\documentclass[twocolumn]{aastex631}

\newcommand\vtan{$V_{\rm tan}$}
\newcommand\kms{km s$^{-1}$}
\usepackage{graphicx, color} % Required for inserting images
\usepackage{wrapfig}
\begin{document}

\title{Parallaxes, Proper Motions, and Near-Infrared Photometry for 173 L and T Dwarfs From The US Naval Observatory Infrared Astrometry Program}

\author{Frederick J. Vrba}
\affil{United States Naval Observatory, Flagstaff Station, 10391 West Naval Observatory Rd., Flagstaff, AZ 86005, USA (retired)}

\author[0000-0002-6294-5937]{Adam C. Schneider}
\affil{United States Naval Observatory, Flagstaff Station, 10391 West Naval Observatory Rd., Flagstaff, AZ 86005, USA}

\author[0000-0002-4603-4834]{Jeffrey A. Munn}
\affil{United States Naval Observatory, Flagstaff Station, 10391 West Naval Observatory Rd., Flagstaff, AZ 86005, USA (retired)}

\author[0009-0008-3683-197X]{Arne A. Henden}
\affil{United States Naval Observatory, Flagstaff Station, 10391 West Naval Observatory Rd., Flagstaff, AZ 86005, USA (retired)}

\author{Christain B. Luginbuhl}
\affil{Flagstaff Dark Skies Coalition, P.O. Box 1892, 86002, Flagstaff, AZ, USA}

\author{Conard C. Dahn}
\affil{United States Naval Observatory, Flagstaff Station, 10391 West Naval Observatory Rd., Flagstaff, AZ 86005, USA (retired)}

\author{Harry H. Guetter}
\affil{United States Naval Observatory, Flagstaff Station, 10391 West Naval Observatory Rd., Flagstaff, AZ 86005, USA (retired)}

\author{Blaise J. Canzian}
\affil{L-3 Communications/Brashear, 615 Epsilon Drive, Pittsburgh, PA 15238-2807, USA}

\author{Trudy M. Tilleman}
\affil{United States Naval Observatory, Flagstaff Station, 10391 West Naval Observatory Rd., Flagstaff, AZ 86005, USA}

\author[0000-0002-2968-2418]{Scott E. Dahm}
\affil{Gemini Observatory/NSF's NOIRLab, 950 N. Cherry Avenue, Tucson, AZ 85719, USA}

\author{Stephen J. Williams}
\affil{United States Naval Observatory, Flagstaff Station, 10391 West Naval Observatory Rd., Flagstaff, AZ 86005, USA}

\author{Justice E. Bruursema}
\affil{United States Naval Observatory, Flagstaff Station, 10391 West Naval Observatory Rd., Flagstaff, AZ 86005, USA (retired)}

\author[0000-0003-4269-260X]{J. Davy Kirkpatrick}
\affil{California Institute of Technology (IPAC), 1200 E. California Blvd., Pasadena, CA 91125, USA}

\author[0000-0002-6523-9536]{Adam J. Burgasser}
\affil{Center for Astrophysics and Space Science, University of California San Diego, La Jolla, CA 92093, USA}

\begin{abstract} 
We present near-infrared parallax and proper motion astrometry for 74 L-dwarfs and 99 T-dwarfs, as single objects or in binary systems, obtained with the ASTROCAM astrometric imager on the USNO, Flagstaff Station 1.55-m telescope over two observing periods. For all 173 objects the median number of observational epochs was 62 with a median time frame of 5.25 years, resulting in median uncertainties of $\sigma$({$\pi_{abs}$}) = 1.51 mas, $\sigma$($\mu_{abs}$) = 1.02 mas yr$^{-1}$, and $\sigma$(\vtan) = 1.01 km s$^{-1}$. Our observations provide the first parallax/proper motion results for 16 objects and the highest precision parallaxes/proper motions for an additional 116 objects. A serendipitous overlap of 40 objects with Gaia DR3 astrometry allows direct comparison and confirmation of our results, along with an investigation on the effects of resolved binarity on astrometric results. We also provide a uniform set of $J$-, $H$-, $K_{S}$-band photometry in the UKIRT/MKO system, most of it being from new observations. We use these results to examine objects included in this study of special-interest populations, consisting of binaries, wide companions, young objects, subdwarfs, and brown dwarf spectral standards.   
\end{abstract}


\section{Introduction} 
Beginning shortly after the early discoveries of significant numbers of brown dwarfs from the Two Micron All Sky Survey (2MASS; \citealt{skrutskie2006}) and the Sloan Sigital Sky Survey (SDSS; \citealt{york2000}) many efforts have been made to obtain high-quality parallaxes and proper motions, as these are essential to understanding fundamental astrophysical properties such as luminosities, effective temperatures, and surface gravities, along with galactic questions such as space densities and velocities. Several larger-scale, facilities-based surveys include USNO \citep{vrba2004,dahn2017}, CFHT \citep{dupuy2012,liu2016, sanghi2023}, UKIRT \citep{marocco2010, best2021}, Palomar \citep{Kirkpatrick2019}, Spitzer \citep{dupuy2013, martin2018, Kirkpatrick2019, kirkpatrick2021}, NTT \citep{smart2013, marocco2013}, CTIO \citep{faherty2012}, ESO 2.2m \citep{smart2018}, VLT \citep{sahlmann2014}, and SOFI \citep{tinney2003}, along with numerous single- or few-object studies. More recently, the Gaia satellite \citep{gaia2016, gaia2023} has provided parallaxes and proper motions of unprecedented quality, but generally for the hotter and/or nearer brown dwarfs.

In this paper we provide the final results of the USNO infrared astrometry program which ran in two epochs between late-2000 to mid-2019. In Section \ref{sec:usnopi} we give a brief history of the USNO parallax and proper motion programs, including early work on brown dwarfs at optical wavelengths. Section \ref{sec:usnoir} presents background on the USNO infrared astrometry program from which results are presented in this paper. In Section \ref{sec:obs} we present detailed descriptions of the observational choices, frame processing and quality control, and astrometric considerations and reductions. Section \ref{sec:absolute} describes our methods for making corrections from relative to absolute parallax and proper motion values and our rationale for choosing these methods. In Section \ref{sec:objects} we describe the sample for which we present results in this work along with selected spectral-typing references, which served as primary selection criteria for inclusion in our program. Section \ref{sec:obssum} gives details of the observations for each object, such as the number of epochs and durations of observations. Section \ref{sec:results} presents the astrometric results including relative and absolute parallax and proper motion determinations and resultant tangential velocities, along with a discussion of the distributions of these values. Section \ref{sec:internal} gives a brief discussion of internal systematics while Section \ref{sec:gaiacomp} gives an extended comparison of results presented here with those of Gaia. New and adopted $JHK$ photometry is presented in Section \ref{sec:JHKphot} for use in the astrophysical analyses in subsequent Sections. In Section \ref{sec:colmag} we present the fundamental near-infrared color vs.~absolute magnitude and spectral type vs.~absolute magnitude diagrams we use for analysis. Section \ref{sec:PopNotes} contains discussions of objects included in this study which are members of populations of interest including binaries, wide companions, young objects, subdwarfs, and brown dwarf spectral standards. Finally, in Section \ref{sec:summary} we summarize some of the main results of this work.

\section{Background of Astrometry at USNO, Flagstaff Station} 
\label{sec:usnopi}
We give here a brief history of ground-based astrometry at the United States Naval Observatory (USNO), Flagstaff Station (NOFS), which began with the commissioning of the 61-inch (1.55-m) telescope in 1963 \citep{strand1964}. This telescope has a ``folded-Newtonian'' optical design employing a parabolic primary mirror and a flat secondary mirror. At a focal ratio of f/7 and a focal length of nearly 50 feet providing a focal plane scale of 13.55 arcsec/mm, it is a physically large telescope for its modest aperture. With a large secondary mirror providing an un-vignetted $\approx$1 deg field of view, the effective open aperture of the telescope is only about 1.25 m. However, at the cost of a large telescope with limited aperture, this optical design provides a uniquely stable astrometric instrument. Both the primary and secondary mirrors are kept fixed and focus is provided by a motor-driven tailpiece instrument mount. Because of this, the focal plane plate scale and point spread function (PSF) are independent of focus. In addition, the telescope has been essentially dedicated to long-term astrometric observations providing many nights of data with few instrument changes or modifications to the telescope itself. NOFS is a generally good observing site with approximately 70 percent of the nights usable and median seeing of 1\farcs0 full-width half-max (FWHM) at zenith, 1\farcs3 FWHM overall, and 0\farcs6 in the best conditions \citep{harris1992}. Together, these circumstances provide the ability for exceptional ground-based long-term astrometric programs.

Between 1964 and 1994 the 1.55-m was used to obtain parallaxes and proper motions for more than 1000 objects using evolving technologies of photographic plates. The results presented in a long series of publications (c.f. \citealt{dahn1988}) were informed by numerous proper motion surveys by Eggen, Giclas, Greenstein, Luyten, and others (c.f. \citealt{luyten1976}), also using small aperture telescopes and photographic technology. Despite the aperture limitations, these combined works helped define the bottom of the stellar main and sub-dwarf sequences almost down to the hydrogen-burning limit. 

With the advent of charge-coupled device (CCD) technology USNO began experimenting with CCD astrometric observations in the mid-1980s, evolving into a full program by the early 1990s (see \citealt{monet1992}). Digital technology allowed for improved parallax accuracies from characteristically 3 mas down to well below 1 mas given a suitable reference frame and target brightness. This program was initially informed for target selection by the same photographic proper motion surveys mentioned above. However, with the advent of the optical Sloan Digital Sky Survey (SDSS; \citealt{york2000}) and the near-infrared Two Micron All Sky Survey (2MASS; \citealt{skrutskie2006}), by 2000 target selection for the USNO program was now being informed by much deeper color-color and approximate color-magnitude studies from these surveys. This allowed the USNO CCD program to extend the low-mass sequence well below the hydrogen burning limit to objects as cool/low-mass as early T spectral types for a few nearby objects. This program continued until 2017 when it was terminated due to the first data release (DR1) from the European Space Agency's astrometric Gaia satellite \citep{gaia2016}, which provides much higher accuracy parallaxes and proper motions than the USNO CCD program and to similar magnitude limits. The final results for this program were given in \cite{dahn2017} which also gives a good retrospective of the program. 

Despite Gaia's outstanding performance, it has limitations in obtaining astrometry for cool objects due to the optical wavelengths at which it operates \citep{gaia2016}. In practice, Gaia reaches few objects cooler than mid-L spectral types due to the general space density of brown dwarfs (e.g., \citealt{theissen2018}). Thus, the USNO infrared astrometry program, which was initiated as an extension of the above programs, can still provide important results which are beyond the reach of Gaia. 

\section{The USNO Infrared Astrometry Program}
\label{sec:usnoir}

Much of the early history, instrumentation, observation, and reduction details of this program have already been given in \cite{vrba2004}, which presented some preliminary astrometric results. Thus, while we give an updated history of the program, we refer the reader to \cite{vrba2004} for many details and we will emphasize changes to observations and reductions since that time. 

Although by the early-1990s the astrometric characteristics of CCD arrays were well known, this was not yet the case for the rapidly-evolving complementary metal-oxide semiconductor (CMOS) array technology, which is necessary for the higher backgrounds encountered even at near infrared wavelengths. Due to this, USNO initiated a test program with a camera built at UCLA and delivered in 1995 using a Rockwell 256$^{2}$ HgCdHe (NICMOS2) as the detector \citep{vrba2000}. Testing over several years with this camera, which was not astrometrically optimized, on the 1.55-m was sufficiently encouraging, both in terms of  immediate astrometry and long term stability to lead USNO to the development of an astrometrically optimized near-infrared camera with a larger field of view.

In the mid-1990s USNO joined with National Optical Astronomical Observatory (NOAO, now NOIRLab) to develop a 1024$^{2}$ InSb array (dubbed ALADDIN) at Santa Barbara Research Corporation (currently part of RTX Corporation). At the time this was the largest CMOS array and most complex integrated chip \citep{fowler1996}. In parallel, USNO joined with the Naval Research Laboratory to design an astrometric camera around the ALADDIN array optimized for use on the 1.55-m telescope. The camera (ASTROCAM) was final-designed and manufactured by Mauna Kea Infrared (MKI; \citealt{fischer2003}). ASTROCAM eliminates refractive optics, other than the entrance window and filters, by the use of an Offner 1:1 re-imaging/apodization system. The ALADDIN array's $\approx$27 $\mu$m pixels thus give a scale of 0\farcs3654/pixel providing nearly-ideal sampling for the site's characteristic 1.0-1.1$\arcsec$ seeing. ASTROCAM together with the 1.55-m telescope's optical design provide an exceptional combination for near-infrared ground-based astrometry, albeit with a limited field of view (FOV) of approximately 6\farcm2$\times$6\farcm2. See \cite{vrba2004} for more details regarding the ALADDIN array and ASTROCAM instrument development.  

ASTROCAM was delivered to NOFS with an engineering-grade array for test observing and characterization in August 1999. In April 2000 a science-grade array was installed and tested prior to the beginning of long-term astrometric observations in August 2000 reported in this work. 

\section{Observations and Data Reduction}
\label{sec:obs}

In this section we briefly discuss the observations and data processing, quality control, and reductions relevant to the astrometric results presented in this work.

\subsection{Series 1 and Series 2 Observations}

Observations were carried out between 2000 and 2017 but with a major interruption resulting in two series of observations as described here.

Series 1: Science astrometric observations commenced on UT 2000 September 09 with 40 initial objects on the program. \cite{vrba2004} reported preliminary results for these objects on very short time baselines of between roughly $\Delta$t $\approx$ 1.3 to 2.0 years, sufficient to give some of the earliest parallax and proper motion results for substellar objects. During this time ASTROCAM was scheduled for observations approximately 14 nights out of each lunar cycle on the 1.55-m telescope with about 65$\%$ of the nights of suitable quality. The number of objects on the program was slowly increased from the original 40 to 65 until the end of Series 1 on UT 2006 June 14 when ASTROCAM was largely destroyed due to an explosion of its LN$_{2}$  pre-charge canister from an uncontrolled warm-up. The latter was the result of a forced evacuation of the site due to a nearby wild fire. Of the 65 objects at Series 1 termination, 13 had insufficient data for even preliminary results. The remaining 52 objects had a median time baseline of $\Delta$t = 5.25 years with a median number of observational epochs of 59.5.

After the ASTROCAM explosion an engineering evaluation study was carried out at MKI to determine if ASTROCAM could be sensibly repaired or if a new camera should be built. Amazingly, the original ALADDIN array suffered only minor damage with fewer than 50 pixels becoming non-functional. In addition, the entrance window and the filters $J$ and $H$ used for this astrometry were not damaged. It was ultimately decided to rebuild ASTROCAM despite most of the structural mechanics needing replacement and a new and optically improved Offner re-imaging  system installed. Between separate contracting for  evaluation and actual ASTROCAM rebuild several years passed until the astrometry program could be re-started. 

Series 2: Astrometric observations were recommenced on UT 2011 May 07 with 50 target objects, including the 13 objects from Series 1 which were incomplete. During the early part of Series 2 nights were evenly split with the optical program each lunar cycle. However, as Gaia astrometric results became available, the optical program was slowly eliminated, such that during the last several years of operation the ASTROCAM program used essentially all nights on the 1.55-m. During this period about 70$\%$ of the nights were of suitable quality. These circumstances allowed the program to increase to roughly 140 objects. The final observations were obtained UT 2019 June 24 due to changing USNO mission requirements and failing ASTROCAM electronics. There are a total of 121 objects in Series 2 which had reasonably complete astrometric results with a median time baseline of $\Delta$t = 5.27 years and a median number of observational epochs of 64. Both of these numbers are serendipitously similar to those of Series 1.

Series 1 and Series 2 combine for a total of 173 objects, the results for which we present in this paper, with total median $\Delta$t = 5.25 years and median observational epochs per target of 62. We note that with both Series being terminated in an uncoordinated manner we have included a few objects, the observations for which were not completed and hence have large error bars, but which we felt still had sufficient value to present. 

\subsection{Observational Procedures}

After the first year of Series 1, during which an observer was required to make target decisions, observations were largely automated. At the beginning of each night biases and dome flats were obtained manually for each filter to be used that night. As a safety precaution it was necessary to start the automated observing manually, after which the program ran until shortly before dawn and closed telescope operations automatically. The program was not a strict queue, but made decisions based on several inputs. First was input from a weather station giving readings on wind, humidity, precipitation, and particulates each of which could terminate observations and close the telescope beyond preset limits. If it was found difficult to find a guide star after several attempts it would assume either clouds or pointing issues and close the telescope. Due to the relatively long exposure times needed, observations were also terminated if seeing degraded to above 2\farcs5. The software also recognized the position of the Moon and twilight sky brightnesses. Due to safety concerns the software was not allowed to restart observations without human intervention. 

The software captured the program observational history during the previous month and prioritized objects with no or few observations during that period. For each target field, filters and exposure times as a function of seeing had been previously determined (see below). All dithered sets of exposures were started such that the central time of the set would be within $\pm$ 15 minutes of time from the meridian to minimize differential color refraction. The optical guide camera position was pre-set to acquire the same guide star for each re-visit and, once guiding, assured that the reference frame would be positioned to better than one pixel, keeping with the differential nature of the astrometry. As has been the policy for all 1.55-m astrometry, it is felt that keeping the reference frame position constant and allowing the target to move via proper motion is optimal. Once the guide star was acquired and properly centered a quick exposure was obtained to measure the seeing and set the exposure time. From previous testing it was determined that three 10$\arcsec$ dithers, in either N-S or E-W directions, were sufficient to prevent PSF overlap and determine background, thus minimizing telescope overhead time; the dither direction being a function of the reference frame stellar positional distribution.        

\subsection{Filter and Initial Reference Frame Selection}

All observations were carried out in either $J$- or $H$-band, giving the highest signal-to-noise when combining the spectra of target objects, minimizing sky/telescope backgrounds, and having the best possible reference frame. Test observations were carried out for each target field in both $J$- and $H$-band to determine the optimum filter to use. As could be expected \textit{a priori} from the spectra of L- and T-dwarfs at these wavelengths, $J$-band observations would typically be optimum for T-dwarfs, while $H$-band observations would be optimum for L-dwarfs. In actuality, 68/74 (92$\%$) of L-dwarfs (or L-dwarf dominated binaries) were observed in $H$-band and 6/74 (8$\%$) in $J$-band, while 81/99 (82$\%$) of T-dwarfs were observed in $J$-band and 18/99 (18$\%$) in $H$-band. The exceptions were typically for objects with unusual blue or red spectra and a general bias toward $J$-band due to somewhat lesser telescope and sky glow backgrounds and typically higher signal for the stellar reference frame.   

The ALADDIN array of 1024$^{2}$ pixels and with 1:1 re-imaging provides a rather small FOV of approximately 6\farcm2$\times$6\farcm2. However, even at this FOV, employing reference stars as near the target object as possible maximizes isoplanarity and astrometric repeatability. Since our target L- and T-dwarfs were already faint for a 1.55-m telescope, few reference frame stars were selected fainter than the target objects. Very bright reference frame stars, although usable with very short exposure times and many internal co-adds at the expense of higher net read noise, are systematically nearby thus stretching the reference frame depth and associated mean error when conversion to absolute parallax (see $\S$\ref{sec:absolute} below). In practice, the initial reference frame stars chosen were typically within 1-2 magnitudes and as near as possible to the target objects. Optimistically, 10 to 30 initial reference frame stars were chosen although the final reference frames would be reduced in numbers due to optimization and inclusion in Gaia Data Release 3 (DR3) \citep{gaia2023} for converting to absolute parallax and proper motions, as explained in $\S$\ref{sec:absolute}.

\subsection{Exposure Time Determinations}

Based on significant testing data the goal was to have a minimum 3000-5000 ADU level of the central pixel of the target object image for each dither. An initial integration time, based on nominal 1\farcs0 seeing, was determined at the time of the filter selection test observations. Since integration time scales as the square of the seeing in order to keep a constant central pixel value, a series of integration times as a function of seeing for each object was developed and used by the automated observing program. While the integration times were calculated up to the seeing limit of 2\farcs5, in practice few observations were obtained at seeing $>$2\farcs0 due to long total integration times. Integration times for each dither were based on the product of a fundamental exposure time and the number of coadds employed to ensure that neither the target object nor any reference frame star would be saturated even at the best seeings. Exposure times ranged from 5 to 60 seconds with the number of coadds determined by the seeing, but ensuring a minimum total integration time of 5 minutes for each of the specified 3 dithers.  

\subsection{Data Processing and Quality Control}
\label{sec:dpqc}

Data processing is the same as described in \cite{vrba2004}, which we briefly summarize here. Biases obtained each night were only used as a test of camera electronics and were not part of data frame processing. Dome flats for each night and each filter used were obtained by differencing three sets of illuminated and un-illuminated flat screen images in order to remove any thermal signature of the flat screen. The resultant difference frames were median-combined and normalized to form a single flat-field image for each filter. All flat field and program frames were linearized, pixel by pixel, in a process described in \cite{luginbuhl1998}. After flat-fielding, each set of dithered target observations was run through a min/max pixel value filter to form a background frame which was subtracted from each frame of the dithered set to form a final frame. 

Each night's data were processed as above by a program that also allowed inspection of each data frame as a first step in quality control. Frames could be removed from further consideration by issues such as trailed images, weak images, images with PSFs larger than 2\farcs5, significant meteor/aircraft trails, or any data wherein all three dithered images in a set were not completed. Only a small fraction of the frames were rejected at this step since the automated observing program performed well, but it did remove data that would have been problematic in later analyses. However, the general philosophy at this point was to not discard frames until detailed scrutiny, as described below.

For the first few nights of observations of a new object on the program, processed frames were inspected carefully to ensure that the initial setup was correct and that observations were being carried out correctly. After that, preliminary astrometric reductions were carried out after every season of observing to ensure that observations had continued correctly and that new issues had not arisen, such as a slightly different bad pixel set after a scheduled ASTROCAM thermal cycle impinging on the target or the reference frame set. This served as the second level of quality control and, equally important, this allowed evaluation of the parallax and proper motion accuracies as more data were being obtained. 

Final astrometry for a given field (described in the next section) was completed after all observations had been collected. At this point, close attention was given as to what frames would be kept in the solution. Frames that were removed were predominantly affected by poorer seeing and weak signal due to clouds that had come in during the long dithered triplet exposures (20$-$60 minutes) or if any crucial reference frame star was affected by a bad pixel. Of the total of more than 36,000 science frames taken during both Series, 16$\%$ were not used in the final astrometry in Series 1 and 17$\%$ were rejected in Series 2. We note that these are very similar rejection rates to those of the $H$-band and $K$-band USNO-led UKIRT Hemisphere Survey (UHS) being carried out with the UKIRT telescope \citep{dye2018, schneider2025}. While this work and the UHS surveys have different goals and observational quality criteria, ground-based, near-infrared surveys will be affected by changing atmospheric conditions and significant quality control is needed to ensure optimal results.  

\subsection{Astrometric Reductions}

The image centroiding technique used in this work for ASTROCAM data was developed by D. Monet in the early 1980s and has been used for all narrow-field CCD astrometry at USNO since that time and thus has a proven record of reliability \citep{monet1983}. Effectively, it is a 2-dimensional Gaussian fit to the PSF and is discussed further in \cite{dahn2017}. The software we use for determining parallax and proper motion, modified for the specifics of ASTROCAM, are again those developed for and used by the USNO CCD astrometry program and are thus well-tested. \cite{monet1992} discusses the philosophy of the reductions in some detail, but we point out some major points here. Dithered frames for an object observation set are treated as individual frames and are not combined in any way. Each frame has unit weight and each reference star within that frame has unit weight.
Ultimately, only linear frame constants are employed due to the nature of the images produced by the 1.55-m telescope and ASTROCAM. Frame scale and rotation are allowed as free parameters. In practice frame scale is effectively a constant due to the nature of the telescope and ASTROCAM, while rotation is solved for due to the slight misalignment of the telescope polar axis.

The data set for an object is solved for proper motion and parallax independently in right ascension ($\alpha$) and declination ($\delta$), relative to the reference frame, by a three-step iteration of least-squares fits to the data. This gives the $\alpha$ and $\delta$ components of the proper motion directly. The parallaxes in $\alpha$ and $\delta$ are corrected for the parallax factor in each component and combined by their weighted averages. Due to the relatively low inclination of the Earth's orbit, the errors in the $\alpha$ determination always have much smaller statistical errors than those derived from the $\delta$ determination such that the $\alpha$ solutions are  the dominant component. For 6 of our 173 target objects, through a combination of the object position in the sky and the distribution of observational epochs, the resultant parallax in $\delta$ was indeterminate and thus not included in the solution. These objects are indicated in Table \ref{tab:astrometry} below.

Once an initial solution for a field has been determined a detailed examination of global  frame errors and individual stellar errors as a function of frame number is initiated. The global frame errors point to frames which were strongly affected by poor seeing, clouds, or trailed images that had not been previously removed. The deletion of frames at this level accounts for nearly all of the 16$\%$ and 17$\%$ deprecation rate for Series 1 and 2, respectively, described in $\S$\ref{sec:dpqc} above. 

\subsection{Reference Frame Determination}

Once a nearly-final data frame set is established a nearly-final set of reference frame stars is determined. During the initial set-up phase for each field, other than checking for resolved binarity or extended PSFs, potential reference frame stars were liberally included as they might prove to be useful and could later be removed at the stage described here. In fact, most of the fainter potential reference frame stars and even some brighter stars which are farther from the target object and shaded by closer useful stars at the same position angle degraded the solutions and were removed. In addition, a few potential reference frame stars were removed due to a variety of reasons, such as routinely landing on a bad pixel in one of the dithers. Since we give unit weight to each star and we require an identical reference for every image frame used in the processing, for each case a decision was made as to whether removing the star from all frames or deleting the frame in question provided the optimal results. We note that these kinds of decisions are a luxury of having a nearly-dedicated telescope producing many epochs/frames of data. There is also one other very minor deletion of potential reference frame stars which is described in $\S$\ref{sec:absolute}.

The final median number of reference frame stars employed was 11 for both Series 1 and 2 with a high of 26 and a low, for one object, of 3. The relatively small number of stars used in the reference frames is directly due to the small 6\farcm2$\times$6\farcm2 FOV of ASTROCAM. However, this gives a tight reference frame and, with Gaia DR3 results available, ensures an accurate correction to absolute parallaxes and proper motions as described in the next section. 

\section{Correction From Relative to Absolute Parallaxes and Proper Motions}
\label{sec:absolute}

The Gaia DR3 catalog \citep{gaia2023} provides both absolute parallax and proper motion data for stars of the general magnitude range used in the reference frames for this work described above. This allows a straightforward method for correcting the parallaxes and proper motions we directly determine relative to the in-field reference frames to absolute values. For parallax corrections we use the unweighted mean value of the Gaia DR3 results for the reference frame stars with an uncertainty calculated as the unweighted standard deviation of the mean values. The corrections are added to our relative parallaxes to determine absolute parallaxes with the absolute parallax uncertainties calculated as the relative parallax errors added in quadrature to the relative to absolute parallax correction errors.  

While USNO has traditionally published only relative proper motion values, we also provide absolute proper motion values. This is similar to the parallax corrections, as we use the unweighted mean Gaia DR3 proper motion of the reference frame stars for each field added to the relative proper motions we determine. The correction uncertainties are taken as the mean of Gaia DR3 proper motion errors for each reference frame field. The absolute proper motion uncertainties are calculated as the relative proper motion errors added in quadrature to the relative to absolute proper motion correction errors. For proper motions the corrections and associated uncertainties are determined for $\alpha$ and $\delta$ components independently.   

There are two reasons why the correction to absolute values may have an affect on the final reference frame employed. First, a small fraction of the desired reference frame stars did not have Gaia DR3 parallaxes and proper motions. Rather than opting for an alternative method of estimating the distances to those stars, we chose to eliminate them in order to have a purely Gaia-based system of reference frame data. The second reason is that we wanted to have a minimal spread in the reference frame distances. Thus, a few closer stars were eliminated to minimize the uncertainty in the correction to absolute parallax, as described above. The mean distance of the reference frames is about  1 kpc and will be discussed in more detail below.

\section{Objects on Program}
\label{sec:objects}
The selection of objects to be put on the near-IR astrometry program evolved considerably over time. All of the Series 1 objects were chosen on the basis of early discoveries from follow-up spectroscopic observations of either 2MASS or SDSS objects: e.g. \cite{burgasser1999}, \cite{kirkpatrick1999}, \cite{geballe2002}. Initially objects were chosen based on discoveries available and thus did not represent systematic choices. In addition, several binaries were included based on apparent brightnesses which were not yet recognized as such by the evolving spectroscopic understanding of brown dwarfs. Nonetheless, such objects were subsequently left on the program. Later in Series 1, objects were selected based on such criteria as the importance of the L-T transition, subdwarf nature, etc. 

By the time Series 2 was initiated the understanding of brown dwarf physics was considerably advanced. In addition to 2MASS and SDSS, many objects were added based on initial discoveries from the Wide-field Infrared Survey Explorer (WISE) (\citealt{wright2010}) such as those from \citep{kirkpatrick2011} and from other surveys such as Pan-STARRS \citep{chambers2016} and UKIDSS \citep{lawrence2007}. A better understanding of brown dwarf ages allowed for the selection of objects more astrophysically interesting, such as potentially young objects and low-metallicity subdwarfs. The program was also tasked with additional practical questions such as testing faintness limits and pushing the limits of astrometric accuracy, both of which influenced target selections. 

Ultimately, a total of 74 L-dwarfs and 99 T-dwarfs were sufficiently observed to produce meaningful results. Of these, numerous targets have special properties such as being in known or suspected binary systems, having a wide companion, having unusual colors or peculiar spectra, are subdwarfs, or objects with metallicity or association indications of youth. These objects will be discussed individually in $\S$\ref{sec:PopNotes}. Serendipitously, 40 objects also have Gaia DR3 astrometric results which we employ below in $\S$\ref{sec:gaiacomp} for checks on astrometric accuracy and systematic errors.

Table 1 presents background information for all 173 single objects or binary pairs. Column (1) gives their CatWISE 2020 identifications \citep{marocco2021}, which we use consistently throughout this paper. Columns (2) and (3) give their discovery names and discovery references, respectively. Columns (4) and (5) give optical spectral typing and associated references, while columns (6) and (7) give infrared spectral typing and associated references. 

\startlongtable
\begin{deluxetable*}{lcccccc}
\label{tab:genprop}
\tabletypesize{\scriptsize}
\tablecaption{Target Sample and Optical and Infrared Wavelength Spectral Types}
\tablehead{
\colhead{Name} & \colhead{Discovery} & \colhead{Ref.} & \colhead{SpT} & \colhead{Ref.} & \colhead{SpT} & \colhead{Ref.} \\
\colhead{(CatWISE)} & \colhead{Name} & & \colhead{(opt)} &  & \colhead{(IR)} \\
\colhead{(1)} & \colhead{(2)} & \colhead{(3)} & \colhead{(4)} & \colhead {(5)} &
\colhead{(6)} & \colhead{(7)} }
\startdata
J000013.51$+$255419.7 & SDSS J000013.54$+$255418.6 & 1 & T5 & 1 & T4.5 & 56 \\ 
J000849.71$-$173925.0 & WISEPC J000849.76$-$173922.6 & 2 & \dots & \dots & T6 & 2 \\
J001502.52$+$295928.7 & 2MASS J00150206$+$2959323 & 3 & L7& 3 & L7.5 pec (blue) & 3 \\
J003030.38$-$145033.8 & 2MASSW J0030300$-$145033 & 4 & L7 & 4 & L6 FLD-G & 57 \\
J003259.67$+$141037.2 & SDSSp J003259.36$+$141036.6 & 5 & \dots & \dots & L9 & 58 \\
\\
J003452.12$+$052307.3 & 2MASS J00345157$+$0523050 & 6 & \dots & \dots & T6.5 & 49 \\
J004121.65$+$354712.5 & 2MASS J00412179$+$3547133 & 6 & sdL0.5 & 59 & sdL? & 6 \\
J004521.80$+$163444.0 & 2MASSW J0045214$+$163445 & 7 & L2$\beta$ & 60 & L2 VL-G & 61 \\
J004928.58$+$044101.0 & CFBDS J004928$+$044058 & 8 & \dots & \dots & L9 & 2 \\
J005911.15$-$011401.4 & CFBDS J005910.90$-$011401.3 & 9 & \dots & \dots & T8.5 & 62 \\
\\
J010332.40$+$193536.5 & 2MASSI J0103320$+$193536 & 4 & L6 & 4 & L6 pec (red) & 63 \\
J010753.11$+$004157.7 & SDSSp J010752.34$+$004156.1  & 10 & L7.5 & 64 & L7 pec (red) & 63 \\
J011912.41$+$240331.7 & SDSS J011912.22$+$240331.6 & 11 & \dots & \dots & T2 & 11 \\
J013836.61$-$032222.8 & WISEPC J013836.59$-$032221.2 & 2 & \dots & \dots & T3 & 2 \\
J015011.41$+$382723.7 & WISEPA J015010.86$+$382724.3 & 2 & \dots & \dots & T0 & 2 \\
\\
J015142.59$+$124428.7 & SDSSp J015141.69$+$124429.6 & 5 & T1 & 5 & T1 & 49 \\
J020625.45$+$264023.3 & WISEPA J020625.26$+$264023.6 & 2 & \dots & \dots & L8 (red) & 57 \\
J020743.00$+$000056.1 & SDSSp J020742.83$+$000056.2 & 5 & \dots & \dots & T4.5 & 5 \\
J023618.06$+$004852.1 & SDSSp J023617.93$+$004855.0 & 5 & L9 & 65 & L6.5 & 5 \\
J024313.38$-$245332.8 & 2MASSI J0243137$-$245329 & 12 & T5.5 & 56 & T6 & 12 \\
\\
J025116.11$-$035315.7 & 2MASS J0251149$-$035246 & 13 & L3 & 16 & L1 & 13 \\
J025409.54$+$022358.7 & WISEPC J025409.45$+$022359.1 & 14 & \dots & \dots & T8 & 66 \\
J030533.66$+$395434.6 & WISEPA J030533.54$+$395434.4 & 2 & \dots & \dots & T6 & 2 \\
J031100.13$+$164815.6 & 2MASSW J0310599$+$164816 & 4 & L8 & 4 & L9.5 & 58 \\
J031326.13$+$780744.6 & WISEPA J031325.96$+$780744.2 & 2 & \dots & \dots & T9 & 67 \\
\\
J032553.05$+$042539.6 & SDSS J032553.17$+$042540.1 & 11 & \dots & \dots & T5.5 & 11 \\
J032842.66$+$230204.1 & 2MASSI J0328426$+$230205 & 4 & L8 & 4 & L9.5 & 5 \\
J035523.53$+$113336.7 & 2MASS J03552337$+$1133437 & 15 & L5$\gamma$ & 60 & L3-L6$\gamma$ & 63 \\
J040709.03$+$151455.0 & 2MASS J04070885$+$1514565 & 6 & \dots & \dots & T5 & 49 \\
J041054.46$+$141130.8 & WISEPA J041054.48$+$141131.6 & 2 & \dots & \dots & T6 & 2 \\
\\
J041522.17$-$093457.1 & 2MASSI J0415195$-$093506 & 12 & T8 & 68 & T8 & 12 \\
J042348.22$-$041402.0 & SDSSp J042348.57$-$041403.5 & 5,10 & L7.5 & 16 & L7.5+T2 & 69 \\
J043900.87$-$235310.7 & 2MASSI J0439010$-$235308 & 16 & L6 & 70 & L4.5 & 58 \\
J044853.70$-$193543.6 & WISEPA J044853.29$-$193548.5 & 2 & \dots & \dots & sdT5 & 82 \\
J050021.02$+$033044.8 & 2MASS J05002100$+$0330501 & 15 & L4  & 71 & \dots & \dots \\
\\
J051317.27$+$060812.3 & WISEPA J051317.28$+$060814.7 & 2 & \dots & \dots & T6.5 & 2 \\
J051609.20$-$044553.3 & 2MASS J05160945$-$0445499 & 17 & \dots & \dots & T6 & 49 \\
J052536.18$+$673951.5 & WISEPA J052536.33$+$673952.3 & 2 & \dots & \dots & T6 pec & 2 \\
J053312.61$+$824617.2 & 2MASS J05325346$+$8246465 & 18 & esdL7 & 72 & esdL8 & 90 \\
J053952.16$-$005856.5 & SDSSp J053951.99$-$005902.0 & 19 & L5 & 19 & L5 (blue) & 58 \\
\\
J054231.20$-$162827.6 & WISEPA J054231.26$-$162829.1 & 2 & \dots & \dots & T6.5 & 2 \\
J055919.85$-$140454.8 & 2MASSI J0559191$-$140448 & 20 & T5 & 56 & T4.5 & 49 \\
J060206.71$+$404355.4 & 2MASS J06020638$+$4043588 & 21 & \dots & \dots & T4.5 & 21 \\
J060738.42$+$242951.2 & WISEP J060738.65$+$242953.4 & 22 & L8 & 73 & L9 & 74 \\
J061407.41$+$391233.2 & WISEPA J061407.49$+$391236.4 & 2 & \dots & \dots & T6 & 2 \\
\\
J062542.19$+$564625.4 & WISEPA J062542.21$+$564625.5 & 2 & \dots & \dots & T6 & 2 \\
J062720.06$-$111429.6 & WISEPA J062720.07$-$111428.8 & 2 & \dots & \dots & T6 & 2 \\
J064626.92$+$793454.5 & HD 46588 B & 23 & \dots & \dots & L9 & 75 \\
J065609.75$+$420532.8 & WISEPA J065609.60$+$420531.0 & 2 & \dots & \dots & T3 & 2 \\
J070036.80$+$315718.2 & 2MASS J07003664$+$3157266 & 24 & L3.5 & 24 & L3+L6.5 & 76 \\
\\
J072226.89$-$054027.5 & UGPS J072227.51$-$054031.2 & 25 & \dots & \dots & T9 & 62 \\
J072719.22$+$170950.1 & 2MASSI J0727182$+$171001 & 12 & T8 & 68 & T7 & 12 \\
J074200.88$+$205515.7 & SDSS J074201.41$+$205520.5 & 1 & \dots & \dots & T5 & 1 \\
J075547.93$+$221212.8 & 2MASSI J0755480$+$221218 & 12 & T6 & 68 & T5 & 49 \\
J075840.07$+$324718.8 & SDSS J075840.33$+$324723.4 & 1 & T3 & 56 & T2 & 1 \\
\\
J081957.98$-$033529.3 & WISEPA J081958.05$-$033529.0 & 2 & T4 & 56 & T4 & 2 \\
J082131.64$+$144317.8 &WISE J0821$+$1443  & 26 & \dots & \dots & T5.5 & 2 \\
J082518.97$+$211546.1 & 2MASSI J0825196$+$211552 & 4 & L7 & 77 & L7 pec (red) & 63 \\
J083006.91$+$482838.3 & SDSSp J083008.12$+$482847.4 & 5 & L8 & 78 & L9.5 & 58 \\
J083541.88$-$081918.0 & 2MASSI 0835425$-$081923 & 27 & L5 & 56 & L4 pec (red) & 63 \\
\\
J083717.18$-$000020.8 & SDSS J083717.21$-$000018.0 & 28 & T0 & 78 & T1 & 55 \\
J085035.75$+$105715.3 & 2MASSs J0850359$+$105716 & 29 & L6 & 29 & L6.5+L8.5 & 76 \\
J085757.60$+$570844.6 & SDSSp J085758.45$+$570851.4 & 5 & L8 & 78 & L8--L9 pec (red) & 63 \\
J085833.72$+$325628.6 & SDSS J085834.42$+$325627.7 & 11 & \dots & \dots & T1 pec (red) & 79 \\
J090023.73$+$253934.1 & SDSS J090023.68$+$253934.3 & 30 & L7 & 77 & \dots & \dots \\
\\
J090900.31$+$652525.8 & SDSS J090900.73$+$652527.2 & 11 & \dots & \dots & T1.5 & 11 \\
J091534.04$+$042204.8 & 2MASS J09153413$+$0422045 & 15 & L7+L7 & 15 & L6+L6 & 80 \\
J092615.40$+$584717.6 & SDSSp J092615.38$+$584720.9 & 5 & T5 & 56 & T4+T5.5 & 81 \\
J092933.34$+$342951.3 & 2MASSW J0929336$+$342952 & 4 & L8 & 4 & L7.5 & 1 \\
J093735.98$+$293120.8 & 2MASSI J0937347$+$293142 & 12 & T7 pec (red) & 68 & sdT6 & 90 \\
\\
J093936.13$-$244844.3 & 2MASS J09393548$-$2448279 & 31 & T8 & 56 & d/sdT8 & 90 \\
J094908.49$-$154548.3 & 2MASS J09490860$-$1545485 & 31 & \dots & \dots & T2 & 49 \\
J095105.35$+$355800.1 & 2MASSW J0951054$+$355801 & 4 & L6 & 4 & L6 & 57 \\ 
J101014.45$-$040650.0 & 2MASS 1010148$+$040649 & 16 & L6 & 16 & L6 & 3 \\
J101905.55$+$652954.5 & WISEPA J101905.63$+$652954.2 & 2 & T7 & 2 & T6 & 2 \\
\\
J102109.51$-$030421.1 & SDSSp J102109.69$-$030420.1 & 28 & T3.5 & 78 & T1+T5.5 & 69 \\
J103931.32$+$325623.7 & SDSS J103931.35$+$325625.5 & 11 & \dots & \dots & T1 & 11 \\
J104307.42$+$222523.3 & 2MASSI 1043075$+$222523 & 32 & L8 & 32 & L8 pec & 63 \\
J104335.10$+$121310.8 & SDSS J104335.08$+$121314.1 & 11 & \dots & \dots & L9 & 3 \\
J104751.75$+$212414.7 & 2MASSI J1047539$+$212423 & 33 & T7 & 68 & T6.5 & 12 \\
\\
J104828.96$+$091940.9 & SDSS J104829.21$+$091937.8 & 11 & \dots & \dots & T2.5 & 11 \\
J110611.58$+$275414.1 & 2MASS J11061197$+$2754225 & 21 & \dots & \dots & T2.5 & 21 \\
J111009.77$+$011608.7 & SDSSp J111010.01$+$011613.1 & 5 & \dots & \dots & T5.5 & 1 \\
J111447.51$-$261830.0 & 2MASS J11145133$-$2618235 & 31 & T8 & 56 & T7.5 & 31 \\
J111812.22$-$085615.0 & 2MASS J11181292$-$0856106 & 3 & L6 & 3 & L6 pec (blue) & 3 \\
\\
J112254.33$+$255020.2 & WISEPC J112254.73$+$255021.5 & 2 & \dots & \dots & T6 & 2 \\
J115553.43$+$055956.6 & SDSS J115553.86$+$055957.5 & 4 & L6 & 77 & L7 pec & 63 \\
J115821.40$+$043446.3 & SDSS J115820.75$+$043501.7 & 30 & sdL7 & 3 & d/sdL8 & 90 \\
J120746.66$+$024426.8 & SDSS J120747.17$+$024424.8 & 34 & L8 & 34 & T0 & 1 \\
J121709.86$-$031111.8 & 2MASSW J1217111$-$031113 & 33 & T7 & 68 & T7.5 & 12 \\
\\
J121757.13$+$162635.1 & WISEPC J121756.91$+$162640.2 & 2 & \dots & \dots & T8.5+Y0 & 83 \\
J122554.84$-$273958.5 & 2MASSW J1225543$-$273947 & 33 & T6 & 68 & T5.5+T8 & 76 \\
J123146.34$+$084717.1 & 2MASS J12314753$+$0847331 & 6 & T6 & 56 & T5.5 & 49 \\
J123737.03$+$652607.2 & 2MASSW J1237392$+$652615 & 33 & T7 & 68 & T6.5 & 12 \\
J125011.59$+$392542.6 & SDSS J125011.65$+$392553.9 & 11 & \dots & \dots & T4 & 11 \\
\\
J125453.41$-$012245.6 & SDSS 125453.90$-$012247.4 & 28 & T2 & 68 & T2 & 5 \\
J131141.74$+$362925.2 & SDSS J131142.11$+$362923.9 & 30 & \dots & \dots & L5 pec (blue) & 2 \\
J132003.78$+$603425.8 & WISEPC J132004.16$+$603426.2 & 2 & \dots & \dots & T6.5 & 2 \\
J132233.51$-$234015.3 & WISEPA J132233.66$-$234017.1 & 2 & \dots & \dots & T8 & 2 \\
J132407.64$+$190625.5 & PSO J201.0320$+$19.1072 & 35 & \dots & \dots & T3.5 & 35 \\
\\
J132434.67$+$635827.1 & 2MASS J13243559$+$6358284 & 21 & T2.5: & 41 & T2:pec (red) & 21 \\
J132605.25$+$120010.1 & ULAS J132605.18$+$120009.9 & 36 & \dots & \dots & T6 pec & 36 \\
J132629.56$-$003833.2 & SDSSp J132629.82$-0$003831.5 & 19 & L8? & 19 & L7 & 58 \\
J133553.35$+$113003.7 & ULAS J133553.45$+$113005.2 & 37 & \dots & \dots & T8.5 & 62 \\
J134645.89$-$003152.2 & 2MASSW J1346464$-$003150 & 33 & \dots & \dots & T6.5 & 49 \\
\\
J134807.08$+$660327.2 & WISEPC J134806.99$+$660327.8 & 2 & \dots & \dots & L9 & 2 \\
J140255.70$+$080053.4 & SDSS J140255.66$+$080055.2 & 11 & \dots & \dots & T2 pec & 84 \\
J140753.30$+$124110.7 & 2MASS J14075361$+$1241099 & 38 & L3 & 77 & L4 & 58 \\
J143517.22$-$004612.9 & SDSS J143517.20$-$004612.9 & 34 & L0 & 34 & \dots & \dots \\
J143535.75$-$004348.6 & SDSS J143535.72$-$004347.0 & 34 & L3 & 34 & L2.5 & 1 \\
\\
J143945.64$+$304218.7 & SDSS J143945.86$+$304220.6 & 11 & \dots & \dots & T2.5 & 11 \\
J144600.78$+$002450.9 & SDSSp J144600.60$+$002452.0 & 5 & L6 & 34 & L5 & 85 \\
J145714.66$+$581509.7 & WISEPC J145715.03$+$581510.2 & 2 & T8 & 2 & T7 & 2 \\
J150319.71$+$252528.3 & 2MASS J15031961$+$2525196 & 27 & T6 & 68 & T5 & 49 \\
J150648.79$+$702741.0 & WISEPC J150649.97$+$702736.0 & 2 & T6 & 56 & T6 & 2 \\
\\
J150653.09$+$132105.8 & 2MASSW J1506544$+$132106 & 39 & L3 & 39 & L3 & 3 \\
J151114.38$+$060739.4 & SDSS J151114.66$+$060742.9 & 11 & \dots & \dots & L5+T5 & 81 \\
J151459.41$+$484803.4 & 2MASS J1515008$+$484742 & 13 & L6 & 32 & L6 & 13 \\
J152040.10$+$354615.7 & SDSS J152039.82$+$354619.8 & 11 & \dots & \dots & L7.5 & 69 \\
J152322.78$+$301453.4 & 2MASSW J152322.6$+$301456 & 40 & L8 & 4 & L8 & 58 \\
\\
J152613.76$+$204334.8 & 2MASSI J1526140$+$204341 & 4 & L7 & 4 & L7 & 6 \\
J154614.94$+$493158.6 & 2MASS 15461461$+$4932114 & 41 & \dots & \dots & T3 & 69 \\
J155301.80$+$153239.5 & 2MASSI J1553$+$1532 & 12 & \dots & \dots & T6.5+T7.5 & 76 \\
J161705.76$+$180713.9 & WISEPC J161705.75$+$180714.0 & 42 & T8 & 2 & T8 & 42 \\
J162413.95$+$002915.6 & SDSS 1624$+$00 & 43 & T6 & 68 & T6 & 12 \\
\\
J162541.25$+$152810.0 & PSO J246.4222$+$15.4698 & 35 & \dots & \dots & T4.5 & 35 \\
J162618.23$+$392523.4 & 2MASS J16262034$+$3925190 & 44 & usdL4 & 59 & sdL4 & 86 \\
J162725.60$+$325522.8 & WISEPA J162725.64$+$325524.1 & 45 & \dots & \dots & T6 & 2 \\
J162918.64$+$033534.8 & PSO J247.3273$+$03.5932 & 35 & T3 & 56 & T2 & 35 \\
J163229.48$+$190439.6 & 2MASSW J1632291$+$190441 & 29 & L8 & 29 & L8 & 3 \\
\\
J164715.47$+$563209.4 & WISEPA J164715.59$+$563208.2 & 2 & L7 & 56 & L9 pec (red) & 2 \\
J165311.00$+$444421.2 & WISEPA J165311.05$+$444423.0 & 45 & T8 & 2 & T8 & 2 \\
J171145.73$+$223204.2 & 2MASSI J1711457$+$223204 & 4 & L6.5 & 4 & L9 & 69 \\
J172811.54$+$394859.0 & 2MASSW J1728114$+$394859 & 4 & L7 & 4 & L5+L6.5 & 87 \\
J174124.05$+$255312.0 & WISEPC J174124.25$+$255319.5 & 14 & T9 & 2 & T9 & 2 \\
\\
J175023.78$+$422238.6 & SDSS J175024.01$+$422237.8 & 1 & \dots & \dots & T1.5 & 69 \\
J175033.14$+$175905.4 & SDSSp J175032.96$+$175903.9 & 5 & T4 & 56 & T3 & 69 \\
J175609.98$+$281516.4 & 2MASS J17561080$+$2815238 & 3 & sdL1 & 3 & L1 pec (blue) & 3 \\
J175805.45$+$463318.2 & SDSS J175805.46$+$463311.9 & 1 & \dots & \dots & T6.5 & 49 \\
J180026.66$+$013450.9 & WISEP J180026.60$+$013453.1 & 46 & L7.5 & 88 & L7.5 & 46 \\
\\
J181210.89$+$272142.8 & WISEPC J181210.85$+$272144.3 & 42 & \dots & \dots & T8.5: & 42 \\
J182128.39$+$141357.2 & 2MASS J18212815$+$1414010 & 47 & L4.5 & 47 & L5 & 47 \\
J183058.60$+$454258.2 & WISEPA J183058.57$+$454257.9 & 3 & \dots & \dots & L9 & 2 \\
J184108.67$+$311728.6 & 2MASSW J1841086$+$311727 & 4 & L4 pec & 4 & \dots & \dots \\
J185215.87$+$353714.9 & WISEPA J185215.78$+$353716.3 & 3 & \dots & \dots & T7 & 2 \\
\\
J190106.23$+$471820.5 & 2MASS J19010601$+$4718136 & 6 & \dots & \dots & T5 & 6 \\
J190624.72$+$450805.2 & WISEPA J190624.75$+$450808.2 & 2 & \dots & \dots & T6 & 2 \\
J190648.66$+$401105.9 & WISEP J190648.47$+$401106.8 & 48 & \dots & \dots & L1 & 48 \\
J195246.34$+$723957.9 & WISEPA J195246.66$+$724000.8 & 2 & \dots & \dots & T4 & 2 \\
J200250.59$-$052154.2 & 2MASSI 2002507$-$052152 & 32 & L5$\beta$ & 89 & L5--L7$\gamma$ & 63 \\
\\
J204749.62$-$071821.6 & SDSS J204749.61$-$071818.3 & 1 & \dots & \dots & L7.5 & 69 \\
J210115.58$+$175656.0 & 2MASSW J2101154$+$175658 & 4 & L7.5 & 4 & L7+L8 & 76 \\
J212414.06$+$010003.8 & SDSS J212413.89$+$010000.3 & 1 & \dots & \dots & T5 & 49 \\
J212702.63$+$761756.8 & 2MASS J21265916$+$7617440 & 3 & L7 & 3 & T0 pec & 3 \\
J213927.29$+$022024.7 & 2MASS J21392676$+$0220226 & 49 & T2 & 56 & T1.5 & 49 \\
\\
J215432.82$+$594211.3 & 2MASS J21543318$+$5942187 & 21 & \dots & \dots & T5.5 & 69 \\
J221354.65$+$091139.0 & WISEPC J221354.69$+$091139.4 & 2 & \dots & \dots & T7 & 2 \\
J222444.39$-$015908.2 & 2MASSW J2224438$-$015852 & 4 & L4.5 & 4 & L3 & 57 \\
J222622.96$+$044001.1 & WISEPC J222623.05$+$044003.9 & 2 & \dots & \dots & T8 & 2 \\
J223937.67$+$161716.8 & WISEPC J223937.55$+$161716.2 & 2 & \dots & \dots & T3 & 2 \\
\\
J224253.65$+$254256.2 & 2MASSJ22425317$+$2542573 & 50 & L3 & 32 & L2 pec & 63 \\
J224431.96$+$204339.1 & 2MASSW J2244316$+$204343 & 51 & L6.5 & 51 & L6-L8$\gamma$ & 63 \\
J225418.98$+$312353.0 & 2MASSI J2254188$+$312349 & 12 & T5 & 56 & T4 & 49 \\
J225529.03$-$003436.4 & SDSS J225529.09$-$003433.4 & 52 & L0 & 77 & \dots & \dots \\
J232123.84$+$135450.3 & ULAS J232123.79$+$135454.8 & 53 & \dots & \dots & T7.5 & 53 \\
\\
J232545.24$+$425143.9 & 2MASS J23254530$+$4251488 & 15 & L8 & 32 & L7 & 87 \\
J232728.85$-$273056.4 & WISEPC J232728.75$-$273056.5 & 2 & \dots & \dots & L9 & 2 \\
J233051.30$-$084455.9 & CFBDS J233051.24$-$084454.6 & 54 & \dots & \dots & T0 & 54 \\
J233910.67$+$135212.9 & 2MASSI J2339101$+$135230 & 12 & \dots & \dots & T5 & 49 \\
J234026.68$-$074509.6 & WISEPC J234026.62$-$074507.2 & 2 & T7 & 2 & T7 & 2 \\
\\
J234841.34$-$102843.2 & WISEPC J234841.10$-$102844.4 & 2 & \dots & \dots & T7 & 2 \\
J235122.32$+$301054.1 & 2MASS J23512200$+$3010540 & 4 & L5.5 & 4 & L5 & 4 \\
J235654.19$-$155322.9 & 2MASSI J2356547$-$155310 & 55 & \dots & \dots & T5 & 49 \\
\enddata
\tablerefs{(1) \cite{knapp2004}; (2) \cite{kirkpatrick2011}; (3) \cite{kirkpatrick2010}; (4) \cite{kirkpatrick2000}; (5) \cite{geballe2002}; (6) \cite{burgasser2004}; (7) \cite{salim2003}; (8) \cite{reyle2010}; (9) \cite{delorme2008}; (10) \cite{schneider2002}; (11) \cite{chiu2006}; (12) \cite{burgasser2002}; (13) \cite{wilson2003}; (14) \cite{scholz2011}; (15) \cite{reid2006b}; (16) \cite{cruz2003}; (17) \cite{burgasser2003c}; (18) \cite{burgasser2003d}; (19) \cite{fan2000}; (20) \cite{burgasser2000}; (21) \cite{looper2007}; (22) \cite{castro2012}; (23) \cite{loutrel2011}; (24) \cite{thorstensen2003}; (25) \cite{lucas2010}; (26) \cite{aberasturi2011}; (27) \cite{burgasser2003}; (28) \cite{leggett2000}; (29) \cite{kirkpatrick1999}; (30) \cite{zhang2009}; (31) \cite{tinney2005}; (32) \cite{cruz2007}; (33) \cite{burgasser1999}; (34) \cite{hawley2002}; (35) \cite{deacon2011}; (36) \cite{burningham2010}; (37) \cite{burningham2008}; (38) \cite{jameson2008}; (39) \cite{gizis2000}; (40) \cite{mclean2000}; (41) \cite{metchev2008}; (42) \cite{burgasser2011b}; (43) \cite{strauss1999}; (44) \cite{burgasser2004b}; (45) \cite{gelino2011}; (46) \cite{gizis2011}; (47) \cite{looper2008b}; (48) \cite{gizis2011b}; (49) \cite{burgasser2006}; (50) \cite{gizis2003}; (51) \cite{dahn2002}; (52) \cite{fan2001}; (53) \cite{scholz2010}; (54) \cite{albert2011}; (55) \cite{mclean2001}; (56) \cite{pineda2016}; (57) \cite{liu2016}; (58) \cite{schneider2014}; (59) \cite{zhang2017}; (60) \cite{cruz2009}; (61) \cite{allers2013}; (62) \cite{cushing2011}; (63) \cite{gagne2015b}; (64) \cite{cruz2018}; (65) \cite{scholz2009}; (66) \cite{liu2011}; (67) \cite{leggett2019}; (68) \cite{burgasser2003b}; (69) \cite{burgasser2010}; (70) \cite{burgasser2007b}; (71) \cite{reid2008b}; (72) \cite{zhang2013}; (73) \cite{castro2013}; (74) \cite{thompson2013}; (75) \cite{loutrel2011}; (76) \cite{dupuy2012}; (77) \cite{schmidt2010}; (78) \cite{kirkpatrick2008}; (79) \cite{faherty2009}; (80) \cite{bardalez2019}; (81) \cite{bardalez2015}; (82) \cite{zhang2019}; (83) \cite{leggett2017}; (84) \cite{kellogg2015}; (85) \cite{marocco2015}; (86) \cite{burgasser2007}; (87) \cite{burgasser2011}; (88) \cite{gizis2015}; (89) \cite{faherty2016}; (90) \cite{burgasser2025}     }
\end{deluxetable*}

\section{Summary of Observations}
\label{sec:obssum}

In this section we give an overview of observations for each object on the program which are summarized in Table \ref{tab:observations}. Column (1) again gives the CatWISE object name, while column (2) gives a shorthand summary of the optical and IR spectral typing for reference. Column (3) shows whether the object was observed during Series 1 (52 objects) or Series 2 (121 objects). Column (4) shows the filter used for the astrometric observations with totals of 87 objects observed in $J$-band and 86 objects observed in $H$-band. Column (5) shows the number of independent observation epochs observed with a range of 11 to 220 and median numbers of 59.5 and 64.0 for Series 1 and Series 2, respectively, and with an overall median of 62 epochs. Column (6) gives the time duration of each set of observations from the first to the last epoch used in the final astrometric reductions with a range of 2.69 to 8.12 with an overall median value of 5.25 years. We note that even the shortest of these time durations encompasses observations during at least parts of four seasons, which USNO has traditionally seen as a minimum to completely separate parallax and proper motion. Column (7) provides the mean epochs of observation calculated from only those frames used in the final astrometric reductions. Column (8) gives the number of reference frame stars used in the final reductions with a median number of 11 for both Series 1 and 2. Column (9) shows whether Gaia DR3 astrometric values are available, which will be used in a consistency and reliability analysis in $\S$\ref{sec:gaiacomp}.

\startlongtable
\begin{deluxetable*}{ccccccccccc}
\label{tab:observations}
\tabletypesize{\scriptsize}
\tablecaption{Parallax and Proper Motion Object Observations and Reference Stars}
\tablewidth{0pt}
\tablehead{
\colhead{Object Name} & \colhead{SpecT} & \colhead{Series} & \colhead{Filter} & \colhead{No. Epochs}  & \colhead{$\Delta$T (yrs)} & \colhead{Mean Epoch} & \colhead{No. Ref. Stars} & \colhead{In Gaia?} \\
\colhead{CatWISE} & \colhead{opt/IR} \\
\colhead{(1)} & \colhead{(2)} & \colhead{(3)} & \colhead{(4)} & \colhead {(5)} &
\colhead{(6)} & \colhead{(7)} & \colhead{(8)} & \colhead{(9)} 
}
\startdata
J000013.51$+$255419.7           & T5/T4.5          & 2 &  $J$  & 90 & 7.34 & 2014.7512 &  9 &      \\
J000849.71$-$173925.0           & $-$/T6           & 2 &  $J$  & 66 & 7.36 & 2014.8949 &  7 &      \\
J001502.52$+$295928.7           & L7/L7.5 pec (blue)& 2 &  $H$  & 38 & 4.30 & 2017.2425 & 18 &      \\
J003030.38$-$145033.8           & L7/L6 FLD-G      & 1 &  $H$  & 65 & 5.27 & 2003.5844 & 13 &      \\
J003259.67$+$141037.2           & $-$/L9           & 1 &  $H$  & 60 & 5.10 & 2003.8004 & 14 & \\
\\
J003452.12$+$052307.3           & $-$/T6.5         & 2 &  $J$  & 73 & 7.36 & 2014.9679 &  5 &      \\ 
J004121.65$+$354712.5           & sdL0.5/sdL?      & 2 &  $H$  & 46 & 5.27 & 2016.3624 & 13 & Gaia \\
J004521.80$+$163444.0           & L2$\beta$/L2 VL-G& 2 &  $H$  & 39 & 5.27 & 2016.4792 &  8 & Gaia \\
J004928.58$+$044101.0           & $-$/L9           & 2 &  $J$  & 66 & 7.26 & 2015.0053 &  6 &      \\
J005911.15$-$011401.4           & $-$/T8.5         & 2 &  $J$  & 11 & 3.16 & 2018.3110 &  9 & \\
\\
J010332.40$+$193536.5           & L6/L6 pec (red)  & 2 &  $H$  & 73 & 4.33 & 2016.7854 & 13 &      \\
J010753.11$+$004157.7           & L7.5/L7 pec (red)& 1 &  $H$  & 90 & 4.95 & 2004.0956 &  7 &      \\
J011912.41$+$240331.7           & $-$/T2           & 2 &  $H$  & 65 & 4.00 & 2017.2812 & 17 &      \\
J013836.61$-$032222.8           & $-$/T3           & 2 &  $H$  & 71 & 3.21 & 2017.6562 &  9 &      \\
J015011.41$+$382723.7           & $-$/T0           & 2 &  $H$  &146 & 5.38 & 2016.1376 & 16 & \\
\\
J015142.59$+$124428.7           & T1/T1            & 1 &  $H$  & 69 & 5.17 & 2003.7852 & 10 &      \\
J020625.45$+$264023.3           & $-$/L8 (red)     & 2 &  $H$  &156 & 5.42 & 2016.3538 & 11 &      \\
J020743.00$+$000056.1           & $-$/T4.5         & 1 &  $H$  & 59 & 5.08 & 2003.9954 & 11 &      \\
J023618.06$+$004852.1           &  L9/L6.5         & 2 &  $H$  & 93 & 4.39 & 2017.0055 &  9 &      \\
J024313.38$-$245332.8           &  T5.5/T6         & 1 &  $J$  & 70 & 5.34 & 2003.6408 &  9 & \\
\\         
J025116.11$-$035315.7           & L3/L1            & 2 &  $H$  & 61 & 5.27 & 2016.0129 &  9 & Gaia \\
J025409.54$+$022358.7           & $-$/T8           & 2 &  $J$  &131 & 7.33 & 2014.4005 & 11 &      \\
J030533.66$+$395434.6           & $-$/T6           & 2 &  $J$  &105 & 7.35 & 2014.4591 & 17 &      \\
J031100.13$+$164815.6           & L8/L9.5          & 1 &  $H$  & 30 & 3.32 & 2004.0835 & 11 &      \\
J031326.13$+$780744.6           & $-$/T9           & 2 &  $J$  & 35 & 3.98 & 2017.1154 & 15 & \\
\\      
J032553.05$+$042539.6           & $-$/T5.5         & 2 &  $J$  & 34 & 3.20 & 2017.4982 &  8 &      \\
J032842.66$+$230204.1           & L8/L9.5          & 1 &  $H$  & 53 & 5.35 & 2003.4117 & 18 &      \\
J035523.53$+$113336.7     &L5$\gamma$/L3-L6$\gamma$& 2 &  $H$  & 95 & 6.05 & 2016.0492 & 13 & Gaia \\
J040709.03$+$151455.0           & $-$/T5           & 2 &  $J$  & 99 & 7.41 & 2015.1326 & 19 &      \\
J041054.46$+$141130.8           & $-$/T6           & 2 &  $J$  &104 & 7.40 & 2015.1590 & 13 & \\
\\
J041522.17$-$093457.1         & T8/T8            & 1 &  $J$  & 62 & 5.34 & 2003.7253 &  7 &      \\
J042348.22$-$041402.0         & L7.5/L7.5+T2     & 1 &  $H$  & 62 & 5.25 & 2003.9656& 11 & Gaia \\
J043900.87$-$235310.7         & L6/L4.5          & 2 &  $H$  & 48 & 4.02 & 2016.9863 & 11 & Gaia \\
J044853.70$-$193543.6         & $-$/sdT5         & 2 &  $H$  & 59 & 4.33 & 2016.8059 & 19 &      \\
J050021.02$+$033044.8         & L4/$-$           & 2 &  $H$  & 90 & 6.05 & 2015.4119 & 11 & Gaia \\
\\    
J051317.27$+$060812.3         & $-$/T6.5         & 2 &  $J$  &149 & 7.40 & 2014.2212 & 15 &      \\
J051609.20$-$044553.3         & $-$/T6           & 1 &  $J$  & 46 & 2.98 & 2004.9613 & 11 &      \\
J052536.18$+$673951.5         & $-$/T6 pec       & 2 &  $J$  & 37 & 4.02 & 2017.2288 & 14 &      \\
J053312.61$+$824617.2         & esdL7/esdL8      & 1 &  $H$  & 37 & 2.98 & 2004.8103 & 12 & Gaia \\
J053952.16$-$005856.5         & L5/L5(blue)      & 1 &  $H$  & 63 & 5.23 & 2003.8234 & 15 & Gaia \\
\\         
J054231.20$-$162827.6         & $-$/T6.5         & 2 &  $J$  & 28 & 3.26 & 2017.3767 & 22 &      \\
J055919.85$-$140454.8         &  T5/T4.5         & 1 &  $J$  & 95 & 5.34 & 2003.8918 & 13 & Gaia \\
J060206.71$+$404355.4         & $-$/T4.5         & 2 &  $J$  & 31 & 3.26 & 2017.5148 & 16 &      \\
J060738.42$+$242951.2         &  L8/L9           & 2 &  $J$  & 93 & 7.27 & 2014.5519 & 22 & Gaia \\
J061407.41$+$391233.2         & $-$/T6           & 2 &  $J$  & 85 & 7.30 & 2014.5799 & 20 &      \\
\\    
J062542.19$+$564625.4         & $-$/T6           & 2 &  $J$  & 88 & 7.28 & 2014.2581 & 15 &      \\
J062720.06$-$111429.6        & $-$/T6            & 2 &  $J$  & 28 & 3.22 & 2017.6912 & 13 &      \\
J064626.92$+$793454.5         & $-$/L9           & 2 &  $H$  & 31 & 3.24 & 2017.7400 & 11 &      \\
J065609.75$+$420532.8         & $-$/T3           & 2 &  $J$  &116 & 7.34 & 2014.6467 & 11 &      \\
J070036.80$+$315718.2         &L3.5/L3+L6.5      & 2 &  $H$  & 91 & 6.14 & 2015.5478 & 12 & Gaia \\
\\     
J072226.89$-$054027.5         & $-$/T9           & 2 &  $J$  & 43 & 4.20 & 2017.1137 & 19 &      \\
J072719.22$+$170950.1         & T8/T7            & 1 &  $J$  & 95 & 5.33 & 2003.7811 & 11 &      \\
J074200.88$+$205515.7         & $-$/T5           & 2 &  $J$  &119 & 7.35 & 2014.7690 & 16 &      \\
J075547.93$+$221212.8         & T6/T5            & 1 &  $J$  & 55 & 3.31 & 2004.4925 & 12 &      \\
J075840.07$+$324718.8         & T3/T2            & 2 &  $J$  &113 & 7.34 & 2014.7385 & 17 & Gaia \\
\\      
J081957.98$-$033529.3         & T4/T4            & 2 &  $J$  & 36 & 3.16 & 2017.7417 & 16 &      \\  
J082131.64$+$144317.8         & $-$/T5.5         & 2 &  $J$  & 99 & 7.37 & 2014.8615 & 16 &      \\
J082518.97$+$211546.1         & L7/L7 pec (red)  & 1 &  $H$  & 54 & 5.25 & 2003.8967 & 15 & Gaia \\
J083006.91$+$482838.3         & L8/L9.5          & 1 &  $H$  & 52 & 5.26 & 2004.3483 & 11 & Gaia \\
J083541.88$-$081918.0         & L5/L4 pec (red)  & 2 &  $H$  & 87 & 6.18 & 2015.8820 & 12 & Gaia \\
\\           
J083717.18$-$000020.8         & T0/T1            & 1 &  $J$  & 39 & 4.95 & 2003.6761 & 12 &   \\
J085035.75$+$105715.3         &L6/L6.5+L8.5      & 1 &  $H$  & 53 & 5.23 & 2003.8130 & 12 &      \\
J085757.60$+$570844.6         &L8/L8-L9 pec (red)& 2 &  $H$  & 70 & 6.19 & 2015.5105 & 11 & Gaia \\
J085833.72$+$325628.6         &$-$/T1 pec (red)  & 2 &  $H$  & 37 & 4.22 & 2017.0858 & 14 &      \\
J090023.73$+$253934.1         & L7/$-$           & 2 &  $H$  & 29 & 4.05 & 2017.1831 & 12 &      \\
\\      
J090900.31$+$656525.8         &  $-$/T1.5        & 2 &  $H$  & 19 & 3.11 & 2017.9158 & 10 &      \\ 
J091534.04$+$042204.8         &L7+L7/L6+L6       & 2 &  $J$  & 91 & 7.42 & 2014.6001 & 10 & Gaia \\
J092615.40$+$584717.6         & T5/T4+T5.5       & 1 &  $J$  & 33 & 3.23 & 2004.5633 & 13 &      \\
J092933.34$+$342951.3         & L8/L7.5          & 2 &  $H$  & 33 & 3.16 & 2017.8105 & 11 &      \\
J093735.98$+$293120.8         & T7 pec (red)/sdT6      & 1 &  $J$  & 44 & 5.28 & 2003.7034 &  8 &      \\
\\     
J093936.13$-$244844.3         & T8/d/sdT8           & 2 &  $J$  & 92 & 7.36 & 2015.0471 & 15 &      \\  
J094908.49$-$154548.3         & $-$/T2           & 2 &  $H$  & 38 & 3.17 & 2017.7768 & 13 &      \\
J095105.35$+$355800.1         & L6/L6            & 1 &  $H$  & 51 & 5.26 & 2003.6711 & 12 & Gaia \\
J101014.45$-$040650.0         & L6/L6            & 2 &  $H$  & 34 & 3.19 & 2017.8427 &  9 &      \\
J101905.55$+$652954.5         & T7/T6            & 2 &  $J$  &109 & 7.95 & 2014.7231 &  7 &      \\
\\   
J102109.51$-$030421.1         &T3.5/T1+T5.5      & 1 &  $J$  & 59 & 5.29 & 2004.1753 & 10 &      \\
J103931.32$+$325623.7         &  $-$/T1          & 2 &  $H$  & 27 & 3.17 & 2017.7852 &  7 &      \\
J104307.42$+$222523.3         & L8/L8 pec        & 2 &  $H$  & 21 & 3.16 & 2017.8359 &  8 &      \\
J104335.10$+$121310.8         & $-$/L9           & 2 &  $H$  & 22 & 3.20 & 2017.5144 &  8 &      \\
J104751.75$+$212414.7         & T7/T6.5          & 1 &  $J$  & 68 & 5.29 & 2003.5886 &  6 &      \\
\\     
J104828.96$+$091940.9         & $-$/T2.5         & 2 &  $H$  & 21 & 3.20 & 2018.0375 & 10 &      \\
J110611.58$+$275414.1         & $-$/T2.5         & 2 &  $J$  & 29 & 3.19 & 2017.9380 &  5 & Gaia \\
J111009.77$+$011608.7         & $-$/T5.5           & 1 &  $J$  & 43 & 3.38 & 2004.3237 &  9 &      \\
J111447.51$-$261830.0         & T8/T7.5          & 2 &  $J$  & 89 & 7.31 & 2015.0406 & 19 &      \\
J111812.22$-$085615.0         &L6/L6 pec (blue)  & 2 &  $H$  & 38 & 3.12 & 2018.1265 &  8 & Gaia \\
\\        
J112254.33$+$255020.2         & $-$/T6           & 2 &  $J$  &110 & 7.95 & 2015.1655 &  7 &      \\
J115553.43$+$055956.6         & L6/L7 pec        & 2 &  $H$  & 72 & 4.29 & 2017.0018 &  7 &      \\
J115821.40$+$043446.3         & sdL7/d/sdL8        & 2 &  $H$  & 37 & 3.13 & 2018.0378 & 10 & Gaia \\
J120746.66$+$024426.8         & L8/T0            & 2 &  $H$  & 55 & 4.09 & 2017.5132 &  7 &      \\
J121709.86$-$031111.8         & T7/T7.5          & 1 &  $J$  & 62 & 5.29 & 2003.8303 &  8 &      \\
\\   
J121757.13$+$162635.1         &$-$/T8.5+Y0       & 2 &  $J$  & 17 & 3.10 & 2017.7509 & 11 &      \\
J122554.84$-$273958.5         & T6/T5.5+T8       & 1 &  $J$  & 46 & 5.25 & 2003.6483 & 11 &      \\
J123146.34$+$084717.1         & T6/T5.5          & 2 &  $J$  &180 & 8.02 & 2014.3637 &  7 &      \\
J123737.03$+$652607.2         & T7/T6.5          & 1 &  $J$  & 53 & 5.33 & 2003.9171 &  5 &      \\
J125011.59$+$392542.6         & $-$/T4           & 2 &  $H$  & 36 & 3.27 & 2017.4639 &  9 &      \\
\\    
J125453.41$-$012245.6         & T2/T2            & 1 &  $J$  & 50 & 5.26 & 2003.7879 &  8 &      \\
J131141.74$+$362925.2         &$-$/L5 pec (blue) & 2 &  $J$  & 65 & 7.98 & 2014.6520 &  9 & Gaia \\
J132003.78$+$603425.8         & $-$/T6.5         & 2 &  $J$  & 60 & 8.00 & 2014.4508 &  7 &      \\
J132233.51$-$234015.3         & $-$/T8           & 2 &  $J$  & 63 & 7.87 & 2014.7101 &  8 &      \\
J132407.64$+$190625.5         & $-$/T3.5         & 2 &  $J$  & 63 & 6.99 & 2015.3268 &  3 &      \\
\\   
J132434.67$+$635827.1         &T2.5:/T2: pec(red)& 2 &  $H$  & 30 & 3.22 & 2018.0138 &  8 &      \\
J132605.25$+$120010.1         & $-$/T6 pec       & 2 &  $J$  & 14 & 3.15 & 2017.7938 & 13 &      \\
J132629.56$-$003833.2         & L8?/L7           & 1 &  $H$  & 85 & 5.33 & 2004.0752 &  6 &      \\
J133553.35$+$113003.7         & $-$/T8.5         & 2 &  $J$  & 16 & 3.15 & 2017.8629 & 12 &      \\
J134645.89$-$003152.2         & $-$/T6.5         & 1 &  $J$  & 92 & 5.36 & 2004.3195 &  8 &      \\
\\   
J134807.08$+$660327.2         & $-$/L9           & 2 &  $H$  & 82 & 5.29 & 2016.8284 & 10 &      \\
J140255.70$+$080053.4         & $-$/T2 pec       & 2 &  $H$  & 67 & 4.10 & 2017.5111 & 10 &      \\
J140753.30$+$124110.7         & L3/L4            & 2 &  $H$  &151 & 6.29 & 2015.1740 &  9 & Gaia \\
J143517.22$-$004612.9         & L0/$-$           & 1 &  $H$  & 78 & 5.35 & 2004.0412 &  6 &      \\
J143535.75$-$004348.6         & L3/L2.5          & 1 &  $H$  & 78 & 5.35 & 2004.0288 &  6 &      \\
\\    
J143945.64$+$304218.7         & $-$/T2.5         & 2 &  $H$  & 76 & 4.31 & 2017.2131 & 11 &      \\
J144600.78$+$002450.9         & L6/L5            & 1 &  $H$  & 69 & 5.33 & 2003.7094 & 10 &      \\
J145714.66$+$581509.7         & T8/T7            & 2 &  $J$  &133 & 8.05 & 2014.8997 & 11 &      \\
J150319.71$+$252528.3         & T6/T5            & 1 &  $J$  & 49 & 3.30 & 2004.5020 &  8 & Gaia \\
J150648.79$+$702741.0         & T6/T6            & 2 &  $J$  &122 & 8.05 & 2014.8189 &  6 & Gaia \\
\\  
J150653.09$+$132105.8         & L3/L3            & 2 &  $H$  & 89 & 6.27 & 2015.6727 & 16 & Gaia \\
J151114.38$+$060739.4         & $-$/L5+T5        & 2 &  $H$  & 38 & 3.98 & 2017.8066 & 11 &      \\
J151459.41$+$484803.4         & L6/L6            & 2 &  $H$  & 39 & 3.98 & 2017.7328 &  7 & Gaia \\
J152040.10$+$354615.7         & $-$/L7.5         & 2 &  $H$  & 50 & 4.06 & 2017.7802 &  8 &      \\
J152322.78$+$301453.4         & L8/L8            & 1 &  $H$  & 79 & 5.32 & 2003.5904 & 14 &      \\
\\     
J152613.76$+$204334.8         & L7/L7            & 2 &  $H$  & 56 & 4.16 & 2017.4366 & 10 & Gaia \\
J154614.94$+$493158.6         & $-$/T3           & 2 &  $J$  &202 & 8.13 & 2014.4620 & 11 &      \\
J155301.80$+$153239.5         & $-$/T6.5+T7.5    & 1 &  $J$  & 42 & 3.31 & 2004.4192 &  9 &      \\
J161705.76$+$180713.9         & T8/T8            & 2 &  $J$  & 40 & 4.03 & 2017.8027 & 10 &      \\
J162413.95$+$002915.6         & T6/T6            & 1 &  $J$  & 55 & 5.25 & 2003.3333 &  9 &      \\
\\   
J162541.25$+$152810.0         & $-$/T4.5         & 2 &  $J$  & 61 & 7.07 & 2015.6299 & 14 &      \\
J162618.23$+$392523.4         & usdL4/sdL4       & 2 &  $H$  & 54 & 6.31 & 2016.3005 & 11 & Gaia \\
J162725.60$+$325522.8         & $-$/T6           & 2 &  $J$  & 84 & 8.05 & 2015.0827 &  9 &      \\
J162918.64$+$033534.8         & T3/T2            & 2 &  $J$  & 55 & 7.11 & 2016.0868 & 15 &      \\
J163229.48$+$190439.6         & L8/L8            & 1 &  $H$  & 59 & 5.25 & 2003.5659 & 16 &      \\
\\  
J164715.47$+$563209.4         & L7/L9 pec (red)  & 2 &  $H$  & 53 & 5.00 & 2016.9886 & 10 &      \\
J165311.00$+$444421.2         & T8/T8            & 2 &  $H$  & 44 & 4.04 & 2017.6767 & 10 &      \\
J171145.73$+$223204.2         & L6.5/L9          & 1 &  $H$  & 84 & 5.18 & 2003.7680 & 17 &      \\
J172811.54$+$394859.0         & L7/L5+L6.5       & 1 &  $H$  & 73 & 5.18 & 2003.6422 & 26 &      \\
J174124.05$+$255312.0         & T9/T9            & 2 &  $J$  & 94 & 8.10 & 2015.4459 & 12 &      \\
\\       
J175023.78$+$422238.6         & $-$/T1.5         & 2 &  $J$  & 80 & 8.12 & 2015.4009 & 12 &      \\
J175033.14$+$175905.4         & T4/T3            & 1 &  $J$  & 71 & 5.04 & 2003.3001 & 25 &      \\
J175609.98$+$281516.4         & sdL1/L1 pec (blue)& 2 &  $H$  & 59 & 6.22 & 2016.3656 & 15 & Gaia \\
J175805.45$+$463318.2         & $-$/T6.5         & 2 &  $J$  & 71 & 8.13 & 2015.0107 & 13 &      \\
J180026.66$+$013450.9         & L7.5/L7.5        & 2 &  $J$  & 64 & 7.90 & 2015.7498 & 13 & Gaia \\
\\      
J181210.89$+$272142.8         & $-$/T8.5:        & 2 &  $J$  & 19 & 4.67 & 2018.0987 & 19 &      \\
J182128.39$+$141357.2         & L4.5/L5          & 2 &  $H$  & 86 & 6.18 & 2016.1783 & 14 & Gaia \\
J183058.60$+$454258.2         & $-$/L9           & 2 &  $H$  & 63 & 5.78 & 2016.9046 & 26 &      \\
J184108.67$+$311728.6         & L4 pec/$-$       & 1 &  $H$  & 91 & 5.75 & 2003.4407 & 26 &      \\
J185215.87$+$353714.9         & $-$/T7           & 2 &  $J$  & 99 & 8.04 & 2015.2472 & 15 &      \\
\\   
J190106.23$+$471820.5         & $-$/T5           & 1 &  $J$  & 45 & 2.69 & 2004.9215 & 21 &      \\
J190624.72$+$450805.2         & $-$/T6           & 2 &  $J$  & 90 & 8.11 & 2015.4566 & 13 &      \\
J190648.66$+$401105.9         & $-$/L1           & 2 &  $J$  & 96 & 7.96 & 2015.2938 & 15 & Gaia \\
J195246.34$+$723957.9         & $-$/T4           & 2 &  $J$  &220 & 8.05 & 2015.2340 & 13 &      \\ 
J200250.59$-$052154.2   &L5$\beta$/L5-L7$\gamma$ & 2 &  $H$  &113 & 4.72 & 2017.4839 & 25 & Gaia \\
\\        
J204749.62$-$071821.6         & $-$/L7.5         & 2 &  $H$  &106 & 4.72 & 2017.5590 & 19 &      \\
J210115.58$+$175656.0         & L7.5/L7+L8       & 1 &  $H$  &119 & 5.76 & 2003.7206 & 17 &      \\
J212414.06$+$010003.8         & $-$/T5           & 2 &  $J$  &155 & 7.94 & 2014.6552 & 17 &      \\ 
J212702.63$+$761756.8         & L7/T0 pec        & 2 &  $H$  & 69 & 4.68 & 2017.0680 & 20 & Gaia \\
J213927.29$+$022024.7         & T2/T1.5          & 2 &  $H$  & 51 & 3.82 & 2017.4431 & 10 &      \\
\\  
J215432.82$+$594211.3         & $-$/T5.5         & 2 &  $J$  & 51 & 3.22 & 2017.3195 & 18 &      \\
J221354.65$+$091139.0         & $-$/T7           & 2 &  $J$  &161 & 7.38 & 2014.5447 &  8 &      \\
J222444.39$-$015908.2         & L4.5/L3          & 1 &  $H$  &118 & 5.19 & 2003.5661 &  5 & Gaia \\
J222622.96$+$044001.1         & $-$/T8           & 2 &  $J$  & 50 & 3.21 & 2017.2994 &  9 &      \\
J223937.67$+$161716.8         & $-$/T3           & 2 &  $H$  & 56 & 3.20 & 2017.3175 &  9 &      \\
\\    
J224253.65$+$254256.2         & L3/L2 pec        & 2 &  $H$  &110 & 5.30 & 2015.6177 &  8 & Gaia \\
J224431.96$+$204339.1         &L6.5/L6-L8$\gamma$ & 1 &  $H$  & 52 & 3.17 & 2004.2541 & 19 &      \\
J225418.98$+$312353.0         & T5/T4            & 1 &  $J$  & 33 & 3.01 & 2004.6663 & 15 &      \\
J225529.03$-$003436.4         & L0/$-$           & 1 &  $H$  & 82 & 5.10 & 2003.2746 & 11 & Gaia \\
J232123.84$+$135450.3         & $-$/T7.5         & 2 &  $J$  & 31 & 3.14 & 2017.4049 & 11 &      \\
\\  
J232545.24$+$425143.9         & L8/L7            & 2 &  $H$  & 22 & 3.19 & 2017.4585 & 16 & Gaia \\
J232728.85$-$273056.4         & $-$/L9           & 2 &  $H$  & 49 & 5.21 & 2016.3766 &  9 &      \\
J233051.30$-$084455.9         & $-$/T0           & 2 &  $H$  & 29 & 3.20 & 2017.3785 &  8 &      \\
J233910.67$+$135212.9         & $-$/T5           & 1 &  $J$  & 56 & 3.08 & 2004.7886 & 10 &      \\
J234026.68$-$074509.6         & T7/T7            & 2 &  $J$  & 74 & 7.23 & 2014.3586 &  9 &      \\
\\                              
J234841.34$-$102843.2         & $-$/T7           & 2 &  $J$  & 74 & 7.29 & 2014.7906 &  7 &      \\
J235122.32$+$301054.1         & L5.5/L5          & 2 &  $H$  & 31 & 3.19 & 2017.4205 & 13 &      \\
J235654.19$-$155322.9         & $-$/T5           & 1 &  $J$  & 88 & 5.26 & 2003.3541 &  8 &   \\
\enddata
\end{deluxetable*}

\section{Astrometric Results}
\label{sec:results}

In this section we present our basic astrometric results along with  brief discussions of these results and associated error distributions.

\subsection{Summary of Astrometric Results}

The main astrometric results of this work are presented in Table \ref{tab:astrometry}. Column (1) again presents the CatWISE object names. Column (2) gives the parallaxes relative to the final reference frame ($\pi$$_{rel}$) and associated uncertainties. Column (3) gives the absolute parallaxes ($\pi$$_{abs}$) after correction for the mean Gaia DR3 distances of the reference frame stars, as described above, along with associated uncertainties, which are calculated by the relative parallax uncertainties added in quadrature with those of the corrections to absolute parallax. Column (4) gives the relative proper motions in the Right Ascension coordinate ($\mu$$\alpha$$_{rel}$) and associated uncertainties. Column (5) gives the absolute proper motions in Right Ascension ($\mu$$\alpha$$_{abs}$) after correction for the mean Gaia DR3 proper motions in Right Ascension of the reference frame stars, as described above, along with associated uncertainties, which are calculated by the relative proper motion errors in Right Ascension added in quadrature with those of the corrections to absolute proper motions in Right Ascension. Similarly, columns (6) and (7) give the relative and absolute proper motions in Declination ($\mu$$\delta$$_{rel}$ and $\mu$$\delta$$_{abs}$, respectively) and associated uncertainties. For convenience, columns (8) and (9) give the total absolute proper motions ($\mu$$_{abs}$) and position angles (P.A.) of proper motion in degrees (in the usual sense of East of North), respectively, and associated uncertainties. Column (10) gives the tangential velocities ($V_{\rm tan}$) based on the combined values of columns (3) and (8). The tangential velocity uncertainties are calculated by combining in quadrature the uncertainties imposed on $V_{\rm tan}$ by the independent uncertainties of $\pi$$_{abs}$ and $\mu$$_{abs}$. All listed uncertainties are 1$\sigma$ values. 

\begin{longrotatetable}
\begin{deluxetable*}{lrrrrrrrrr}
\label{tab:astrometry}
\tabletypesize{\scriptsize}
\tablecaption{Astrometric Results}
\tablewidth{0pt}
\tablehead{
\colhead{Object Name} & \colhead{$\pi_{rel}$} & \colhead{$\pi_{abs}$} & \colhead{$\mu\alpha_{rel}$} &
\colhead{$\mu\alpha_{abs}$} & \colhead{$\mu\delta_{rel}$} &
\colhead{$\mu\delta_{abs}$} & \colhead{$\mu_{abs}$} & \colhead{P.A.} & \colhead{$V_{\rm tan}$}  \\
\colhead{CatWISE} & \colhead{(mas)} & \colhead{(mas)} & \colhead{(mas yr$^{-1}$)} & \colhead{(mas yr$^{-1}$)} & \colhead{(mas yr$^{-1}$)} & \colhead{(mas yr$^{-1}$)}
& \colhead{(mas yr$^{-1}$)} & \colhead{(deg)} & \colhead{(km s$^{-1}$)}\\
\colhead{(1)} & \colhead{(2)} & \colhead{(3)} & \colhead{(4)} & \colhead{(5)} & \colhead{(6)} &
\colhead{(7)} & \colhead{(8)} & \colhead{(9)} & \colhead{(10)}
}
\startdata
J000013.54$+$255419.7  & 74.25 $\pm$ 1.00 & 75.54 $\pm$ 1.06 & -51.94 $\pm$ 0.26 & -38.49 $\pm$ 0.27 & 130.33 $\pm$ 0.37 & 126.68 $\pm$ 0.37 & 132.40 $\pm$ 0.46 & 343.10 $\pm$ 0.10 & 8.42 $\pm$ 0.16 \\
J000849.76$-$173925.0\tablenotemark{a} & 36.57 $\pm$ 3.91 & 37.64 $\pm$ 3.92 & -177.36 $\pm$ 0.92 & -166.64 $\pm$ 0.93 & -361.40 $\pm$ 1.08 & -370.53 $\pm$ 1.08 & 406.28 $\pm$ 1.43 & 204.22 $\pm$ 0.10 & 51.17 $\pm$ 5.39 \\
J001502.52$+$295928.7   & 31.86 $\pm$ 1.65 & 33.05 $\pm$ 1.66 & 371.22 $\pm$ 0.91 & 374.28 $\pm$ 0.91 & -211.74 $\pm$ 1.05 & -223.02 $\pm$ 1.05 & 435.69 $\pm$ 1.39 & 120.79 $\pm$ 0.09 & 62.49 $\pm$ 3.16 \\
J003030.38$-$145033.8 &  34.14 $\pm$ 1.75 & 35.64 $\pm$ 1.78 & 244.20 $\pm$ 0.58 & 252.36 $\pm$ 0.59 & -27.90 $\pm$ 0.74 & -41.70 $\pm$ 0.74 & 255.78 $\pm$ 0.95 & 99.38 $\pm$ 0.11 & 34.02 $\pm$ 1.70 \\
J003259.67$+$141037.2   &  29.79 $\pm$ 1.89 & 30.61 $\pm$ 1.89 & 271.67 $\pm$ 0.74 & 281.65 $\pm$ 0.75 & 34.79 $\pm$ 1.06 & 27.32 $\pm$ 1.06 & 282.97 $\pm$ 1.30 & 84.46 $\pm$ 0.14 & 43.82 $\pm$ 2.73 \\
\\
J003452.12$+$052307.3   &  115.47 $\pm$ 1.22 & 116.34 $\pm$ 1.28 & 670.32 $\pm$ 0.31 & 666.59 $\pm$ 0.33 & 183.50 $\pm$ 0.35 & 172.57 $\pm$ 0.36 & 688.57 $\pm$ 0.49 & 75.49 $\pm$ 0.04 & 28.07 $\pm$ 0.31 \\
J004121.65$+$354712.5\tablenotemark{a}   &  8.50 $\pm$ 1.56 & 9.38 $\pm$ 1.56 & -113.74 $\pm$ 0.54 & -114.80 $\pm$ 0.54 & -54.77 $\pm$ 1.18 & -57.94 $\pm$ 1.18 & 128.59 $\pm$ 1.30 & 243.22 $\pm$ 0.29 & 64.98 $\pm$ 11.14 \\
\enddata
\tablenotetext{a}{$\pi$ measured in $\alpha$ only due to small $\pi$ factor in $\delta$}
\end{deluxetable*}
\end{longrotatetable}

\subsection{Astrometric Results Discussion}
 
Table \ref{tab:summary} presents median values of the parallax results for Series 1 (52 objects), Series 2 (121 objects), and combined Series 1+2. Column (2) gives the median relative parallaxes and median relative parallax uncertainties. We note that the median uncertainty for Series 2 is larger than for Series 1, despite similar numbers of frames and $\Delta$T, which is likely due to the selection of somewhat more difficult objects due to programmatic decisions as described in $\S$\ref{sec:objects}. Column (3) gives the median parallax values for the reference frames, and hence the median corrections from relative to absolute parallax, along with the median reference frame distances in parsecs in parentheses. Column (4) gives the program object median absolute parallaxes, median absolute parallax uncertainties, and, parenthetically, the median program object distances. A typical reference frame is at a distance of 900 parsecs, compared to a typical program object distance of 18 parsecs; roughly a factor of 50. This distance ratio is sufficient to cause only a small increase in uncertainty in the conversion to absolute parallax.   

\startlongtable
\begin{deluxetable}{cccc}
\label{tab:summary}
\tabletypesize{\scriptsize}
\tablecaption{Summary: Object and Reference Frame Parallaxes and Distances.}
\tablewidth{0pt}
\tablehead{
\colhead{Series} & \colhead{$\pi$$_{rel}$} & $\pi$$_{rel}$$\rightarrow$$\pi$$_{abs}$ & \colhead{$\pi$$_{abs}$} \\
& \colhead{(mas)} & \colhead{(mas/pc)}  & \colhead{(mas/pc)} \\
\colhead{(1)} & \colhead{(2)} & \colhead{(3)} & \colhead{(4)}  
}
\startdata
Series 1 &   56.75$\pm$1.27 & 0.98 & 57.73$\pm$1.28 \\
         &                & (1020 pc) & (17.32 pc)\\
\\         
Series 2 &   54.15$\pm$1.57 & 1.16 & 55.31$\pm$1.62 \\
         &                & (862 pc)  & (18.08 pc) \\
\\         
Series 1+2 & 54.49$\pm$1.47 & 1.10 & 55.59$\pm$1.51 \\
         &                & (909 pc)  & (17.99 pc)
\enddata
\end{deluxetable}

\begin{figure}
\epsscale{1.2}
\plotone{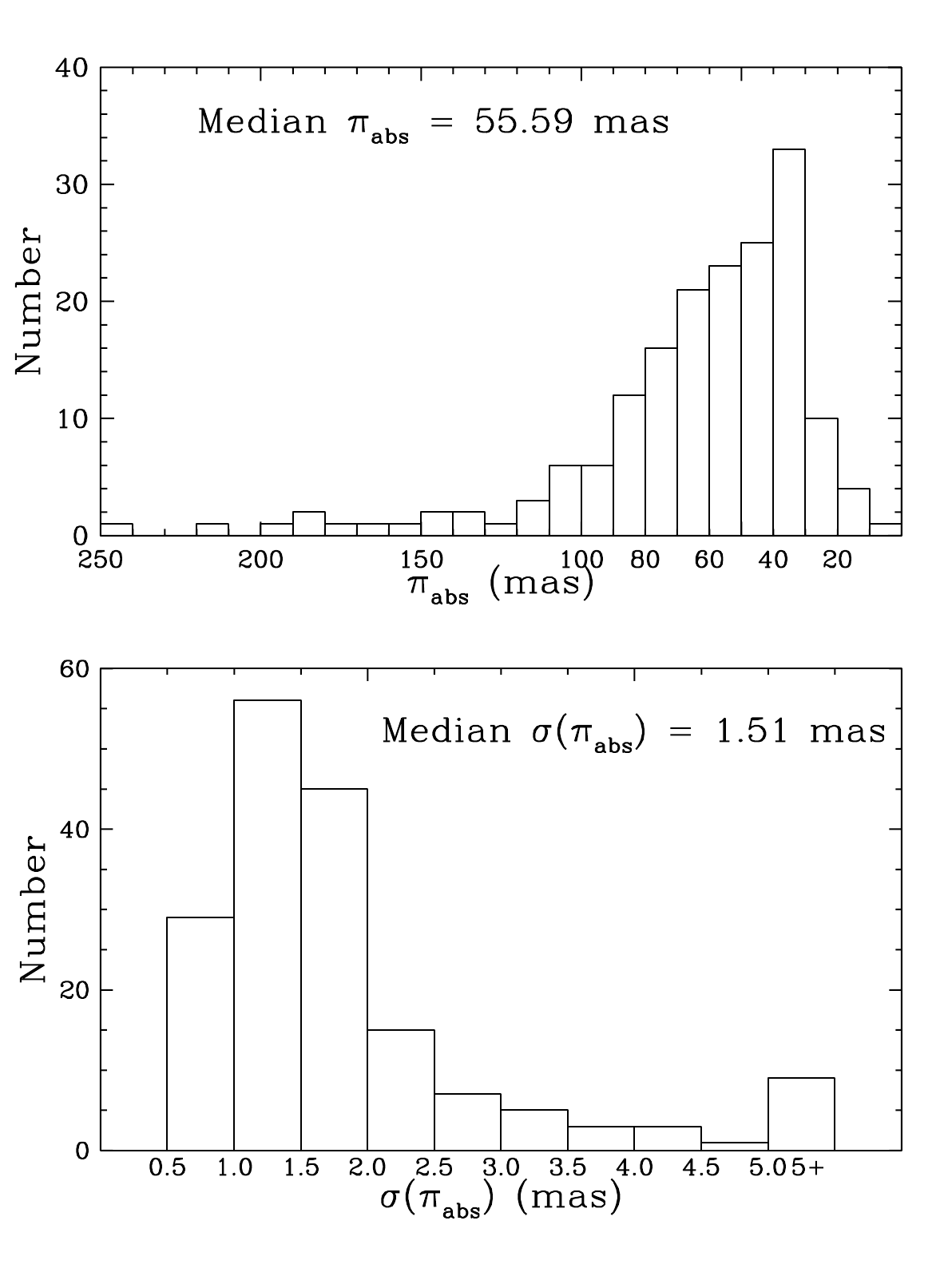}
\caption{Distributions of $\pi_{abs}$ and $\sigma$($\pi_{abs}$) along with median values for the combined Series 1 and Series 2 results.}
\label{fig:piabs}
\end{figure}

Figure \ref{fig:piabs} expands on the absolute parallax results showing the distribution of absolute parallaxes, the distribution of absolute uncertainties, and their associated median values. The distance distribution, peaking at about 30 parsecs, is less physical and more a product of the relatively small number of brown dwarfs known at the time within 10 parsecs and our desire to observe astrophysically interesting objects. The distribution of absolute parallax uncertainties, while having a median value of 1.51 mas, shows a tail of larger errors both due to selection of interesting, but faint, objects and the fact that both Series 1 and Series 2 observations had unplanned, abrupt endings leaving some objects without coverage we would otherwise have provided. Nonetheless, to our knowledge, our results provide the first published parallaxes for 16 objects and the highest precision parallaxes available for an additional 106 objects, along with concomitant proper motions. 

\begin{figure}
\epsscale{1.2}
\plotone{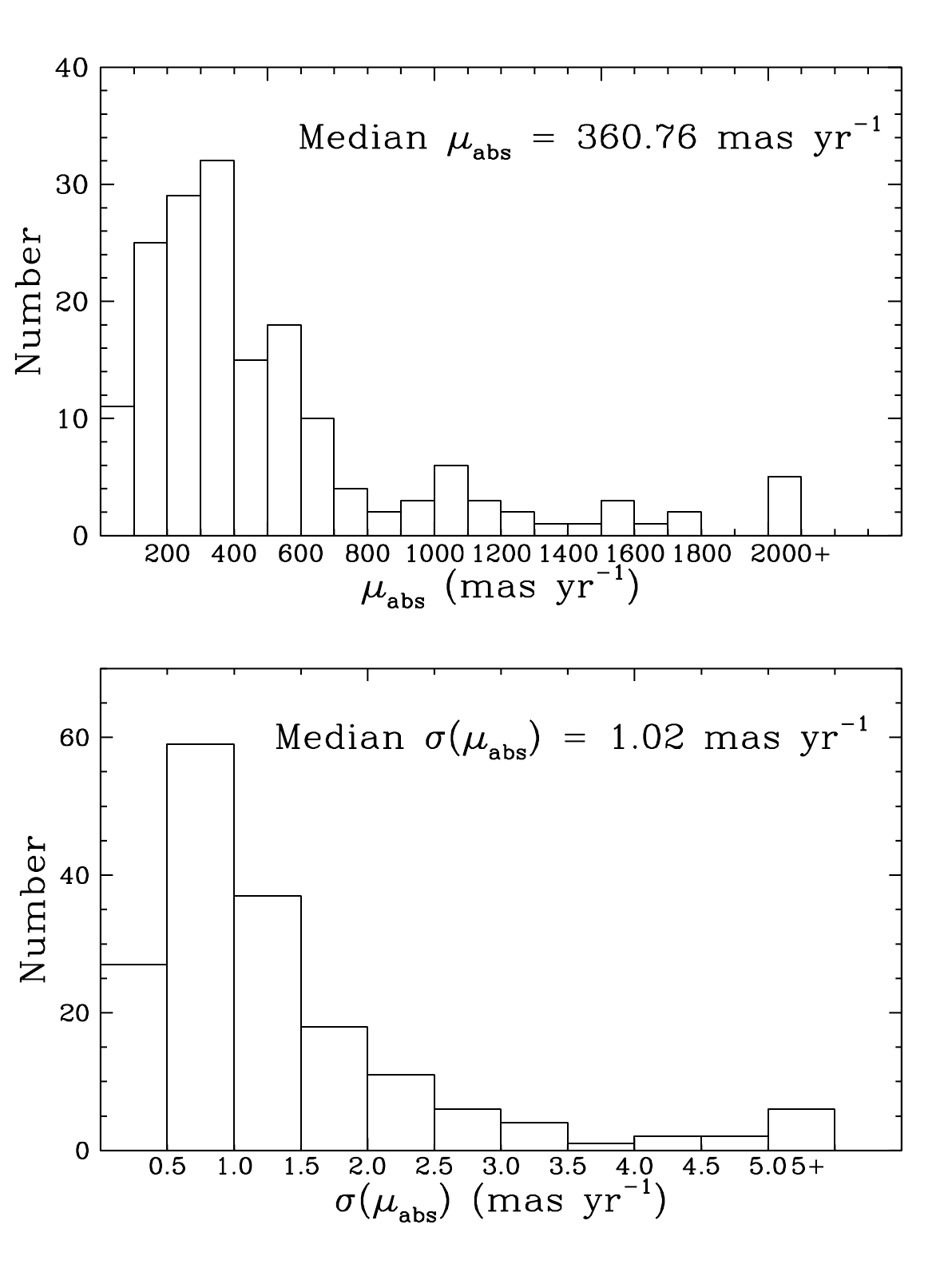}
\caption{Distributions of $\mu$$_{abs}$ and $\sigma$($\mu$$_{abs}$) along with median values for combined Series 1 and Series 2 results.}
\label{fig:muabs}
\end{figure}

Figure \ref{fig:muabs} shows similar diagrams for absolute total proper motion values, the distribution of absolute uncertainties, and their associated median values. As with Figure \ref{fig:piabs}, the distribution of absolute proper motion uncertainties, with a median value of 1.02 mas yr$^{-1}$, has a long tail toward larger uncertainties for similar reasons.

\begin{figure}
\epsscale{1.2}
\plotone{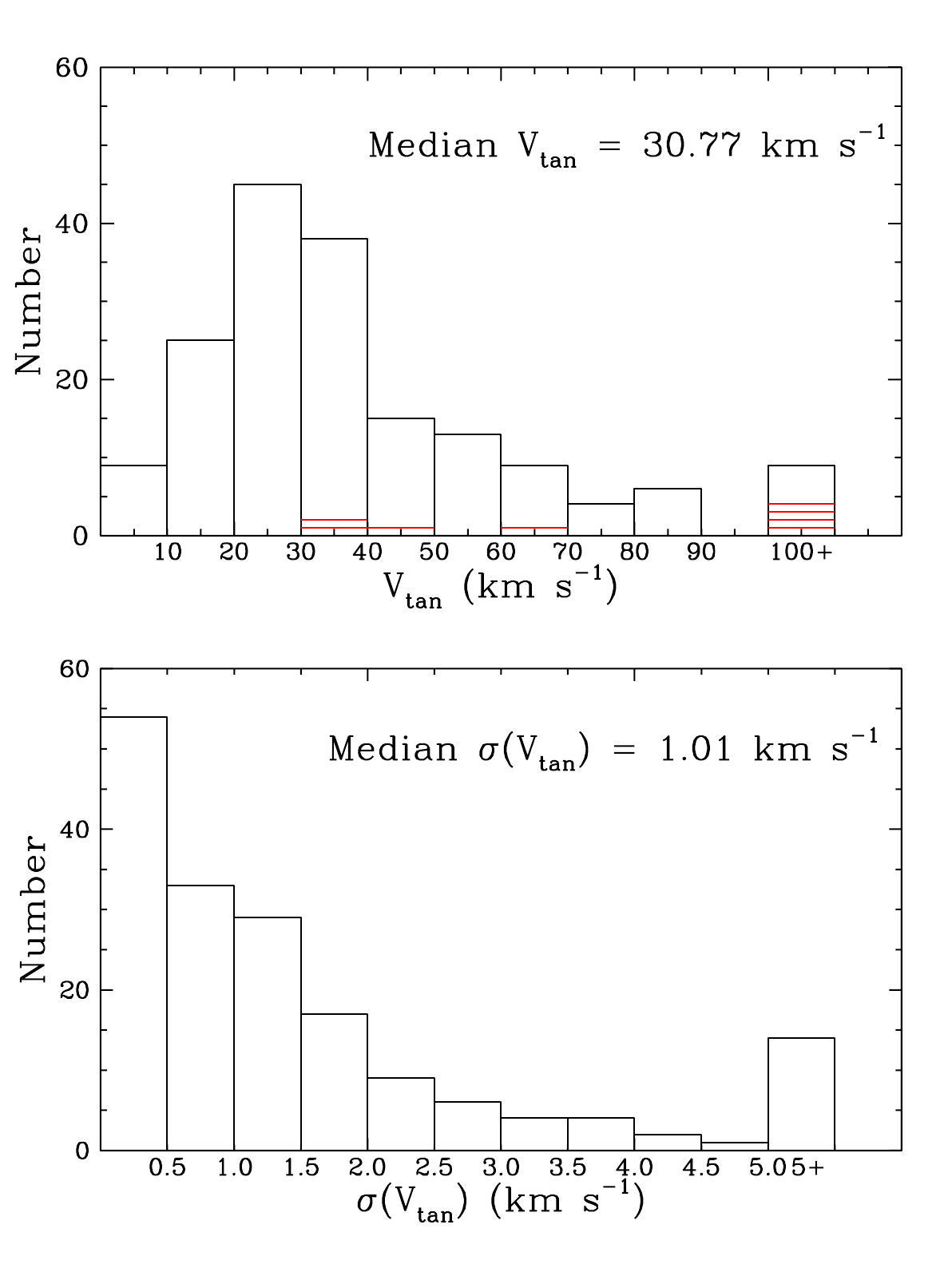}
\caption{Distributions of $V_{\rm tan}$ and $\sigma$($V_{\rm tan}$) along with median values. The red lines correspond to the $V_{\rm tan}$ velocities of the eight verified subdwarfs in this study (see text).}
\label{fig:vtanall}
\end{figure}

The panels in Figure \ref{fig:vtanall} show the distribution of tangential velocities and tangential velocity uncertainties, along with mean values for both, based on the absolute parallax and absolute proper motion determinations. The red lines depict the tangential velocities of the eight verified subdwarfs included in this study (J0041$+$35, J0448$-$19, J0533$+$82, J0937$+$29, J0939$-$24, J1158$+$04, J1626$+$39, J1756$+$28), reflecting their large range in tangential velocities. See $\S$\ref{sec:subdwarfs} for brief reviews of the certain and likely subdwarfs included in this study. As described above, the tangential velocity uncertainties are based on a combination of absolute parallax and absolute proper motion uncertainties and thus have a tail toward larger values. Nonetheless, the median tangential velocity uncertainty is about 1 km s$^{-1}$, with a plurality having uncertainties of $<$0.5 km s$^{-1}$. We will re-visit the validity of the uncertainties presented in Table \ref{tab:astrometry} and in this section in the comparison with Gaia astrometric results below.

\section{Systematic Internal Analysis}
\label{sec:internal}

\begin{figure}
\epsscale{1.2}
\plotone{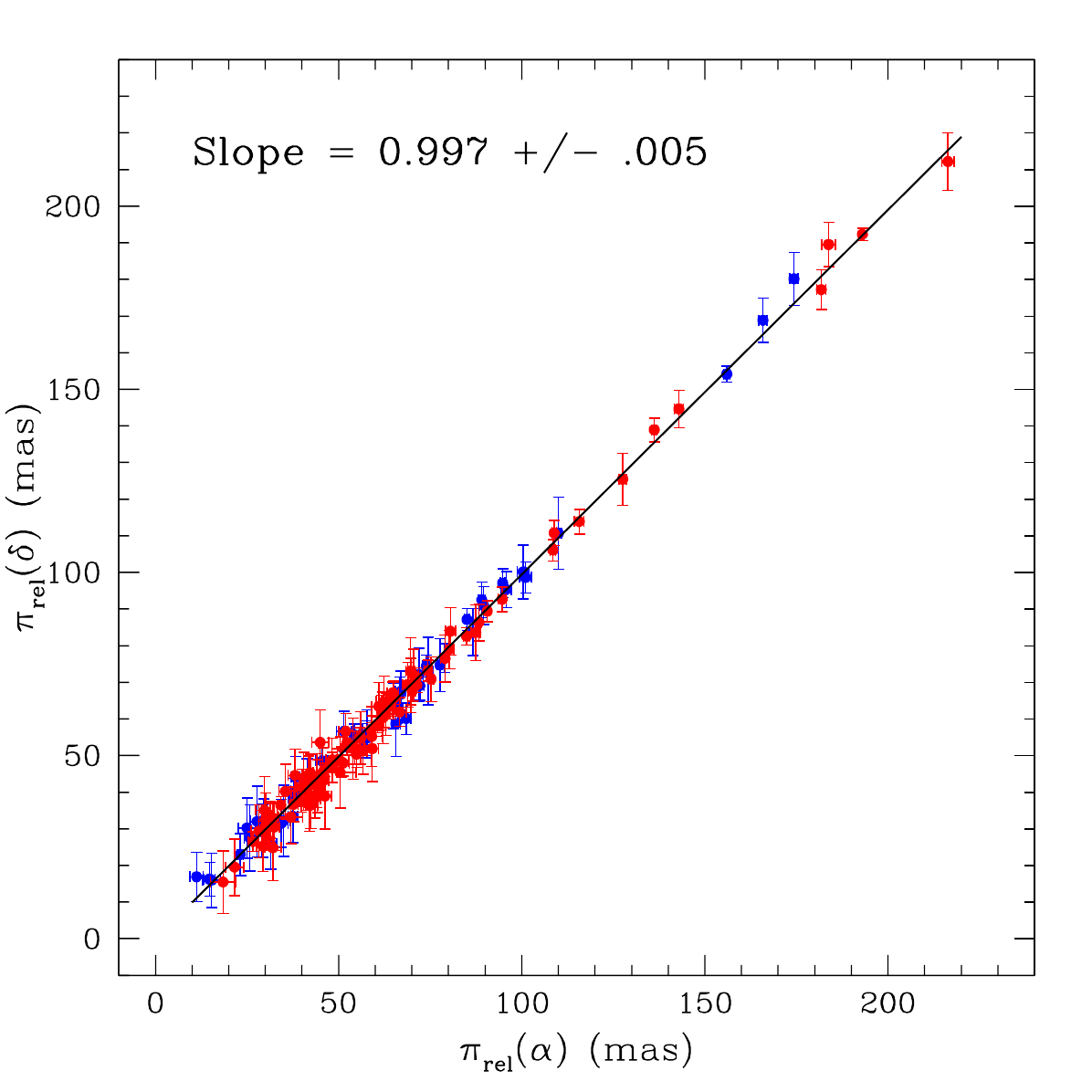}
\caption{Comparison of relative parallaxes determined in the $\alpha$ and $\delta$ directions independently. Blue points display Series 1 data while red points display Series 2 data.}
\label{fig:pixvpiy}
\end{figure}

While every effort has been made to properly calculate parallaxes and proper motions in this study, an internal consistency evaluation is still of value. In our previous study \citep{vrba2004}, with fewer observations and a time baseline of no more than $\Delta$T = 2.0 years, the data were insufficient to base determinations of parallaxes on more than the parallactic factor in Right Ascension and, hence, were deemed ``preliminary". A general rule of thumb for USNO narrow-field astrometry has long been to consider any result as ``preliminary" unless based on a minimum time baseline of $\Delta$T$\ge$3 years (i.e., a minimum of 4 observing seasons) in order to properly separate parallax and proper motions. The current study has sufficient data to allow combined weighted means of parallaxes determined independently in Right Ascension and Declination, as described in \S4.6, except for six objects with indeterminate Declination solutions. For the remaining 167 objects, in Figure \ref{fig:pixvpiy}  we plot the relative parallax solutions in Right Ascension versus those in Declination, with blue and red points representing Series 1 and Series 2 results, respectively. Error bars in both coordinates are plotted, however, only those in Declination are easily seen. A weighted fit to the data is shown with a slope of 0.997$\pm$0.005, indicating no systematic differences in the Right Ascension and Declination solutions and justifying the combined results presented in Column (2) of Table \ref{tab:astrometry}.

\section{Comparison With Gaia Astrometric Results}
\label{sec:gaiacomp}

The European Space Agency's Gaia satellite mission \citep{gaia2016} has provided astrometry at optical wavelengths with unprecedented accuracy. The International Astronomical Union has recognized Gaia products as the optical realization of the International Celestial Reference Frame 3 (ICRF3) via Resolution B3 at its XXXIst General Assembly. The latest data update from Gaia is Data Release 3 (DR3) \citep{gaia2023}, providing both parallax and proper motion values. There are 40 objects serendipitously in common between this study and Gaia DR3. In this section we compare our parallax and proper motion results with those of the Gaia/ICRF standard values.

\subsection{Parallax Comparison}

\begin{figure}
\epsscale{1.15}
\plotone{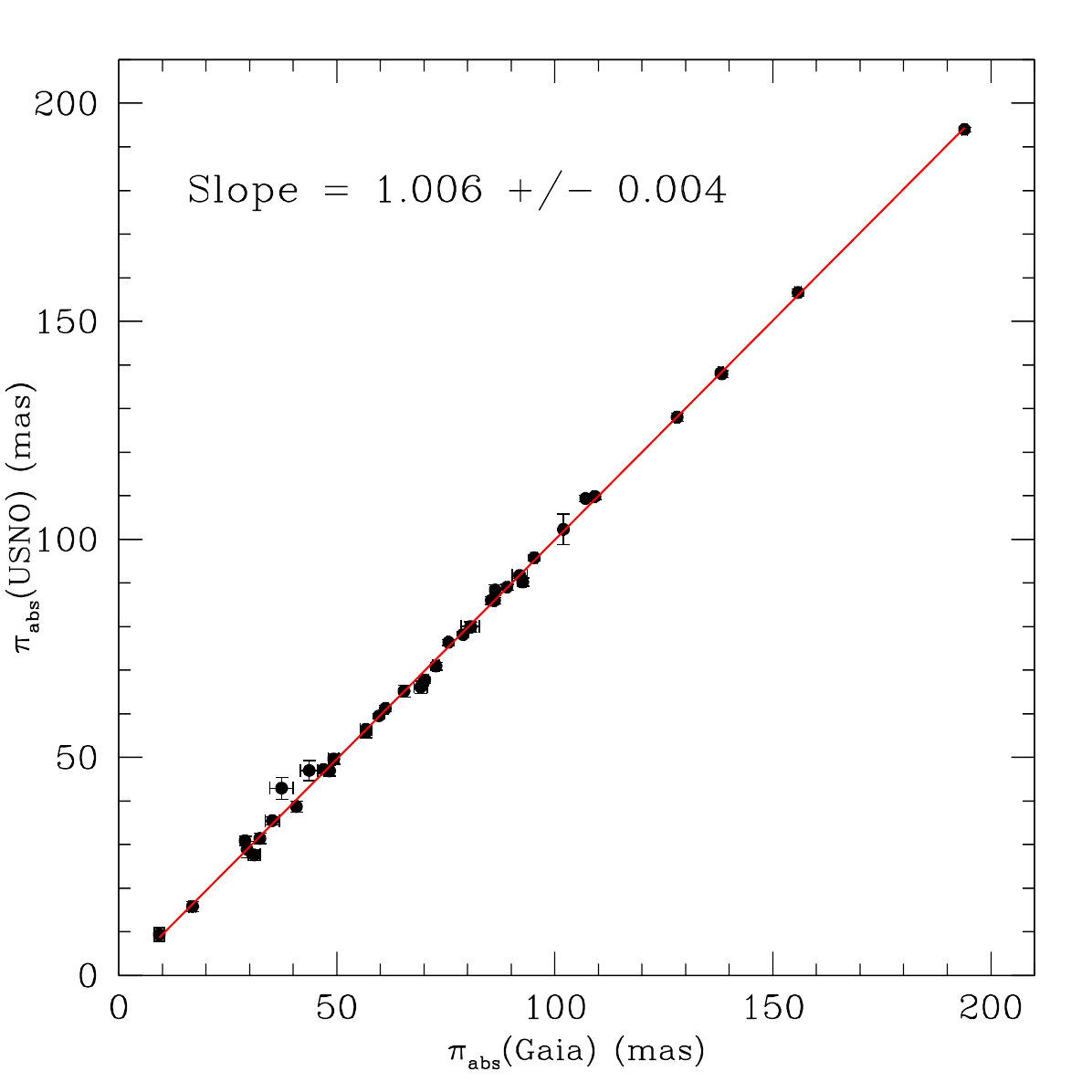}
\caption{A plot of USNO versus Gaia $\pi_{abs}$ values for the 40 objects in common. Error bars for both sets of data are plotted along with the solid red line being the best fit to the data weighted by ($\sigma$($\pi_{abs}$ USNO$))^{-2}$, and a resulting slope of 1.006.}
\label{fig:usnovgaia}
\end{figure}

We compare our $\pi$$_{abs}$ values from Column 3 of Table \ref{tab:astrometry} of this study to the values presented in Gaia DR3. Figure \ref{fig:usnovgaia} plots our values against those of Gaia. The best-fit straight line, weighted by ($\sigma$($\pi_{abs}$ USNO$))^{-2}$, has a slope of 1.006 $\pm$ 0.004. Error bars for both the USNO and Gaia results are shown in this figure, although typically too small to be apparent at this scale.

\begin{figure}
\epsscale{1.2}
\plotone{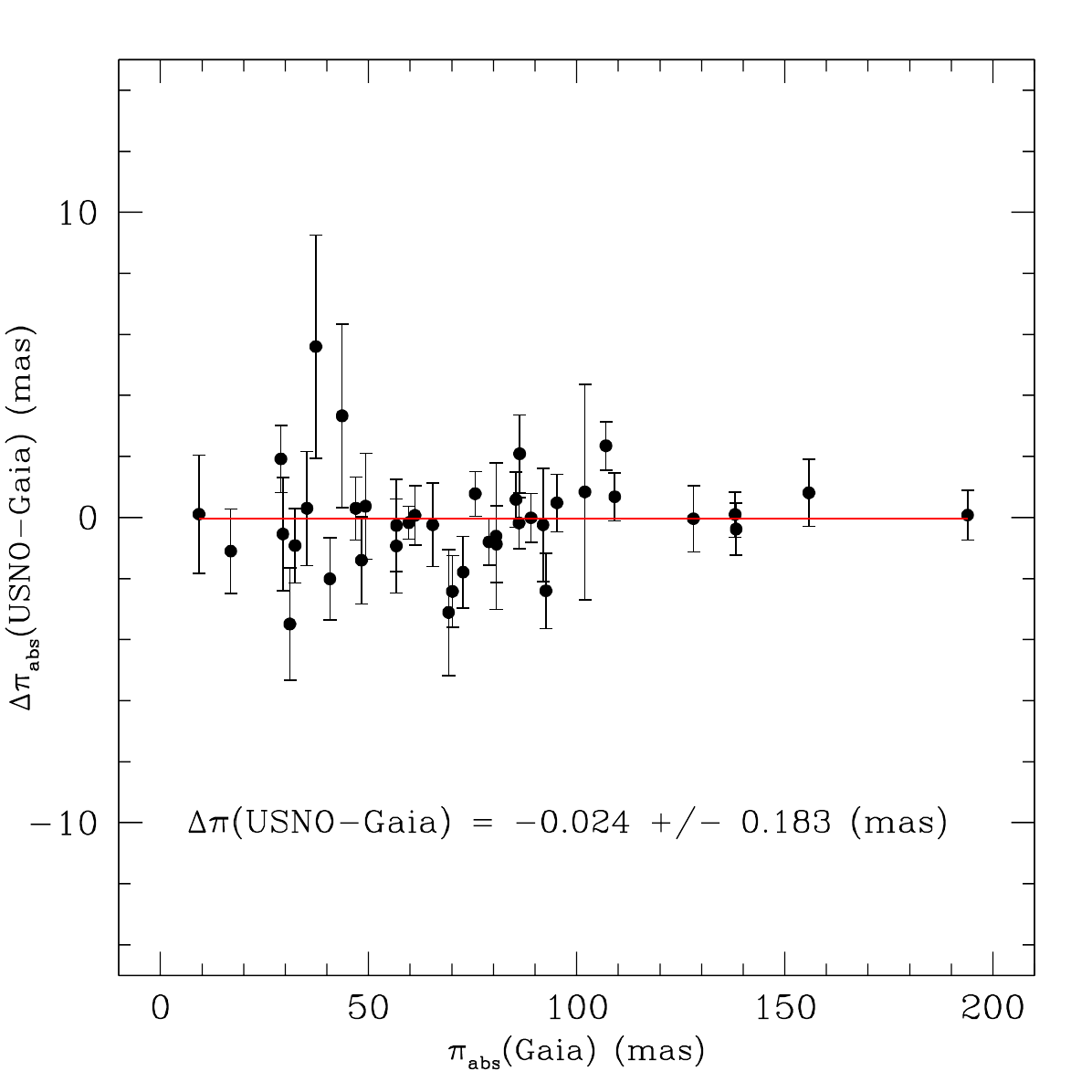}
\caption{The differences in USNO minus Gaia $\pi$$_{abs}$ values plotted versus Gaia $\pi_{abs}$ for the the 40 objects in common. The error bars displayed are the USNO and Gaia errors combined in quadrature. The solid red line shows the net difference between the two sets of results of -0.024$\pm$0.183 mas, weighted by the inverse squared sum of the errors.}
\label{fig:usnovgaia2}
\end{figure}

In Figure \ref{fig:usnovgaia2} we plot the differences in $\pi_{abs}$ in the sense of (USNO-Gaia) versus Gaia $\pi_{abs}$. The error bars shown are the USNO and Gaia errors combined in quadrature. The solid red line shows the net difference between the two sets of results of -0.024 $\pm$ 0.183 mas, weighted by the inverse squared sum of the errors. We note that there is only one object with $\Delta\pi_{abs}$ at the $\approx$ 3$\sigma$ level (J042348.22$-$041402.0 -- see discussion of this object in Section \ref{sec:pmcomp}), which can be expected for 40 sets of measures with a normal error distribution.

Based on the results shown in Figures \ref{fig:usnovgaia} and \ref{fig:usnovgaia2} we find no significant offsets or systematic differences as a function of object distances, including conversions to absolute parallax, between the $\pi_{abs}$ values determined in this work and those from Gaia DR3.  

\subsection{Proper Motion Comparison}
\label{sec:pmcomp}
\begin{figure}
\epsscale{1.2}
\plotone{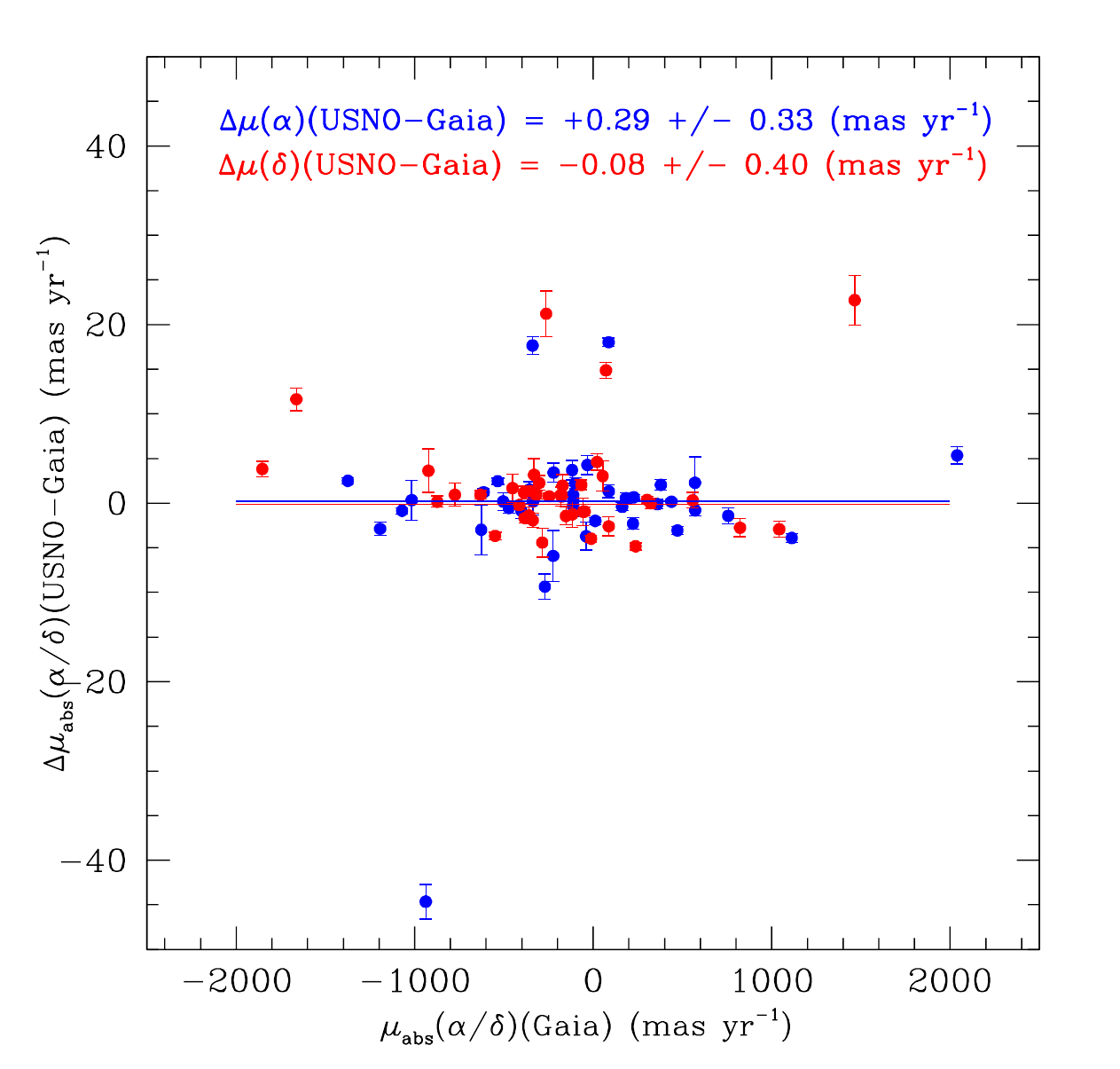}
\caption{The USNO-Gaia differences in absolute proper motions ($\Delta$$\mu$$_{abs}$) in $\alpha$ (blue) and $\delta$ (red) versus the absolute Gaia proper motion in $\alpha$/$\delta$ for the 40 objects in common. The error bars displayed are the USNO and Gaia errors combined in quadrature. The several outlying data points are for five objects with resolved or nearly-resolved binarity likely affecting the Gaia results, as discussed in $\S$\ref{sec:pmcomp}. For the remaining 35 objects the blue and red lines show the mean differences in $\Delta$$\mu$$_{abs}$ in $\alpha$ and $\delta$, respectively, of +0.29 $\pm$ 0.33 mas yr$^{-1}$ and -0.08 $\pm$ 0.40 mas yr$^{-1}$. }
\label{fig:usnomuvgaia}
\end{figure}

The comparison of USNO and Gaia proper motion determinations is a bit more complicated due to Gaia's enhanced resolution of closely separated binaries, which did not appear to be an issue in the comparison of parallaxes in the previous section. In Figure \ref{fig:usnomuvgaia}, we plot the USNO-Gaia differences in absolute proper motions in $\alpha$ (blue) and $\delta$ (red) versus the absolute Gaia proper motion in $\alpha$ and $\delta$. Clearly, there are a number of outlying data points in this figure. In Table \ref{tab:comparison} we list the USNO-Gaia differences (U-G) in $\mu$$_{abs}$ in both $\alpha$ and $\delta$, along with total errors
added in quadrature, in columns (2) and (3), respectively, for the 40 objects in common. We also list the Gaia DR3 values in columns (4) and (5), respectively, for the Renormalized Unit Weight Error (RUWE) and Image Parameter Determination (gofha) (IPD), which we take to be independent indicators of binarity for values significantly above RUWE $>$ 1.4 and IPD $>$ 0.1 \citep{fabricius2021}. Objects for which their RUWE or IPD values exceed these limits are noted with a tablenote in Table \ref{tab:comparison}, along with J151459.41$+$484803.4 as discussed below.  See \cite{castro2024} for a discussion of Gaia-based discovery of otherwise unresolved binaries based on RUWE values. Column (6) provides comments on potentially relevant properties such as binarity, high proper motions, or unusual spectral types. 

\begin{figure}
\label{fig8}
\epsscale{1.2}
\plotone{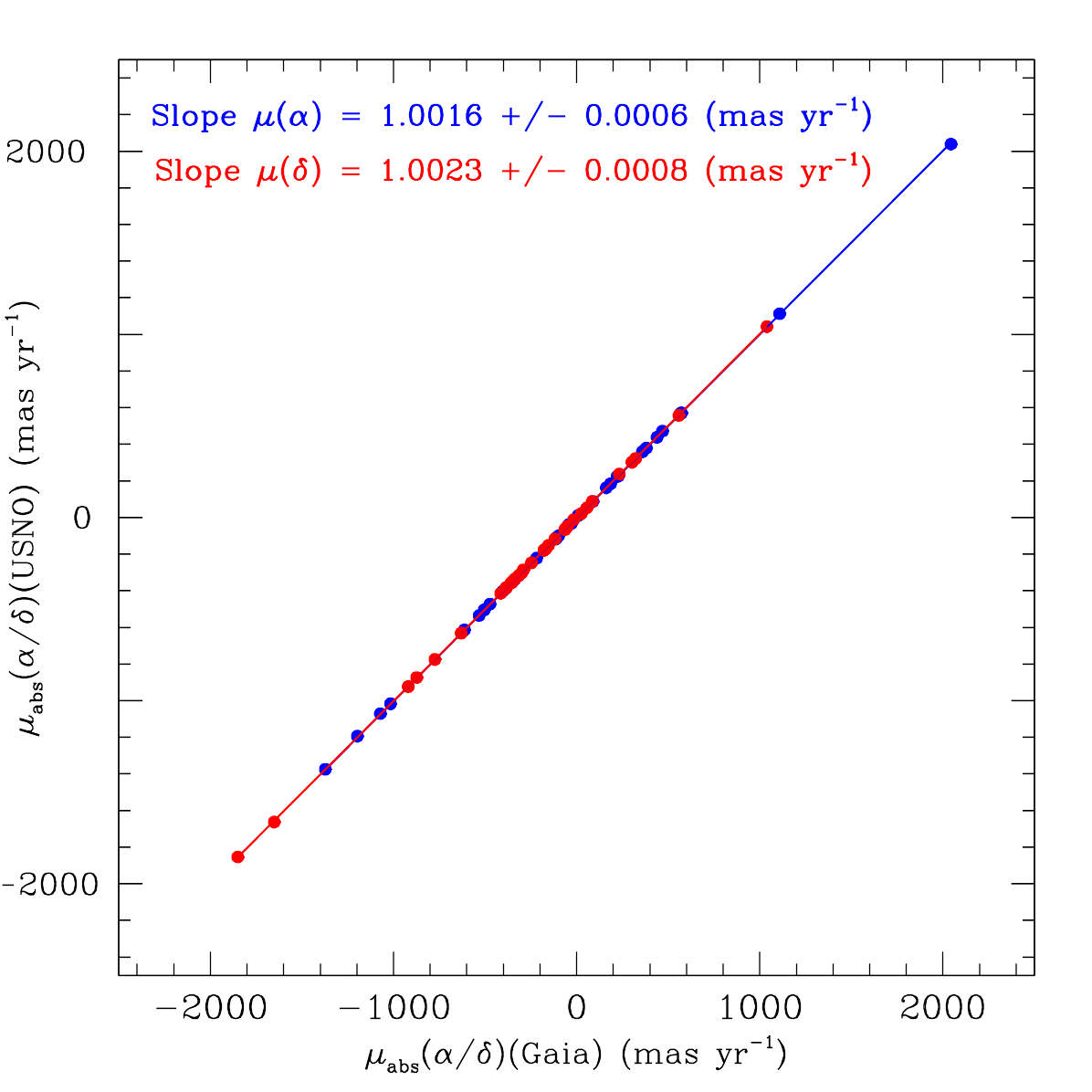}
\caption{We plot the Gaia versus USNO values of $\mu$$_{abs}$ in $\alpha$ (blue) and $\delta$ (red), respectively, for the 35 not affected by likely binarity, as discussed in $\S$\ref{sec:pmcomp}. The blue and red lines represent the best fits to the $\alpha$ and $\delta$ data with slopes of 1.0016 $\pm$ 0.0006 and 1.0023 $\pm$ 0.0008, respectively.}
\label{fig:usnomuvgaia2}
\end{figure}

\startlongtable
\begin{deluxetable*}{lrrccc}
\label{tab:comparison}
\tabletypesize{\scriptsize}
\tablecaption{Comparison of USNO and Gaia Proper Motions and Gaia DR3 Indicators of Binarity}
\tablewidth{0pt}
\tablehead{
\colhead{Object Name} & \colhead{$\Delta\mu(\alpha)({\rm U}-{\rm G})$} & \colhead{$\Delta\mu(\delta)({\rm U}-{\rm G})$} & \colhead{RUWE} & \colhead{IPD}  & \colhead{Comments} \\
\colhead{CatWISE} & \colhead{(mas yr$^{-1}$)} & \colhead{(mas yr$^{-1}$)} \\
\colhead{(1)} & \colhead{(2)} & \colhead{(3)} & \colhead{(4)} & \colhead {(5)} &
\colhead{(6)} 
}
\startdata
J004121.65$+$354712.5 & $+$3.69 $\pm$ 1.09 & $-$0.97 $\pm$ 1.51 & 1.147 & 0.0470  &  Subdwarf \\
J004521.80$+$163444.0 & $-$0.11 $\pm$ 0.50 & $-$0.94 $\pm$ 0.51 & 0.925 & 0.0159  &  \\  
J025116.11$-$035315.7 & $-$3.90 $\pm$ 0.49 & $+$3.82 $\pm$ 0.87 & 1.178 & 0.0152  &  Large $\mu$ (2\farcs1 yr$^{-1}$) \\
J035523.53$+$113336.7 & $-$2.30 $\pm$ 0.64 & $+$0.93 $\pm$ 0.46 & 1.210 & 0.0182  &  \\
J042348.22$-$041402.0\tablenotemark{a}  &$+$17.65 $\pm$ 1.01 & $+$14.87 $\pm$ 0.86 & 1.573 & 0.1940  & 0\farcs16 binary (HST)  \\
J043900.87$-$235310.7 & $-$0.33 $\pm$ 0.74 & $-$1.44 $\pm$ 1.00 & 1.166 & 0.0071  &  \\
J050021.02$+$033044.8 & $-$2.00 $\pm$ 0.42 & $+$1.34 $\pm$ 0.45 & 1.009 & 0.0084  &  \\
J053312.61$+$824617.2 & $+$5.34 $\pm$ 0.98 & $+$11.63 $\pm$ 1.28 & 1.119 & 0.0145 & Subdwarf, Large $\mu$ (2\farcs6 yr$^{-1}$) \\
J053952.16$-$005856.5 & $-$0.41 $\pm$ 0.44 & $-$0.04 $\pm$ 0.59 & 1.176 & 0.0406  &  \\
J055919.85$-$140454.8 & $-$0.80 $\pm$ 0.67 & $-$1.93 $\pm$ 0.78 & 1.303 & 0.0256  &  \\
J060738.42$+$242951.2 & $-$0.54 $\pm$ 0.54 & $+$0.92 $\pm$ 0.47 & 1.128 & 0.0023  &  \\
J070036.80$+$315718.2\tablenotemark{a}  &$+$18.02 $\pm$ 0.46 & $-$3.67 $\pm$ 0.40 & 1.995 & 0.1370  & 0\farcs2 binary (HST) \\ 
J075840.07$+$324718.8\tablenotemark{a}  & $-$5.91 $\pm$ 2.88 & $+$3.17 $\pm$ 1.83 & 1.137 & 0.1210  & Spectral binary \\
J082518.97$+$211546.1 & $+$0.18 $\pm$ 0.97 & $+$2.25 $\pm$ 0.82 & 1.111 & 0.0188  & \\
J083006.91$+$482838.3 & $+$0.33 $\pm$ 2.23 & $+$0.93 $\pm$ 1.04 & 1.090 & 0.0161  & \\
J083541.88$-$081918.0 & $+$2.46 $\pm$ 0.35 & $+$0.37 $\pm$ 0.39 & 1.418 & 0.0237  & \\
J085757.60$+$570844.6 & $-$0.84 $\pm$ 0.84 & $+$1.18 $\pm$ 0.72 & 1.244 & 0.0265  & \\
J091534.04$+$042204.8 & $+$0.91 $\pm$ 1.04 & $+$4.61 $\pm$ 0.94 & 1.048 & 0.0653  & 0\farcs73 binary (HST) \\
J095105.35$+$355800.1 & $+$2.22 $\pm$ 0.63 & $+$1.94 $\pm$ 1.24 & 1.289 & 0.0360  & \\
J110611.58$+$275414.1\tablenotemark{a}  & $-$9.35 $\pm$ 1.40 & $+$1.68 $\pm$ 1.57 & 1.733 & 0.0116  & Over-luminous, spectral binary \\
J111812.22$-$085615.0 & $-$2.99 $\pm$ 2.80 &$+$21.21 $\pm$ 2.58 & 1.104 & 0.1510   & Unusually blue\\
J115821.40$+$043446.3 & $+$2.28 $\pm$ 2.88 & $+$3.63 $\pm$ 2.45 & 1.225 & 0.0173  & Subdwarf \\
J131141.74$+$362925.2 & $+$1.53 $\pm$ 0.87 & $-$2.60 $\pm$ 1.05 & 1.156 & 0.0116  & Spectral binary \\
J140753.30$+$124110.7 & $+$0.19 $\pm$ 1.47 & $+$3.03 $\pm$ 1.70 & 1.225 & 0.0217  & \\
J150319.71$+$252528.3 & $+$1.31 $\pm$ 0.73 & $+$0.31 $\pm$ 0.88 & 1.274 & 0.0184  & \\
J150648.79$+$702741.0 & $-$2.87 $\pm$ 0.73 & $-$2.93 $\pm$ 0.86 & 1.311 & 0.0182  & \\
J150653.09$+$132105.8 & $-$0.85 $\pm$ 0.35 & $-$3.98 $\pm$ 0.39 & 0.979 & 0.0251  & \\
J151459.41$+$484803.4\tablenotemark{a}  &$-$44.64 $\pm$ 1.93 & $+$22.73 $\pm$ 2.78 & 1.203 & 0.0289  & Large $\mu$ (1\farcs8 yr$^{-1}$) \\
J152613.76$+$204334.8 & $+$3.42 $\pm$ 1.08 & $-$1.38 $\pm$ 0.93 & 1.018 & 0.0160  & \\
J162618.23$+$392523.4 & $+$2.50 $\pm$ 0.34 & $-$4.84 $\pm$ 0.40 & 1.088 & 0.0045  & Subdwarf, Large $\mu$ (1\farcs4 yr$^{-1}$) \\
J175609.98$+$281516.4 & $+$1.22 $\pm$ 0.39 & $-$0.22 $\pm$ 0.46 & 1.162 & 0.0160  & Subdwarf \\
J180026.66$+$013450.9 & $+$0.66 $\pm$ 0.58 & $-$1.68 $\pm$ 0.50 & 1.183 & 0.0252  & \\
J182128.39$+$141357.2 & $+$0.65 $\pm$ 0.28 & $+$0.77 $\pm$ 0.30 & 1.156 & 0.0105  & \\
J190648.66$+$401105.9 & $-$0.16 $\pm$ 0.18 & $+$0.20 $\pm$ 0.20 & 1.121 & 0.0073  & \\
J200250.59$-$052154.2 & $+$0.32 $\pm$ 1.87 & $-$1.24 $\pm$ 1.47 & 1.197 & 0.0336  & \\ 
J212702.63$+$761756.8 & $-$1.42 $\pm$ 0.93 & $-$2.75 $\pm$ 1.01 & 1.465 & 0.0468  & Spectral binary \\
J222444.39$-$015908.2 & $-$3.05 $\pm$ 0.44 & $+$0.15 $\pm$ 0.62 & 1.023 & 0.1000  & Astrometric binary candidate\\
J224253.65$+$254256.2 & $+$2.04 $\pm$ 0.58 & $+$2.03 $\pm$ 0.56 & 1.190 & 0.0466  & \\
J225529.03$-$003436.4 & $+$4.27 $\pm$ 1.06 & $+$0.74 $\pm$ 1.00 & 1.039 & 0.1140  & \\
J232545.24$+$425143.9 & $-$3.71 $\pm$ 1.54 & $-$4.42 $\pm$ 1.59 & 1.225 & 0.0110  & \\
\enddata
\tablenotetext{a}{Objects which have indications of potential Gaia-resolved binarity as discussed in $\S$\ref{sec:pmcomp}. See also $\S$\ref{sec:pmcomp} for definitions of (U$-$G), RUWE, and IPD.}
\end{deluxetable*}

From Table \ref{tab:comparison} we note that the objects with increased values of either RUWE or IPD have significantly different $\mu_{abs}$ in either $\alpha$ or $\delta$ between this study and Gaia DR3, and most have some independent indication of binarity. J042348.22$-$041402.0, with a RUWE value of 1.573 and IPD value of 0.1940, is a resolved binary with a separation of 0\farcs16 found using Hubble Space Telescope (HST) imaging \citep{burgasser2005}. J070036.80+315718.2, with RUWE = 1.995 and IPD = 0.137, is another HST resolved 0\farcs2 binary \citep{reid2006b}. J075840.07+324718.8 has a RUWE indicative of a single object (1.137), but an IPD value indicative of binarity (0.121).  This object has been identified as a spectral binary candidate \citep{burgasser2010, bardalez2015}, but is unresolved in high-resolution imaging \citep{radigan2013}. J110611.58$+$275414.1 has a high RUWE value (1.733), has been noted as over-luminous \citep{manjavacas2013}, shows radial velocity variations \citep{hsu2021}, and is considered as strong spectral binary candidate in \cite{burgasser2010}, though unresolved \citep{looper2008, bardalez2015}. We remove these objects from our comparison of USNO and Gaia results along with the blue L dwarf J111812.22$-$085615.0 due to its high value of IPD (0.151). We also remove the high proper motion object J151459.41$+$484803.4 which, while having large proper motion discrepancies between USNO and Gaia in both $\alpha$ and $\delta$, has no previous indications of binarity but is noted as a blue L dwarf contaminant in \cite{bardalez2014} with stable radial velocity measurements \citep{wilson2003}. We note, however, that \cite{dahn2017} also find proper motion values for this object that are significantly discrepant with both this study and Gaia, while all three studies show consistent parallax results. Thus, a high-accuracy proper motion for J151459.41$+$484803.4 is indeterminate at this time. We leave in the mix J091534.04$+$042204.8, an HST resolved 0\farcs73 binary \citep{reid2006b} we take to have a wide enough separation so as to not affect Gaia proper motion astrometry. 

For the five objects discussed above with significantly discrepant USNO and Gaia values of proper motion we cannot say with certainty which set of values are more reliable. However, we do note that the projected spatial separations are all well within the typical USNO PSFs, there were no issues found with the USNO data reductions for these objects, and that the data time baselines were considerably longer for the USNO data than for those of Gaia. The latter point is also relevant for considering the somewhat enhanced differences in USNO and Gaia proper motion values for high proper motion objects in Table \ref{tab:comparison}.

For the remaining 35 objects in common with Gaia we return to Figure \ref{fig:usnomuvgaia} where the blue and red lines show the mean differences in $\Delta$$\mu$$_{abs}$ in $\alpha$ and $\delta$, respectively, of +0.29 $\pm$ 0.33 mas yr$^{-1}$ and -0.08 $\pm$ 0.40 mas yr$^{-1}$. In Figure \ref{fig:usnomuvgaia2} we plot the Gaia versus USNO values of $\mu$$_{abs}$ in $\alpha$ (blue) and $\delta$ (red), respectively. The blue and red lines represent the best fits to the $\alpha$ and $\delta$ data with slopes of 1.0016 $\pm$ 0.0006 and 1.0023 $\pm$ 0.0008, respectively. We conclude that the proper motion values determined in this study do not differ systematically from those of Gaia DR3 for those objects not affected by resolved or nearly-resolved binarity.

\section{New UKIRT and Adopted MKO $JHK$ Photometry}
\label{sec:JHKphot}

For the following science sections we found it optimal to obtain or utilize $J$, $H$, and $K_{\rm s}$ photometry based on as uniform of a system as possible. We chose the UKIRT (MKO) photometry system \citep{simons2002, hodgkin2009} due to its widespread use and USNO involvement in the UKIRT Hemisphere Survey project (UHS) \citep{dye2018, schneider2025}. While we obtained considerable new photometry for Series 2 objects using WFCAM \citep{casali2007} at UKIRT, we also employ previously published data from other sources, most of it obtained with WFCAM. Table \ref{tab:photometry} presents the collected $J$, $H$, and $K_{\rm s}$ photometry used in this study. Column (1) gives the CatWISE name while columns (2),(3), and (4) give the $J$, $H$, and $K_{\rm s}$ photometry and associated uncertainties based on photon statistics, respectively. Column (5) gives the source of the photometry, which needs some clarification. In order of $J$/$H$/$K_{\rm s}$, numbers show the number of independent new observations obtained in this study to give the photometry listed for the object and represent a large majority of the entries. These targeted UKIRT/WFCAM observations employed exposure times customized for the object and were run through the ususal WFCAM photometric pipeline \citep{dye2018}.
Otherwise, letters represent the source of the photometry. Specifically, U represents data from the on-going UHS survey data  \citep{dye2018, schneider2025}. Most of the other data were also taken with the WFCAM instrument on UKIRT,
 the primary exceptions being data from the 2MASS \citep{skrutskie2006} and VISTA \citep{mcmahon2013, sutherland2015} surveys, which use very similar filter systems. A full explanation of column (5) symbols is given in the footnotes to Table \ref{tab:photometry}. Thus, for the purposes of this work, all photometry in Table \ref{tab:photometry} is on the same photometric system.  

For completeness sake, we also present in Appendix A additional $J$, $H$, and $K_{\rm s}$ photometry obtained using the USNO 1.55-m telescope and ASTROCAM in the CIT photometric system, based on the standards of \cite{guetter2003}, primarily for Series 1 objects, however, this photometry is not used in this study.

\startlongtable
\begin{deluxetable*}{lcccc}
\tabletypesize{\scriptsize}
\tablecaption{New UKIRT and Adopted $JHK$$_{\rm S}$ Photometry.}
\label{tab:photometry}
\tablewidth{0pt}
\tablehead{
\colhead{Name} & \colhead{$J$} & \colhead{$H$} & \colhead{$K_{\rm S}$} &
\colhead{Source}  \\
\colhead{(CatWISE)} & \colhead{(mag)} & \colhead{(mag)} & \colhead{(mag)} & \colhead{($J$/$H$/$K_{\rm s}$: No. Obs./other)} \\
\colhead{(1)} & \colhead{(2)} & \colhead{(3)} & \colhead{(4)} & \colhead{(5)}
}
\startdata
J000013.51+255419.7   & 14.864 $\pm$ 0.003 & 14.817 $\pm$ 0.005 & 14.970 $\pm$ 0.007 & 2/2/2 \\
J000849.71$-$173925.0 & 16.872 $\pm$ 0.010 & 17.168 $\pm$ 0.013 & 17.304 $\pm$ 0.020 & 1/1/1 \\
J001502.52+295928.7   & 16.165 $\pm$ 0.009 & 15.209 $\pm$ 0.006 & 14.535 $\pm$ 0.006 & 1/1/1 \\
J003030.38$-$145033.8 & 16.471 $\pm$ 0.010 & 15.362 $\pm$ 0.006 & 14.490 $\pm$ 0.009 & 6/6/5 \\
J003259.67+141037.2   & 16.645 $\pm$ 0.016 & 15.689 $\pm$ 0.011 & 15.002 $\pm$ 0.013 & L/L/L \\
                      &                    &                    &                    &       \\
J003452.12+052307.3   & 15.202 $\pm$ 0.005 & 15.575 $\pm$ 0.008 & 16.093 $\pm$ 0.023 & 1/1/1 \\
J004121.65+354712.5   & 16.072 $\pm$ 0.008 & 15.509 $\pm$ 0.008 & 15.126 $\pm$ 0.011 & 1/1/1 \\
\enddata
\tablecomments{\\$\#$/$\#$/$\#$: This paper. Number of observations. See text.\\
L: UKIDSS Large Area Survey\\
G: UKIDSS Galactic Clusters Survey\\
M: 2 Micron All Sky Survey (2MASS)\\
V: VISTA Hemisphere Survey\\
U: UKIRT Hemisphere Survey\\
S: UltracoolSheet -- ``Unpublished work/analysis done specifically for the UltracoolSheet version 2.0 release in 2024''\\
K: \cite{kirkpatrick2011}\\
B: \cite{best2021}\\
T: \cite{leggett2002}\\
C: \cite{chiu2006}\\
P: \cite{knapp2004}\\
A: \cite{liu2016}\\
D: \cite{leggett2010}\\
R: \cite{strauss1999}}
\end{deluxetable*}

\section{Colors and Absolute Magnitudes}
\label{sec:colmag}

\begin{figure*}
\plotone{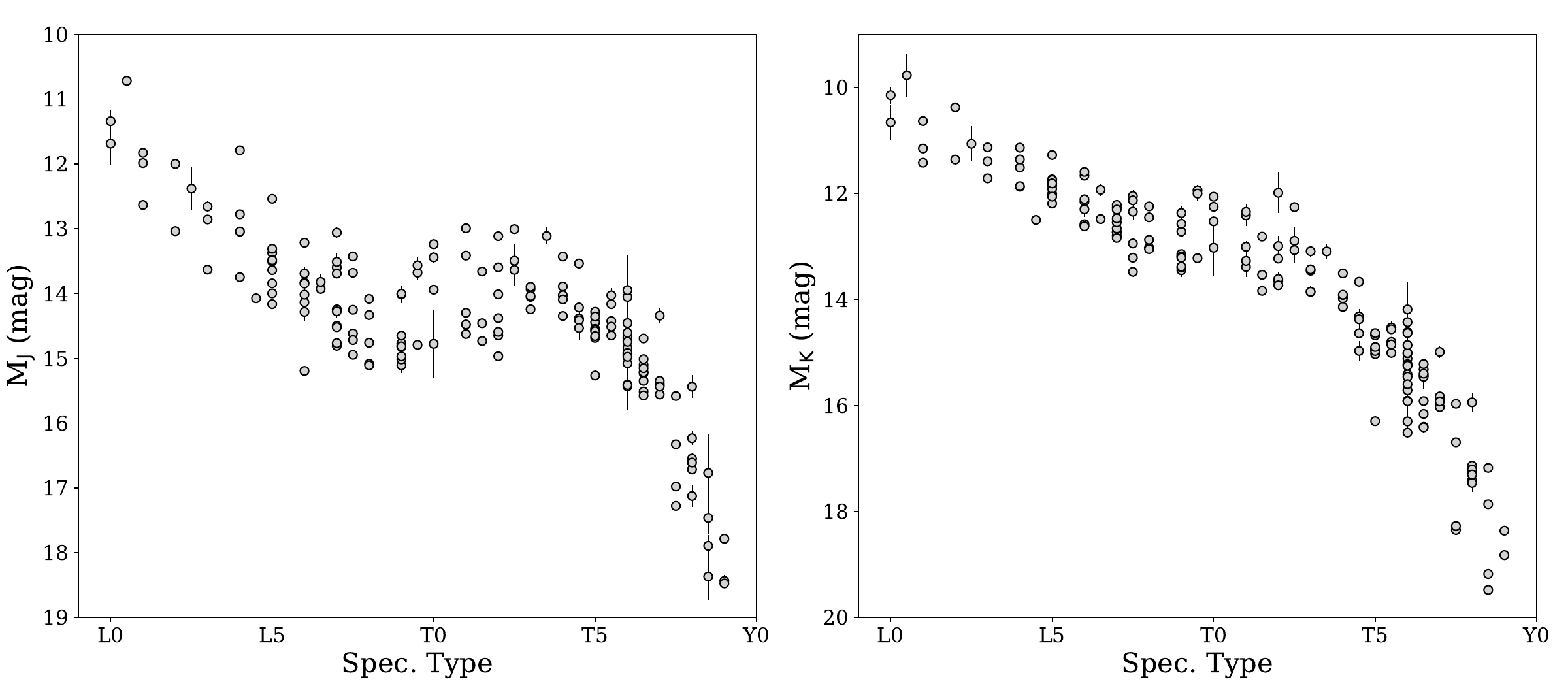}
\caption{Absolute $J$-band (left) and absolute $K$-band (right) magnitudes as a function of spectral type for the sample presented in this work.  Near-infrared spectral types from Table \ref{tab:genprop} are used for both panels except in cases where no near-infrared spectral type exists, then optical types are used.}
\label{fig:abspt1}
\end{figure*}

\begin{figure}
\plotone{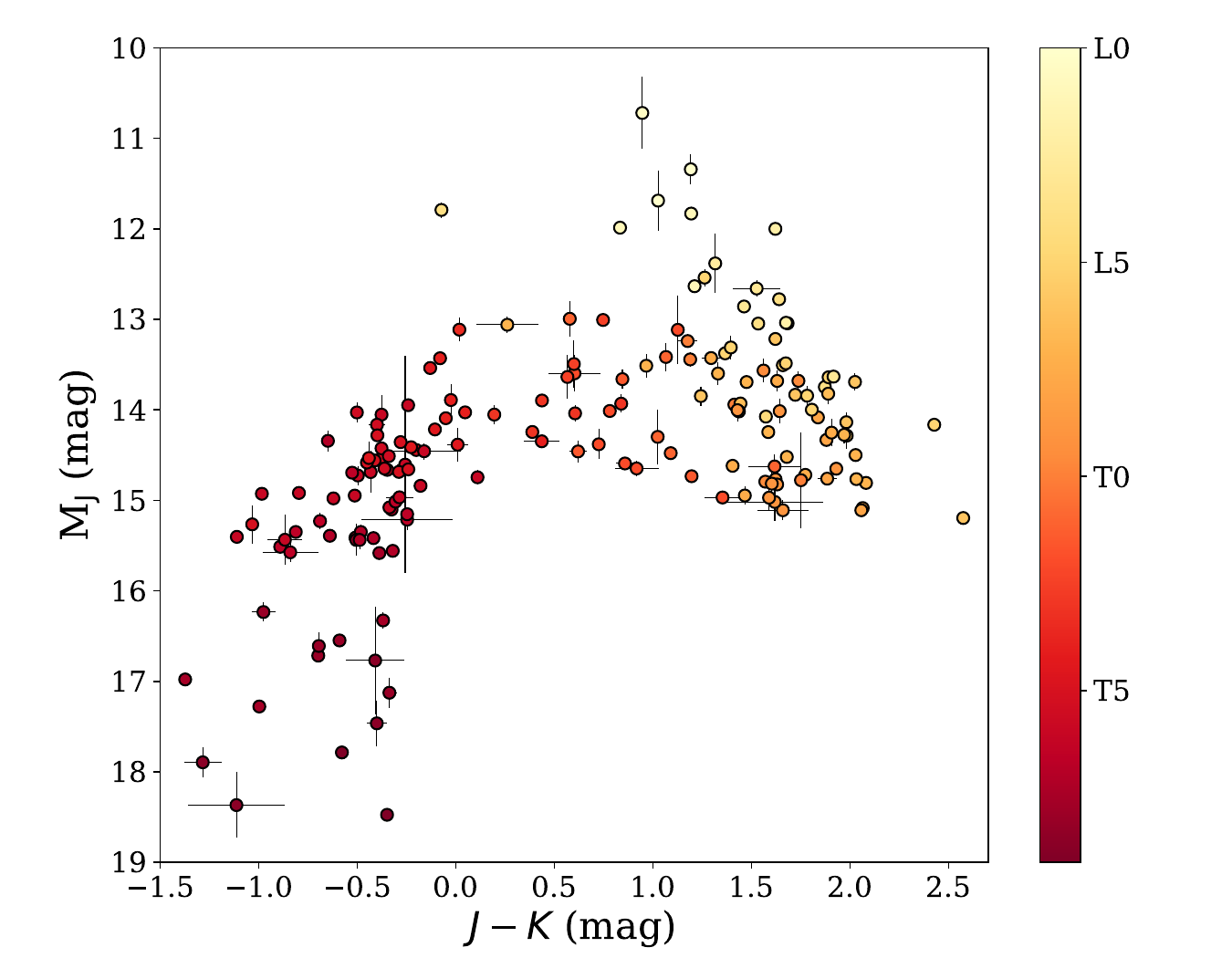}
\caption{Absolute $J$-band magnitude as a function of $J-K$ color for the sample presented in this work.  Near-infrared spectral types from Table \ref{tab:genprop} are used for both panels except in cases where no near-infrared spectral type exists, then optical types are used.  Symbols are colored by spectral type.}
\label{fig:cmd1}
\end{figure}

In Figure \ref{fig:abspt1} we show the absolute magnitudes in $J$- and $K$- bands as a function of spectral type for all
objects in this study. The absolute magnitudes are based on the astrometry from Table \ref{tab:astrometry} and infrared photometry given in Table \ref{tab:photometry}. Near-infrared spectral types from Table \ref{tab:genprop} are used for both panels except in cases where no near-infrared spectral type exists, then optical types are used. In Figure \ref{fig:cmd1} we show the absolute $J$-band magnitudes of this sample as a function of their $J-K$ colors with spectral types encoded by color, again using the data in Tables \ref{tab:genprop}, \ref{tab:astrometry}, \ref{tab:photometry} as described above. For both Figures all photometric error bars are shown, while spectral types are assumed to be known to one spectral subclass or better and are not specifically shown.

Rather than presenting numerous color-absolute magnitude and color-color diagrams, we have opted to use the three basic diagrams shown in Figures \ref{fig:abspt1} and \ref{fig:cmd1} which are sufficient to demonstrate the points we make about various brown dwarf populations in the next Section. These diagrams will be re-visited, in turn, as we discuss each of the populations of interest. 

\section{Notes on Various Populations}
\label{sec:PopNotes}

In this section we discuss objects in this study which are or may be members of populations of special interest. These include binaries, wide companions, young objects, subdwarfs, and spectral standards. While we give some background for each object considered and point out
where our new results may help in resolving some issues, these are brief comments and we acknowledge that we have not included all useful references to previous relevant work.

\subsection{Binaries}

We first review objects in this study that are binaries or suspected binaries. This includes resolved
binaries, spectral binaries, and objects which might be binary due to their positions in spectral type vs. absolute magnitude diagrams.

\subsubsection{Resolved Binaries}
\label{sec:resbin}

We note that for the objects in this subsection, although resolved with larger aperture ground-based telescopes or HST, separations are typically a small fraction of an arcsecond, well within the $\sim$1\farcs1 FWHM of a typical frame taken with ASTROCAM on the 1.55-m telescope at NOFS. Thus, while these binaries likely had some affect on the final astrometric quality, we find no systematic effects. 

{\it J031100.13$+$164815.6:} Early spectral typing of this object classified it as an L8 at both optical \citep{kirkpatrick2000} and near-infrared \citep{reid2001} wavelengths. It was later typed as L9 using optical and near-infrared spectra by \cite{geballe2002}, while \cite{schneider2014} adopted a near-infrared spectral type of L9.5, and \cite{manjavacas2019} determined a near-infrared type of T2. It was resolved as an approximately equal brightness binary with a separation of 204.3$\pm$0.4 mas in \cite{stumpf2010}, who also estimated a distance of 25$\pm$4 pc for the pair.  
A $\pi_{\rm abs}$ of 36.9$\pm$3.4 mas was presented in \cite{smart2013} which is consistent with our result of $\pi_{\rm abs}$ of 35.89$\pm$1.73 mas.    

{\it J042348.22$-$041402.0:} \cite{cruz2003} noted a discrepancy between optical and near-infrared spectral types for this object (also discussed in \citealt{burgasser2003b}).  \cite{vrba2004} and \cite{knapp2004} also noted the discrepant optical/infrared types and suggest binarity or youth could be an explanation.  \cite{burgasser2005} is devoted to this object, where {\it HST} data were used to resolve this as an L6+T2 binary with a separation of 0\farcs16. \cite{burgasser2010} used spectral decomposition to find types L7.5$\pm$0.3 and T2.0$\pm$0.2.  \cite{dupuy2012} derived spectral component types of L6.5$\pm$1 and T2$\pm$0.5. \cite{dupuy2017} monitored this binary with {\it HST} and found a period of 12.30$\pm$0.06 years and a total mass of 83$\pm$3 $M_{\rm Jup}$. 

{\it J070036.80$+$315718.2:} This object was found to be a binary by \cite{reid2006b} using {\it HST} images, who estimate a spectral type of L6 for the secondary and a separation of $\sim$0\farcs2. \cite{dupuy2012} used spectral decomposition to estimate types of L3$\pm$1 and L6.5$\pm$1.5.  \cite{dupuy2017} measured a full orbit for the pair (23.9$\pm$0.5 years) and suggested that the secondary (2M0700+3157B) is itself an unresolved binary, because they find that it is more massive than the L3 primary.

{\it J085035.75$+$105715.3:} \cite{reid2001} resolved this object as a 0\farcs16 binary in {\it HST} imaging and suggested the fainter companion is a very late L or early T dwarf.  \cite{konopacky2010} found an orbital period of 24$^{+69}_{-6}$ years.  \cite{faherty2011} found this to be a hierarchical system containing a young ($\lesssim$1.5 Gyr) M5$+$M6 primary (NLTT 20346).  \cite{burgasser2011} determined L6+L7 component types but suggested that this may be a triple system.  \cite{dupuy2012} determined component types of L6.5$\pm$1 and L8.5$\pm$1.  However, \cite{dupuy2012} suggested that companionship with the M+M pair NLTT 20346 is a chance alignment, based on their updated astrometry, and find that there is no need to invoke additional multiplicity to either of the L components. Based on Gaia results for NLTT 20346AB and the astrometry from this work (which is consistent with \cite{dupuy2012, dupuy2017}), this is likely a chance alignment.  \cite{dupuy2017} give an orbital period of 48$^{+7}_{-6}$ years (though state that the solution is poorly constrained), while also measuring a low total mass for the system, suggesting that it may be young.

{\it J091534.04$+$042204.8:} This object was discovered by \cite{reid2006b} as 2MASS J09153413$+$0422045, who resolved this as a 0\farcs73 separation candidate binary from {\it HST} images. With flux ratios close to equal they gave optical spectral types of L7 to both components.  \cite{reid2006b} also state that this object was independently identified as a binary by M.~Liu (private communication).  Several authors have provided updated spectral typing with similar results: \cite{reid2008b} gave component types of L6: and L6:, \cite{liu2010} gave component types of L6$\pm$1 and L7$^{+3}_{-1.5}$, and \cite{bardalez2019} gave component types of L6 and L6.

{\it J092615.40$+$584717.6:} \cite{burgasser2006c} found that this source was slightly elongated in {\it HST} images ($a$ = 0\farcs07) and suggested component spectral types of T4+T4 based on their approximately equal brightness. Several authors
have provided updated spectral typing giving similar results: \cite{liu2010} gave component types of T4.5$\pm$0.5 and T3$^{+2.5}_{-4.0}$, \cite{geissler2011} estimate component types of T3--T4+T6 based on spectral template fitting, \cite{dupuy2012} find component types of T3.5$\pm$1 and T5$\pm$1, and \cite{bardalez2015} found component types of T4$\pm$0.1 and T5.3$\pm$0.7.  A poorly constrained orbit was given in \cite{dupuy2017}, who found a period of 12.9$^{+1.3}_{-1.9}$ years.

{\it J102109.51$-$030421.1:} This object was identified by \cite{burgasser2006c} as a binary from {\it HST} images with component types of T1 and T5 based on spectral decomposition and a separation of 0\farcs17 (one of the first flux-reversal binaries, maybe the first).  \cite{burgasser2010} found component types of T1.0$\pm$0.4 and T5.5$\pm$0.7, while \cite{dupuy2012} found component types of T0$\pm$1 and T5$\pm$0.5 and suggested, based on color-magnitude diagram positions, that the A component of this system may in fact be an unresolved binary.  \cite{dupuy2017} provided a poorly constrained orbit with a period of 86$^{+13}_{-17}$ years.

{\it J121757.13$+$162635.1:}  \cite{liu2012} resolved this object as an $\sim$0\farcs76 binary and estimated component spectral types of T9$\pm$0.5 and Y0$\pm$0.5.  \cite{leggett2014} presented resolved spectroscopy of each component and found spectral types of T8.5 and Y0--Y0.5.  \cite{zhang2019} suggested an sdT9 spectral type for the primary.

{\it J122554.84$-$273958.5:}  This object was resolved by \cite{burgasser2003d} as a 0\farcs28 binary with spectral component types of T6 and T8.  \cite{dupuy2012} determined similar spectral component types of T5.5$\pm$0.5 and T8$\pm$0.5.

{\it J151114.38$+$060739.4:} \cite{burgasser2010} found this object to be a strong spectral binary candidate with component types of L5.5$\pm$0.8 and T5$\pm$0.4.  It was resolved in Keck/NIRC2 images by \cite{bardalez2015} with a separation of 108$\pm$11 mas, who determined similar spectral component types of L5$\pm$1 and T5$\pm$0.5.

{\it J155301.80$+$153239.5:} The first published mention of this object being a binary came from \cite{knapp2004} who listed it as 2MASS J1553+1532AB, citing A.~Burgasser (private communication). \cite{burgasser2006} suggested this object as an alternate T7 standard, although mentioning it as an approximately equal magnitude binary citing Burgasser in prep. \cite{burgasser2006c} resolved this object as a 0\farcs35 binary in both {\it HST} and Keck/NIRC images, estimating component spectral types of T6.5 and T7.  \cite{dupuy2012} found similar component types of T6.5$\pm$0.5 and T7.5$\pm$0.5. See also $\S$\ref{sec:young} for evidence that this
object may be young, as a member of the Carina-Near Moving Group.

{\it J172811.54$+$394859.0:} This object was resolved as a binary in {\it HST} images by \cite{gizis2003} and  \cite{bouy2003} with a separation of $\sim$130 mas.  Orbital parameters were determined by \cite{konopacky2010}, who found an orbital period of 31.3$\pm$12.7 yrs.  \cite{burgasser2011} presented a detailed investigation of {\it HST} data and found component spectral types of L5 and L6.5, with the primary component having thick condensate clouds to account for its colors and low luminosity on CMDs.  \cite{dupuy2012} found component types of L5$\pm$1 and L7$\pm$1 and also noted that the primary is unusually red.  \cite{dupuy2017} calculated orbital parameters including a period of 40.8$^{+1.6}_{-1.2}$ years.

{\it J210115.58$+$175656.0:} This object was found to be a binary in {\it HST} images by \cite{bouy2003} and \cite{gizis2003} with a separation of $\sim$235 mas, with \cite{gizis2003} estimating the secondary having a spectral type of L8 or later.  \cite{liu2010} estimated component spectral types of L7.5$\pm$0.5 and T1$^{+2}_{-2.5}$, while \cite{dupuy2012} found component spectral types of L7$\pm$1 and L8$\pm$1.

\begin{figure*}
\plotone{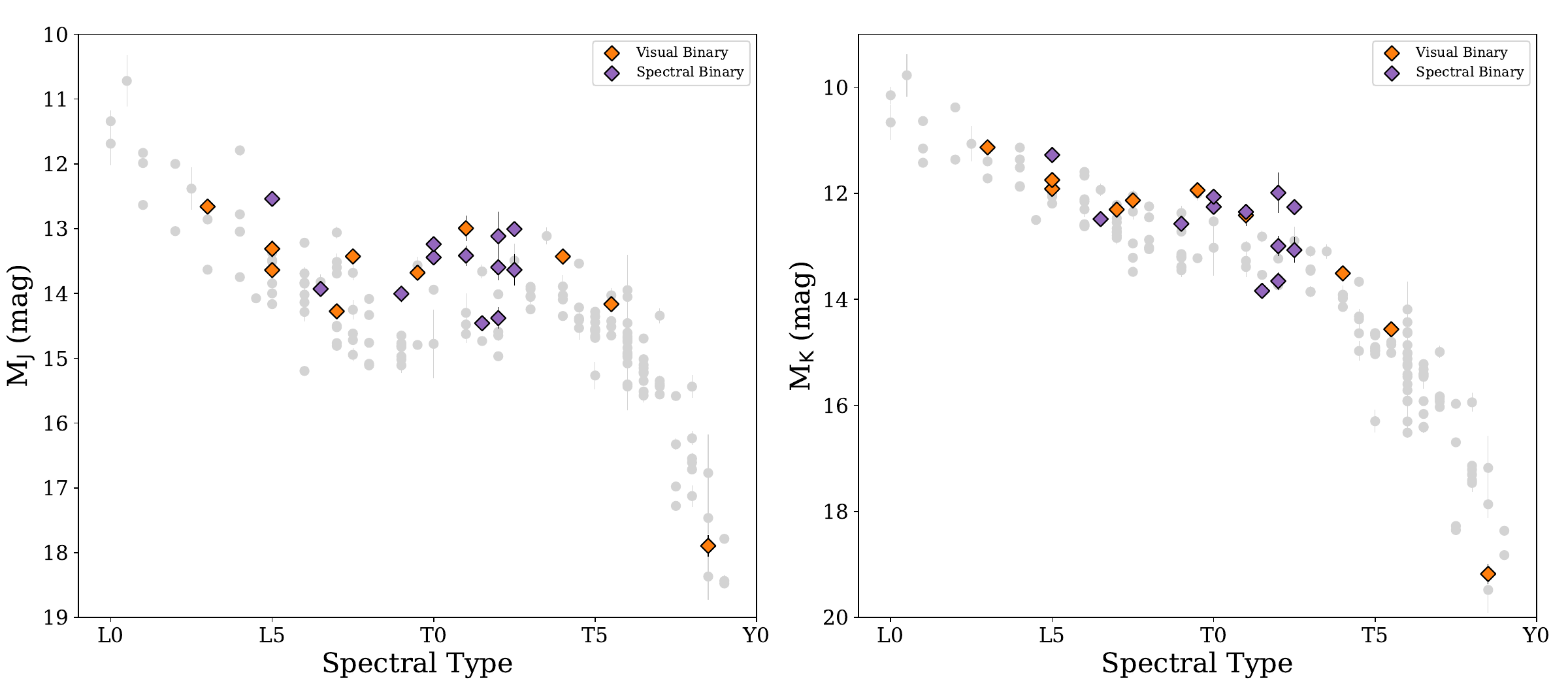}
\caption{Same as \ref{fig:abspt1} but with the positions of visual and spectral binaries highlighted.}
\label{fig:abspt2}
\end{figure*}

\begin{figure}
\plotone{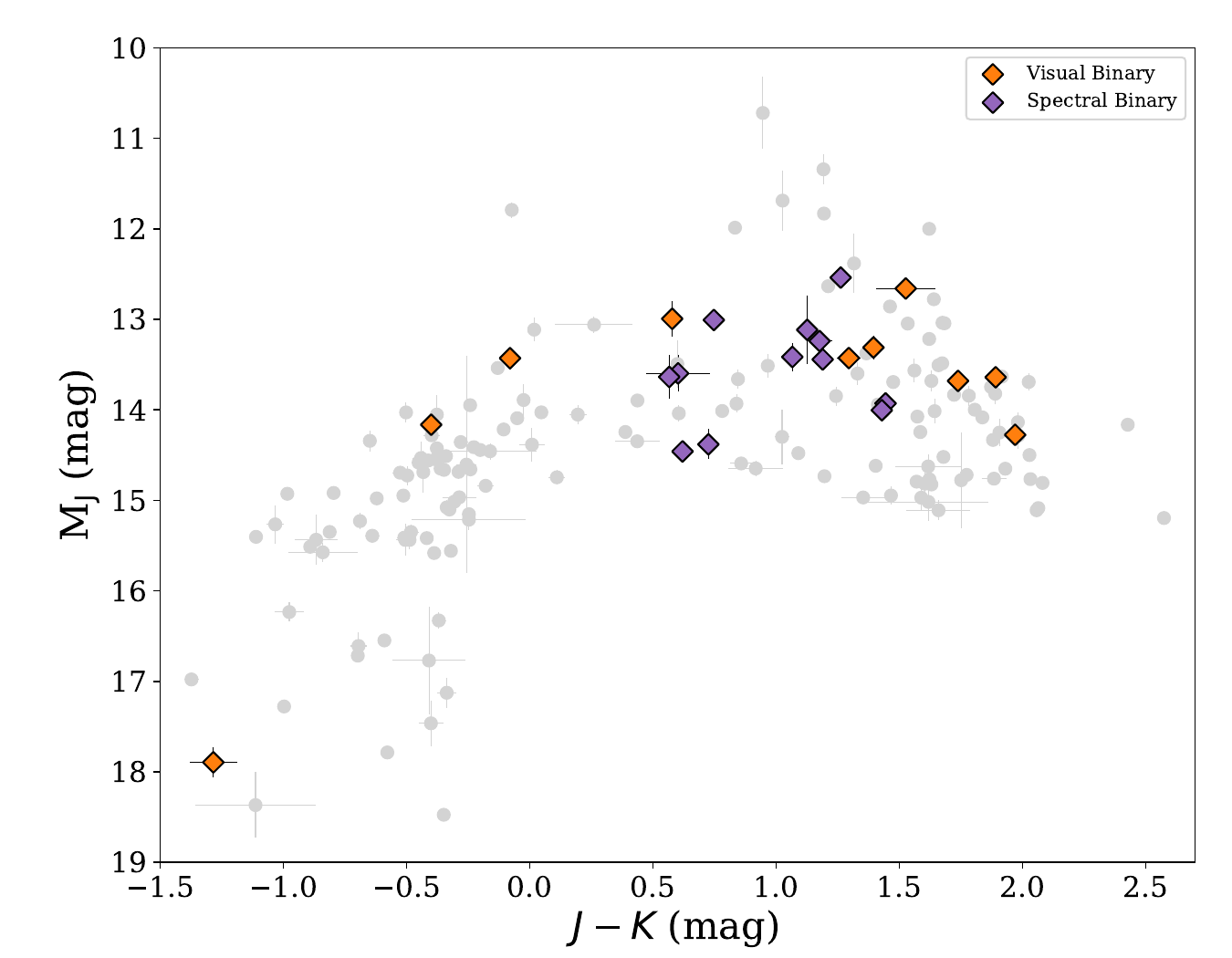}
\caption{Same as \ref{fig:cmd1} but with the positions of visual and spectral binaries highlighted.}
\label{fig:cmd2}
\end{figure}

The absolute magnitudes of these resolved binaries compared to the rest of the sample in this work are shown on Figures \ref{fig:abspt2} and \ref{fig:cmd2}.

\subsubsection{Spectral Binaries}
\label{sec:specbin}

In this section we give brief summaries of previous work on spectrally-detected binaries, along with our own estimations
of binary likelihood based on template-fitting.

{\it J003259.67$+$141037.2:} This object was classified as L9 by \cite{schneider2014} based on near-infrared spectra. It was suggested as a visual spectroscopic binary by \cite{bardalez2015} with component types L6.2$\pm$0.7 and T2.4$\pm$1.9.  

{\it J011912.41$+$240331.7:} \cite{chiu2006} classified this object as a T2 based on near-infrared spectra. It was suggested as an unresolved binary with spectral types T0$\pm$0.7 and T4$\pm$0.4 by \cite{burgasser2010}.  We note that this object, with an absolute $J$-band magnitude of $\sim$14.4 and an absolute $K$-band magnitude of $\sim$13.7, does not appear overluminous in either of the panels of Figure \ref{fig:abspt2}. 

{\it J023618.06$+$004852.1:} This object was classified as an L9 from visual spectra by \cite{scholz2009} and as an L6.5
from near-infrared spectra by \cite{geballe2002}. It was selected as a strong spectral binary candidate by \cite{bardalez2014}, with individual components of L5.0$\pm$0.6 and T1.9$\pm$1.1.  

{\it J075840.07$+$324718.8:} \cite{pineda2016} classified this object as a T3 based on visual spectra while \cite{knapp2004} classified it as a T2 based on near-infrared spectra. It was identified as a weak near-infrared spectroscopic binary candidate in \cite{burgasser2010} with individual components of L7.5 and T4. \cite{bardalez2015} selected it as a visual spectral binary with components of T2.3 and T2.2.  

{\it J090900.31$+$652525.8:} This object was classified as a T1.5 based on near-infrared spectra by \cite{chiu2006}. It was labeled as a weak spectroscopic binary candidate based on near-infrared spectra by \cite{burgasser2010}, with component types of T1.5 and T2.5. We note that this object, with an absolute $J$-band magnitude of $\sim$14.5 and an absolute $K$-band magnitude of $\sim$13.8, does not appear overluminous in either of the panels of Figure \ref{fig:abspt2}. 

{\it J094908.49$-$154548.3:} This object was discovered as 2MASS J09490860$-$1545485 in \cite{tinney2005}, who found a near-infrared type of T1.  It was found to be a weak spectroscopic binary candidate in \cite{burgasser2010}, who found component types of T2 and T1.5$\pm$1.

{\it J103931.32$+$325623.7:} This object was classified as a T1 based on near-infrared spectra by \cite{chiu2006}. It was identified as a strong binary candidate based on near-infrared spectra by \cite{burgasser2010}, with component types of L7$\pm$0.2 and T4$\pm$0.2.  

{\it J110611.58$+$275414.1:} \cite{looper2007} discovered this object and classified it as a T2.5 based on near-infrared spectra. It is a strong binary candidate based on near-infrared spectra by \cite{burgasser2010}, who found component types of T0$\pm$0.2 and T4.5$\pm$0.2.  

{\it J120746.66$+$024426.8:} Discovered as SDSS J120747.17$+$024424.8 in \cite{hawley2002} with an optical spectral type of L8.  Suggested as a weak binary candidate in \cite{burgasser2010} with component types of L6.5$\pm$0.7 and T2.5$\pm$0.5.

{\it J131141.74$+$362925.2:} This object was found by \cite{kirkpatrick2011} to have a near-infrared spectral type of L5 pec (blue) and who suggested its peculiar spectroscopic features may be indicative of unresolved binarity, with component types of L3.5$\pm$0.7 and T2$\pm$0.5.  \cite{bardalez2014} also identified it as a probable binary based on near-infrared spectra, with component types of L4.8$\pm$0.6 and T4.1$\pm$2.7. 

{\it J140255.70$+$080053.4:} \cite{kellogg2015} found this object to have a near infrared type of T2 pec, but also suggested it is a binary with L8+T5 components.  

{\it J143945.64$+$304218.7:} \cite{chiu2006} classified this object as a T2.5 from near-infrared spectra. It was deemed a strong binary candidate based on near-infrared spectra by \cite{burgasser2010} with component types of T1$\pm$0.2 and T5$\pm$0.6. 

{\it J171145.73$+$223204.2:} This object was discovered by \cite{burgasser2010} who gave it a near-infrared type of L9, but suggested it to be a strong binary candidate with component types of L5$\pm$0.4 and T5.5$\pm$1.2.  \cite{bardalez2014} found component types of L1.5$\pm$0.6 and T2.5$\pm$1.0 based on near-infrared spectra. \cite{bardalez2015} later suggested component types of L5.5$\pm$0.5 and T5.3$\pm$1.0 and showed that this system is not resolved in Keck/NIRC2 images. \cite{manjavacas2019} found component types of L6 and T3 based on $HST$ imaging.  

{\it J183058.60$+$454258.2:} This object was originally classified as an L9 by \cite{kirkpatrick2011}. However, it was also mentioned in \cite{best2015} to show signs of methane absorption and meet several of the criteria outlined in \cite{burgasser2010} for binarity. 

{\it J212702.63$+$761756.8:} \cite{kirkpatrick2010} discovered this object and measured optical and near-infrared spectral types of L7 and T0 pec, respectively, suggesting that this is an L+T binary with component types of L7 and T3.5.  \cite{bardalez2014} gave component types of L8.5$\pm$1.0 and T4.5$\pm$2.0, based on near-infrared spectra.

As a check on the above binarity indications, we use the binary template fitting procedure outlined in \cite{bravo2023}, available spectra from the Spex Prism Archive \citep{burgasser2017}, and the parallax distance determinations of this paper to make an estimate of whether the objects are strong, marginal, or poor binary candidates based on comparison of fits to binary or single-object templates.

We confirm these objects as strong spectral binary candidates: J103931.32$+$325623.7 (L7+T4), J110611.58$+$275414.1 (T1+T3), J131141.74$+$362925.2 (L4+T2), J140255.70$+$080053.4 (L8+T3), J171145.73$+$223204.2 (L6+T7), and J120746.66$+$024426.8 (T0+T1).

These objects are marginally better fit by a binary template than a single template (2 $\leq$ $\Delta$$\chi^{2}$ $\leq$ 3): J011912.41$+$240331.7 (T2+T8), J023618.06$+$004852.1 (L5+T1), J090900.31$+$652525.8 (T1+T3), and J094908.49$-$154548.3 (T1+T4). 

We find that these objects do not have binary fits that are significantly better than single object fits: J003259.67$+$141037.2 (L9), J075840.07$+$324718.8 (T2), J143945.64$+$304218.7 (T3), and J183058.60$+$454258.2 (L9).

The absolute magnitudes of spectral binaries compared to the rest of the sample in this work are shown on Figures \ref{fig:abspt2} and \ref{fig:cmd2}.

\subsubsection{Binary Objects Which may Have Been Missed}
\label{sec:potbin}

\begin{figure}
\plotone{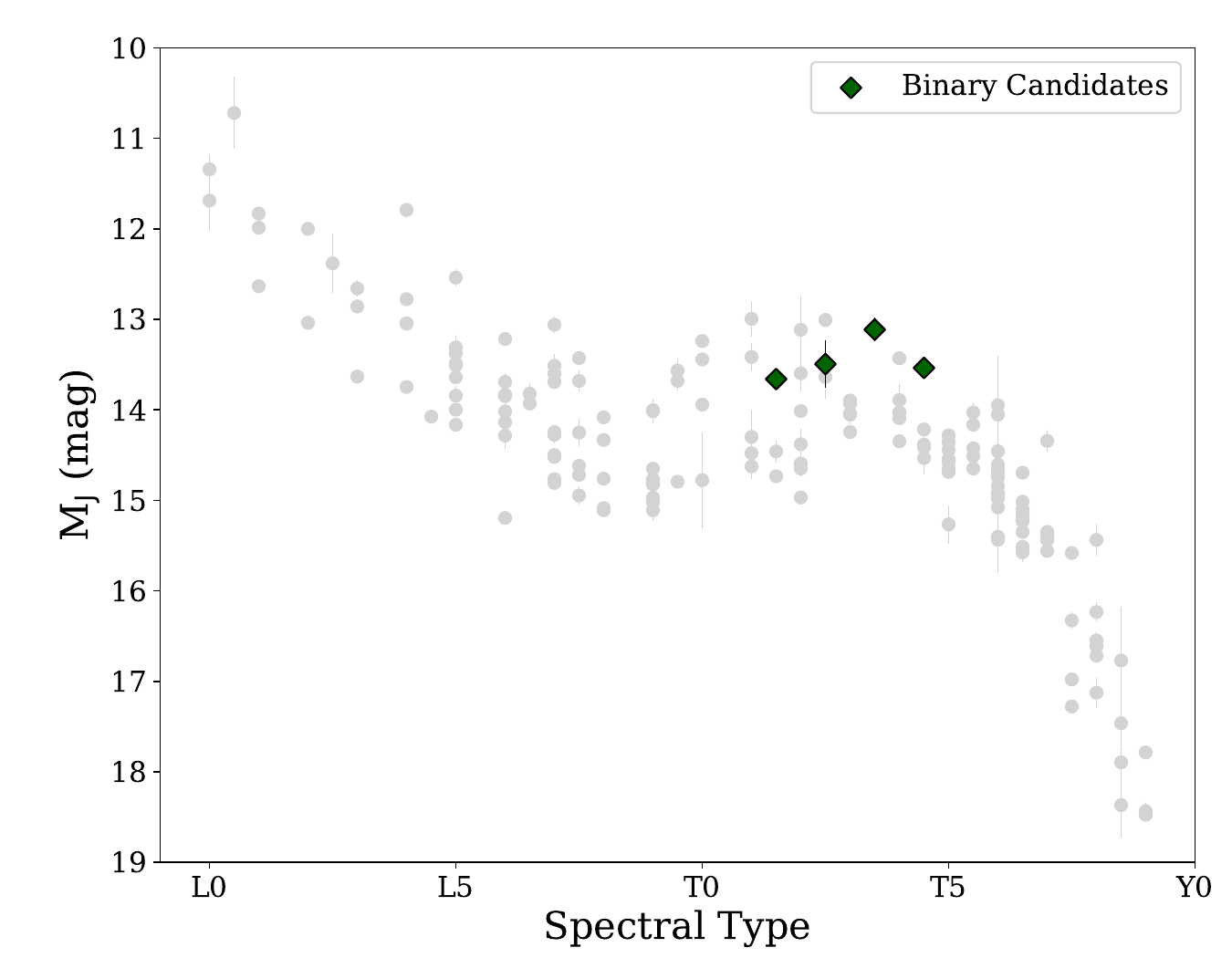}
\caption{Absolute $J$-band magnitudes as a function of spectral type with the positions of potential binary objects highlighted.}
\label{fig:abspt5}
\end{figure}

Despite best efforts to find binary brown dwarfs by high spatial resolution imaging and spectral de-convolution, it is likely that there remain some number of undiscovered binary objects, including those in this study. A difficult class for binary discovery would be those for which the components have very nearly equal spectral types, remain unresolved due to close orbits and/or distance, but for which a combination of low masses and orbital parameters give little indication of velocity spectral line broadening. Due to having similar spectral types, such objects might still be revealed by being roughly a factor of two ($\approx$0.75 mag) brighter in absolute magnitude than single objects of similar spectral type. This brightness difference can still be difficult to detect without accurate parallaxes and photometry and due to the width and steepness of the general cooling curve as seen in absolute magnitude versus spectral type diagrams. Fortunately, the ``J-band bump'' produces a flat region from late-L to mid-T spectral types in the M$_J$ versus spectral type diagram. Inspection of the left panel of Figure \ref{fig:abspt2} shows a clear locus of known non-subdwarf binaries roughly a factor of two above the single object locus for this spectral range. We note that there are several other objects in the binary locus which are not currently-known binaries. While not a proof of binarity we list these objects for further scrutiny: J055919.85$-$140454.8, J104828.96$+$091940.9, J132407.64$+$190625.5, and J175023.78$+$422238.6 and plot them in Figure \ref{fig:abspt5}.  We note that J055919.85$-$140454.8 has previously been mentioned as being overluminous in \cite{dahn2002} and \cite{burgasser2007b}.

\subsection{Wide Companions}
\label{sec:widecomp}

In this section we discuss brown dwarfs included in this work that have previously been suggested as being widely separated companions to stars. In several cases we parenthetically add the phrase ``Astrometry is consistent'', which refers to a general consistency of the brown dwarf proper motions and parallaxes determined in this work to those of the Gaia DR3 data for the putative companion star, although a rigorous comparison has not been carried out. For two
objects we provide somewhat more extended discussions due to their circumstances.

{\it J001502.52$+$295928.7:} This object was suggested as a possible wide ($\sim$5070 AU) companion to NLTT 730 in \cite{deacon2014}.  (Astrometry is consistent.)  

{\it J064626.92$+$793454.5:} This object was discovered as HD 46588 B in \cite{loutrel2011} with a near-infrared spectral type of L9$\pm$1 and a separation of $\sim$80\arcsec. (Astrometry is consistent.)

{\it J091534.04$+$042204.8AB:} We present the possibility that J091534.04$+$042204.8AB is a wide ($\sim$660\arcsec) companion to HD 79555AB, which is itself a K4 spectroscopic binary. See $\S$\ref{sec:resbin} for a discussion of the discovery and binary properties of J091534.04$+$042204.8AB. We first consider the Gaia astrometry for both objects, but we note the difficulties found for Gaia proper motions for resolved binaries, including J091534.04$+$042204.8AB, which we discuss in $\S$\ref{sec:pmcomp}. Gaia astrometry for HD 79555 has $\pi$=55.9338$\pm$0.1607 mas, $\mu_{\alpha}$ = -75.445$\pm$0.161 mas/yr, $\mu_{\delta}$ = 19.604$\pm$0.182 mas/yr which is largely consistent with its astrometry for J091534.04$+$042204.8AB  of $\pi$=56.6678$\pm$0.9493 mas, $\mu_{\alpha}$ = -113.180$\pm$1.004 mas/yr, $\mu_{\delta}$ = 21.278$\pm$0.890 mas/yr. While the parallaxes and $\mu_{\delta}$'s are consistent, the $\mu{\alpha}$'s are similar, but statistically significantly different. This may not be surprising considering the binary nature of both objects and noting that HD~79555AB itself has an elevated  RUWE value of 5.021. However, we note that the proper motion derived for HD 79555AB from \cite{hog2000} ($\mu_{\alpha}$ = -114.4$\pm$1.3 mas/yr, $\mu_{\delta}$ = 26.8$\pm$1.3 mas/yr) agrees much better with the value for J091534.04$+$042204.8AB we find in this paper ($\mu_{\alpha}$ = -112.27$\pm$0.28 mas/yr, $\mu_{\delta}$ = 25.89$\pm$0.31 mas/yr), which strongly suggests these objects are wide companions and would be the first time these objects have been associated.  The putative primary has been suggested to be part of the Castor Moving Group \citep{anosova1991} in \cite{maldonado2010} with a young age determination confirmed in \cite{brandt2014}, thus making this a new benchmark system if physical association can be confirmed.  J091534.04$+$042204.8AB is assigned an INT-G gravity classification in \cite{bouy2022}, so the ages of the components are generally consistent. However, this object was rejected as a low-gravity candidate from visual inspection in \cite{bardalez2019}.  The separation is 11,845 AU at the distance of HD 79555 from Gaia.  

{\it J095105.35$+$355800.1:}  This object is noted as a wide separation (13\farcs3) companion to LP 261-75 (M:) in \cite{burgasser2005b}.  The system is listed as young (100-200 Myr) with a primary spectral type of M4.5 in \cite{zhang2020}. 
(Astrometry is consistent.)

{\it J112254.33$+$255020.2:} \cite{kirkpatrick2011} discovered this object as WISEPC J112254.73$+$255021.5 with a near-infrared spectral type of T6.  They further noted that it is near the M5 star LHS 302, with a similar distance estimate and proper motion. (Astrometry is consistent.)

{\it J152322.78$+$301453.4:} \cite{mclean2000} first published this object as 2MASSW J152322.6$+$301456 with a near-infrared spectral type of L8/L9.  \cite{kirkpatrick2000} gave an optical spectral type of L8 and note that this object is also known as Gl 584C. \cite{kirkpatrick2001} confirmed the wide companionship of this source to the G+G dwarf primary Gliese 584AB. (Astrometry is consistent.)

{\it J175805.45$+$463318.2:} \cite{faherty2010} suggested this object as a companion to G 204$-$39 (M3) with a separation of 198\arcsec, suggesting it is young and metal-rich. (Astrometry is consistent.)

{\it J235122.32$+$301054.1:} We note the possibility that this object is a wide ($\sim$930\arcsec) companion to BD+29 5007.  Gaia astrometry for BD+29 5007 ($\pi$=42.2919$\pm$0.0166 mas, $\mu_{\alpha}$ = 255.381$\pm$0.018 mas/yr, $\mu_{\delta}$ = 10.628$\pm$0.011 mas/yr) is consistent with its astrometry for the putative companion from this work ($\pi$=40.79$\pm$2.06 mas, $\mu_{\alpha}$ = 254.86$\pm$1.27 mas/yr, $\mu_{\delta}$ = 6.89$\pm$1.48 mas/yr).  This would be the first time that these objects have been associated. The separation is 22,103 AU at the distance of BD+29 5007. (Astrometry is consistent.)

\subsection{Young Objects}
\label{sec:young}

The astrometry for every object in this study was input into BANYAN $\Sigma$ \citep{gagne2018} to look for kinematic matches to known moving groups.  Objects with BANYAN $\Sigma$ probabilities greater than 80\% are considered for further discussion below, where we review previous evidence for youth.  

{\it J004521.80$+$163444.0:} This object has a 98.4\% probability from BANYAN $\Sigma$ of belonging to the Argus association \citep{torres2008, zuckerman2019}.  \cite{schmidt2007} measured an H$\alpha$ equivalent width for this object of -10.06~\AA.  \cite{reid2008b} provide an optical spectral type of (L2), the parenthesis indicating signs of low gravity and measure an H$\alpha$ equivalent width of -14~\AA.  \cite{cruz2009} provide optical and near-infrared types of L2$\beta$ and L2$\gamma$, respectively, and give an H$\alpha$ equivalent width of -13.1$\pm$2.2~\AA.   \cite{allers2013} give this object a VL-G gravity classification.  \cite{gagne2014} suggest that this object is a high-probability member of the Argus association.  \cite{zapatero2014} detect lithium in this object's spectrum.  Using updated parallax and radial velocity information, \cite{gagne2015b} reaffirm membership in the Argus association.  \cite{faherty2016} measure $\pi_{\rm abs}$ = 62.5$\pm$3.7 mas, measured a new radial velocity, adopt a spectral type of L0$\beta$, give a gravity score of INT-G, and suggest that this object is a bonafide member of Argus.  \cite{liu2016} find $\pi_{\rm abs}$ = 65.9$\pm$1.3 mas and also conclude Argus membership.  This object was proposed as the low-gravity L1 spectral standard in \cite{piscarreta2024}.

{\it J005911.15$-$011401.4:} This object has a 96.6\% probability from BANYAN $\Sigma$ of belonging to the Carina-Near Moving Group \citep{zuckerman2006} and has no mention of youth elsewhere in the literature.  A young age for this source is unlikely considering the T8.5 spectral type of this object \citep{cushing2011}.

{\it J020743.00$+$000056.1:} This object has a 98.4\% probability from BANYAN $\Sigma$ of belonging to the Argus association \citep{torres2008, zuckerman2019}.  This object has been known to have odd colors in \cite{marley2002}, and was suggested to be metal-rich in \cite{marocco2010}.  \cite{zhang2021} also suggest Argus membership for this object, but was ultimately ruled-out as member. Youth is not evident by its normal T4.5 spectral type \citep{geballe2002}. 

{\it J023618.06$+$004852.1:} This object has a 98.2\% probability from BANYAN $\Sigma$ of belonging to the AB Doradus Moving Group \citep{zuckerman2004}.  It is suggested as a possible member of the Pleiades Moving Group in \cite{casewell2008} and listed as an AB Dor member in \cite{hurt2024}.  It is discussed as being a spectral binary candidate in $\S$\ref{sec:specbin}.

{\it J035523.53$+$113336.7:} This object has a 99.7\% probability from BANYAN $\Sigma$ of belonging to the AB Doradus Moving Group \citep{zuckerman2004}. \cite{reid2008b} give an optical spectral type of L5 and note the presence of lithium in this object's spectrum and its exceptionally red $J-K$ color.  It was given an L5$\gamma$ spectral type by \cite{cruz2009} who suggest it may be a planetary mass field object. \cite{faherty2013} discussed this object extensively and propose it to be a member of the AB Dor association.  Likewise, \cite{liu2013} suggest AB Dor membership. \cite{allers2013} classify this object as L3 VL-G in the near-infrared. It is considered a bona fide AB Dor member by \cite{gagne2014}, \cite{faherty2016}, and \cite{liu2016}.

{\it J062720.06$-$111429.6:} This object has a 99.1\% probability from BANYAN $\Sigma$ of belonging to the AB Doradus Moving Group \citep{zuckerman2004} and is also listed as candidate AB Dor member by \cite{zhang2021}. The available near-infrared spectral type of T6 by \cite{kirkpatrick2011} was not noted to have any indications of youth.

{\it J075840.07$+$324718.8:} This object has a 88.3\% probability from BANYAN $\Sigma$ of belonging to the Carina-Near Moving Group \citep{zuckerman2006} and was also noted as high-probability Carina-Near member by \cite{ashraf2022}. It was suggested as having low-gravity and being young (0.04-0.4 Gyr) by \cite{stephens2009}.  Strong brightness variations were seen by \cite{radigan2014}, who found a period of 4.9$\pm$0.2 hrs.  

{\it J081957.98$-$033529.3:} This object has a 74.9\% probability from BANYAN $\Sigma$ of belonging to the $\beta$ Pictoris Moving Group \citep{zuckerman2001}.  However, this probability increases to 85\% when its radial velocity, measured by \cite{hsu2021}, is included.  It is also listed as a candidate $\beta$ Pictoris Moving Group member in \cite{zhang2021} and has been shown to have significant photometric variability \citep{liu2024}.   

{\it J101014.45$-$040650.0:} This object has a 97.0\% probability from BANYAN $\Sigma$ of belonging to the Carina-Near Moving Group \citep{zuckerman2006} and is also considered a Carina-Near Moving Group member by \cite{sanghi2023} and \cite{hurt2024}. Large amplitude variability was noted by \cite{wilson2014} and confirmed by \cite{radigan2014b}. It has also been used as the L6 near-infrared standard \citep{kirkpatrick2010, cruz2018} (see $\S$\ref{sec:standards}).

{\it J111009.77$+$011608.7:}  This object has a 98.8\% probability from BANYAN $\Sigma$ of belonging to the AB Doradus Moving Group \citep{zuckerman2004}. \cite{knapp2004} found a spectral type of T5.5 and suggested that this object may be young and have a planetary mass based on evidence of low surface gravity in its spectrum.  \cite{stephens2009} presented a {\it Spitzer} IRS mid-IR spectrum and noted this object as very red.  \cite{gagne2015c} measured a radial velocity of 7.5$\pm$3.8 km s$^{-1}$ and firmly establish that this object is a planetary mass member of the AB Dor association.  This object was proposed as the T5 low-gravity spectral standard in \cite{piscarreta2024}.

{\it J132434.67$+$635827.1:} This object has a 98.2\% probability from BANYAN $\Sigma$ of belonging to the AB Doradus Moving Group \citep{zuckerman2004}. Discovered as 2MASS J13243559$+$6358284 by \cite{looper2007} who gave it a near-infrared spectral type of T2:pec due to being noticeably redder than the T2 standard.  \cite{looper2007} further suggested binarity or youth may be the cause of this object's unusual spectrum.  This object was labeled as a red photometric outlier in \cite{faherty2009}. \cite{yang2016} also found significant variability with {\it Spitzer} monitoring.  Significant banding was inferred by \cite{apai2017}.  \cite{gagne2018b} measured $\pi$ = 78.7$\pm$9.0 mas and RV = -23.7$^{+0.4}_{-0.2}$ and deduced that this object is a young, planetary-mass T dwarf in the AB Dor association.

{\it J155301.80$+$153239.5:} This object has a 92.8\% probability from BANYAN $\Sigma$ of belonging to the Carina-Near Moving Group \citep{zuckerman2006}.  It is also considered a Carina-Near candidate member by \cite{zhang2021}, with kinematics confirmed by \cite{hsu2021}. It is a resolved binary in Keck AO images from \cite{dupuy2012} who determined T6.5+T7.5 components. (see $\S$\ref{sec:resbin}). 

{\it J162413.95$+$002915.6:} This object has a 98.3\% probability from BANYAN $\Sigma$ of belonging to the Carina-Near Moving Group \citep{zuckerman2006}.  It is also suggested as a Carina-Near Moving Group candidate by \cite{zhang2021}, although ruled out as Carina-Near moving group member by \cite{hsu2021}.  

{\it J213927.29$+$022024.7:} This object has a 81.5\% probability from BANYAN $\Sigma$ of belonging to the Carina-Near Moving Group \citep{zuckerman2006} and is also considered to be a Carina-Near Moving Group member by \cite{zhang2021}. It is found to be a variable object in several studies, including \cite{radigan2012} who found large-amplitude variations ($\sim$26\% in $J-$band) with a period of 7.721$\pm$0.005 hr and note this object is single based on {\it HST} imaging.  \cite{apai2013} presented time-resolved {\it HST} spectroscopy again confirming the variability and finding a period of 7.83$\pm$0.1 hr, while \cite{vos2023} found evidence of patchy high-altitude fosterite clouds in this object's atmosphere. 

{\it J224431.96$+$204339.1:} This object has a 99.7\% probability from BANYAN $\Sigma$ of belonging to the AB Doradus Moving Group \citep{zuckerman2004}.  \cite{dahn2002} note this object's very red $J-K$ color, while \cite{mclean2003} presented a near-infrared spectrum and noted this object's unusual red color, a peaked $H-$band, and weak potassium lines and suggest low-gravity or low metallicity.  \cite{kirkpatrick2008} discussed at length why this object looks normal in the optical but unusual in the infrared.  \cite{allers2013} type this object as L6 VL-G in the near-infrared. \cite{gagne2014} found that this object is a strong candidate member of AB Dor.  It is given a near-infrared type of L6--L8$\gamma$ in \cite{gagne2015b} and  was considered to be a likely AB Dor member in \cite{faherty2016}, albeit with no RV or parallax data. \cite{liu2016} measured a $\pi_{\rm abs}$ of 58.7$\pm$1.0 mas which strengthened the case for AB Dor membership. \cite{vos2018} measured an RV of -16$^{+0.8}_{-0.9}$ km s$^{-1}$, confirming AB Dor membership, and found {\it Spitzer} variability with a period of 11$\pm$2 hr.

\begin{figure*}
\plotone{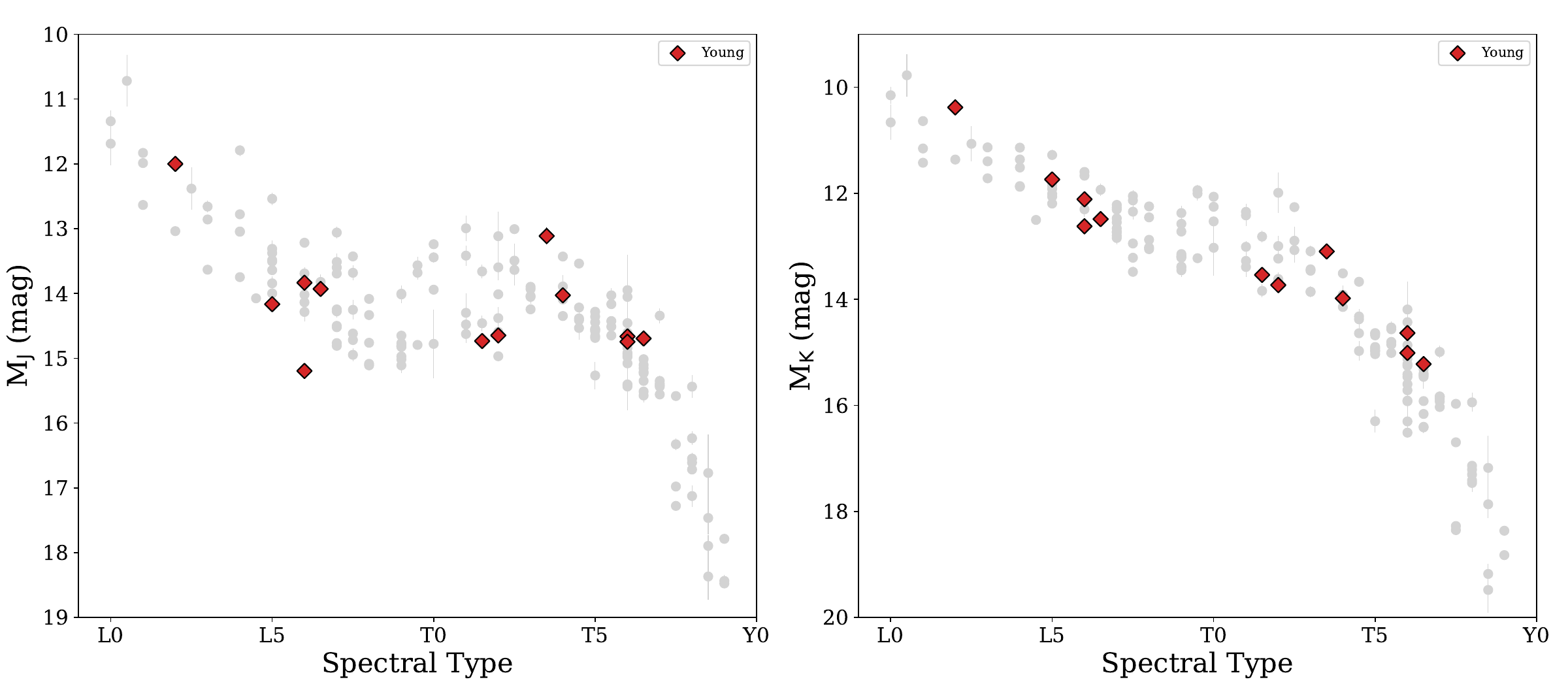}
\caption{Same as \ref{fig:abspt1} but with the positions of young objects highlighted.}
\label{fig:abspt3}
\end{figure*}

\begin{figure}
\plotone{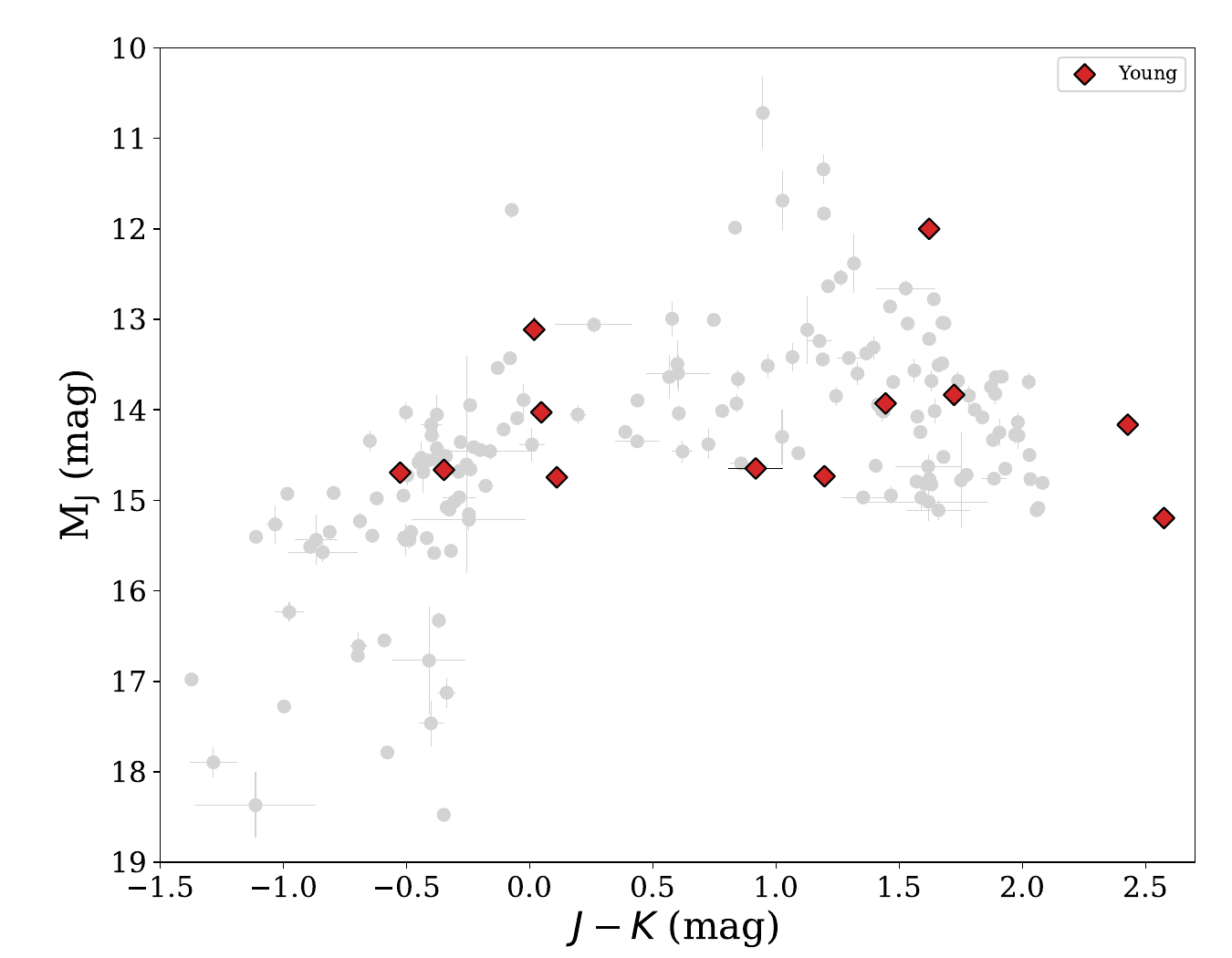}
\caption{Same as \ref{fig:cmd1} but with the positions of young objects highlighted.}
\label{fig:cmd3}
\end{figure}

The absolute magnitudes of the young objects identified in this work are shown in Figures \ref{fig:abspt3} and \ref{fig:cmd3}.

\subsection{Subdwarfs}
\label{sec:subdwarfs}

In this section we give brief summaries of the known and suspected subdwarfs which were part of
this study. We also list parenthetically the values of \vtan\ we determined for each object.

{\it J004121.65$+$354712.5:} Discovered as 2MASS J00412179$+$3547133 by \cite{burgasser2004} who typed it as sdL? because it exhibited spectral features similar to the L1 near-infrared standard, but also showed stronger FeH absorption, weaker CO absorption, and a bluer spectral slope. \cite{zhang2017} provide an optical spectrum and classify this object as sdL0.5.  (\vtan\ = 64.98$\pm$11.14 \kms)

{\it J044853.70$-$193543.6:} Discovered as WISEPA J044853.29$-$193548.5 by \cite{kirkpatrick2011}, who gave a near-infrared spectral type of T5 pec because of enhanced Y band flux and suppressed K band flux in its spectrum.  \cite{kirkpatrick2011} speculate that it may be metal-poor.  It is included in the list of metal-poor T dwarfs in \cite{zhang2019} and classified as an sdT5 in \cite{burgasser2025}. (\vtan\ = 103.61$\pm$9.99 \kms)

{\it J052536.18$+$673951.5:} Discovered as WISEPA J052536.33$+$673952.3 by \cite{kirkpatrick2011}, who determined a 
near-infrared type of T6 pec and suggested low metallicity. It is also included in the list of metal-poor T dwarfs in \cite{zhang2019}. (\vtan\ = 31.08$\pm$3.92 \kms)

{\it J053312.61$+$824617.2:} Discovered by \cite{burgasser2003b} as 2MASS J05325346$+$8246465, they classified it as the
first known L-type subdwarf, based on both optical and near-infrared spectra; although no exact spectral type was given, late-L was suggested. \cite{burgasser2007} later classified this object as sdL7.  It is typed as esdL7 in \cite{zhang2013} and as an esdL8 standard in \cite{burgasser2025}. (\vtan\ = 321.85$\pm$10.57 \kms)

{\it J061407.41$+$391233.2:} Discovered as WISEPA J061407.49$+$391236.4 by \cite{kirkpatrick2011} who found a near-infrared spectral type of T6.  Suggested to be a T subdwarf in \cite{zhang2019}. (\vtan\ = 50.17$\pm$1.40 \kms).

{\it J093735.98$+$293120.8:} Discovered as 2MASSI J0937347$+$293142 by \cite{burgasser2002} who noted this object as a peculiar T dwarf with very blue near-infrared colors (the bluest T dwarf at that time), possibly suggestive of low-metallicity and give a spectral type T6p. \cite{burgasser2003e} presented an optical spectrum, finding a type of T7 and a very red slope, suggesting that this object may be a metal-poor halo/thick disk brown dwarf.  A $\pi_{\rm abs}$ of 162.84$\pm$3.88 was presented in \cite{vrba2004} and noted as subluminous in all near-infrared bands. A {\it Spitzer} IRS spectrum was presented in \cite{cushing2006} who again suggested low-metallicity and/or high log g. \cite{burgasser2025} take this object as an sdT6 standard. (\vtan\ = 46.06$\pm$0.34 \kms)

{\it J093936.13$-$244844.3:} Discovered as 2MASS J09393548$-$2448279 by \cite{tinney2005} who gave it an initial spectral type of T8 and noted its blue near-infrared color. An optical spectral type of T8 is given by \cite{pineda2016}. 
It was suggested as having low metallicity by \cite{leggett2007}.  \cite{burgasser2008} argue that model fits require a highly inflated radius, in conflict with brown dwarf models, and suggest that this source is an unresolved equal-mass binary with low-metallicity ([M/H] $\approx$ -0.3).  Used in retrieval analysis of \cite{line2017}, who also find significant low-metallcity.  \cite{zhang2019} suggest an sdT7.5 spectral type while \cite{burgasser2025} consider it to be an d/sdT8 standard. (\vtan\ = 30.77$\pm$0.21 \kms)

{\it J111447.51$-$261830.0:} Discovered as 2MASS J11145133$-$2618235 by \cite{tinney2005} who classified it as T7.5$\pm$0.5 from near-infrared spectra, estimated a distance of 7 pc (possibly the coldest brown dwarf known at that time), and noted it as having a large space velocity.  \cite{burgasser2006} suggested this object may have subsolar metallicity while \cite{leggett2007} suggest that value is [m/H] $\approx$ -0.3.  Used in retrieval analysis of \cite{line2017} who derive a near-solar metallicity. An optical spectral type of T8 is given by \cite{pineda2016}. (\vtan\ = 77.83$\pm$0.76 \kms)

{\it J115821.40$+$043446.3:} Discovered as SDSS J115820.75$+$043501.7 by \cite{zhang2009}. It was spectral-typed by \cite{kirkpatrick2010} as sdL7 in both the near-infrared and the optical based on enhanced abundances of TiO, FeH, and CrH and a blue near-infrared spectrum.  It was suggested as an sdL7 standard by \cite{greco2019} and adopted as a d/sdL8 standard by \cite{burgasser2025}. (\vtan\ = 119.64$\pm$7.02 \kms)

{\it J143535.75$-$004348.6:} Discovered as SDSS J143535.72$-$004347.0 by \cite{hawley2002} who classified it as an L3 based on optical spectra, while \citep{knapp2004} provided an L2.5 classification from near-infrared spectra. It has been suggested as metal-poor by \cite{lodieu2017} based on various color indices. (\vtan\ = 30.22$\pm$4.45 \kms)

{\it J162618.23$+$392523.4:}  Discovered as 2MASS J16262034$+$3925190 by \cite{burgasser2004b} and determined to be a high-proper motion, halo, low-metallicity L dwarf based on a near-infrared spectrum (the 2nd known L subdwarf). An optical spectrum presented by \cite{gizis2006} indicated this object to be a mid-L subdwarf. \cite{burgasser2007} suggested an sdL4 classification based on its near-infrared spectrum,  while \cite{zhang2013} gave an optical spectral type of esdL4 (referencing a Zhang et al.~2013 conference proceeding). It is classified as usdL4 by \cite{zhang2017} and determined to be part of the ``halo transition zone'' in \cite{zhang2017b}. \cite{dahn2017} find $\pi_{\rm abs}$ = 31.08$\pm$0.55 mas and found residuals possibly indicative of a binary system.   \cite{gonzales2018} provided 8.0 $\mu$m {\it Spitzer} photometry and derived numerous physical parameters based on this object's SED.  It is listed as the sdL4 optical/near-infrared standard in \cite{greco2019}. (\vtan\ = 113.63$\pm$3.96 \kms)

{\it J175609.98$+$281516.4:} Discovered as 2MASS J17561080$+$2815238 by \cite{kirkpatrick2010} who also determined optical and near-infrared spectral types of sdL1 and L1 (pec), respectively.  \cite{greco2019} suggest this object to be an optical/near-infrared sdL1 standard. We note that the relatively high value of \vtan\ is consistent with the subdwarf nature of this object. (\vtan\ = 113.63$\pm$3.96 \kms)

\begin{figure*}
\plotone{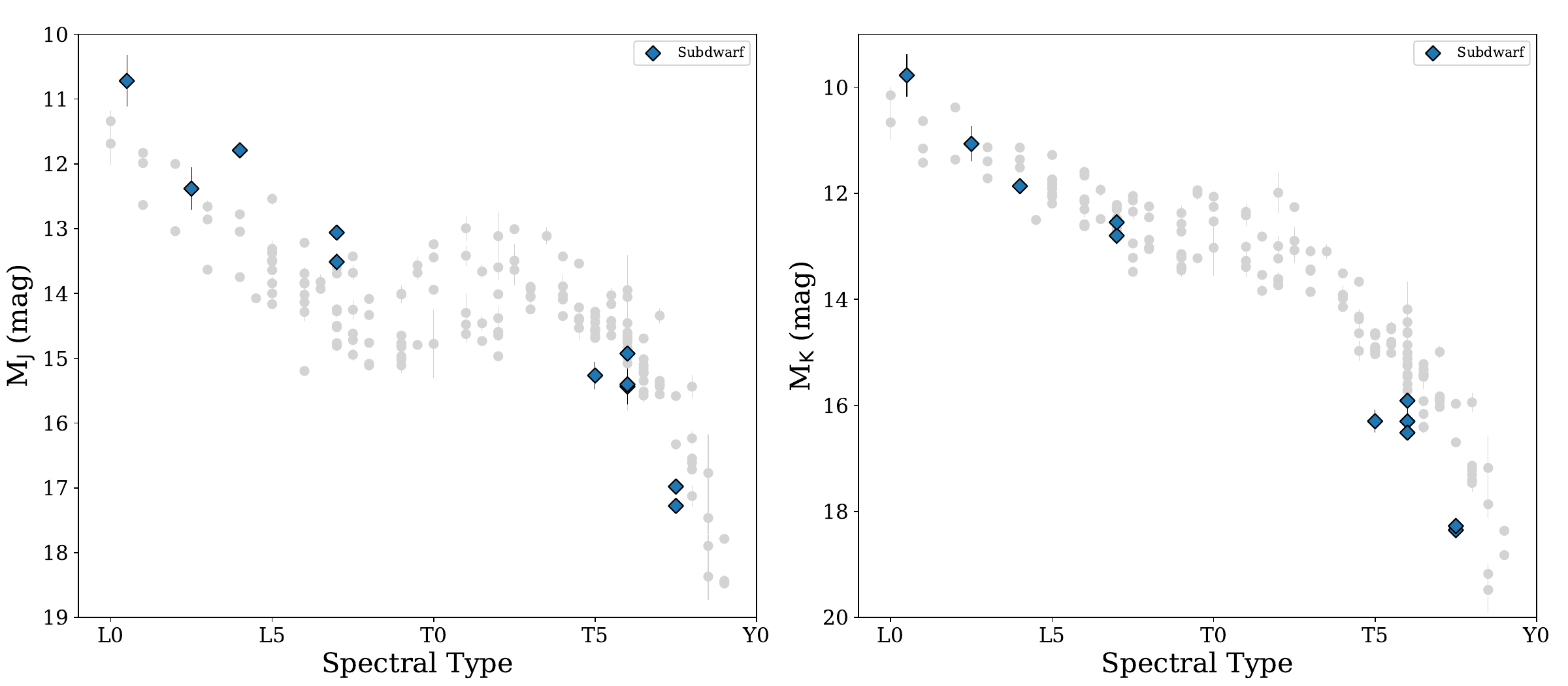}
\caption{Same as \ref{fig:abspt1} but with the positions of subdwarfs highlighted.}
\label{fig:abspt4}
\end{figure*}

\begin{figure}
\plotone{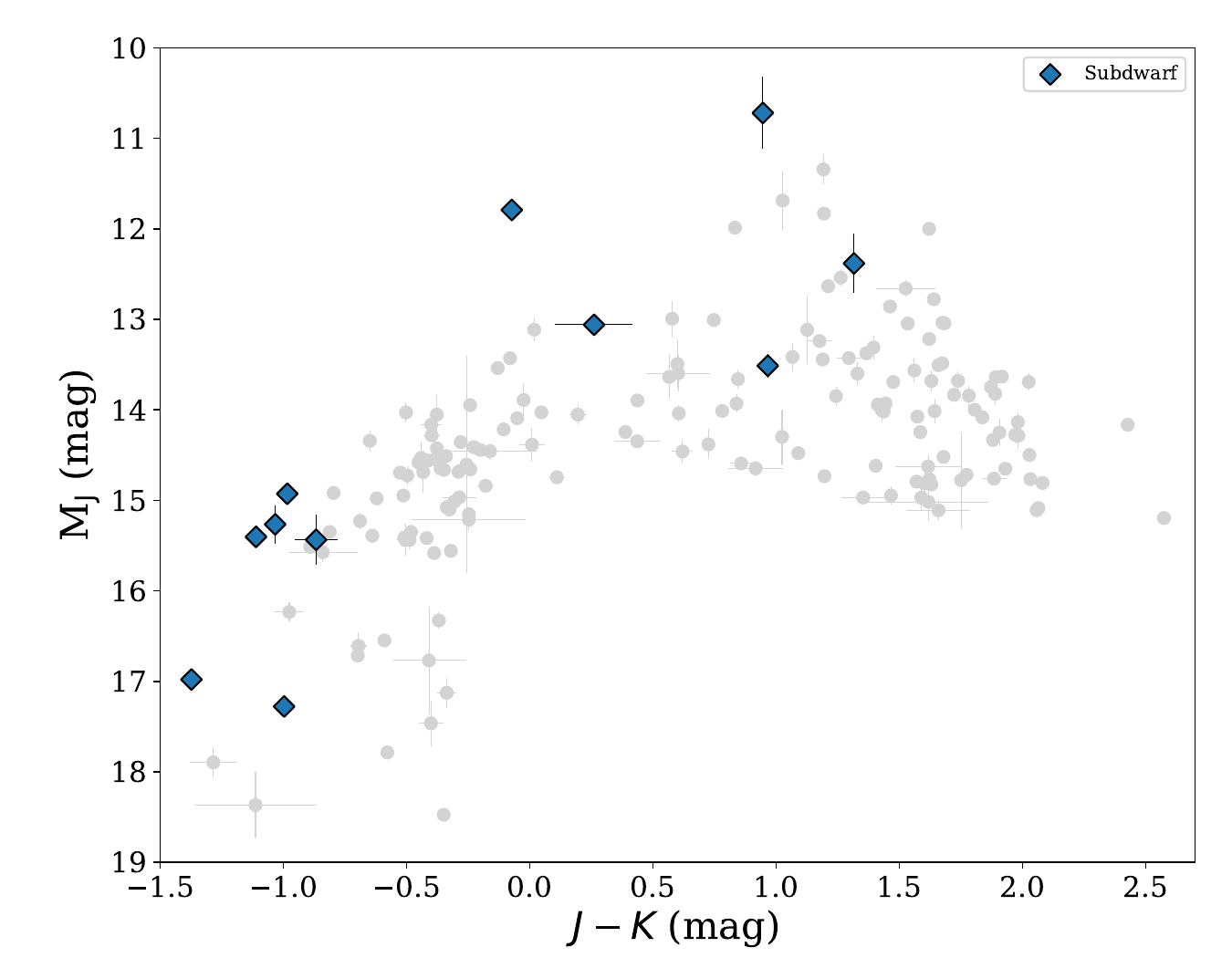}
\caption{Same as \ref{fig:cmd1} but with the positions of subdwarfs highlighted.}
\label{fig:cmd4}
\end{figure}

The absolute magnitudes of the subdwarfs included in this parallax program are shown in Figures \ref{fig:abspt4} and \ref{fig:cmd4}.

\subsection{Brown Dwarf Spectral Standards}
\label{sec:standards}

Several L- and T-type optical and near-infrared standards were observed as part of this program. While some of these have
been mentioned in the previous section, here we give a complete list of the objects which are considered standards and for
which we provide parallax and proper motion results. We note that, for twelve of the twenty standards listed below, the parallax results in this work are the most precise yet measured.

The near-infrared L dwarf standards from \cite{kirkpatrick2010} include J150653.09$+$132105.8 (L3), J101014.45$-$040650.0 (L6), J010332.40$+$193536.5 (L7), and J163229.48$+$190439.6 (L8).  J101014.45$-$040650.0 and J163229.48$+$190439.6 are also the L6 and L8 optical standards, respectively \citep{kirkpatrick2005, geissler2011}. J010332.40$+$193536.5 was later typed as L6$\beta$ \citep{faherty2012} and has been considered a possible member of the Argus association \citep{torres2008, zuckerman2019} by \cite{gagne2014, gagne2015b}. (We also note that J010332.40$+$193536.5 did not meet our BANYAN $\Sigma$ criterion for association with any moving group and thus was left out of $\S$\ref{sec:young}.) \cite{gagne2015} described this object as a red brown dwarf with no clear signs of low gravity, and retype it as L6 pec.  \cite{cruz2018} no longer consider this object a near-infrared standard, and instead proposed J082518.97$+$211546.1 as the L7.5 near-infrared standard, which is included in this program.

The T dwarf near-infrared standards are given in \cite{burgasser2006}, most of which are included in this parallax program, including J120746.66$+$024426.8 (T0), J083717.18$-$000020.8 (T1), J125453.41$-$012245.6 (T2), J225418.98$+$312353.0 (T4), J150319.71$+$252528.3 (T5), J162413.95$+$002915.6 (T6), J072719.22$+$170950.1 (T7), and J041522.17$-$093457.1 (T8).  The T9 near-infrared standard J072226.89$-$054027.5 (\citealt{lucas2010, cushing2011}) was also included in our program. 

Several subdwarf standards were also included as part of this program.  The near-infrared subdwarf standards from \cite{greco2019} include J175609.98$+$281516.4 (sdL1), J162618.23$+$392523.4 (sdL4), and J115821.40$+$043446.3 (sdL7). J115821.40$+$043446.3 was later updated to be the d/sdL8 standard in \cite{burgasser2025}. We note that \cite{greco2019} consider J175609.98$+$281516.4 to also be an optical sdL1 standard. Other subdwarf standards from \cite{burgasser2025} include J053312.61$+$824617.2 (esdL8), J093735.98$+$293120.8 (sdT6), and J093936.13$-$244844.3 (d/sdT8). 

\section{Summary}
\label{sec:summary}

We have measured parallaxes and proper motions with high accuracy for a vast majority of the 173 L- and T-type dwarfs and subdwarfs in this study with median accuracies of $\sigma$({$\pi_{abs}$}) = 1.51 mas and $\sigma$($\mu_{abs}$) = 1.02 mas yr$^{-1}$, with target object median distances of $\approx$18 pc and median reference frame distances of $\approx$1 kpc. These results provide the first parallaxes and proper motions for 16 objects and the current highest precision parallaxes and proper motions for an additional 106 objects. Comparison of our results with 40 objects with parallaxes and proper motions in Gaia DR3 showed no systematic differences, confirmed our quoted precisions, and allowed for an investigation of resolved binarity on astrometric results. We also provide a uniform set of $J,H,K_S$ photometry in the UKIRT/MKO system for all objects in our study. Although quality parallax and proper motion determinations are important for all brown dwarfs, we further investigate several brown dwarf populations included in our sample. While for all binary objects in the study we provide new absolute magnitudes, for the spectral binary subset we also include new spectral template fitting to help determine binarity. For wide companion candidates our proper motions results are used for confirmation. Proper motions are also used for confirmation of potential young objects with stellar moving groups. Several subdwarfs are discussed in some detail. Finally, our parallaxes and new photometry provide reliable near-infrared absolute magnitudes for several brown dwarf spectral standards. 

\section{Appendix A}

We present in Table \ref{tab:cit} additional $JHK$ photometry obtained with ASTROCAM at the 1.55-m telescope, primarily of Series 1 objects. The observations, obtained in the early 2000's, are on the CIT photometric system using the standards of \cite{guetter2003}. Column (1) gives the CatWISE object names, column (2) the $J$-band magnitudes and associated error bars, columns (3) and (4) the ($J-H$) and ($J-K$) colors and associated error bars, respectively, and column (5) the number of independent nights of observation. The results are given in terms of J magnitude and ($J-H$) and ($J-K$) colors, as that was the fashion in which the data were reduced. This photometry is presented for purposes of completeness only and is not used in any of the science discussions in this paper, as our subsequent UKIRT $JHK$ photometry presented in this paper supersedes in accuracy and uniformity this earlier photometry.

\startlongtable
\begin{deluxetable*}{lcccc}
\label{tab:cit}
\tabletypesize{\scriptsize}
\tablecaption{New $JHK$ Photometry in the CIT System}
\tablewidth{0pt}
\tablehead{
\colhead{Name} & \colhead{$J$ $\pm$ $\sigma$($J$)} & 
\colhead{($J-H$) $\pm$ $\sigma$($J-H$)}  & \colhead{($J-K$) $\pm$ $\sigma$($J-K$)} & \colhead{Nights} \\
\colhead{(CatWISE)}   & \colhead{(mag)}   & \colhead{(mag)}  &   
\colhead{(mag)} &  \colhead{(Number)}  \\
\colhead{(1)} & \colhead{(2)}   & \colhead{(3)}   &  \colhead{(4)}  & 
\colhead{(5)}
}
\startdata
J003030.38$-$145033.8    & 16.51 $\pm$ 0.03 & 1.18 $\pm$ 0.02  & 2.01 $\pm$ 0.03 & 5    \\
J003259.67$+$141037.2    & 16.65 $\pm$ 0.03 & 0.98 $\pm$ 0.03  & 1.60 $\pm$ 0.03 & 5    \\
J010753.11$+$004157.7    & 15.75 $\pm$ 0.03 & 1.25 $\pm$ 0.03  & 2.05 $\pm$ 0.07 & 1    \\
J015142.59$+$124428.7    & 16.30 $\pm$ 0.02 & 0.80 $\pm$ 0.02  & 1.00 $\pm$ 0.03 & 4    \\
J020743.00$+$000056.1    & 16.73 $\pm$ 0.02 &$-$0.06 $\pm$ 0.03&$-$0.08 $\pm$ 0.04 & 4    \\
J024313.38$-$245332.8    & 15.08 $\pm$ 0.03 &$-$0.34 $\pm$ 0.02&$-$0.12 $\pm$ 0.03 & 4    \\
J031100.13$+$164815.6    & 15.90 $\pm$ 0.06 & 0.96 $\pm$ 0.04  & 1.61 $\pm$ 0.03 & 2      \\
J032842.66$+$230204.1    & 16.51 $\pm$ 0.02 & 0.98 $\pm$ 0.02  & 1.49 $\pm$ 0.02 & 4     \\
J041522.17$-$093457.1    & 15.19 $\pm$ 0.02 &$-$0.55 $\pm$ 0.02&$-$0.45 $\pm$ 0.06 & 5  \\
J042348.22$-$041402.0    & 14.26 $\pm$ 0.03 & 0.82 $\pm$ 0.02  & 1.32 $\pm$ 0.02 & 2   \\
J051609.20$-$044553.3    & 15.49 $\pm$ 0.03 &$-$0.32 $\pm$ 0.02&$-$0.19 $\pm$ 0.03 & 4    \\
J053312.61$+$824617.2    & 15.02 $\pm$ 0.03 & 0.13 $\pm$ 0.02  & 0.11 $\pm$ 0.02 & 3      \\
J053952.16$-$005856.5    & 13.88 $\pm$ 0.05 & 0.83 $\pm$ 0.03  & 1.40 $\pm$ 0.03 & 1    \\
J055919.85$-$140454.8    & 13.55 $\pm$ 0.02 &$-$0.12 $\pm$ 0.02&$-$0.21 $\pm$ 0.03 & 2   \\
J072719.22$+$170950.1    & 15.10 $\pm$ 0.03 &$-$0.56 $\pm$ 0.05&$-$0.31 $\pm$ 0.03 & 3    \\
J075547.93$+$221212.8    & 15.39 $\pm$ 0.04 &$-$0.33 $\pm$ 0.02&$-$0.39 $\pm$ 0.05 & 4    \\
J083006.91$+$482838.3    & 15.27 $\pm$ 0.02 & 0.90 $\pm$ 0.02  & 1.47 $\pm$ 0.02 & 2       \\
J083717.18$-$000020.8    & 17.03 $\pm$ 0.02 & 0.67 $\pm$ 0.02  & 1.05 $\pm$ 0.03 & 3      \\
J092615.40$+$584717.6    & 15.41 $\pm$ 0.02 &$-$0.03 $\pm$ 0.03&$-$0.05 $\pm$ 0.04 & 4    \\
J093735.98$+$293120.8    & 14.28 $\pm$ 0.02 &$-$0.47 $\pm$ 0.02&$-$1.03 $\pm$ 0.02 & 2   \\
J093936.13$-$244844.3   & 15.60 $\pm$ 0.02 &$-$0.55 $\pm$ 0.02&$-$1.30 $\pm$ 0.04 & 3   \\
J095105.35$+$355800.1   & 17.13 $\pm$ 0.05 &   1.14 $\pm$ 0.02&   1.89 $\pm$ 0.03 & 2   \\
J102109.51$-$030421.1   & 15.94 $\pm$ 0.02 &   0.29 $\pm$ 0.02  & 0.49 $\pm$ 0.03 & 2   \\
J104751.75$+$212414.7   & 15.37 $\pm$ 0.02 &$-$0.49 $\pm$ 0.02&$-$0.80 $\pm$ 0.03 & 4   \\
J111009.77$+$011608.7   & 16.11 $\pm$ 0.04 &$-$0.17 $\pm$ 0.03& 0.06 $\pm$ 0.04 & 3     \\
J111447.51$-$261830.0   & 15.59 $\pm$ 0.02 &$-$0.56 $\pm$ 0.02&$-$1.04 $\pm$ 0.04 & 4   \\
J121709.86$-$031111.8     & 15.42 $\pm$ 0.02 &$-$0.56 $\pm$ 0.03&$-$0.31 $\pm$ 0.03 & 2   \\
J122554.84$-$273958.5   & 14.83 $\pm$ 0.02 &$-$0.38 $\pm$ 0.03&$-$0.39 $\pm$ 0.02 & 2   \\
J123737.03$+$652607.2   & 15.59 $\pm$ 0.02 &$-$0.52 $\pm$ 0.02&$-$0.99 $\pm$ 0.04 & 3   \\
J125453.41$-$012245.6   & 14.71 $\pm$ 0.04 & 0.59 $\pm$ 0.03  & 0.75 $\pm$ 0.05 & 1     \\
J134645.89$-$003152.2   & 15.53 $\pm$ 0.03 &$-$0.47 $\pm$ 0.02&$-$0.27 $\pm$ 0.03 & 4   \\
J143517.22$-$004612.9   & 16.45 $\pm$ 0.04 & 0.66 $\pm$ 0.03  & 1.12 $\pm$ 0.04 & 4     \\
J143535.75$-$004348.6   & 16.52 $\pm$ 0.03 & 0.81 $\pm$ 0.03  & 1.30 $\pm$ 0.03 & 4     \\
J144600.78$+$002450.9   & 15.57 $\pm$ 0.02 & 1.02 $\pm$ 0.04  & 1.63 $\pm$ 0.06 & 2     \\
J150319.71$+$252528.3   & 13.51 $\pm$ 0.03 &$-$0.36 $\pm$ 0.03&$-$0.44 $\pm$ 0.03 & 3   \\
J152322.78$+$301453.4   & 16.11 $\pm$ 0.04 & 1.07 $\pm$ 0.02  & 1.72 $\pm$ 0.02 & 3     \\
J155301.80$+$153239.5   & 15.34 $\pm$ 0.02 &$-$0.51 $\pm$ 0.02&$-$0.44 $\pm$ 0.02 & 2    \\
J162413.95$+$002915.6   & 15.07 $\pm$ 0.02 &$-$0.42 $\pm$ 0.03&$-$0.57 $\pm$ 0.05 & 4    \\
J163229.48$+$190439.6       & 15.73 $\pm$ 0.03 & 1.10 $\pm$ 0.02  & 1.77 $\pm$ 0.04 & 2       \\
J171145.73$+$223204.2   & 16.72 $\pm$ 0.04 & 0.82 $\pm$ 0.02  & 1.40 $\pm$ 0.04 & 5      \\
J172811.54$+$394859.0   & 15.79 $\pm$ 0.03 & 1.11 $\pm$ 0.03  & 1.88 $\pm$ 0.03 & 2     \\
J175023.78$+$422238.6   & 16.13 $\pm$ 0.03 & 0.59 $\pm$ 0.02  & 0.77 $\pm$ 0.02 & 3     \\
J175033.14$+$175905.4   & 16.12 $\pm$ 0.05 & 0.22 $\pm$ 0.03  & 0.05 $\pm$ 0.03 & 3    \\
J175805.45$+$463318.2   & 15.80 $\pm$ 0.04 &$-$0.44 $\pm$ 0.02&$-$0.21 $\pm$ 0.02 & 4    \\
J184108.67$+$311728.6   & 15.94 $\pm$ 0.02 & 1.01 $\pm$ 0.04  & 1.59 $\pm$ 0.02 & 5     \\
J190106.23$+$471820.5   & 15.47 $\pm$ 0.02 &$-$0.21 $\pm$ 0.02&$-$0.31 $\pm$ 0.02 & 7   \\  
J210115.58$+$175656.0   & 16.87 $\pm$ 0.04 & 1.19 $\pm$ 0.04  & 1.91 $\pm$ 0.06 & 3     \\
J212414.06$+$010003.8   & 15.72 $\pm$ 0.03 &$-$0.24 $\pm$ 0.02&$-$0.21 $\pm$ 0.04 & 4   \\
J222444.39$-$015908.2   & 13.96 $\pm$ 0.03 & 1.14 $\pm$ 0.06  & 1.91 $\pm$ 0.05 & 2     \\
J224253.65$+$254256.2   & 14.64 $\pm$ 0.02 & 1.00 $\pm$ 0.02  & 1.68 $\pm$ 0.02 & 3      \\
J224431.96$+$204339.1   & 16.40 $\pm$ 0.03 & 1.39 $\pm$ 0.03  & 2.42 $\pm$ 0.03 & 4     \\
J225418.98$+$312353.0   & 14.95 $\pm$ 0.02 & 0.05 $\pm$ 0.02  &-0.03 $\pm$ 0.03 & 5     \\
J225529.03$-$003436.4   & 15.57 $\pm$ 0.02 & 0.77 $\pm$ 0.02  & 1.21 $\pm$ 0.02 & 3     \\
J233910.67$+$135212.9   & 15.74 $\pm$ 0.03 &$-$0.28 $\pm$ 0.03&$-$0.30 $\pm$ 0.04 & 2   \\
J235654.19$-$155322.9   & 15.44 $\pm$ 0.02 &$-$0.28 $\pm$ 0.03&$-$0.17 $\pm$ 0.03 & 2   \\
\enddata
\end{deluxetable*}

\clearpage

\section{Acknowledgements}

\begin{acknowledgments}
FJV would like to thank C.C. Dahn, apart from his direct contributions to this paper, for his many personal discussions over the years on how to correctly do ground-based astrometry and optical photometry. FJV would also like to acknowledge D.G. Monet for developing the software, used in this and previous USNO/NOFS astrometry, which included the tools necessary for detailed analyses of the astrometric data.

This work has made use of data from the European Space Agency (ESA) mission Gaia (https://www.cosmos.esa.int/gaia), processed by the Gaia Data Processing and Analysis Consortium (DPAC, https://www.cosmos.esa.int/web/gaia/dpac/consortium). Funding for the DPAC has been provided by national institutions, in particular the institutions participating in the Gaia Multilateral Agreement.

This publication makes use of data products from the Two Micron All Sky Survey, which is a joint project of the University of Massachusetts and the Infrared Processing and Analysis Center/California Institute of Technology, funded by the National Aeronautics and Space Administration and the National Science Foundation.

This publication makes use of data products from the Wide-field Infrared Survey Explorer, which is a joint project of the University of California, Los Angeles, and the Jet Propulsion Laboratory/California Institute of Technology, funded by the National Aeronautics and Space Administration.  This publication also makes use of data products from NEOWISE, which is a project of the Jet Propulsion Laboratory/California Institute of Technology, funded by the Planetary Science Division of the National Aeronautics and Space Administration.

This publication makes use of data products from the UKIRT Hemisphere Survey, which is a joint project of the United States Naval Observatory, the University of Hawaii Institute for Astronomy, the Cambridge University Cambridge Astronomy Survey Unit, and the University of Edinburgh Wide-Field Astronomy Unit (WFAU). The WFAU gratefully acknowledges support for this work from the Science and Technology Facilities Council (STFC) through ST/T002956/1 and previous grants. 

This work has benefited from The UltracoolSheet at http://bit.ly/UltracoolSheet, maintained by Will Best, Trent Dupuy, Michael Liu, Aniket Sanghi, Rob Siverd, and Zhoujian Zhang, and developed from compilations by \cite{dupuy2012, dupuy2013, deacon2014, liu2016, best2018, best2021, sanghi2023, schneider2023}. 

\end{acknowledgments}

\bibliography{references}{}

@ARTICLE{vrba2004,
       author = {{Vrba}, F.~J. and {Henden}, A.~A. and {Luginbuhl}, C.~B. and {Guetter}, H.~H. and {Munn}, J.~A. and {Canzian}, B. and {Burgasser}, A.~J. and {Kirkpatrick}, J. Davy and {Fan}, X. and {Geballe}, T.~R. and {Golimowski}, D.~A. and {Knapp}, G.~R. and {Leggett}, S.~K. and {Schneider}, D.~P. and {Brinkmann}, J.},
        title = "{Preliminary Parallaxes of 40 L and T Dwarfs from the US Naval Observatory Infrared Astrometry Program}",
      journal = {\aj},
         year = 2004,
        month = may,
       volume = {127},
       number = {5},
        pages = {2948-2968},
          doi = {10.1086/383554},
archivePrefix = {arXiv},
       eprint = {astro-ph/0402272},
 primaryClass = {astro-ph},
       adsurl = {https://ui.adsabs.harvard.edu/abs/2004AJ....127.2948V}
}

@ARTICLE{dupuy2012,
       author = {{Dupuy}, Trent J. and {Liu}, Michael C.},
        title = "{The Hawaii Infrared Parallax Program. I. Ultracool Binaries and the L/T Transition}",
      journal = {\apjs},
         year = 2012,
        month = aug,
       volume = {201},
       number = {2},
          eid = {19},
        pages = {19},
          doi = {10.1088/0067-0049/201/2/19},
archivePrefix = {arXiv},
       eprint = {1201.2465},
 primaryClass = {astro-ph.SR},
       adsurl = {https://ui.adsabs.harvard.edu/abs/2012ApJS..201...19D}
}

@ARTICLE{dahn2017,
       author = {{Dahn}, Conard C. and {Harris}, Hugh C. and {Subasavage}, John P. and {Ables}, Harold D. and {Canzian}, Blaise J. and {Guetter}, Harry H. and {Harris}, Fred H. and {Henden}, Arne H. and {Leggett}, S.~K. and {Levine}, Stephen E. and {Luginbuhl}, Christian B. and {Monet}, Alice B. and {Monet}, David G. and {Munn}, Jeffrey A. and {Pier}, Jeffrey R. and {Stone}, Ronald C. and {Vrba}, Frederick J. and {Walker}, Richard L. and {Tilleman}, Trudy M.},
        title = "{CCD Parallaxes for 309 Late-type Dwarfs and Subdwarfs}",
      journal = {\aj},
         year = 2017,
        month = oct,
       volume = {154},
       number = {4},
          eid = {147},
        pages = {147},
          doi = {10.3847/1538-3881/aa880b},
archivePrefix = {arXiv},
       eprint = {1709.02729},
 primaryClass = {astro-ph.SR},
       adsurl = {https://ui.adsabs.harvard.edu/abs/2017AJ....154..147D}
}

@ARTICLE{dahn2002,
       author = {{Dahn}, Conard C. and {Harris}, Hugh C. and {Vrba}, Frederick J. and {Guetter}, Harry H. and {Canzian}, Blaise and {Henden}, Arne A. and {Levine}, Stephen E. and {Luginbuhl}, Christian B. and {Monet}, Alice K.~B. and {Monet}, David G. and {Pier}, Jeffrey R. and {Stone}, Ronald C. and {Walker}, Richard L. and {Burgasser}, Adam J. and {Gizis}, John E. and {Kirkpatrick}, J. Davy and {Liebert}, James and {Reid}, I. Neill},
        title = "{Astrometry and Photometry for Cool Dwarfs and Brown Dwarfs}",
      journal = {\aj},
         year = 2002,
        month = aug,
       volume = {124},
       number = {2},
        pages = {1170-1189},
          doi = {10.1086/341646},
archivePrefix = {arXiv},
       eprint = {astro-ph/0205050},
 primaryClass = {astro-ph},
       adsurl = {https://ui.adsabs.harvard.edu/abs/2002AJ....124.1170D}
}

@ARTICLE{tinney2003,
       author = {{Tinney}, C.~G. and {Burgasser}, Adam J. and {Kirkpatrick}, J. Davy},
        title = "{Infrared Parallaxes for Methane T Dwarfs}",
      journal = {\aj},
         year = 2003,
        month = aug,
       volume = {126},
       number = {2},
        pages = {975-992},
          doi = {10.1086/376481},
archivePrefix = {arXiv},
       eprint = {astro-ph/0304339},
 primaryClass = {astro-ph},
       adsurl = {https://ui.adsabs.harvard.edu/abs/2003AJ....126..975T}
}

@ARTICLE{faherty2011,
       author = {{Faherty}, Jacqueline K. and {Burgasser}, Adam J. and {Bochanski}, John J. and {Looper}, Dagny L. and {West}, Andrew A. and {van der Bliek}, Nicole S.},
        title = "{Identification of a Wide, Low-Mass Multiple System Containing the Brown Dwarf 2MASS J0850359+105716}",
      journal = {\aj},
         year = 2011,
        month = mar,
       volume = {141},
       number = {3},
          eid = {71},
        pages = {71},
          doi = {10.1088/0004-6256/141/3/71},
archivePrefix = {arXiv},
       eprint = {1012.4232},
 primaryClass = {astro-ph.SR},
       adsurl = {https://ui.adsabs.harvard.edu/abs/2011AJ....141...71F}
}

@ARTICLE{faherty2012,
       author = {{Faherty}, Jacqueline K. and {Burgasser}, Adam J. and {Walter}, Frederick M. and {Van der Bliek}, Nicole and {Shara}, Michael M. and {Cruz}, Kelle L. and {West}, Andrew A. and {Vrba}, Frederick J. and {Anglada-Escud{\'e}}, Guillem},
        title = "{The Brown Dwarf Kinematics Project (BDKP). III. Parallaxes for 70 Ultracool Dwarfs}",
      journal = {\apj},
         year = 2012,
        month = jun,
       volume = {752},
       number = {1},
          eid = {56},
        pages = {56},
          doi = {10.1088/0004-637X/752/1/56},
archivePrefix = {arXiv},
       eprint = {1203.5543},
 primaryClass = {astro-ph.SR},
       adsurl = {https://ui.adsabs.harvard.edu/abs/2012ApJ...752...56F}
}

@ARTICLE{kirkpatrick2000,
       author = {{Kirkpatrick}, J. Davy and {Reid}, I. Neill and {Liebert}, James and {Gizis}, John E. and {Burgasser}, Adam J. and {Monet}, David G. and {Dahn}, Conard C. and {Nelson}, Brant and {Williams}, Rik J.},
        title = "{67 Additional L Dwarfs Discovered by the Two Micron All Sky Survey}",
      journal = {\aj},
         year = 2000,
        month = jul,
       volume = {120},
       number = {1},
        pages = {447-472},
          doi = {10.1086/301427},
archivePrefix = {arXiv},
       eprint = {astro-ph/0003317},
 primaryClass = {astro-ph},
       adsurl = {https://ui.adsabs.harvard.edu/abs/2000AJ....120..447K}
}

@ARTICLE{geballe2002,
       author = {{Geballe}, T.~R. and {Knapp}, G.~R. and {Leggett}, S.~K. and {Fan}, X. and {Golimowski}, D.~A. and {Anderson}, S. and {Brinkmann}, J. and {Csabai}, I. and {Gunn}, J.~E. and {Hawley}, S.~L. and {Hennessy}, G. and {Henry}, T.~J. and {Hill}, G.~J. and {Hindsley}, R.~B. and {Ivezi{\'c}}, {\v{Z}}. and {Lupton}, R.~H. and {McDaniel}, A. and {Munn}, J.~A. and {Narayanan}, V.~K. and {Peng}, E. and {Pier}, J.~R. and {Rockosi}, C.~M. and {Schneider}, D.~P. and {Smith}, J. Allyn and {Strauss}, M.~A. and {Tsvetanov}, Z.~I. and {Uomoto}, A. and {York}, D.~G. and {Zheng}, W.},
        title = "{Toward Spectral Classification of L and T Dwarfs: Infrared and Optical Spectroscopy and Analysis}",
      journal = {\apj},
         year = 2002,
        month = jan,
       volume = {564},
       number = {1},
        pages = {466-481},
          doi = {10.1086/324078},
archivePrefix = {arXiv},
       eprint = {astro-ph/0108443},
 primaryClass = {astro-ph},
       adsurl = {https://ui.adsabs.harvard.edu/abs/2002ApJ...564..466G}
}

@ARTICLE{burgasser2002,
       author = {{Burgasser}, Adam J. and {Kirkpatrick}, J. Davy and {Brown}, Michael E. and {Reid}, I. Neill and {Burrows}, Adam and {Liebert}, James and {Matthews}, Keith and {Gizis}, John E. and {Dahn}, Conard C. and {Monet}, David G. and {Cutri}, Roc M. and {Skrutskie}, Michael F.},
        title = "{The Spectra of T Dwarfs. I. Near-Infrared Data and Spectral Classification}",
      journal = {\apj},
         year = 2002,
        month = jan,
       volume = {564},
       number = {1},
        pages = {421-451},
          doi = {10.1086/324033},
archivePrefix = {arXiv},
       eprint = {astro-ph/0108452},
 primaryClass = {astro-ph},
       adsurl = {https://ui.adsabs.harvard.edu/abs/2002ApJ...564..421B}
}

@ARTICLE{hawley2002,
       author = {{Hawley}, Suzanne L. and {Covey}, Kevin R. and {Knapp}, Gillian R. and {Golimowski}, David A. and {Fan}, Xiaohui and {Anderson}, Scott F. and {Gunn}, James E. and {Harris}, Hugh C. and {Ivezi{\'c}}, {\v{Z}}eljko and {Long}, Gary M. and {Lupton}, Robert H. and {McGehee}, Peregrine M. and {Narayanan}, Vijay and {Peng}, Eric and {Schlegel}, David and {Schneider}, Donald P. and {Spahn}, Emily Y. and {Strauss}, Michael A. and {Szkody}, Paula and {Tsvetanov}, Zlatan and {Walkowicz}, Lucianne M. and {Brinkmann}, J. and {Harvanek}, Michael and {Hennessy}, Gregory S. and {Kleinman}, S.~J. and {Krzesinski}, Jurek and {Long}, Dan and {Neilsen}, Eric H. and {Newman}, Peter R. and {Nitta}, Atsuko and {Snedden}, Stephanie A. and {York}, Donald G.},
        title = "{Characterization of M, L, and T Dwarfs in the Sloan Digital Sky Survey}",
      journal = {\aj},
         year = 2002,
        month = jun,
       volume = {123},
       number = {6},
        pages = {3409-3427},
          doi = {10.1086/340697},
archivePrefix = {arXiv},
       eprint = {astro-ph/0204065},
 primaryClass = {astro-ph},
       adsurl = {https://ui.adsabs.harvard.edu/abs/2002AJ....123.3409H}
}

@ARTICLE{burgasser2003,
       author = {{Burgasser}, Adam J. and {McElwain}, Michael W. and {Kirkpatrick}, J. Davy},
        title = "{The 2MASS Wide-Field T Dwarf Search. II. Discovery of Three T Dwarfs in the Southern Hemisphere}",
      journal = {\aj},
         year = 2003,
        month = nov,
       volume = {126},
       number = {5},
        pages = {2487-2494},
          doi = {10.1086/378608},
archivePrefix = {arXiv},
       eprint = {astro-ph/0307374},
 primaryClass = {astro-ph},
       adsurl = {https://ui.adsabs.harvard.edu/abs/2003AJ....126.2487B}
}

@ARTICLE{burgasser2003b,
       author = {{Burgasser}, Adam J. and {Kirkpatrick}, J. Davy and {Burrows}, Adam and {Liebert}, James and {Reid}, I. Neill and {Gizis}, John E. and {McGovern}, Mark R. and {Prato}, L. and {McLean}, Ian S.},
        title = "{The First Substellar Subdwarf? Discovery of a Metal-poor L Dwarf with Halo Kinematics}",
      journal = {\apj},
         year = 2003,
        month = aug,
       volume = {592},
       number = {2},
        pages = {1186-1192},
          doi = {10.1086/375813},
archivePrefix = {arXiv},
       eprint = {astro-ph/0304174},
 primaryClass = {astro-ph},
       adsurl = {https://ui.adsabs.harvard.edu/abs/2003ApJ...592.1186B}
}

@ARTICLE{fan2000,
       author = {{Fan}, Xiaohui and {Knapp}, G.~R. and {Strauss}, Michael A. and {Gunn}, James E. and {Lupton}, Robert H. and {Ivezi{\'c}}, {\v{Z}}eljko and {Rockosi}, Constance M. and {Yanny}, Brian and {Kent}, Stephen and {Schneider}, Donald P. and {Kirkpatrick}, J. Davy and {Annis}, James and {Bastian}, Steven and {Berman}, Eileen and {Brinkmann}, J. and {Csabai}, Istv{\'a}n and {Federwitz}, Glenn R. and {Fukugita}, Masataka and {Gurbani}, Vijay K. and {Hennessy}, G.~S. and {Hindsley}, Robert B. and {Ichikawa}, Takashi and {Lamb}, D.~Q. and {Lindenmeyer}, Carl and {Mantsch}, P.~M. and {McKay}, Timothy A. and {Munn}, Jeffrey A. and {Nash}, Thomas and {Okamura}, Sadanori and {Pauls}, A. George and {Pier}, Jeffrey R. and {Rechenmacher}, Ron and {Rivetta}, Claudio H. and {Sergey}, Gary and {Stoughton}, Chris and {Szalay}, Alexander S. and {Szokoly}, Gyula P. and {Tucker}, Douglas L. and {York}, Donald G. and {SDSS Collaboration}},
        title = "{L Dwarfs Found in Sloan Digital Sky Survey Commissioning Imaging Data}",
      journal = {\aj},
         year = 2000,
        month = feb,
       volume = {119},
       number = {2},
        pages = {928-935},
          doi = {10.1086/301224},
archivePrefix = {arXiv},
       eprint = {astro-ph/9909263},
 primaryClass = {astro-ph},
       adsurl = {https://ui.adsabs.harvard.edu/abs/2000AJ....119..928F}
}

@ARTICLE{burgasser2000,
       author = {{Burgasser}, Adam J. and {Wilson}, John C. and {Kirkpatrick}, J. Davy and {Skrutskie}, Michael F. and {Colonno}, Michael R. and {Enos}, Alan T. and {Smith}, J.~D. and {Henderson}, Charles P. and {Gizis}, John E. and {Brown}, Michael E. and {Houck}, James R.},
        title = "{Discovery of a Bright Field Methane (T-Type) Brown Dwarf by 2MASS}",
      journal = {\aj},
         year = 2000,
        month = aug,
       volume = {120},
       number = {2},
        pages = {1100-1105},
          doi = {10.1086/301475},
archivePrefix = {arXiv},
       eprint = {astro-ph/0004239},
 primaryClass = {astro-ph},
       adsurl = {https://ui.adsabs.harvard.edu/abs/2000AJ....120.1100B}
}

@ARTICLE{leggett2000,
       author = {{Leggett}, S.~K. and {Geballe}, T.~R. and {Fan}, Xiaohui and {Schneider}, Donald P. and {Gunn}, James E. and {Lupton}, Robert H. and {Knapp}, G.~R. and {Strauss}, Michael A. and {McDaniel}, Alex and {Golimowski}, David A. and {Henry}, Todd J. and {Peng}, Eric and {Tsvetanov}, Zlatan I. and {Uomoto}, Alan and {Zheng}, Wei and {Hill}, G.~J. and {Ramsey}, L.~W. and {Anderson}, Scott F. and {Annis}, James A. and {Bahcall}, Neta A. and {Brinkmann}, J. and {Chen}, Bing and {Csabai}, Istv{\'a}n and {Fukugita}, Masataka and {Hennessy}, G.~S. and {Hindsley}, Robert B. and {Ivezi{\'c}}, {\v{Z}}eljko and {Lamb}, D.~Q. and {Munn}, Jeffrey A. and {Pier}, Jeffrey R. and {Schlegel}, David J. and {Smith}, J. Allyn and {Stoughton}, Chris and {Thakar}, A.~R. and {York}, Donald G.},
        title = "{The Missing Link: Early Methane (``T'') Dwarfs in the Sloan Digital Sky Survey}",
      journal = {\apjl},
         year = 2000,
        month = jun,
       volume = {536},
       number = {1},
        pages = {L35-L38},
          doi = {10.1086/312728},
archivePrefix = {arXiv},
       eprint = {astro-ph/0004408},
 primaryClass = {astro-ph},
       adsurl = {https://ui.adsabs.harvard.edu/abs/2000ApJ...536L..35L}
}

@ARTICLE{kirkpatrick1999,
       author = {{Kirkpatrick}, J. Davy and {Reid}, I. Neill and {Liebert}, James and {Cutri}, Roc M. and {Nelson}, Brant and {Beichman}, Charles A. and {Dahn}, Conard C. and {Monet}, David G. and {Gizis}, John E. and {Skrutskie}, Michael F.},
        title = "{Dwarfs Cooler than ``M``: The Definition of Spectral Type ``L'' Using Discoveries from the 2 Micron All-Sky Survey (2MASS)}",
      journal = {\apj},
         year = 1999,
        month = jul,
       volume = {519},
       number = {2},
        pages = {802-833},
          doi = {10.1086/307414},
       adsurl = {https://ui.adsabs.harvard.edu/abs/1999ApJ...519..802K}
}

@ARTICLE{burgasser1999,
       author = {{Burgasser}, Adam J. and {Kirkpatrick}, J. Davy and {Brown}, Michael E. and {Reid}, I. Neill and {Gizis}, John E. and {Dahn}, Conard C. and {Monet}, David G. and {Beichman}, Charles A. and {Liebert}, James and {Cutri}, Roc M. and {Skrutskie}, Michael F.},
        title = "{Discovery of Four Field Methane (T-Type) Dwarfs with the Two Micron All-Sky Survey}",
      journal = {\apjl},
         year = 1999,
        month = sep,
       volume = {522},
       number = {1},
        pages = {L65-L68},
          doi = {10.1086/312221},
archivePrefix = {arXiv},
       eprint = {astro-ph/9907019},
 primaryClass = {astro-ph},
       adsurl = {https://ui.adsabs.harvard.edu/abs/1999ApJ...522L..65B}
}

@ARTICLE{burgasser2003c,
       author = {{Burgasser}, Adam J. and {Kirkpatrick}, J. Davy and {McElwain}, Michael W. and {Cutri}, Roc M. and {Burgasser}, Albert J. and {Skrutskie}, Michael F.},
        title = "{The 2Mass Wide-Field T Dwarf Search. I. Discovery of a Bright T Dwarf within 10 Parsecs of the Sun}",
      journal = {\aj},
         year = 2003,
        month = feb,
       volume = {125},
       number = {2},
        pages = {850-857},
          doi = {10.1086/345975},
archivePrefix = {arXiv},
       eprint = {astro-ph/0211117},
 primaryClass = {astro-ph},
       adsurl = {https://ui.adsabs.harvard.edu/abs/2003AJ....125..850B}
}

@ARTICLE{strauss1999,
       author = {{Strauss}, Michael A. and {Fan}, Xiaohui and {Gunn}, James E. and {Leggett}, S.~K. and {Geballe}, T.~R. and {Pier}, Jeffrey R. and {Lupton}, Robert H. and {Knapp}, G.~R. and {Annis}, James and {Brinkmann}, J. and {Crocker}, J.~H. and {Csabai}, Istv{\'a}n and {Fukugita}, Masataka and {Golimowski}, David A. and {Harris}, Frederick H. and {Hennessy}, G.~S. and {Hindsley}, Robert B. and {Ivezi{\'c} }, {\v{Z}}eljko and {Kent}, Stephen and {Lamb}, D.~Q. and {Munn}, Jeffrey A. and {Newberg}, Heidi Jo and {Rechenmacher}, Ron and {Schneider}, Donald P. and {Smith}, J. Allyn and {Stoughton}, Chris and {Tucker}, Douglas L. and {Waddell}, Patrick and {York}, Donald G.},
        title = "{The Discovery of a Field Methane Dwarf from Sloan Digital Sky Survey Commissioning Data}",
      journal = {\apjl},
         year = 1999,
        month = sep,
       volume = {522},
       number = {1},
        pages = {L61-L64},
          doi = {10.1086/312218},
archivePrefix = {arXiv},
       eprint = {astro-ph/9905391},
 primaryClass = {astro-ph},
       adsurl = {https://ui.adsabs.harvard.edu/abs/1999ApJ...522L..61S}
}

@ARTICLE{burgasser2004,
       author = {{Burgasser}, Adam J. and {McElwain}, Michael W. and {Kirkpatrick}, J. Davy and {Cruz}, Kelle L. and {Tinney}, Chris G. and {Reid}, I. Neill},
        title = "{The 2MASS Wide-Field T Dwarf Search. III. Seven New T Dwarfs and Other Cool Dwarf Discoveries}",
      journal = {\aj},
         year = 2004,
        month = may,
       volume = {127},
       number = {5},
        pages = {2856-2870},
          doi = {10.1086/383549},
archivePrefix = {arXiv},
       eprint = {astro-ph/0402325},
 primaryClass = {astro-ph},
       adsurl = {https://ui.adsabs.harvard.edu/abs/2004AJ....127.2856B}
}

@ARTICLE{schneider2002,
       author = {{Schneider}, Donald P. and {Knapp}, Gillian R. and {Hawley}, Suzanne L. and {Covey}, Kevin R. and {Fan}, Xiaohui and {Ramsey}, Lawrence W. and {Richards}, Gordon T. and {Strauss}, Michael A. and {Gunn}, James E. and {Hill}, Gary J. and {MacQueen}, Phillip J. and {Adams}, Mark T. and {Hill}, Grant M. and {Ivezi{\'c}}, {\v{Z}}eljko and {Lupton}, Robert H. and {Pier}, Jeffrey R. and {Saxe}, David H. and {Shetrone}, Matthew and {Tufts}, Joseph R. and {Wolf}, Marsha J. and {Brinkmann}, J. and {Csabai}, Istv{\'a}n and {Hennessy}, G.~S. and {York}, Donald G.},
        title = "{L Dwarfs Found in Sloan Digital Sky Survey Commissioning Data. II. Hobby-Eberly Telescope Observations}",
      journal = {\aj},
         year = 2002,
        month = jan,
       volume = {123},
       number = {1},
        pages = {458-465},
          doi = {10.1086/338095},
archivePrefix = {arXiv},
       eprint = {astro-ph/0110273},
 primaryClass = {astro-ph},
       adsurl = {https://ui.adsabs.harvard.edu/abs/2002AJ....123..458S}
}

@ARTICLE{tinney2005,
       author = {{Tinney}, C.~G. and {Burgasser}, Adam J. and {Kirkpatrick}, J. Davy and {McElwain}, Michael W.},
        title = "{The 2MASS Wide-Field T Dwarf Search. IV. Hunting Out T Dwarfs with Methane Imaging}",
      journal = {\aj},
         year = 2005,
        month = nov,
       volume = {130},
       number = {5},
        pages = {2326-2346},
          doi = {10.1086/491734},
archivePrefix = {arXiv},
       eprint = {astro-ph/0508150},
 primaryClass = {astro-ph},
       adsurl = {https://ui.adsabs.harvard.edu/abs/2005AJ....130.2326T}
}

@ARTICLE{knapp2004,
       author = {{Knapp}, G.~R. and {Leggett}, S.~K. and {Fan}, X. and {Marley}, M.~S. and {Geballe}, T.~R. and {Golimowski}, D.~A. and {Finkbeiner}, D. and {Gunn}, J.~E. and {Hennawi}, J. and {Ivezi{\'c}}, Z. and {Lupton}, R.~H. and {Schlegel}, D.~J. and {Strauss}, M.~A. and {Tsvetanov}, Z.~I. and {Chiu}, K. and {Hoversten}, E.~A. and {Glazebrook}, K. and {Zheng}, W. and {Hendrickson}, M. and {Williams}, C.~C. and {Uomoto}, A. and {Vrba}, F.~J. and {Henden}, A.~A. and {Luginbuhl}, C.~B. and {Guetter}, H.~H. and {Munn}, J.~A. and {Canzian}, B. and {Schneider}, Donald P. and {Brinkmann}, J.},
        title = "{Near-Infrared Photometry and Spectroscopy of L and T Dwarfs: The Effects of Temperature, Clouds, and Gravity}",
      journal = {\aj},
         year = 2004,
        month = jun,
       volume = {127},
       number = {6},
        pages = {3553-3578},
          doi = {10.1086/420707},
archivePrefix = {arXiv},
       eprint = {astro-ph/0402451},
 primaryClass = {astro-ph},
       adsurl = {https://ui.adsabs.harvard.edu/abs/2004AJ....127.3553K}
}

@ARTICLE{gizis2003,
       author = {{Gizis}, John E. and {Reid}, I. Neill and {Knapp}, Gillian R. and {Liebert}, James and {Kirkpatrick}, J. Davy and {Koerner}, David W. and {Burgasser}, Adam J.},
        title = "{Hubble Space Telescope Observations of Binary Very Low Mass Stars and Brown Dwarfs}",
      journal = {\aj},
         year = 2003,
        month = jun,
       volume = {125},
       number = {6},
        pages = {3302-3310},
          doi = {10.1086/374991},
archivePrefix = {arXiv},
       eprint = {astro-ph/0302526},
 primaryClass = {astro-ph},
       adsurl = {https://ui.adsabs.harvard.edu/abs/2003AJ....125.3302G}
}

@ARTICLE{marocco2010,
       author = {{Marocco}, F. and {Smart}, R.~L. and {Jones}, H.~R.~A. and {Burningham}, B. and {Lattanzi}, M.~G. and {Leggett}, S.~K. and {Lucas}, P.~W. and {Tinney}, C.~G. and {Adamson}, A. and {Evans}, D.~W. and {Lodieu}, N. and {Murray}, D.~N. and {Pinfield}, D.~J. and {Tamura}, M.},
        title = "{Parallaxes and physical properties of 11 mid-to-late T dwarfs}",
      journal = {\aap},
         year = 2010,
        month = dec,
       volume = {524},
          eid = {A38},
        pages = {A38},
          doi = {10.1051/0004-6361/201015394},
archivePrefix = {arXiv},
       eprint = {1010.1135},
 primaryClass = {astro-ph.SR},
       adsurl = {https://ui.adsabs.harvard.edu/abs/2010A&A...524A..38M}
}

@ARTICLE{smart2013,
       author = {{Smart}, R.~L. and {Tinney}, C.~G. and {Bucciarelli}, B. and {Marocco}, F. and {Abbas}, U. and {Andrei}, A. and {Bernardi}, G. and {Burningham}, B. and {Cardoso}, C. and {Costa}, E. and {Crosta}, M.~T. and {Dapr{\'a}}, M. and {Day-Jones}, A. and {Goldman}, B. and {Jones}, H.~R.~A. and {Lattanzi}, M.~G. and {Leggett}, S.~K. and {Lucas}, P. and {Mendez}, R. and {Penna}, J.~L. and {Pinfield}, D. and {Smith}, L. and {Sozzetti}, A. and {Vecchiato}, A.},
        title = "{NPARSEC: NTT Parallaxes of Southern Extremely Cool objects. Goals, targets, procedures and first results}",
      journal = {\mnras},
         year = 2013,
        month = aug,
       volume = {433},
       number = {3},
        pages = {2054-2063},
          doi = {10.1093/mnras/stt876},
archivePrefix = {arXiv},
       eprint = {1306.4527},
 primaryClass = {astro-ph.SR},
       adsurl = {https://ui.adsabs.harvard.edu/abs/2013MNRAS.433.2054S}
}

@ARTICLE{burgasser2008,
       author = {{Burgasser}, Adam J. and {Tinney}, C.~G. and {Cushing}, Michael C. and {Saumon}, Didier and {Marley}, Mark S. and {Bennett}, Clara S. and {Kirkpatrick}, J. Davy},
        title = "{2MASS J09393548-2448279: The Coldest and Least Luminous Brown Dwarf Binary Known?}",
      journal = {\apjl},
         year = 2008,
        month = dec,
       volume = {689},
       number = {1},
        pages = {L53},
          doi = {10.1086/595747},
       adsurl = {https://ui.adsabs.harvard.edu/abs/2008ApJ...689L..53B}
}

@ARTICLE{manjavacas2013,
       author = {{Manjavacas}, E. and {Goldman}, B. and {Reffert}, S. and {Henning}, T.},
        title = "{Parallax measurements of cool brown dwarfs}",
      journal = {\aap},
         year = 2013,
        month = dec,
       volume = {560},
          eid = {A52},
        pages = {A52},
          doi = {10.1051/0004-6361/201321720},
archivePrefix = {arXiv},
       eprint = {1310.2191},
 primaryClass = {astro-ph.SR},
       adsurl = {https://ui.adsabs.harvard.edu/abs/2013A&A...560A..52M}
}

@ARTICLE{burgasser2006,
       author = {{Burgasser}, Adam J. and {Geballe}, T.~R. and {Leggett}, S.~K. and {Kirkpatrick}, J. Davy and {Golimowski}, David A.},
        title = "{A Unified Near-Infrared Spectral Classification Scheme for T Dwarfs}",
      journal = {\apj},
         year = 2006,
        month = feb,
       volume = {637},
       number = {2},
        pages = {1067-1093},
          doi = {10.1086/498563},
archivePrefix = {arXiv},
       eprint = {astro-ph/0510090},
 primaryClass = {astro-ph},
       adsurl = {https://ui.adsabs.harvard.edu/abs/2006ApJ...637.1067B}
}

@ARTICLE{burgasser2005,
       author = {{Burgasser}, Adam J. and {Reid}, I. Neill and {Leggett}, S.~K. and {Kirkpatrick}, J. Davy and {Liebert}, James and {Burrows}, Adam},
        title = "{SDSS J042348.57-041403.5AB: A Brown Dwarf Binary Straddling the L/T Transition}",
      journal = {\apjl},
         year = 2005,
        month = dec,
       volume = {634},
       number = {2},
        pages = {L177-L180},
          doi = {10.1086/498866},
archivePrefix = {arXiv},
       eprint = {astro-ph/0510580},
 primaryClass = {astro-ph},
       adsurl = {https://ui.adsabs.harvard.edu/abs/2005ApJ...634L.177B}
}

@ARTICLE{burgasser2007,
       author = {{Burgasser}, Adam J. and {Cruz}, Kelle L. and {Kirkpatrick}, J. Davy},
        title = "{Optical Spectroscopy of 2MASS Color-selected Ultracool Subdwarfs}",
      journal = {\apj},
         year = 2007,
        month = mar,
       volume = {657},
       number = {1},
        pages = {494-510},
          doi = {10.1086/510148},
archivePrefix = {arXiv},
       eprint = {astro-ph/0610096},
 primaryClass = {astro-ph},
       adsurl = {https://ui.adsabs.harvard.edu/abs/2007ApJ...657..494B}
}

@ARTICLE{kirkpatrick2008,
       author = {{Kirkpatrick}, J. Davy and {Cruz}, Kelle L. and {Barman}, Travis S. and {Burgasser}, Adam J. and {Looper}, Dagny L. and {Tinney}, C.~G. and {Gelino}, Christopher R. and {Lowrance}, Patrick J. and {Liebert}, James and {Carpenter}, John M. and {Hillenbrand}, Lynne A. and {Stauffer}, John R.},
        title = "{A Sample of Very Young Field L Dwarfs and Implications for the Brown Dwarf ``Lithium Test'' at Early Ages}",
      journal = {\apj},
         year = 2008,
        month = dec,
       volume = {689},
       number = {2},
        pages = {1295-1326},
          doi = {10.1086/592768},
archivePrefix = {arXiv},
       eprint = {0808.3153},
 primaryClass = {astro-ph},
       adsurl = {https://ui.adsabs.harvard.edu/abs/2008ApJ...689.1295K}
}

@ARTICLE{reid2001,
       author = {{Reid}, I. Neill and {Gizis}, John E. and {Kirkpatrick}, J. Davy and {Koerner}, D.~W.},
        title = "{A Search for L Dwarf Binary Systems}",
      journal = {\aj},
         year = 2001,
        month = jan,
       volume = {121},
       number = {1},
        pages = {489-502},
          doi = {10.1086/318023},
archivePrefix = {arXiv},
       eprint = {astro-ph/0010202},
 primaryClass = {astro-ph},
       adsurl = {https://ui.adsabs.harvard.edu/abs/2001AJ....121..489R}
}

@ARTICLE{burgasser2003d,
       author = {{Burgasser}, Adam J. and {Kirkpatrick}, J. Davy and {Reid}, I. Neill and {Brown}, Michael E. and {Miskey}, Cherie L. and {Gizis}, John E.},
        title = "{Binarity in Brown Dwarfs: T Dwarf Binaries Discovered with the Hubble Space Telescope Wide Field Planetary Camera 2}",
      journal = {\apj},
         year = 2003,
        month = mar,
       volume = {586},
       number = {1},
        pages = {512-526},
          doi = {10.1086/346263},
archivePrefix = {arXiv},
       eprint = {astro-ph/0211470},
 primaryClass = {astro-ph},
       adsurl = {https://ui.adsabs.harvard.edu/abs/2003ApJ...586..512B}
}

@ARTICLE{cruz2007,
       author = {{Cruz}, Kelle L. and {Reid}, I. Neill and {Kirkpatrick}, J. Davy and {Burgasser}, Adam J. and {Liebert}, James and {Solomon}, Adam R. and {Schmidt}, Sarah J. and {Allen}, Peter R. and {Hawley}, Suzanne L. and {Covey}, Kevin R.},
        title = "{Meeting the Cool Neighbors. IX. The Luminosity Function of M7-L8 Ultracool Dwarfs in the Field}",
      journal = {\aj},
         year = 2007,
        month = feb,
       volume = {133},
       number = {2},
        pages = {439-467},
          doi = {10.1086/510132},
archivePrefix = {arXiv},
       eprint = {astro-ph/0609648},
 primaryClass = {astro-ph},
       adsurl = {https://ui.adsabs.harvard.edu/abs/2007AJ....133..439C}
}

@ARTICLE{leggett2002,
       author = {{Leggett}, S.~K. and {Golimowski}, David A. and {Fan}, Xiaohui and {Geballe}, T.~R. and {Knapp}, G.~R. and {Brinkmann}, J. and {Csabai}, Istv{\'a}n and {Gunn}, James E. and {Hawley}, Suzanne L. and {Henry}, Todd J. and {Hindsley}, Robert and {Ivezi{\'c}}, {\v{Z}}eljko and {Lupton}, Robert H. and {Pier}, Jeffrey R. and {Schneider}, Donald P. and {Smith}, J. Allyn and {Strauss}, Michael A. and {Uomoto}, Alan and {York}, D.~G.},
        title = "{Infrared Photometry of Late-M, L, and T Dwarfs}",
      journal = {\apj},
         year = 2002,
        month = jan,
       volume = {564},
       number = {1},
        pages = {452-465},
          doi = {10.1086/324037},
archivePrefix = {arXiv},
       eprint = {astro-ph/0108435},
 primaryClass = {astro-ph},
       adsurl = {https://ui.adsabs.harvard.edu/abs/2002ApJ...564..452L}
}

@ARTICLE{skrutskie2006,
       author = {{Skrutskie}, M.~F. and {Cutri}, R.~M. and {Stiening}, R. and {Weinberg}, M.~D. and {Schneider}, S. and {Carpenter}, J.~M. and {Beichman}, C. and {Capps}, R. and {Chester}, T. and {Elias}, J. and {Huchra}, J. and {Liebert}, J. and {Lonsdale}, C. and {Monet}, D.~G. and {Price}, S. and {Seitzer}, P. and {Jarrett}, T. and {Kirkpatrick}, J.~D. and {Gizis}, J.~E. and {Howard}, E. and {Evans}, T. and {Fowler}, J. and {Fullmer}, L. and {Hurt}, R. and {Light}, R. and {Kopan}, E.~L. and {Marsh}, K.~A. and {McCallon}, H.~L. and {Tam}, R. and {Van Dyk}, S. and {Wheelock}, S.},
        title = "{The Two Micron All Sky Survey (2MASS)}",
      journal = {\aj},
         year = 2006,
        month = feb,
       volume = {131},
       number = {2},
        pages = {1163-1183},
          doi = {10.1086/498708},
       adsurl = {https://ui.adsabs.harvard.edu/abs/2006AJ....131.1163S}
}

@ARTICLE{kirkpatrick2011,
       author = {{Kirkpatrick}, J. Davy and {Cushing}, Michael C. and {Gelino}, Christopher R. and {Griffith}, Roger L. and {Skrutskie}, Michael F. and {Marsh}, Kenneth A. and {Wright}, Edward L. and {Mainzer}, A. and {Eisenhardt}, Peter R. and {McLean}, Ian S. and {Thompson}, Maggie A. and {Bauer}, James M. and {Benford}, Dominic J. and {Bridge}, Carrie R. and {Lake}, Sean E. and {Petty}, Sara M. and {Stanford}, S.~A. and {Tsai}, Chao-Wei and {Bailey}, Vanessa and {Beichman}, Charles A. and {Bloom}, Joshua S. and {Bochanski}, John J. and {Burgasser}, Adam J. and {Capak}, Peter L. and {Cruz}, Kelle L. and {Hinz}, Philip M. and {Kartaltepe}, Jeyhan S. and {Knox}, Russell P. and {Manohar}, Swarnima and {Masters}, Daniel and {Morales-Calder{\'o}n}, Maria and {Prato}, Lisa A. and {Rodigas}, Timothy J. and {Salvato}, Mara and {Schurr}, Steven D. and {Scoville}, Nicholas Z. and {Simcoe}, Robert A. and {Stapelfeldt}, Karl R. and {Stern}, Daniel and {Stock}, Nathan D. and {Vacca}, William D.},
        title = "{The First Hundred Brown Dwarfs Discovered by the Wide-field Infrared Survey Explorer (WISE)}",
      journal = {\apjs},
         year = 2011,
        month = dec,
       volume = {197},
       number = {2},
          eid = {19},
        pages = {19},
          doi = {10.1088/0067-0049/197/2/19},
archivePrefix = {arXiv},
       eprint = {1108.4677},
 primaryClass = {astro-ph.SR},
       adsurl = {https://ui.adsabs.harvard.edu/abs/2011ApJS..197...19K}
}

@ARTICLE{best2021,
       author = {{Best}, William M.~J. and {Liu}, Michael C. and {Magnier}, Eugene A. and {Dupuy}, Trent J.},
        title = "{A Volume-limited Sample of Ultracool Dwarfs. I. Construction, Space Density, and a Gap in the L/T Transition}",
      journal = {\aj},
         year = 2021,
        month = jan,
       volume = {161},
       number = {1},
          eid = {42},
        pages = {42},
          doi = {10.3847/1538-3881/abc893},
archivePrefix = {arXiv},
       eprint = {2010.15853},
 primaryClass = {astro-ph.SR},
       adsurl = {https://ui.adsabs.harvard.edu/abs/2021AJ....161...42B}
}

@ARTICLE{chiu2006,
       author = {{Chiu}, K. and {Fan}, X. and {Leggett}, S.~K. and {Golimowski}, D.~A. and {Zheng}, W. and {Geballe}, T.~R. and {Schneider}, D.~P. and {Brinkmann}, J.},
        title = "{Seventy-One New L and T Dwarfs from the Sloan Digital Sky Survey}",
      journal = {\aj},
         year = 2006,
        month = jun,
       volume = {131},
       number = {5},
        pages = {2722-2736},
          doi = {10.1086/501431},
archivePrefix = {arXiv},
       eprint = {astro-ph/0601089},
 primaryClass = {astro-ph},
       adsurl = {https://ui.adsabs.harvard.edu/abs/2006AJ....131.2722C}
}

@ARTICLE{liu2016,
       author = {{Liu}, Michael C. and {Dupuy}, Trent J. and {Allers}, Katelyn N.},
        title = "{The Hawaii Infrared Parallax Program. II. Young Ultracool Field Dwarfs}",
      journal = {\apj},
         year = 2016,
        month = dec,
       volume = {833},
       number = {1},
          eid = {96},
        pages = {96},
          doi = {10.3847/1538-4357/833/1/96},
archivePrefix = {arXiv},
       eprint = {1612.02426},
 primaryClass = {astro-ph.SR},
       adsurl = {https://ui.adsabs.harvard.edu/abs/2016ApJ...833...96L}
}

@ARTICLE{leggett2010,
       author = {{Leggett}, S.~K. and {Burningham}, Ben and {Saumon}, D. and {Marley}, M.~S. and {Warren}, S.~J. and {Smart}, R.~L. and {Jones}, H.~R.~A. and {Lucas}, P.~W. and {Pinfield}, D.~J. and {Tamura}, Motohide},
        title = "{Mid-Infrared Photometry of Cold Brown Dwarfs: Diversity in Age, Mass, and Metallicity}",
      journal = {\apj},
         year = 2010,
        month = feb,
       volume = {710},
       number = {2},
        pages = {1627-1640},
          doi = {10.1088/0004-637X/710/2/1627},
archivePrefix = {arXiv},
       eprint = {1001.0762},
 primaryClass = {astro-ph.SR},
       adsurl = {https://ui.adsabs.harvard.edu/abs/2010ApJ...710.1627L}
}

@ARTICLE{kirkpatrick2010,
       author = {{Kirkpatrick}, J. Davy and {Looper}, Dagny L. and {Burgasser}, Adam J. and {Schurr}, Steven D. and {Cutri}, Roc M. and {Cushing}, Michael C. and {Cruz}, Kelle L. and {Sweet}, Anne C. and {Knapp}, Gillian R. and {Barman}, Travis S. and {Bochanski}, John J. and {Roellig}, Thomas L. and {McLean}, Ian S. and {McGovern}, Mark R. and {Rice}, Emily L.},
        title = "{Discoveries from a Near-infrared Proper Motion Survey Using Multi-epoch Two Micron All-Sky Survey Data}",
      journal = {\apjs},
         year = 2010,
        month = sep,
       volume = {190},
       number = {1},
        pages = {100-146},
          doi = {10.1088/0067-0049/190/1/100},
archivePrefix = {arXiv},
       eprint = {1008.3591},
 primaryClass = {astro-ph.SR},
       adsurl = {https://ui.adsabs.harvard.edu/abs/2010ApJS..190..100K}
}

@ARTICLE{salim2003,
       author = {{Salim}, Samir and {L{\'e}pine}, S{\'e}bastien and {Rich}, R. Michael and {Shara}, Michael M.},
        title = "{LSR 0602+3910: Discovery of a Bright Nearby L-Type Brown Dwarf}",
      journal = {\apjl},
         year = 2003,
        month = apr,
       volume = {586},
       number = {2},
        pages = {L149-L152},
          doi = {10.1086/374794},
archivePrefix = {arXiv},
       eprint = {astro-ph/0301282},
 primaryClass = {astro-ph},
       adsurl = {https://ui.adsabs.harvard.edu/abs/2003ApJ...586L.149S}
}

@ARTICLE{reyle2010,
       author = {{Reyl{\'e}}, C. and {Delorme}, P. and {Willott}, C.~J. and {Albert}, L. and {Delfosse}, X. and {Forveille}, T. and {Artigau}, E. and {Malo}, L. and {Hill}, G.~J. and {Doyon}, R.},
        title = "{The ultracool-field dwarf luminosity-function and space density from the Canada-France Brown Dwarf Survey}",
      journal = {\aap},
         year = 2010,
        month = nov,
       volume = {522},
          eid = {A112},
        pages = {A112},
          doi = {10.1051/0004-6361/200913234},
archivePrefix = {arXiv},
       eprint = {1008.2301},
 primaryClass = {astro-ph.SR},
       adsurl = {https://ui.adsabs.harvard.edu/abs/2010A&A...522A.112R}
}

@ARTICLE{delorme2008,
       author = {{Delorme}, P. and {Delfosse}, X. and {Albert}, L. and {Artigau}, E. and {Forveille}, T. and {Reyl{\'e}}, C. and {Allard}, F. and {Homeier}, D. and {Robin}, A.~C. and {Willott}, C.~J. and {Liu}, M.~C. and {Dupuy}, T.~J.},
        title = "{CFBDS J005910.90-011401.3: reaching the T-Y brown dwarf transition?}",
      journal = {\aap},
         year = 2008,
        month = may,
       volume = {482},
       number = {3},
        pages = {961-971},
          doi = {10.1051/0004-6361:20079317},
archivePrefix = {arXiv},
       eprint = {0802.4387},
 primaryClass = {astro-ph},
       adsurl = {https://ui.adsabs.harvard.edu/abs/2008A&A...482..961D}
}

@ARTICLE{scholz2011,
       author = {{Scholz}, R. -D. and {Bihain}, G. and {Schnurr}, O. and {Storm}, J.},
        title = "{Two very nearby (d \raisebox{-0.5ex}\textasciitilde 5 pc) ultracool brown dwarfs detected by their large proper motions from WISE, 2MASS, and SDSS data}",
      journal = {\aap},
         year = 2011,
        month = aug,
       volume = {532},
          eid = {L5},
        pages = {L5},
          doi = {10.1051/0004-6361/201117297},
archivePrefix = {arXiv},
       eprint = {1105.4059},
 primaryClass = {astro-ph.GA},
       adsurl = {https://ui.adsabs.harvard.edu/abs/2011A&A...532L...5S}
}

@ARTICLE{reid2006b,
       author = {{Reid}, I. Neill and {Lewitus}, E. and {Allen}, P.~R. and {Cruz}, Kelle L. and {Burgasser}, Adam J.},
        title = "{A Search for Binary Systems among the Nearest L Dwarfs}",
      journal = {\aj},
         year = 2006,
        month = aug,
       volume = {132},
       number = {2},
        pages = {891-901},
          doi = {10.1086/505626},
archivePrefix = {arXiv},
       eprint = {astro-ph/0606331},
 primaryClass = {astro-ph},
       adsurl = {https://ui.adsabs.harvard.edu/abs/2006AJ....132..891R}
}

@ARTICLE{cruz2003,
       author = {{Cruz}, Kelle L. and {Reid}, I. Neill and {Liebert}, James and {Kirkpatrick}, J. Davy and {Lowrance}, Patrick J.},
        title = "{Meeting the Cool Neighbors. V. A 2MASS-Selected Sample of Ultracool Dwarfs}",
      journal = {\aj},
         year = 2003,
        month = nov,
       volume = {126},
       number = {5},
        pages = {2421-2448},
          doi = {10.1086/378607},
archivePrefix = {arXiv},
       eprint = {astro-ph/0307429},
 primaryClass = {astro-ph},
       adsurl = {https://ui.adsabs.harvard.edu/abs/2003AJ....126.2421C}
}

@ARTICLE{looper2007,
       author = {{Looper}, Dagny L. and {Kirkpatrick}, J. Davy and {Burgasser}, Adam J.},
        title = "{Discovery of 11 New T Dwarfs in the Two Micron All Sky Survey, Including a Possible L/T Transition Binary}",
      journal = {\aj},
         year = 2007,
        month = sep,
       volume = {134},
       number = {3},
        pages = {1162-1182},
          doi = {10.1086/520645},
archivePrefix = {arXiv},
       eprint = {0706.1601},
 primaryClass = {astro-ph},
       adsurl = {https://ui.adsabs.harvard.edu/abs/2007AJ....134.1162L}
}

@ARTICLE{loutrel2011,
       author = {{Loutrel}, N.~P. and {Luhman}, K.~L. and {Lowrance}, P.~J. and {Bochanski}, J.~J.},
        title = "{Discovery of a Companion at the L/T Transition with the Wide-field Infrared Survey Explorer}",
      journal = {\apj},
         year = 2011,
        month = oct,
       volume = {739},
       number = {2},
          eid = {81},
        pages = {81},
          doi = {10.1088/0004-637X/739/2/81},
archivePrefix = {arXiv},
       eprint = {1107.1812},
 primaryClass = {astro-ph.GA},
       adsurl = {https://ui.adsabs.harvard.edu/abs/2011ApJ...739...81L}
}

@INPROCEEDINGS{wilson2003,
       author = {{Wilson}, J.~C. and {Miller}, Neal A. and {Gizis}, J.~E. and {Skrutskie}, M.~F. and {Houck}, J.~R. and {Kirkpatrick}, J. Davy and {Burgasser}, A.~J. and {Monet}, D.~G.},
        title = "{New M and L Dwarfs Confirmed with CorMASS}",
    booktitle = {Brown Dwarfs},
         year = 2003,
       editor = {{Mart{\'\i}n}, Eduardo},
       series = {IAU Symposium},
       volume = {211},
        month = jun,
        pages = {197},
       adsurl = {https://ui.adsabs.harvard.edu/abs/2003IAUS..211..197W}
}

@ARTICLE{thorstensen2003,
       author = {{Thorstensen}, John R. and {Kirkpatrick}, J. Davy},
        title = "{Serendipitous Discovery and Parallax of a Nearby L Dwarf}",
      journal = {\pasp},
         year = 2003,
        month = oct,
       volume = {115},
       number = {812},
        pages = {1207-1210},
          doi = {10.1086/378080},
archivePrefix = {arXiv},
       eprint = {astro-ph/0307295},
 primaryClass = {astro-ph},
       adsurl = {https://ui.adsabs.harvard.edu/abs/2003PASP..115.1207T}
}

@ARTICLE{castro2012,
       author = {{Castro}, Philip J. and {Gizis}, John E.},
        title = "{Discovery of a Late L Dwarf: WISEP J060738.65+242953.4}",
      journal = {\apj},
         year = 2012,
        month = feb,
       volume = {746},
       number = {1},
          eid = {3},
        pages = {3},
          doi = {10.1088/0004-637X/746/1/3},
archivePrefix = {arXiv},
       eprint = {1110.4351},
 primaryClass = {astro-ph.SR},
       adsurl = {https://ui.adsabs.harvard.edu/abs/2012ApJ...746....3C}
}

@ARTICLE{lucas2010,
       author = {{Lucas}, P.~W. and {Tinney}, C.~G. and {Burningham}, Ben and {Leggett}, S.~K. and {Pinfield}, David J. and {Smart}, Richard and {Jones}, Hugh R.~A. and {Marocco}, Federico and {Barber}, Robert J. and {Yurchenko}, Sergei N. and {Tennyson}, Jonathan and {Ishii}, Miki and {Tamura}, Motohide and {Day-Jones}, Avril C. and {Adamson}, Andrew and {Allard}, France and {Homeier}, Derek},
        title = "{The discovery of a very cool, very nearby brown dwarf in the Galactic plane}",
      journal = {\mnras},
         year = 2010,
        month = oct,
       volume = {408},
       number = {1},
        pages = {L56-L60},
          doi = {10.1111/j.1745-3933.2010.00927.x},
archivePrefix = {arXiv},
       eprint = {1004.0317},
 primaryClass = {astro-ph.SR},
       adsurl = {https://ui.adsabs.harvard.edu/abs/2010MNRAS.408L..56L}
}

@ARTICLE{aberasturi2011,
       author = {{Aberasturi}, M. and {Solano}, E. and {Mart{\'\i}n}, E.~L.},
        title = "{WISE/2MASS-SDSS brown dwarfs candidates using Virtual Observatory tools}",
      journal = {\aap},
         year = 2011,
        month = oct,
       volume = {534},
          eid = {L7},
        pages = {L7},
          doi = {10.1051/0004-6361/201117822},
archivePrefix = {arXiv},
       eprint = {1108.1724},
 primaryClass = {astro-ph.SR},
       adsurl = {https://ui.adsabs.harvard.edu/abs/2011A&A...534L...7A}
}

@ARTICLE{zhang2009,
       author = {{Zhang}, Z.~H. and {Pokorny}, R.~S. and {Jones}, H.~R.~A. and {Pinfield}, D.~J. and {Chen}, P.~S. and {Han}, Z. and {Chen}, D. and {G{\'a}lvez-Ortiz}, M.~C. and {Burningham}, B.},
        title = "{Ultra-cool dwarfs: new discoveries, proper motions, and improved spectral typing from SDSS and 2MASS photometric colors}",
      journal = {\aap},
         year = 2009,
        month = apr,
       volume = {497},
       number = {2},
        pages = {619-633},
          doi = {10.1051/0004-6361/200810314},
archivePrefix = {arXiv},
       eprint = {0902.2798},
 primaryClass = {astro-ph.GA},
       adsurl = {https://ui.adsabs.harvard.edu/abs/2009A&A...497..619Z}
}

@ARTICLE{deacon2011,
       author = {{Deacon}, Niall R. and {Liu}, Michael C. and {Magnier}, Eugene A. and {Bowler}, Brendan P. and {Goldman}, Bertrand and {Redstone}, Joshua A. and {Burgett}, W.~S. and {Chambers}, K.~C. and {Flewelling}, H. and {Kaiser}, N. and {Lupton}, R.~H. and {Morgan}, J.~S. and {Price}, P.~A. and {Sweeney}, W.~E. and {Tonry}, J.~L. and {Wainscoat}, R.~J. and {Waters}, C.},
        title = "{Four New T Dwarfs Identified in Pan-STARRS 1 Commissioning Data}",
      journal = {\aj},
         year = 2011,
        month = sep,
       volume = {142},
       number = {3},
          eid = {77},
        pages = {77},
          doi = {10.1088/0004-6256/142/3/77},
archivePrefix = {arXiv},
       eprint = {1106.3105},
 primaryClass = {astro-ph.SR},
       adsurl = {https://ui.adsabs.harvard.edu/abs/2011AJ....142...77D}
}

@ARTICLE{burningham2010,
       author = {{Burningham}, Ben and {Pinfield}, D.~J. and {Lucas}, P.~W. and {Leggett}, S.~K. and {Deacon}, N.~R. and {Tamura}, M. and {Tinney}, C.~G. and {Lodieu}, N. and {Zhang}, Z.~H. and {Huelamo}, N. and {Jones}, H.~R.~A. and {Murray}, D.~N. and {Mortlock}, D.~J. and {Patel}, M. and {Barrado Y Navascu{\'e}s}, D. and {Zapatero Osorio}, M.~R. and {Ishii}, M. and {Kuzuhara}, M. and {Smart}, R.~L.},
        title = "{47 new T dwarfs from the UKIDSS Large Area Survey}",
      journal = {\mnras},
         year = 2010,
        month = aug,
       volume = {406},
       number = {3},
        pages = {1885-1906},
          doi = {10.1111/j.1365-2966.2010.16800.x},
archivePrefix = {arXiv},
       eprint = {1004.1912},
 primaryClass = {astro-ph.GA},
       adsurl = {https://ui.adsabs.harvard.edu/abs/2010MNRAS.406.1885B}
}

@ARTICLE{burningham2008,
       author = {{Burningham}, Ben and {Pinfield}, D.~J. and {Leggett}, S.~K. and {Tamura}, M. and {Lucas}, P.~W. and {Homeier}, D. and {Day-Jones}, A. and {Jones}, H.~R.~A. and {Clarke}, J.~R.~A. and {Ishii}, M. and {Kuzuhara}, M. and {Lodieu}, N. and {Zapatero Osorio}, M.~R. and {Venemans}, B.~P. and {Mortlock}, D.~J. and {Barrado Y Navascu{\'e}s}, D. and {Martin}, E.~L. and {Magazz{\`u}}, A.},
        title = "{Exploring the substellar temperature regime down to \raisebox{-0.5ex}\textasciitilde550K}",
      journal = {\mnras},
         year = 2008,
        month = nov,
       volume = {391},
       number = {1},
        pages = {320-333},
          doi = {10.1111/j.1365-2966.2008.13885.x},
archivePrefix = {arXiv},
       eprint = {0806.0067},
 primaryClass = {astro-ph},
       adsurl = {https://ui.adsabs.harvard.edu/abs/2008MNRAS.391..320B}
}

@ARTICLE{jameson2008,
       author = {{Jameson}, R.~F. and {Casewell}, S.~L. and {Bannister}, N.~P. and {Lodieu}, N. and {Keresztes}, K. and {Dobbie}, P.~D. and {Hodgkin}, S.~T.},
        title = "{Proper motions of field L and T dwarfs}",
      journal = {\mnras},
         year = 2008,
        month = mar,
       volume = {384},
       number = {4},
        pages = {1399-1413},
          doi = {10.1111/j.1365-2966.2007.12637.x},
archivePrefix = {arXiv},
       eprint = {0710.4786},
 primaryClass = {astro-ph},
       adsurl = {https://ui.adsabs.harvard.edu/abs/2008MNRAS.384.1399J}
}

@ARTICLE{gizis2000,
       author = {{Gizis}, John E. and {Monet}, David G. and {Reid}, I. Neill and {Kirkpatrick}, J. Davy and {Liebert}, James and {Williams}, Rik J.},
        title = "{New Neighbors from 2MASS: Activity and Kinematics at the Bottom of the Main Sequence}",
      journal = {\aj},
         year = 2000,
        month = aug,
       volume = {120},
       number = {2},
        pages = {1085-1099},
          doi = {10.1086/301456},
archivePrefix = {arXiv},
       eprint = {astro-ph/0004361},
 primaryClass = {astro-ph},
       adsurl = {https://ui.adsabs.harvard.edu/abs/2000AJ....120.1085G}
}

@ARTICLE{mclean2000,
       author = {{McLean}, Ian S. and {Wilcox}, Mavourneen K. and {Becklin}, E.~E. and {Figer}, Donald F. and {Gilbert}, Andrea M. and {Graham}, James R. and {Larkin}, James E. and {Levenson}, N.~A. and {Teplitz}, Harry I. and {Kirkpatrick}, J. Davy},
        title = "{J-Band Infrared Spectroscopy of a Sample of Brown Dwarfs Using NIRSPEC on Keck II}",
      journal = {\apjl},
         year = 2000,
        month = apr,
       volume = {533},
       number = {1},
        pages = {L45-L48},
          doi = {10.1086/312600},
archivePrefix = {arXiv},
       eprint = {astro-ph/0003035},
 primaryClass = {astro-ph},
       adsurl = {https://ui.adsabs.harvard.edu/abs/2000ApJ...533L..45M}
}

@ARTICLE{metchev2008,
       author = {{Metchev}, Stanimir A. and {Kirkpatrick}, J. Davy and {Berriman}, G. Bruce and {Looper}, Dagny},
        title = "{A Cross-Match of 2MASS and SDSS: Newly Found L and T Dwarfs and an Estimate of the Space Density of T Dwarfs}",
      journal = {\apj},
         year = 2008,
        month = apr,
       volume = {676},
       number = {2},
        pages = {1281-1306},
          doi = {10.1086/524721},
archivePrefix = {arXiv},
       eprint = {0710.4157},
 primaryClass = {astro-ph},
       adsurl = {https://ui.adsabs.harvard.edu/abs/2008ApJ...676.1281M}
}

@ARTICLE{burgasser2011,
       author = {{Burgasser}, Adam J. and {Bardalez-Gagliuffi}, Daniella C. and {Gizis}, John E.},
        title = "{Hubble Space Telescope Imaging and Spectral Analysis of Two Brown Dwarf Binaries at the L Dwarf/T Dwarf Transition}",
      journal = {\aj},
         year = 2011,
        month = mar,
       volume = {141},
       number = {3},
          eid = {70},
        pages = {70},
          doi = {10.1088/0004-6256/141/3/70},
archivePrefix = {arXiv},
       eprint = {1011.0798},
 primaryClass = {astro-ph.SR},
       adsurl = {https://ui.adsabs.harvard.edu/abs/2011AJ....141...70B}
}

@ARTICLE{burgasser2011b,
       author = {{Burgasser}, Adam J. and {Cushing}, Michael C. and {Kirkpatrick}, J. Davy and {Gelino}, Christopher R. and {Griffith}, Roger L. and {Looper}, Dagny L. and {Tinney}, Christopher and {Simcoe}, Robert A. and {Bochanski}, John J. and {Skrutskie}, Michael F. and {Mainzer}, A. and {Thompson}, Maggie A. and {Marsh}, Kenneth A. and {Bauer}, James M. and {Wright}, Edward L.},
        title = "{Fire Spectroscopy of Five Late-type T Dwarfs Discovered with the Wide-field Infrared Survey Explorer}",
      journal = {\apj},
         year = 2011,
        month = jul,
       volume = {735},
       number = {2},
          eid = {116},
        pages = {116},
          doi = {10.1088/0004-637X/735/2/116},
archivePrefix = {arXiv},
       eprint = {1104.2537},
 primaryClass = {astro-ph.SR},
       adsurl = {https://ui.adsabs.harvard.edu/abs/2011ApJ...735..116B}
}

@ARTICLE{burgasser2004b,
       author = {{Burgasser}, Adam J.},
        title = "{Discovery of a Second L Subdwarf in the Two Micron All Sky Survey}",
      journal = {\apjl},
         year = 2004,
        month = oct,
       volume = {614},
       number = {1},
        pages = {L73-L76},
          doi = {10.1086/425418},
archivePrefix = {arXiv},
       eprint = {astro-ph/0409179},
 primaryClass = {astro-ph},
       adsurl = {https://ui.adsabs.harvard.edu/abs/2004ApJ...614L..73B}
}

@ARTICLE{gelino2011,
       author = {{Gelino}, Christopher R. and {Kirkpatrick}, J. Davy and {Cushing}, Michael C. and {Eisenhardt}, Peter R. and {Griffith}, Roger L. and {Mainzer}, Amanda K. and {Marsh}, Kenneth A. and {Skrutskie}, Michael F. and {Wright}, Edward L.},
        title = "{WISE Brown Dwarf Binaries: The Discovery of a T5+T5 and a T8.5+T9 System}",
      journal = {\aj},
         year = 2011,
        month = aug,
       volume = {142},
       number = {2},
          eid = {57},
        pages = {57},
          doi = {10.1088/0004-6256/142/2/57},
archivePrefix = {arXiv},
       eprint = {1106.3142},
 primaryClass = {astro-ph.SR},
       adsurl = {https://ui.adsabs.harvard.edu/abs/2011AJ....142...57G}
}

@ARTICLE{gizis2011b,
       author = {{Gizis}, John E. and {Burgasser}, Adam J. and {Faherty}, Jacqueline K. and {Castro}, Philip J. and {Shara}, Michael M.},
        title = "{WISEP J180026.60+013453.1: A nearby late-L Dwarf near the Galactic Plane}",
      journal = {\aj},
         year = 2011,
        month = nov,
       volume = {142},
       number = {5},
          eid = {171},
        pages = {171},
          doi = {10.1088/0004-6256/142/5/171},
archivePrefix = {arXiv},
       eprint = {1109.0054},
 primaryClass = {astro-ph.SR},
       adsurl = {https://ui.adsabs.harvard.edu/abs/2011AJ....142..171G}
}

@ARTICLE{looper2008,
       author = {{Looper}, Dagny L. and {Gelino}, Christopher R. and {Burgasser}, Adam J. and {Kirkpatrick}, J. Davy},
        title = "{Discovery of a T Dwarf Binary with the Largest Known J-Band Flux Reversal}",
      journal = {\apj},
         year = 2008,
        month = oct,
       volume = {685},
       number = {2},
        pages = {1183-1192},
          doi = {10.1086/590382},
archivePrefix = {arXiv},
       eprint = {0803.0544},
 primaryClass = {astro-ph},
       adsurl = {https://ui.adsabs.harvard.edu/abs/2008ApJ...685.1183L}
}

@ARTICLE{looper2008b,
       author = {{Looper}, Dagny L. and {Kirkpatrick}, J. Davy and {Cutri}, Roc M. and {Barman}, Travis and {Burgasser}, Adam J. and {Cushing}, Michael C. and {Roellig}, Thomas and {McGovern}, Mark R. and {McLean}, Ian S. and {Rice}, Emily and {Swift}, Brandon J. and {Schurr}, Steven D.},
        title = "{Discovery of Two Nearby Peculiar L Dwarfs from the 2MASS Proper-Motion Survey: Young or Metal-Rich?}",
      journal = {\apj},
         year = 2008,
        month = oct,
       volume = {686},
       number = {1},
        pages = {528-541},
          doi = {10.1086/591025},
archivePrefix = {arXiv},
       eprint = {0806.1059},
 primaryClass = {astro-ph},
       adsurl = {https://ui.adsabs.harvard.edu/abs/2008ApJ...686..528L}
}

@ARTICLE{gizis2011,
       author = {{Gizis}, John E. and {Troup}, Nicholas W. and {Burgasser}, Adam J.},
        title = "{A Very High Proper Motion Star and the First L Dwarf in the Kepler Field}",
      journal = {\apjl},
         year = 2011,
        month = aug,
       volume = {736},
       number = {2},
          eid = {L34},
        pages = {L34},
          doi = {10.1088/2041-8205/736/2/L34},
archivePrefix = {arXiv},
       eprint = {1106.4526},
 primaryClass = {astro-ph.SR},
       adsurl = {https://ui.adsabs.harvard.edu/abs/2011ApJ...736L..34G}
}

@ARTICLE{fan2001,
       author = {{Fan}, Xiaohui and {Narayanan}, Vijay K. and {Lupton}, Robert H. and {Strauss}, Michael A. and {Knapp}, Gillian R. and {Becker}, Robert H. and {White}, Richard L. and {Pentericci}, Laura and {Leggett}, S.~K. and {Haiman}, Zolt{\'a}n and {Gunn}, James E. and {Ivezi{\'c}}, {\v{Z}}eljko and {Schneider}, Donald P. and {Anderson}, Scott F. and {Brinkmann}, J. and {Bahcall}, Neta A. and {Connolly}, Andrew J. and {Csabai}, Istv{\'a}n and {Doi}, Mamoru and {Fukugita}, Masataka and {Geballe}, Tom and {Grebel}, Eva K. and {Harbeck}, Daniel and {Hennessy}, Gregory and {Lamb}, Don Q. and {Miknaitis}, Gajus and {Munn}, Jeffrey A. and {Nichol}, Robert and {Okamura}, Sadanori and {Pier}, Jeffrey R. and {Prada}, Francisco and {Richards}, Gordon T. and {Szalay}, Alex and {York}, Donald G.},
        title = "{A Survey of z>5.8 Quasars in the Sloan Digital Sky Survey. I. Discovery of Three New Quasars and the Spatial Density of Luminous Quasars at z\raisebox{-0.5ex}\textasciitilde6}",
      journal = {\aj},
         year = 2001,
        month = dec,
       volume = {122},
       number = {6},
        pages = {2833-2849},
          doi = {10.1086/324111},
archivePrefix = {arXiv},
       eprint = {astro-ph/0108063},
 primaryClass = {astro-ph},
       adsurl = {https://ui.adsabs.harvard.edu/abs/2001AJ....122.2833F}
}

@ARTICLE{scholz2010,
       author = {{Scholz}, R. -D.},
        title = "{Hip 63510C, Hip 73786B, and nine new isolated high proper motion T dwarf candidates from UKIDSS DR6 and SDSS DR7}",
      journal = {\aap},
         year = 2010,
        month = jun,
       volume = {515},
          eid = {A92},
        pages = {A92},
          doi = {10.1051/0004-6361/201014264},
archivePrefix = {arXiv},
       eprint = {1002.3073},
 primaryClass = {astro-ph.GA},
       adsurl = {https://ui.adsabs.harvard.edu/abs/2010A&A...515A..92S}
}

@ARTICLE{albert2011,
       author = {{Albert}, Lo{\"\i}c and {Artigau}, {\'E}tienne and {Delorme}, Philippe and {Reyl{\'e}}, C{\'e}line and {Forveille}, Thierry and {Delfosse}, Xavier and {Willott}, Chris J.},
        title = "{37 New T-type Brown Dwarfs in the Canada-France Brown Dwarfs Survey}",
      journal = {\aj},
         year = 2011,
        month = jun,
       volume = {141},
       number = {6},
          eid = {203},
        pages = {203},
          doi = {10.1088/0004-6256/141/6/203},
       adsurl = {https://ui.adsabs.harvard.edu/abs/2011AJ....141..203A}
}

@ARTICLE{mclean2001,
       author = {{McLean}, Ian S. and {Prato}, L. and {Kim}, Sungsoo S. and {Wilcox}, M.~K. and {Kirkpatrick}, J. Davy and {Burgasser}, Adam},
        title = "{Near-Infrared Spectroscopy of Brown Dwarfs: Methane and the Transition between the L and T Spectral Types}",
      journal = {\apjl},
         year = 2001,
        month = nov,
       volume = {561},
       number = {1},
        pages = {L115-L118},
          doi = {10.1086/324380},
archivePrefix = {arXiv},
       eprint = {astro-ph/0109390},
 primaryClass = {astro-ph},
       adsurl = {https://ui.adsabs.harvard.edu/abs/2001ApJ...561L.115M}
}

@ARTICLE{pineda2016,
       author = {{Pineda}, J. Sebastian and {Hallinan}, Gregg and {Kirkpatrick}, J. Davy and {Cotter}, Garret and {Kao}, Melodie M. and {Mooley}, Kunal},
        title = "{A Survey for H{\ensuremath{\alpha}} Emission from Late L Dwarfs and T Dwarfs}",
      journal = {\apj},
         year = 2016,
        month = jul,
       volume = {826},
       number = {1},
          eid = {73},
        pages = {73},
          doi = {10.3847/0004-637X/826/1/73},
archivePrefix = {arXiv},
       eprint = {1604.03941},
 primaryClass = {astro-ph.SR},
       adsurl = {https://ui.adsabs.harvard.edu/abs/2016ApJ...826...73P}
}

@ARTICLE{schneider2014,
       author = {{Schneider}, Adam C. and {Cushing}, Michael C. and {Kirkpatrick}, J. Davy and {Mace}, Gregory N. and {Gelino}, Christopher R. and {Faherty}, Jacqueline K. and {Fajardo-Acosta}, Sergio and {Sheppard}, Scott S.},
        title = "{Discovery of the Young L Dwarf WISE J174102.78-464225.5}",
      journal = {\aj},
         year = 2014,
        month = feb,
       volume = {147},
       number = {2},
          eid = {34},
        pages = {34},
          doi = {10.1088/0004-6256/147/2/34},
archivePrefix = {arXiv},
       eprint = {1311.5941},
 primaryClass = {astro-ph.SR},
       adsurl = {https://ui.adsabs.harvard.edu/abs/2014AJ....147...34S}
}

@ARTICLE{zhang2017,
       author = {{Zhang}, Z.~H. and {Pinfield}, D.~J. and {G{\'a}lvez-Ortiz}, M.~C. and {Burningham}, B. and {Lodieu}, N. and {Marocco}, F. and {Burgasser}, A.~J. and {Day-Jones}, A.~C. and {Allard}, F. and {Jones}, H.~R.~A. and {Homeier}, D. and {Gomes}, J. and {Smart}, R.~L.},
        title = "{Primeval very low-mass stars and brown dwarfs - I. Six new L subdwarfs, classification and atmospheric properties}",
      journal = {\mnras},
         year = 2017,
        month = jan,
       volume = {464},
       number = {3},
        pages = {3040-3059},
          doi = {10.1093/mnras/stw2438},
archivePrefix = {arXiv},
       eprint = {1609.07181},
 primaryClass = {astro-ph.SR},
       adsurl = {https://ui.adsabs.harvard.edu/abs/2017MNRAS.464.3040Z}
}

@ARTICLE{cruz2009,
       author = {{Cruz}, Kelle L. and {Kirkpatrick}, J. Davy and {Burgasser}, Adam J.},
        title = "{Young L Dwarfs Identified in the Field: A Preliminary Low-Gravity, Optical Spectral Sequence from L0 to L5}",
      journal = {\aj},
         year = 2009,
        month = feb,
       volume = {137},
       number = {2},
        pages = {3345-3357},
          doi = {10.1088/0004-6256/137/2/3345},
archivePrefix = {arXiv},
       eprint = {0812.0364},
 primaryClass = {astro-ph},
       adsurl = {https://ui.adsabs.harvard.edu/abs/2009AJ....137.3345C}
}

@ARTICLE{allers2013,
       author = {{Allers}, K.~N. and {Liu}, Michael C.},
        title = "{A Near-infrared Spectroscopic Study of Young Field Ultracool Dwarfs}",
      journal = {\apj},
         year = 2013,
        month = aug,
       volume = {772},
       number = {2},
          eid = {79},
        pages = {79},
          doi = {10.1088/0004-637X/772/2/79},
archivePrefix = {arXiv},
       eprint = {1305.4418},
 primaryClass = {astro-ph.SR},
       adsurl = {https://ui.adsabs.harvard.edu/abs/2013ApJ...772...79A}
}

@ARTICLE{cushing2011,
       author = {{Cushing}, Michael C. and {Kirkpatrick}, J. Davy and {Gelino}, Christopher R. and {Griffith}, Roger L. and {Skrutskie}, Michael F. and {Mainzer}, A. and {Marsh}, Kenneth A. and {Beichman}, Charles A. and {Burgasser}, Adam J. and {Prato}, Lisa A. and {Simcoe}, Robert A. and {Marley}, Mark S. and {Saumon}, D. and {Freedman}, Richard S. and {Eisenhardt}, Peter R. and {Wright}, Edward L.},
        title = "{The Discovery of Y Dwarfs using Data from the Wide-field Infrared Survey Explorer (WISE)}",
      journal = {\apj},
         year = 2011,
        month = dec,
       volume = {743},
       number = {1},
          eid = {50},
        pages = {50},
          doi = {10.1088/0004-637X/743/1/50},
archivePrefix = {arXiv},
       eprint = {1108.4678},
 primaryClass = {astro-ph.SR},
       adsurl = {https://ui.adsabs.harvard.edu/abs/2011ApJ...743...50C}
}

@ARTICLE{gagne2015,
       author = {{Gagn{\'e}}, Jonathan and {Faherty}, Jacqueline K. and {Cruz}, Kelle L. and {Lafreni{\'e}re}, David and {Doyon}, Ren{\'e} and {Malo}, Lison and {Burgasser}, Adam J. and {Naud}, Marie-Eve and {Artigau}, {\'E}tienne and {Bouchard}, Sandie and {Gizis}, John E. and {Albert}, Lo{\"\i}c},
        title = "{BANYAN. VII. A New Population of Young Substellar Candidate Members of Nearby Moving Groups from the BASS Survey}",
      journal = {\apjs},
         year = 2015,
        month = aug,
       volume = {219},
       number = {2},
          eid = {33},
        pages = {33},
          doi = {10.1088/0067-0049/219/2/33},
archivePrefix = {arXiv},
       eprint = {1506.07712},
 primaryClass = {astro-ph.SR},
       adsurl = {https://ui.adsabs.harvard.edu/abs/2015ApJS..219...33G}
}

@ARTICLE{cruz2018,
       author = {{Cruz}, Kelle L. and {N{\'u}{\~n}ez}, Alejandro and {Burgasser}, Adam J. and {Abrahams}, Ellianna and {Rice}, Emily L. and {Reid}, I. Neill and {Looper}, Dagny},
        title = "{Meeting the Cool Neighbors. XII. An Optically Anchored Analysis of the Near-infrared Spectra of L Dwarfs}",
      journal = {\aj},
         year = 2018,
        month = jan,
       volume = {155},
       number = {1},
          eid = {34},
        pages = {34},
          doi = {10.3847/1538-3881/aa9d8a},
       adsurl = {https://ui.adsabs.harvard.edu/abs/2018AJ....155...34C}
}

@ARTICLE{scholz2009,
       author = {{Scholz}, R. -D. and {Storm}, J. and {Knapp}, G.~R. and {Zinnecker}, H.},
        title = "{Extremely faint high proper motion objects from SDSS stripe 82. Optical classification spectroscopy of about 40 new objects}",
      journal = {\aap},
         year = 2009,
        month = feb,
       volume = {494},
       number = {3},
        pages = {949-967},
          doi = {10.1051/0004-6361:200811053},
archivePrefix = {arXiv},
       eprint = {0812.1495},
 primaryClass = {astro-ph},
       adsurl = {https://ui.adsabs.harvard.edu/abs/2009A&A...494..949S}
}

@ARTICLE{liu2011,
       author = {{Liu}, Michael C. and {Deacon}, Niall R. and {Magnier}, Eugene A. and {Dupuy}, Trent J. and {Aller}, Kimberly M. and {Bowler}, Brendan P. and {Redstone}, Joshua and {Goldman}, Bertrand and {Burgett}, W.~S. and {Chambers}, K.~C. and {Hodapp}, K.~W. and {Kaiser}, N. and {Kudritzki}, R. -P. and {Morgan}, J.~S. and {Price}, P.~A. and {Tonry}, J.~L. and {Wainscoat}, R.~J.},
        title = "{A Search for High Proper Motion T Dwarfs with PAN-STARRS1 + 2MASS + WISE}",
      journal = {\apjl},
         year = 2011,
        month = oct,
       volume = {740},
       number = {2},
          eid = {L32},
        pages = {L32},
          doi = {10.1088/2041-8205/740/2/L32},
archivePrefix = {arXiv},
       eprint = {1107.4608},
 primaryClass = {astro-ph.SR},
       adsurl = {https://ui.adsabs.harvard.edu/abs/2011ApJ...740L..32L}
}

@ARTICLE{leggett2019,
       author = {{Leggett}, S.~K. and {Dupuy}, Trent J. and {Morley}, Caroline V. and {Marley}, Mark S. and {Best}, William M.~J. and {Liu}, Michael C. and {Apai}, D. and {Casewell}, S.~L. and {Geballe}, T.~R. and {Gizis}, John E. and {Pineda}, J. Sebastian and {Rieke}, Marcia and {Wright}, G.~S.},
        title = "{3.8 {\ensuremath{\mu}}m Imaging of 400-600 K Brown Dwarfs and Orbital Constraints for WISEP J045853.90+643452.6AB}",
      journal = {\apj},
         year = 2019,
        month = sep,
       volume = {882},
       number = {2},
          eid = {117},
        pages = {117},
          doi = {10.3847/1538-4357/ab3393},
archivePrefix = {arXiv},
       eprint = {1907.07798},
 primaryClass = {astro-ph.SR},
       adsurl = {https://ui.adsabs.harvard.edu/abs/2019ApJ...882..117L}
}

@ARTICLE{burgasser2010,
       author = {{Burgasser}, Adam J. and {Cruz}, Kelle L. and {Cushing}, Michael and {Gelino}, Christopher R. and {Looper}, Dagny L. and {Faherty}, Jacqueline K. and {Kirkpatrick}, J. Davy and {Reid}, I. Neill},
        title = "{SpeX Spectroscopy of Unresolved Very Low Mass Binaries. I. Identification of 17 Candidate Binaries Straddling the L Dwarf/T Dwarf Transition}",
      journal = {\apj},
         year = 2010,
        month = feb,
       volume = {710},
       number = {2},
        pages = {1142-1169},
          doi = {10.1088/0004-637X/710/2/1142},
archivePrefix = {arXiv},
       eprint = {0912.3808},
 primaryClass = {astro-ph.SR},
       adsurl = {https://ui.adsabs.harvard.edu/abs/2010ApJ...710.1142B}
}

@ARTICLE{burgasser2007b,
       author = {{Burgasser}, Adam J.},
        title = "{Binaries and the L Dwarf/T Dwarf Transition}",
      journal = {\apj},
         year = 2007,
        month = apr,
       volume = {659},
       number = {1},
        pages = {655-674},
          doi = {10.1086/511027},
archivePrefix = {arXiv},
       eprint = {astro-ph/0611505},
 primaryClass = {astro-ph},
       adsurl = {https://ui.adsabs.harvard.edu/abs/2007ApJ...659..655B}
}

@ARTICLE{reid2008b,
       author = {{Reid}, I. Neill and {Cruz}, Kelle L. and {Kirkpatrick}, J. Davy and {Allen}, Peter R. and {Mungall}, F. and {Liebert}, James and {Lowrance}, Patrick and {Sweet}, Anne},
        title = "{Meeting the Cool Neighbors. X. Ultracool Dwarfs from the 2MASS All-Sky Data Release}",
      journal = {\aj},
         year = 2008,
        month = sep,
       volume = {136},
       number = {3},
        pages = {1290-1311},
          doi = {10.1088/0004-6256/136/3/1290},
       adsurl = {https://ui.adsabs.harvard.edu/abs/2008AJ....136.1290R}
}

@ARTICLE{zhang2013,
       author = {{Zhang}, Z.~H. and {Pinfield}, D.~J. and {Burningham}, B. and {Jones}, H.~R.~A. and {G{\'a}lvez-Ortiz}, M.~C. and {Catal{\'a}n}, S. and {Smart}, R.~L. and {L{\'e}pine}, S. and {Clarke}, J.~R.~A. and {Pavlenko}, Ya. V. and {Murray}, D.~N. and {Kuznetsov}, M.~K. and {Day-Jones}, A.~C. and {Gomes}, J. and {Marocco}, F. and {Sip{\H{o}}cz}, B.},
        title = "{A spectroscopic and proper motion search of Sloan Digital Sky Survey: red subdwarfs in binary systems}",
      journal = {\mnras},
         year = 2013,
        month = sep,
       volume = {434},
       number = {2},
        pages = {1005-1027},
          doi = {10.1093/mnras/stt1030},
archivePrefix = {arXiv},
       eprint = {1306.3060},
 primaryClass = {astro-ph.GA},
       adsurl = {https://ui.adsabs.harvard.edu/abs/2013MNRAS.434.1005Z}
}

@ARTICLE{castro2013,
       author = {{Castro}, Philip J. and {Gizis}, John E. and {Harris}, Hugh C. and {Mace}, Gregory N. and {Kirkpatrick}, J. Davy and {McLean}, Ian S. and {Pattarakijwanich}, Petchara and {Skrutskie}, Michael F.},
        title = "{Discovery of Four High Proper Motion L Dwarfs, Including a 10 pc L Dwarf at the L/T Transition}",
      journal = {\apj},
         year = 2013,
        month = oct,
       volume = {776},
       number = {2},
          eid = {126},
        pages = {126},
          doi = {10.1088/0004-637X/776/2/126},
archivePrefix = {arXiv},
       eprint = {1308.5252},
 primaryClass = {astro-ph.SR},
       adsurl = {https://ui.adsabs.harvard.edu/abs/2013ApJ...776..126C}
}

@ARTICLE{thompson2013,
       author = {{Thompson}, Maggie A. and {Kirkpatrick}, J. Davy and {Mace}, Gregory N. and {Cushing}, Michael C. and {Gelino}, Christopher R. and {Griffith}, Roger L. and {Skrutskie}, Michael F. and {Eisenhardt}, Peter R.~M. and {Wright}, Edward L. and {Marsh}, Kenneth A. and {Mix}, Katholeen J. and {Beichman}, Charles A. and {Faherty}, Jacqueline K. and {Toloza}, Odette and {Ferrara}, Jocelyn and {Apodaca}, Brian and {McLean}, Ian S. and {Bloom}, Joshua S.},
        title = "{Nearby M, L, and T Dwarfs Discovered by the Wide-field Infrared Survey Explorer (WISE)}",
      journal = {\pasp},
         year = 2013,
        month = jul,
       volume = {125},
       number = {929},
        pages = {809},
          doi = {10.1086/671426},
archivePrefix = {arXiv},
       eprint = {1305.4590},
 primaryClass = {astro-ph.SR},
       adsurl = {https://ui.adsabs.harvard.edu/abs/2013PASP..125..809T}
}

@ARTICLE{schmidt2010,
       author = {{Schmidt}, Sarah J. and {West}, Andrew A. and {Hawley}, Suzanne L. and {Pineda}, J. Sebastian},
        title = "{Colors and Kinematics of L Dwarfs from the Sloan Digital Sky Survey}",
      journal = {\aj},
         year = 2010,
        month = may,
       volume = {139},
       number = {5},
        pages = {1808-1821},
          doi = {10.1088/0004-6256/139/5/1808},
archivePrefix = {arXiv},
       eprint = {1001.3402},
 primaryClass = {astro-ph.SR},
       adsurl = {https://ui.adsabs.harvard.edu/abs/2010AJ....139.1808S}
}

@ARTICLE{faherty2009,
       author = {{Faherty}, Jacqueline K. and {Burgasser}, Adam J. and {Cruz}, Kelle L. and {Shara}, Michael M. and {Walter}, Frederick M. and {Gelino}, Christopher R.},
        title = "{The Brown Dwarf Kinematics Project I. Proper Motions and Tangential Velocities for a Large Sample of Late-Type M, L, and T Dwarfs}",
      journal = {\aj},
         year = 2009,
        month = jan,
       volume = {137},
       number = {1},
        pages = {1-18},
          doi = {10.1088/0004-6256/137/1/1},
archivePrefix = {arXiv},
       eprint = {0809.3008},
 primaryClass = {astro-ph},
       adsurl = {https://ui.adsabs.harvard.edu/abs/2009AJ....137....1F}
}

@ARTICLE{bardalez2019,
       author = {{Bardalez Gagliuffi}, Daniella C. and {Burgasser}, Adam J. and {Schmidt}, Sarah J. and {Theissen}, Christopher and {Gagn{\'e}}, Jonathan and {Gillon}, Michael and {Sahlmann}, Johannes and {Faherty}, Jacqueline K. and {Gelino}, Christopher and {Cruz}, Kelle L. and {Skrzypek}, Nathalie and {Looper}, Dagny},
        title = "{The Ultracool SpeXtroscopic Survey. I. Volume-limited Spectroscopic Sample and Luminosity Function of M7-L5 Ultracool Dwarfs}",
      journal = {\apj},
         year = 2019,
        month = oct,
       volume = {883},
       number = {2},
          eid = {205},
        pages = {205},
          doi = {10.3847/1538-4357/ab253d},
archivePrefix = {arXiv},
       eprint = {1906.04166},
 primaryClass = {astro-ph.SR},
       adsurl = {https://ui.adsabs.harvard.edu/abs/2019ApJ...883..205B}
}

@ARTICLE{bardalez2015,
       author = {{Bardalez Gagliuffi}, Daniella C. and {Gelino}, Christopher R. and {Burgasser}, Adam J.},
        title = "{High Resolution Imaging of Very Low Mass Spectral Binaries: Three Resolved Systems and Detection of Orbital Motion in an L/T Transition Binary}",
      journal = {\aj},
         year = 2015,
        month = nov,
       volume = {150},
       number = {5},
          eid = {163},
        pages = {163},
          doi = {10.1088/0004-6256/150/5/163},
archivePrefix = {arXiv},
       eprint = {1510.00392},
 primaryClass = {astro-ph.SR},
       adsurl = {https://ui.adsabs.harvard.edu/abs/2015AJ....150..163B}
}

@ARTICLE{zhang2019,
       author = {{Zhang}, Z.~H. and {Burgasser}, A.~J. and {G{\'a}lvez-Ortiz}, M.~C. and {Lodieu}, N. and {Zapatero Osorio}, M.~R. and {Pinfield}, D.~J. and {Allard}, F.},
        title = "{Primeval very low-mass stars and brown dwarfs - VI. Population properties of metal-poor degenerate brown dwarfs}",
      journal = {\mnras},
         year = 2019,
        month = jun,
       volume = {486},
       number = {1},
        pages = {1260-1282},
          doi = {10.1093/mnras/stz777},
archivePrefix = {arXiv},
       eprint = {1903.05536},
 primaryClass = {astro-ph.SR},
       adsurl = {https://ui.adsabs.harvard.edu/abs/2019MNRAS.486.1260Z}
}

@ARTICLE{leggett2017,
       author = {{Leggett}, S.~K. and {Tremblin}, P. and {Esplin}, T.~L. and {Luhman}, K.~L. and {Morley}, Caroline V.},
        title = "{The Y-type Brown Dwarfs: Estimates of Mass and Age from New Astrometry, Homogenized Photometry, and Near-infrared Spectroscopy}",
      journal = {\apj},
         year = 2017,
        month = jun,
       volume = {842},
       number = {2},
          eid = {118},
        pages = {118},
          doi = {10.3847/1538-4357/aa6fb5},
archivePrefix = {arXiv},
       eprint = {1704.03573},
 primaryClass = {astro-ph.SR},
       adsurl = {https://ui.adsabs.harvard.edu/abs/2017ApJ...842..118L}
}

@ARTICLE{kellogg2015,
       author = {{Kellogg}, Kendra and {Metchev}, Stanimir and {Gei{\ss}ler}, Kerstin and {Hicks}, Shannon and {Kirkpatrick}, J. Davy and {Kurtev}, Radostin},
        title = "{A Targeted Search for Peculiarly Red L and T Dwarfs in SDSS, 2MASS, and WISE: Discovery of a Possible L7 Member of the TW Hydrae Association}",
      journal = {\aj},
         year = 2015,
        month = dec,
       volume = {150},
       number = {6},
          eid = {182},
        pages = {182},
          doi = {10.1088/0004-6256/150/6/182},
archivePrefix = {arXiv},
       eprint = {1510.08464},
 primaryClass = {astro-ph.SR},
       adsurl = {https://ui.adsabs.harvard.edu/abs/2015AJ....150..182K}
}

@ARTICLE{marocco2015,
       author = {{Marocco}, F. and {Jones}, H.~R.~A. and {Day-Jones}, A.~C. and {Pinfield}, D.~J. and {Lucas}, P.~W. and {Burningham}, B. and {Zhang}, Z.~H. and {Smart}, R.~L. and {Gomes}, J.~I. and {Smith}, L.},
        title = "{A large spectroscopic sample of L and T dwarfs from UKIDSS LAS: peculiar objects, binaries, and space density}",
      journal = {\mnras},
         year = 2015,
        month = jun,
       volume = {449},
       number = {4},
        pages = {3651-3692},
          doi = {10.1093/mnras/stv530},
archivePrefix = {arXiv},
       eprint = {1503.05082},
 primaryClass = {astro-ph.SR},
       adsurl = {https://ui.adsabs.harvard.edu/abs/2015MNRAS.449.3651M}
}

@ARTICLE{gizis2015,
       author = {{Gizis}, John E. and {Burgasser}, Adam J. and {Vrba}, Frederick J.},
        title = "{Properties of the nearby Brown Dwarf WISEP J180026.60+013453.1}",
      journal = {\aj},
         year = 2015,
        month = dec,
       volume = {150},
       number = {6},
          eid = {179},
        pages = {179},
          doi = {10.1088/0004-6256/150/6/179},
archivePrefix = {arXiv},
       eprint = {1509.04690},
 primaryClass = {astro-ph.SR},
       adsurl = {https://ui.adsabs.harvard.edu/abs/2015AJ....150..179G}
}

@ARTICLE{faherty2016,
       author = {{Faherty}, Jacqueline K. and {Riedel}, Adric R. and {Cruz}, Kelle L. and {Gagne}, Jonathan and {Filippazzo}, Joseph C. and {Lambrides}, Erini and {Fica}, Haley and {Weinberger}, Alycia and {Thorstensen}, John R. and {Tinney}, C.~G. and {Baldassare}, Vivienne and {Lemonier}, Emily and {Rice}, Emily L.},
        title = "{Population Properties of Brown Dwarf Analogs to Exoplanets}",
      journal = {\apjs},
         year = 2016,
        month = jul,
       volume = {225},
       number = {1},
          eid = {10},
        pages = {10},
          doi = {10.3847/0067-0049/225/1/10},
archivePrefix = {arXiv},
       eprint = {1605.07927},
 primaryClass = {astro-ph.SR},
       adsurl = {https://ui.adsabs.harvard.edu/abs/2016ApJS..225...10F}
}

@ARTICLE{gagne2015b,
       author = {{Gagn{\'e}}, Jonathan and {Lafreni{\`e}re}, David and {Doyon}, Ren{\'e} and {Malo}, Lison and {Artigau}, {\'E}tienne},
        title = "{BANYAN. V. A Systematic All-sky Survey for New Very Late-type Low-mass Stars and Brown Dwarfs in Nearby Young Moving Groups}",
      journal = {\apj},
         year = 2015,
        month = jan,
       volume = {798},
       number = {2},
          eid = {73},
        pages = {73},
          doi = {10.1088/0004-637X/798/2/73},
archivePrefix = {arXiv},
       eprint = {1410.4864},
 primaryClass = {astro-ph.SR},
       adsurl = {https://ui.adsabs.harvard.edu/abs/2015ApJ...798...73G}
}

@ARTICLE{strand1964,
       author = {{Strand}, K. Aa.},
        title = "{Determination of Stellar Distances}",
      journal = {Science},
         year = 1964,
        month = jun,
       volume = {144},
       number = {3624},
        pages = {1299-1309},
          doi = {10.1126/science.144.3624.1299},
       adsurl = {https://ui.adsabs.harvard.edu/abs/1964Sci...144.1299S}
}

@ARTICLE{harris1992,
       author = {{Harris}, Hugh C. and {Vrba}, Frederick J.},
        title = "{Seeing Measurements and Observing Statistics at the U.S. Naval Observatory, Flagstaff Station}",
      journal = {\pasp},
         year = 1992,
        month = feb,
       volume = {104},
        pages = {140},
          doi = {10.1086/132969},
       adsurl = {https://ui.adsabs.harvard.edu/abs/1992PASP..104..140H}
}

@ARTICLE{dahn1988,
       author = {{Dahn}, C.~C. and {Harrington}, R.~S. and {Kallarakal}, V.~V. and {Guetter}, H.~H. and {Luginbuhl}, C.~B. and {Riepe}, B.~Y. and {Walker}, R.~L. and {Pier}, J.~R. and {Vrba}, F.~J. and {Monet}, D.~G. and {Ables}, H.~D.},
        title = "{U.S. Naval Observatory Parallaxes of Faint Stars, List VIII}",
      journal = {\aj},
         year = 1988,
        month = jan,
       volume = {95},
        pages = {237},
          doi = {10.1086/114633},
       adsurl = {https://ui.adsabs.harvard.edu/abs/1988AJ.....95..237D}
}

@ARTICLE{luyten1976,
       author = {{Luyten}, W.~J.},
        title = "{A Catalogue of 1849 Stars with Proper Motions greater than 0''5 annually (LHS)}",
      journal = {Univ. Minnesota},
         year = 1976,
        month = jan,
        pages = {0},
       adsurl = {https://ui.adsabs.harvard.edu/abs/1976LHS...C......0L}
}

@ARTICLE{monet1992,
       author = {{Monet}, David G. and {Dahn}, Conard C. and {Vrba}, Frederick J. and {Harris}, Hugh C. and {Pier}, Jeffrey R. and {Luginbuhl}, Christian B. and {Ables}, Harold D.},
        title = "{U.S. Naval Observatory CCD Parallaxes of Faint Stars. I. Program Description and First Results}",
      journal = {\aj},
         year = 1992,
        month = feb,
       volume = {103},
        pages = {638},
          doi = {10.1086/116091},
       adsurl = {https://ui.adsabs.harvard.edu/abs/1992AJ....103..638M}
}

@ARTICLE{gaia2016,
       author = {{Gaia Collaboration} and {Brown}, A.~G.~A. and {Vallenari}, A. and {Prusti}, T. and {de Bruijne}, J.~H.~J. and {Mignard}, F. and {Drimmel}, R. and {Babusiaux}, C. and {Bailer-Jones}, C.~A.~L. and {Bastian}, U. and {Biermann}, M. and {Evans}, D.~W. and {Eyer}, L. and {Jansen}, F. and {Jordi}, C. and {Katz}, D. and {Klioner}, S.~A. and {Lammers}, U. and {Lindegren}, L. and {Luri}, X. and {O'Mullane}, W. and {Panem}, C. and {Pourbaix}, D. and {Randich}, S. and {Sartoretti}, P. and {Siddiqui}, H.~I. and {Soubiran}, C. and {Valette}, V. and {van Leeuwen}, F. and {Walton}, N.~A. and {Aerts}, C. and {Arenou}, F. and {Cropper}, M. and {H{\o}g}, E. and {Lattanzi}, M.~G. and {Grebel}, E.~K. and {Holland}, A.~D. and {Huc}, C. and {Passot}, X. and {Perryman}, M. and {Bramante}, L. and {Cacciari}, C. and {Casta{\~n}eda}, J. and {Chaoul}, L. and {Cheek}, N. and {De Angeli}, F. and {Fabricius}, C. and {Guerra}, R. and {Hern{\'a}ndez}, J. and {Jean-Antoine-Piccolo}, A. and {Masana}, E. and {Messineo}, R. and {Mowlavi}, N. and {Nienartowicz}, K. and {Ord{\'o}{\~n}ez-Blanco}, D. and {Panuzzo}, P. and {Portell}, J. and {Richards}, P.~J. and {Riello}, M. and {Seabroke}, G.~M. and {Tanga}, P. and {Th{\'e}venin}, F. and {Torra}, J. and {Els}, S.~G. and {Gracia-Abril}, G. and {Comoretto}, G. and {Garcia-Reinaldos}, M. and {Lock}, T. and {Mercier}, E. and {Altmann}, M. and {Andrae}, R. and {Astraatmadja}, T.~L. and {Bellas-Velidis}, I. and {Benson}, K. and {Berthier}, J. and {Blomme}, R. and {Busso}, G. and {Carry}, B. and {Cellino}, A. and {Clementini}, G. and {Cowell}, S. and {Creevey}, O. and {Cuypers}, J. and {Davidson}, M. and {De Ridder}, J. and {de Torres}, A. and {Delchambre}, L. and {Dell'Oro}, A. and {Ducourant}, C. and {Fr{\'e}mat}, Y. and {Garc{\'\i}a-Torres}, M. and {Gosset}, E. and {Halbwachs}, J. -L. and {Hambly}, N.~C. and {Harrison}, D.~L. and {Hauser}, M. and {Hestroffer}, D. and {Hodgkin}, S.~T. and {Huckle}, H.~E. and {Hutton}, A. and {Jasniewicz}, G. and {Jordan}, S. and {Kontizas}, M. and {Korn}, A.~J. and {Lanzafame}, A.~C. and {Manteiga}, M. and {Moitinho}, A. and {Muinonen}, K. and {Osinde}, J. and {Pancino}, E. and {Pauwels}, T. and {Petit}, J. -M. and {Recio-Blanco}, A. and {Robin}, A.~C. and {Sarro}, L.~M. and {Siopis}, C. and {Smith}, M. and {Smith}, K.~W. and {Sozzetti}, A. and {Thuillot}, W. and {van Reeven}, W. and {Viala}, Y. and {Abbas}, U. and {Abreu Aramburu}, A. and {Accart}, S. and {Aguado}, J.~J. and {Allan}, P.~M. and {Allasia}, W. and {Altavilla}, G. and {{\'A}lvarez}, M.~A. and {Alves}, J. and {Anderson}, R.~I. and {Andrei}, A.~H. and {Anglada Varela}, E. and {Antiche}, E. and {Antoja}, T. and {Ant{\'o}n}, S. and {Arcay}, B. and {Bach}, N. and {Baker}, S.~G. and {Balaguer-N{\'u}{\~n}ez}, L. and {Barache}, C. and {Barata}, C. and {Barbier}, A. and {Barblan}, F. and {Barrado y Navascu{\'e}s}, D. and {Barros}, M. and {Barstow}, M.~A. and {Becciani}, U. and {Bellazzini}, M. and {Bello Garc{\'\i}a}, A. and {Belokurov}, V. and {Bendjoya}, P. and {Berihuete}, A. and {Bianchi}, L. and {Bienaym{\'e}}, O. and {Billebaud}, F. and {Blagorodnova}, N. and {Blanco-Cuaresma}, S. and {Boch}, T. and {Bombrun}, A. and {Borrachero}, R. and {Bouquillon}, S. and {Bourda}, G. and {Bouy}, H. and {Bragaglia}, A. and {Breddels}, M.~A. and {Brouillet}, N. and {Br{\"u}semeister}, T. and {Bucciarelli}, B. and {Burgess}, P. and {Burgon}, R. and {Burlacu}, A. and {Busonero}, D. and {Buzzi}, R. and {Caffau}, E. and {Cambras}, J. and {Campbell}, H. and {Cancelliere}, R. and {Cantat-Gaudin}, T. and {Carlucci}, T. and {Carrasco}, J.~M. and {Castellani}, M. and {Charlot}, P. and {Charnas}, J. and {Chiavassa}, A. and {Clotet}, M. and {Cocozza}, G. and {Collins}, R.~S. and {Costigan}, G. and {Crifo}, F. and {Cross}, N.~J.~G. and {Crosta}, M. and {Crowley}, C. and {Dafonte}, C. and {Damerdji}, Y. and {Dapergolas}, A. and {David}, P. and {David}, M. and {De Cat}, P.},
        title = "{Gaia Data Release 1. Summary of the astrometric, photometric, and survey properties}",
      journal = {\aap},
         year = 2016,
        month = nov,
       volume = {595},
          eid = {A2},
        pages = {A2},
          doi = {10.1051/0004-6361/201629512},
archivePrefix = {arXiv},
       eprint = {1609.04172},
 primaryClass = {astro-ph.IM},
       adsurl = {https://ui.adsabs.harvard.edu/abs/2016A&A...595A...2G}
}

@INPROCEEDINGS{vrba2000,
       author = {{Vrba}, F.~J. and {Henden}, A.~A. and {Luginbuhl}, C.~B. and {Guetter}, H.~H. and {Monet}, D.~G.},
        title = "{Evaluation of the Astrometric Potential of NIR Focal Plane Arrays for Determination of Parallaxes and Proper Motions of L and T Dwarfs}",
    booktitle = {American Astronomical Society Meeting Abstracts \#196},
         year = 2000,
       series = {American Astronomical Society Meeting Abstracts},
       volume = {196},
        month = may,
          eid = {03.08},
        pages = {03.08},
       adsurl = {https://ui.adsabs.harvard.edu/abs/2000AAS...196.0308V}
}

@INPROCEEDINGS{fowler1996,
       author = {{Fowler}, Albert M. and {Gatley}, Ian and {McIntyre}, Paul and {Vrba}, Frederick J. and {Hoffman}, Alan W.},
        title = "{ALADDIN: the 1024x1024 InSb array--design, description, and results}",
    booktitle = {Infrared Detectors for Remote Sensing: Physics, Materials, and Devices},
         year = 1996,
       editor = {{Longshore}, Randolph E. and {Baars}, Jan W.},
       series = {Society of Photo-Optical Instrumentation Engineers (SPIE) Conference Series},
       volume = {2816},
        month = oct,
        pages = {150-160},
          doi = {10.1117/12.255162},
       adsurl = {https://ui.adsabs.harvard.edu/abs/1996SPIE.2816..150F}
}

@INPROCEEDINGS{fischer2003,
       author = {{Fischer}, Jacqueline and {Vrba}, Frederick J. and {Toomey}, Douglas W. and {Lucke}, Bob L. and {Wang}, Shu-i. and {Henden}, Arne A. and {Robichaud}, Joseph L. and {Onaka}, Peter M. and {Hicks}, Brian and {Harris}, Frederick H. and {Stahlberger}, Werner E. and {Kosakowski}, Kris E. and {Dudley}, Charles C. and {Johnston}, Kenneth J.},
        title = "{ASTROCAM: offner re-imaging 1024 X 1024 InSb camera for near-infrared astrometry on the USNO 1.55-m telescope}",
    booktitle = {Instrument Design and Performance for Optical/Infrared Ground-based Telescopes},
         year = 2003,
       editor = {{Iye}, Masanori and {Moorwood}, Alan F.~M.},
       series = {Society of Photo-Optical Instrumentation Engineers (SPIE) Conference Series},
       volume = {4841},
        month = mar,
        pages = {564-577},
          doi = {10.1117/12.461033},
       adsurl = {https://ui.adsabs.harvard.edu/abs/2003SPIE.4841..564F}
}

@ARTICLE{gaia2023,
       author = {{Gaia Collaboration} and {Vallenari}, A. and {Brown}, A.~G.~A. and {Prusti}, T. and {de Bruijne}, J.~H.~J. and {Arenou}, F. and {Babusiaux}, C. and {Biermann}, M. and {Creevey}, O.~L. and {Ducourant}, C. and {Evans}, D.~W. and {Eyer}, L. and {Guerra}, R. and {Hutton}, A. and {Jordi}, C. and {Klioner}, S.~A. and {Lammers}, U.~L. and {Lindegren}, L. and {Luri}, X. and {Mignard}, F. and {Panem}, C. and {Pourbaix}, D. and {Randich}, S. and {Sartoretti}, P. and {Soubiran}, C. and {Tanga}, P. and {Walton}, N.~A. and {Bailer-Jones}, C.~A.~L. and {Bastian}, U. and {Drimmel}, R. and {Jansen}, F. and {Katz}, D. and {Lattanzi}, M.~G. and {van Leeuwen}, F. and {Bakker}, J. and {Cacciari}, C. and {Casta{\~n}eda}, J. and {De Angeli}, F. and {Fabricius}, C. and {Fouesneau}, M. and {Fr{\'e}mat}, Y. and {Galluccio}, L. and {Guerrier}, A. and {Heiter}, U. and {Masana}, E. and {Messineo}, R. and {Mowlavi}, N. and {Nicolas}, C. and {Nienartowicz}, K. and {Pailler}, F. and {Panuzzo}, P. and {Riclet}, F. and {Roux}, W. and {Seabroke}, G.~M. and {Sordo}, R. and {Th{\'e}venin}, F. and {Gracia-Abril}, G. and {Portell}, J. and {Teyssier}, D. and {Altmann}, M. and {Andrae}, R. and {Audard}, M. and {Bellas-Velidis}, I. and {Benson}, K. and {Berthier}, J. and {Blomme}, R. and {Burgess}, P.~W. and {Busonero}, D. and {Busso}, G. and {C{\'a}novas}, H. and {Carry}, B. and {Cellino}, A. and {Cheek}, N. and {Clementini}, G. and {Damerdji}, Y. and {Davidson}, M. and {de Teodoro}, P. and {Nu{\~n}ez Campos}, M. and {Delchambre}, L. and {Dell'Oro}, A. and {Esquej}, P. and {Fern{\'a}ndez-Hern{\'a}ndez}, J. and {Fraile}, E. and {Garabato}, D. and {Garc{\'\i}a-Lario}, P. and {Gosset}, E. and {Haigron}, R. and {Halbwachs}, J. -L. and {Hambly}, N.~C. and {Harrison}, D.~L. and {Hern{\'a}ndez}, J. and {Hestroffer}, D. and {Hodgkin}, S.~T. and {Holl}, B. and {Jan{\ss}en}, K. and {Jevardat de Fombelle}, G. and {Jordan}, S. and {Krone-Martins}, A. and {Lanzafame}, A.~C. and {L{\"o}ffler}, W. and {Marchal}, O. and {Marrese}, P.~M. and {Moitinho}, A. and {Muinonen}, K. and {Osborne}, P. and {Pancino}, E. and {Pauwels}, T. and {Recio-Blanco}, A. and {Reyl{\'e}}, C. and {Riello}, M. and {Rimoldini}, L. and {Roegiers}, T. and {Rybizki}, J. and {Sarro}, L.~M. and {Siopis}, C. and {Smith}, M. and {Sozzetti}, A. and {Utrilla}, E. and {van Leeuwen}, M. and {Abbas}, U. and {{\'A}brah{\'a}m}, P. and {Abreu Aramburu}, A. and {Aerts}, C. and {Aguado}, J.~J. and {Ajaj}, M. and {Aldea-Montero}, F. and {Altavilla}, G. and {{\'A}lvarez}, M.~A. and {Alves}, J. and {Anders}, F. and {Anderson}, R.~I. and {Anglada Varela}, E. and {Antoja}, T. and {Baines}, D. and {Baker}, S.~G. and {Balaguer-N{\'u}{\~n}ez}, L. and {Balbinot}, E. and {Balog}, Z. and {Barache}, C. and {Barbato}, D. and {Barros}, M. and {Barstow}, M.~A. and {Bartolom{\'e}}, S. and {Bassilana}, J. -L. and {Bauchet}, N. and {Becciani}, U. and {Bellazzini}, M. and {Berihuete}, A. and {Bernet}, M. and {Bertone}, S. and {Bianchi}, L. and {Binnenfeld}, A. and {Blanco-Cuaresma}, S. and {Blazere}, A. and {Boch}, T. and {Bombrun}, A. and {Bossini}, D. and {Bouquillon}, S. and {Bragaglia}, A. and {Bramante}, L. and {Breedt}, E. and {Bressan}, A. and {Brouillet}, N. and {Brugaletta}, E. and {Bucciarelli}, B. and {Burlacu}, A. and {Butkevich}, A.~G. and {Buzzi}, R. and {Caffau}, E. and {Cancelliere}, R. and {Cantat-Gaudin}, T. and {Carballo}, R. and {Carlucci}, T. and {Carnerero}, M.~I. and {Carrasco}, J.~M. and {Casamiquela}, L. and {Castellani}, M. and {Castro-Ginard}, A. and {Chaoul}, L. and {Charlot}, P. and {Chemin}, L. and {Chiaramida}, V. and {Chiavassa}, A. and {Chornay}, N. and {Comoretto}, G. and {Contursi}, G. and {Cooper}, W.~J. and {Cornez}, T. and {Cowell}, S. and {Crifo}, F. and {Cropper}, M. and {Crosta}, M. and {Crowley}, C. and {Dafonte}, C. and {Dapergolas}, A. and {David}, M. and {David}, P. and {de Laverny}, P. and {De Luise}, F. and {De March}, R.},
        title = "{Gaia Data Release 3. Summary of the content and survey properties}",
      journal = {\aap},
         year = 2023,
        month = jun,
       volume = {674},
          eid = {A1},
        pages = {A1},
          doi = {10.1051/0004-6361/202243940},
archivePrefix = {arXiv},
       eprint = {2208.00211},
 primaryClass = {astro-ph.GA},
       adsurl = {https://ui.adsabs.harvard.edu/abs/2023A&A...674A...1G}
}

@INPROCEEDINGS{luginbuhl1998,
       author = {{Luginbuhl}, Christian B. and {Henden}, Arne A. and {Vrba}, Frederick J. and {Guetter}, Harry H.},
        title = "{Photometric linearization of the NICMOS-3 array}",
    booktitle = {Infrared Astronomical Instrumentation},
         year = 1998,
       editor = {{Fowler}, Albert M.},
       series = {Society of Photo-Optical Instrumentation Engineers (SPIE) Conference Series},
       volume = {3354},
        month = aug,
        pages = {240-246},
          doi = {10.1117/12.317305},
       adsurl = {https://ui.adsabs.harvard.edu/abs/1998SPIE.3354..240L}
}

@ARTICLE{monet1983,
       author = {{Monet}, D.~G. and {Dahn}, C.~C.},
        title = "{CCD astrometry. I. Preliminary results from the KPNO 4-m/CCD parallaxprogram.}",
      journal = {\aj},
         year = 1983,
        month = oct,
       volume = {88},
        pages = {1489-1507},
          doi = {10.1086/113438},
       adsurl = {https://ui.adsabs.harvard.edu/abs/1983AJ.....88.1489M}
}

@ARTICLE{marocco2021,
       author = {{Marocco}, Federico and {Eisenhardt}, Peter R.~M. and {Fowler}, John W. and {Kirkpatrick}, J. Davy and {Meisner}, Aaron M. and {Schlafly}, Edward F. and {Stanford}, S.~A. and {Garcia}, Nelson and {Caselden}, Dan and {Cushing}, Michael C. and {Cutri}, Roc M. and {Faherty}, Jacqueline K. and {Gelino}, Christopher R. and {Gonzalez}, Anthony H. and {Jarrett}, Thomas H. and {Koontz}, Renata and {Mainzer}, Amanda and {Marchese}, Elijah J. and {Mobasher}, Bahram and {Schlegel}, David J. and {Stern}, Daniel and {Teplitz}, Harry I. and {Wright}, Edward L.},
        title = "{The CatWISE2020 Catalog}",
      journal = {\apjs},
         year = 2021,
        month = mar,
       volume = {253},
       number = {1},
          eid = {8},
        pages = {8},
          doi = {10.3847/1538-4365/abd805},
archivePrefix = {arXiv},
       eprint = {2012.13084},
 primaryClass = {astro-ph.IM},
       adsurl = {https://ui.adsabs.harvard.edu/abs/2021ApJS..253....8M}
}

@ARTICLE{castro2024,
       author = {{Castro-Ginard}, Alfred and {Penoyre}, Zephyr and {Casey}, Andrew R. and {Brown}, Anthony G.~A. and {Belokurov}, Vasily and {Cantat-Gaudin}, Tristan and {Drimmel}, Ronald and {Fouesneau}, Morgan and {Khanna}, Shourya and {Kurbatov}, Evgeny P. and {Price-Whelan}, Adrian M. and {Rix}, Hans-Walter and {Smart}, Richard L.},
        title = "{Gaia DR3 detectability of unresolved binary systems}",
      journal = {\aap},
         year = 2024,
        month = aug,
       volume = {688},
          eid = {A1},
        pages = {A1},
          doi = {10.1051/0004-6361/202450172},
archivePrefix = {arXiv},
       eprint = {2404.14127},
 primaryClass = {astro-ph.GA},
       adsurl = {https://ui.adsabs.harvard.edu/abs/2024A&A...688A...1C}
}

@ARTICLE{sutherland2015,
       author = {{Sutherland}, Will and {Emerson}, Jim and {Dalton}, Gavin and {Atad-Ettedgui}, Eli and {Beard}, Steven and {Bennett}, Richard and {Bezawada}, Naidu and {Born}, Andrew and {Caldwell}, Martin and {Clark}, Paul and {Craig}, Simon and {Henry}, David and {Jeffers}, Paul and {Little}, Bryan and {McPherson}, Alistair and {Murray}, John and {Stewart}, Malcolm and {Stobie}, Brian and {Terrett}, David and {Ward}, Kim and {Whalley}, Martin and {Woodhouse}, Guy},
        title = "{The Visible and Infrared Survey Telescope for Astronomy (VISTA): Design, technical overview, and performance}",
      journal = {\aap},
         year = 2015,
        month = mar,
       volume = {575},
          eid = {A25},
        pages = {A25},
          doi = {10.1051/0004-6361/201424973},
archivePrefix = {arXiv},
       eprint = {1409.4780},
 primaryClass = {astro-ph.IM},
       adsurl = {https://ui.adsabs.harvard.edu/abs/2015A&A...575A..25S}
}

@ARTICLE{simons2002,
       author = {{Simons}, D.~A. and {Tokunaga}, A.},
        title = "{The Mauna Kea Observatories Near-Infrared Filter Set. I. Defining Optimal 1-5 Micron Bandpasses}",
      journal = {\pasp},
         year = 2002,
        month = feb,
       volume = {114},
       number = {792},
        pages = {169-179},
          doi = {10.1086/338544},
archivePrefix = {arXiv},
       eprint = {astro-ph/0110594},
 primaryClass = {astro-ph},
       adsurl = {https://ui.adsabs.harvard.edu/abs/2002PASP..114..169S}
}

@ARTICLE{casali2007,
       author = {{Casali}, M. and {Adamson}, A. and {Alves de Oliveira}, C. and {Almaini}, O. and {Burch}, K. and {Chuter}, T. and {Elliot}, J. and {Folger}, M. and {Foucaud}, S. and {Hambly}, N. and {Hastie}, M. and {Henry}, D. and {Hirst}, P. and {Irwin}, M. and {Ives}, D. and {Lawrence}, A. and {Laidlaw}, K. and {Lee}, D. and {Lewis}, J. and {Lunney}, D. and {McLay}, S. and {Montgomery}, D. and {Pickup}, A. and {Read}, M. and {Rees}, N. and {Robson}, I. and {Sekiguchi}, K. and {Vick}, A. and {Warren}, S. and {Woodward}, B.},
        title = "{The UKIRT wide-field camera}",
      journal = {\aap},
         year = 2007,
        month = may,
       volume = {467},
       number = {2},
        pages = {777-784},
          doi = {10.1051/0004-6361:20066514},
       adsurl = {https://ui.adsabs.harvard.edu/abs/2007A&A...467..777C}
}

@ARTICLE{dye2018,
       author = {{Dye}, S. and {Lawrence}, A. and {Read}, M.~A. and {Fan}, X. and {Kerr}, T. and {Varricatt}, W. and {Furnell}, K.~E. and {Edge}, A.~C. and {Irwin}, M. and {Hambly}, N. and {Lucas}, P. and {Almaini}, O. and {Chambers}, K. and {Green}, R. and {Hewett}, P. and {Liu}, M.~C. and {McGreer}, I. and {Best}, W. and {Zhang}, Z. and {Sutorius}, E. and {Froebrich}, D. and {Magnier}, E. and {Hasinger}, G. and {Lederer}, S.~M. and {Bold}, M. and {Tedds}, J.~A.},
        title = "{The UKIRT Hemisphere Survey: definition and J-band data release}",
      journal = {\mnras},
         year = 2018,
        month = feb,
       volume = {473},
       number = {4},
        pages = {5113-5125},
          doi = {10.1093/mnras/stx2622},
archivePrefix = {arXiv},
       eprint = {1707.09975},
 primaryClass = {astro-ph.IM},
       adsurl = {https://ui.adsabs.harvard.edu/abs/2018MNRAS.473.5113D}
}

@ARTICLE{mcmahon2013,
       author = {{McMahon}, R.~G. and {Banerji}, M. and {Gonzalez}, E. and {Koposov}, S.~E. and {Bejar}, V.~J. and {Lodieu}, N. and {Rebolo}, R. and {VHS Collaboration}},
        title = "{First Scientific Results from the VISTA Hemisphere Survey (VHS)}",
      journal = {The Messenger},
         year = 2013,
        month = dec,
       volume = {154},
        pages = {35-37},
       adsurl = {https://ui.adsabs.harvard.edu/abs/2013Msngr.154...35M}
}

@ARTICLE{york2000,
       author = {{York}, Donald G. and {Adelman}, J. and {Anderson}, Jr., John E. and {Anderson}, Scott F. and {Annis}, James and {Bahcall}, Neta A. and {Bakken}, J.~A. and {Barkhouser}, Robert and {Bastian}, Steven and {Berman}, Eileen and {Boroski}, William N. and {Bracker}, Steve and {Briegel}, Charlie and {Briggs}, John W. and {Brinkmann}, J. and {Brunner}, Robert and {Burles}, Scott and {Carey}, Larry and {Carr}, Michael A. and {Castander}, Francisco J. and {Chen}, Bing and {Colestock}, Patrick L. and {Connolly}, A.~J. and {Crocker}, J.~H. and {Csabai}, Istv{\'a}n and {Czarapata}, Paul C. and {Davis}, John Eric and {Doi}, Mamoru and {Dombeck}, Tom and {Eisenstein}, Daniel and {Ellman}, Nancy and {Elms}, Brian R. and {Evans}, Michael L. and {Fan}, Xiaohui and {Federwitz}, Glenn R. and {Fiscelli}, Larry and {Friedman}, Scott and {Frieman}, Joshua A. and {Fukugita}, Masataka and {Gillespie}, Bruce and {Gunn}, James E. and {Gurbani}, Vijay K. and {de Haas}, Ernst and {Haldeman}, Merle and {Harris}, Frederick H. and {Hayes}, J. and {Heckman}, Timothy M. and {Hennessy}, G.~S. and {Hindsley}, Robert B. and {Holm}, Scott and {Holmgren}, Donald J. and {Huang}, Chi-hao and {Hull}, Charles and {Husby}, Don and {Ichikawa}, Shin-Ichi and {Ichikawa}, Takashi and {Ivezi{\'c}}, {\v{Z}}eljko and {Kent}, Stephen and {Kim}, Rita S.~J. and {Kinney}, E. and {Klaene}, Mark and {Kleinman}, A.~N. and {Kleinman}, S. and {Knapp}, G.~R. and {Korienek}, John and {Kron}, Richard G. and {Kunszt}, Peter Z. and {Lamb}, D.~Q. and {Lee}, B. and {Leger}, R. French and {Limmongkol}, Siriluk and {Lindenmeyer}, Carl and {Long}, Daniel C. and {Loomis}, Craig and {Loveday}, Jon and {Lucinio}, Rich and {Lupton}, Robert H. and {MacKinnon}, Bryan and {Mannery}, Edward J. and {Mantsch}, P.~M. and {Margon}, Bruce and {McGehee}, Peregrine and {McKay}, Timothy A. and {Meiksin}, Avery and {Merelli}, Aronne and {Monet}, David G. and {Munn}, Jeffrey A. and {Narayanan}, Vijay K. and {Nash}, Thomas and {Neilsen}, Eric and {Neswold}, Rich and {Newberg}, Heidi Jo and {Nichol}, R.~C. and {Nicinski}, Tom and {Nonino}, Mario and {Okada}, Norio and {Okamura}, Sadanori and {Ostriker}, Jeremiah P. and {Owen}, Russell and {Pauls}, A. George and {Peoples}, John and {Peterson}, R.~L. and {Petravick}, Donald and {Pier}, Jeffrey R. and {Pope}, Adrian and {Pordes}, Ruth and {Prosapio}, Angela and {Rechenmacher}, Ron and {Quinn}, Thomas R. and {Richards}, Gordon T. and {Richmond}, Michael W. and {Rivetta}, Claudio H. and {Rockosi}, Constance M. and {Ruthmansdorfer}, Kurt and {Sandford}, Dale and {Schlegel}, David J. and {Schneider}, Donald P. and {Sekiguchi}, Maki and {Sergey}, Gary and {Shimasaku}, Kazuhiro and {Siegmund}, Walter A. and {Smee}, Stephen and {Smith}, J. Allyn and {Snedden}, S. and {Stone}, R. and {Stoughton}, Chris and {Strauss}, Michael A. and {Stubbs}, Christopher and {SubbaRao}, Mark and {Szalay}, Alexander S. and {Szapudi}, Istvan and {Szokoly}, Gyula P. and {Thakar}, Anirudda R. and {Tremonti}, Christy and {Tucker}, Douglas L. and {Uomoto}, Alan and {Vanden Berk}, Dan and {Vogeley}, Michael S. and {Waddell}, Patrick and {Wang}, Shu-i. and {Watanabe}, Masaru and {Weinberg}, David H. and {Yanny}, Brian and {Yasuda}, Naoki and {SDSS Collaboration}},
        title = "{The Sloan Digital Sky Survey: Technical Summary}",
      journal = {\aj},
         year = 2000,
        month = sep,
       volume = {120},
       number = {3},
        pages = {1579-1587},
          doi = {10.1086/301513},
archivePrefix = {arXiv},
       eprint = {astro-ph/0006396},
 primaryClass = {astro-ph},
       adsurl = {https://ui.adsabs.harvard.edu/abs/2000AJ....120.1579Y}
}

@ARTICLE{guetter2003,
       author = {{Guetter}, H.~H. and {Vrba}, F.~J. and {Henden}, A.~A. and {Luginbuhl}, C.~B.},
        title = "{JHK Standard Stars on the CIT Photometric System}",
      journal = {\aj},
         year = 2003,
        month = jun,
       volume = {125},
       number = {6},
        pages = {3344-3348},
          doi = {10.1086/375305},
       adsurl = {https://ui.adsabs.harvard.edu/abs/2003AJ....125.3344G}
}

@ARTICLE{kirkpatrick2019,
       author = {{Kirkpatrick}, J. Davy and {Martin}, Emily C. and {Smart}, Richard L. and {Cayago}, Alfred J. and {Beichman}, Charles A. and {Marocco}, Federico and {Gelino}, Christopher R. and {Faherty}, Jacqueline K. and {Cushing}, Michael C. and {Schneider}, Adam C. and {Mace}, Gregory N. and {Tinney}, Christopher G. and {Wright}, Edward L. and {Lowrance}, Patrick J. and {Ingalls}, James G. and {Vrba}, Frederick J. and {Munn}, Jeffrey A. and {Dahm}, Scott E. and {McLean}, Ian S.},
        title = "{Preliminary Trigonometric Parallaxes of 184 Late-T and Y Dwarfs and an Analysis of the Field Substellar Mass Function into the {\textquotedblleft}Planetary{\textquotedblright} Mass Regime}",
      journal = {\apjs},
         year = 2019,
        month = feb,
       volume = {240},
       number = {2},
          eid = {19},
        pages = {19},
          doi = {10.3847/1538-4365/aaf6af},
archivePrefix = {arXiv},
       eprint = {1812.01208},
 primaryClass = {astro-ph.SR},
       adsurl = {https://ui.adsabs.harvard.edu/abs/2019ApJS..240...19K}
}

@ARTICLE{kirkpatrick2021,
       author = {{Kirkpatrick}, J. Davy and {Marocco}, Federico and {Caselden}, Dan and {Meisner}, Aaron M. and {Faherty}, Jacqueline K. and {Schneider}, Adam C. and {Kuchner}, Marc J. and {Casewell}, S.~L. and {Gelino}, Christopher R. and {Cushing}, Michael C. and {Eisenhardt}, Peter R. and {Wright}, Edward L. and {Schurr}, Steven D.},
        title = "{The Enigmatic Brown Dwarf WISEA J153429.75-104303.3 (a.k.a. ``The Accident'')}",
      journal = {\apjl},
         year = 2021,
        month = jul,
       volume = {915},
       number = {1},
          eid = {L6},
        pages = {L6},
          doi = {10.3847/2041-8213/ac0437},
archivePrefix = {arXiv},
       eprint = {2106.13408},
 primaryClass = {astro-ph.SR},
       adsurl = {https://ui.adsabs.harvard.edu/abs/2021ApJ...915L...6K}
}

@ARTICLE{marocco2013,
       author = {{Marocco}, F. and {Andrei}, A.~H. and {Smart}, R.~L. and {Jones}, H.~R.~A. and {Pinfield}, D.~J. and {Day-Jones}, A.~C. and {Clarke}, J.~R.~A. and {Sozzetti}, A. and {Lucas}, P.~W. and {Bucciarelli}, B. and {Penna}, J.~L.},
        title = "{Parallaxes of Southern Extremely Cool Objects (PARSEC). II. Spectroscopic Follow-up and Parallaxes of 52 Targets}",
      journal = {\aj},
         year = 2013,
        month = dec,
       volume = {146},
       number = {6},
          eid = {161},
        pages = {161},
          doi = {10.1088/0004-6256/146/6/161},
archivePrefix = {arXiv},
       eprint = {1309.6525},
 primaryClass = {astro-ph.SR},
       adsurl = {https://ui.adsabs.harvard.edu/abs/2013AJ....146..161M}
}

@ARTICLE{wright2010,
       author = {{Wright}, Edward L. and {Eisenhardt}, Peter R.~M. and {Mainzer}, Amy K. and {Ressler}, Michael E. and {Cutri}, Roc M. and {Jarrett}, Thomas and {Kirkpatrick}, J. Davy and {Padgett}, Deborah and {McMillan}, Robert S. and {Skrutskie}, Michael and {Stanford}, S.~A. and {Cohen}, Martin and {Walker}, Russell G. and {Mather}, John C. and {Leisawitz}, David and {Gautier}, III, Thomas N. and {McLean}, Ian and {Benford}, Dominic and {Lonsdale}, Carol J. and {Blain}, Andrew and {Mendez}, Bryan and {Irace}, William R. and {Duval}, Valerie and {Liu}, Fengchuan and {Royer}, Don and {Heinrichsen}, Ingolf and {Howard}, Joan and {Shannon}, Mark and {Kendall}, Martha and {Walsh}, Amy L. and {Larsen}, Mark and {Cardon}, Joel G. and {Schick}, Scott and {Schwalm}, Mark and {Abid}, Mohamed and {Fabinsky}, Beth and {Naes}, Larry and {Tsai}, Chao-Wei},
        title = "{The Wide-field Infrared Survey Explorer (WISE): Mission Description and Initial On-orbit Performance}",
      journal = {\aj},
         year = 2010,
        month = dec,
       volume = {140},
       number = {6},
        pages = {1868-1881},
          doi = {10.1088/0004-6256/140/6/1868},
archivePrefix = {arXiv},
       eprint = {1008.0031},
 primaryClass = {astro-ph.IM},
       adsurl = {https://ui.adsabs.harvard.edu/abs/2010AJ....140.1868W}
}

@ARTICLE{lawrence2007,
       author = {{Lawrence}, A. and {Warren}, S.~J. and {Almaini}, O. and {Edge}, A.~C. and {Hambly}, N.~C. and {Jameson}, R.~F. and {Lucas}, P. and {Casali}, M. and {Adamson}, A. and {Dye}, S. and {Emerson}, J.~P. and {Foucaud}, S. and {Hewett}, P. and {Hirst}, P. and {Hodgkin}, S.~T. and {Irwin}, M.~J. and {Lodieu}, N. and {McMahon}, R.~G. and {Simpson}, C. and {Smail}, I. and {Mortlock}, D. and {Folger}, M.},
        title = "{The UKIRT Infrared Deep Sky Survey (UKIDSS)}",
      journal = {\mnras},
         year = 2007,
        month = aug,
       volume = {379},
       number = {4},
        pages = {1599-1617},
          doi = {10.1111/j.1365-2966.2007.12040.x},
archivePrefix = {arXiv},
       eprint = {astro-ph/0604426},
 primaryClass = {astro-ph},
       adsurl = {https://ui.adsabs.harvard.edu/abs/2007MNRAS.379.1599L}
}

@ARTICLE{chambers2016,
       author = {{Chambers}, K.~C. and {Magnier}, E.~A. and {Metcalfe}, N. and {Flewelling}, H.~A. and {Huber}, M.~E. and {Waters}, C.~Z. and {Denneau}, L. and {Draper}, P.~W. and {Farrow}, D. and {Finkbeiner}, D.~P. and {Holmberg}, C. and {Koppenhoefer}, J. and {Price}, P.~A. and {Rest}, A. and {Saglia}, R.~P. and {Schlafly}, E.~F. and {Smartt}, S.~J. and {Sweeney}, W. and {Wainscoat}, R.~J. and {Burgett}, W.~S. and {Chastel}, S. and {Grav}, T. and {Heasley}, J.~N. and {Hodapp}, K.~W. and {Jedicke}, R. and {Kaiser}, N. and {Kudritzki}, R. -P. and {Luppino}, G.~A. and {Lupton}, R.~H. and {Monet}, D.~G. and {Morgan}, J.~S. and {Onaka}, P.~M. and {Shiao}, B. and {Stubbs}, C.~W. and {Tonry}, J.~L. and {White}, R. and {Ba{\~n}ados}, E. and {Bell}, E.~F. and {Bender}, R. and {Bernard}, E.~J. and {Boegner}, M. and {Boffi}, F. and {Botticella}, M.~T. and {Calamida}, A. and {Casertano}, S. and {Chen}, W. -P. and {Chen}, X. and {Cole}, S. and {Deacon}, N. and {Frenk}, C. and {Fitzsimmons}, A. and {Gezari}, S. and {Gibbs}, V. and {Goessl}, C. and {Goggia}, T. and {Gourgue}, R. and {Goldman}, B. and {Grant}, P. and {Grebel}, E.~K. and {Hambly}, N.~C. and {Hasinger}, G. and {Heavens}, A.~F. and {Heckman}, T.~M. and {Henderson}, R. and {Henning}, T. and {Holman}, M. and {Hopp}, U. and {Ip}, W. -H. and {Isani}, S. and {Jackson}, M. and {Keyes}, C.~D. and {Koekemoer}, A.~M. and {Kotak}, R. and {Le}, D. and {Liska}, D. and {Long}, K.~S. and {Lucey}, J.~R. and {Liu}, M. and {Martin}, N.~F. and {Masci}, G. and {McLean}, B. and {Mindel}, E. and {Misra}, P. and {Morganson}, E. and {Murphy}, D.~N.~A. and {Obaika}, A. and {Narayan}, G. and {Nieto-Santisteban}, M.~A. and {Norberg}, P. and {Peacock}, J.~A. and {Pier}, E.~A. and {Postman}, M. and {Primak}, N. and {Rae}, C. and {Rai}, A. and {Riess}, A. and {Riffeser}, A. and {Rix}, H.~W. and {R{\"o}ser}, S. and {Russel}, R. and {Rutz}, L. and {Schilbach}, E. and {Schultz}, A.~S.~B. and {Scolnic}, D. and {Strolger}, L. and {Szalay}, A. and {Seitz}, S. and {Small}, E. and {Smith}, K.~W. and {Soderblom}, D.~R. and {Taylor}, P. and {Thomson}, R. and {Taylor}, A.~N. and {Thakar}, A.~R. and {Thiel}, J. and {Thilker}, D. and {Unger}, D. and {Urata}, Y. and {Valenti}, J. and {Wagner}, J. and {Walder}, T. and {Walter}, F. and {Watters}, S.~P. and {Werner}, S. and {Wood-Vasey}, W.~M. and {Wyse}, R.},
        title = "{The Pan-STARRS1 Surveys}",
      journal = {arXiv e-prints},
         year = 2016,
        month = dec,
          eid = {arXiv:1612.05560},
        pages = {arXiv:1612.05560},
          doi = {10.48550/arXiv.1612.05560},
archivePrefix = {arXiv},
       eprint = {1612.05560},
 primaryClass = {astro-ph.IM},
       adsurl = {https://ui.adsabs.harvard.edu/abs/2016arXiv161205560C}
}

@ARTICLE{fabricius2021,
       author = {{Fabricius}, C. and {Luri}, X. and {Arenou}, F. and {Babusiaux}, C. and {Helmi}, A. and {Muraveva}, T. and {Reyl{\'e}}, C. and {Spoto}, F. and {Vallenari}, A. and {Antoja}, T. and {Balbinot}, E. and {Barache}, C. and {Bauchet}, N. and {Bragaglia}, A. and {Busonero}, D. and {Cantat-Gaudin}, T. and {Carrasco}, J.~M. and {Diakit{\'e}}, S. and {Fabrizio}, M. and {Figueras}, F. and {Garcia-Gutierrez}, A. and {Garofalo}, A. and {Jordi}, C. and {Kervella}, P. and {Khanna}, S. and {Leclerc}, N. and {Licata}, E. and {Lambert}, S. and {Marrese}, P.~M. and {Masip}, A. and {Ramos}, P. and {Robichon}, N. and {Robin}, A.~C. and {Romero-G{\'o}mez}, M. and {Rubele}, S. and {Weiler}, M.},
        title = "{Gaia Early Data Release 3. Catalogue validation}",
      journal = {\aap},
         year = 2021,
        month = may,
       volume = {649},
          eid = {A5},
        pages = {A5},
          doi = {10.1051/0004-6361/202039834},
archivePrefix = {arXiv},
       eprint = {2012.06242},
 primaryClass = {astro-ph.GA},
       adsurl = {https://ui.adsabs.harvard.edu/abs/2021A&A...649A...5F}
}

@ARTICLE{radigan2013,
       author = {{Radigan}, Jacqueline and {Jayawardhana}, Ray and {Lafreni{\`e}re}, David and {Dupuy}, Trent J. and {Liu}, Michael C. and {Scholz}, Alexander},
        title = "{Discovery of a Visual T-dwarf Triple System and Binarity at the L/T Transition}",
      journal = {\apj},
         year = 2013,
        month = nov,
       volume = {778},
       number = {1},
          eid = {36},
        pages = {36},
          doi = {10.1088/0004-637X/778/1/36},
archivePrefix = {arXiv},
       eprint = {1308.5702},
 primaryClass = {astro-ph.SR},
       adsurl = {https://ui.adsabs.harvard.edu/abs/2013ApJ...778...36R}
}

@ARTICLE{hodgkin2009,
       author = {{Hodgkin}, S.~T. and {Irwin}, M.~J. and {Hewett}, P.~C. and {Warren}, S.~J.},
        title = "{The UKIRT wide field camera ZYJHK photometric system: calibration from 2MASS}",
      journal = {\mnras},
         year = 2009,
        month = apr,
       volume = {394},
       number = {2},
        pages = {675-692},
          doi = {10.1111/j.1365-2966.2008.14387.x},
archivePrefix = {arXiv},
       eprint = {0812.3081},
 primaryClass = {astro-ph},
       adsurl = {https://ui.adsabs.harvard.edu/abs/2009MNRAS.394..675H}
}

@ARTICLE{geissler2011,
       author = {{Gei{\ss}ler}, Kerstin and {Metchev}, Stanimir and {Kirkpatrick}, J. Davy and {Berriman}, G. Bruce and {Looper}, Dagny},
        title = "{A Cross-match of 2MASS and SDSS. II. Peculiar L Dwarfs, Unresolved Binaries, and the Space Density of T Dwarf Secondaries}",
      journal = {\apj},
         year = 2011,
        month = may,
       volume = {732},
       number = {1},
          eid = {56},
        pages = {56},
          doi = {10.1088/0004-637X/732/1/56},
archivePrefix = {arXiv},
       eprint = {1103.1160},
 primaryClass = {astro-ph.SR},
       adsurl = {https://ui.adsabs.harvard.edu/abs/2011ApJ...732...56G}
}

@ARTICLE{kirkpatrick2005,
       author = {{Kirkpatrick}, J. Davy},
        title = "{New Spectral Types L and T}",
      journal = {\araa},
         year = 2005,
        month = sep,
       volume = {43},
       number = {1},
        pages = {195-245},
          doi = {10.1146/annurev.astro.42.053102.134017},
       adsurl = {https://ui.adsabs.harvard.edu/abs/2005ARA&A..43..195K}
}

@ARTICLE{zuckerman2019,
       author = {{Zuckerman}, B.},
        title = "{The Nearby, Young, Argus Association: Membership, Age, and Dusty Debris Disks}",
      journal = {\apj},
         year = 2019,
        month = jan,
       volume = {870},
       number = {1},
          eid = {27},
        pages = {27},
          doi = {10.3847/1538-4357/aaee66},
archivePrefix = {arXiv},
       eprint = {1811.01508},
 primaryClass = {astro-ph.SR},
       adsurl = {https://ui.adsabs.harvard.edu/abs/2019ApJ...870...27Z}
}

@INCOLLECTION{torres2008,
       author = {{Torres}, C.~A.~O. and {Quast}, G.~R. and {Melo}, C.~H.~F. and {Sterzik}, M.~F.},
        title = "{Young Nearby Loose Associations}",
    booktitle = {Handbook of Star Forming Regions, Volume II},
         year = 2008,
       editor = {{Reipurth}, B.},
       volume = {5},
        pages = {757},
          doi = {10.48550/arXiv.0808.3362},
       adsurl = {https://ui.adsabs.harvard.edu/abs/2008hsf2.book..757T}
}

@ARTICLE{gagne2014,
       author = {{Gagn{\'e}}, Jonathan and {Lafreni{\`e}re}, David and {Doyon}, Ren{\'e} and {Malo}, Lison and {Artigau}, {\'E}tienne},
        title = "{BANYAN. II. Very Low Mass and Substellar Candidate Members to Nearby, Young Kinematic Groups with Previously Known Signs of Youth}",
      journal = {\apj},
         year = 2014,
        month = mar,
       volume = {783},
       number = {2},
          eid = {121},
        pages = {121},
          doi = {10.1088/0004-637X/783/2/121},
archivePrefix = {arXiv},
       eprint = {1312.5864},
 primaryClass = {astro-ph.SR},
       adsurl = {https://ui.adsabs.harvard.edu/abs/2014ApJ...783..121G}
}

@ARTICLE{greco2019,
       author = {{Greco}, Jennifer J. and {Schneider}, Adam C. and {Cushing}, Michael C. and {Kirkpatrick}, J. Davy and {Burgasser}, Adam J.},
        title = "{Spectroscopic Follow-up of Discoveries from the NEOWISE Proper Motion Survey}",
      journal = {\aj},
         year = 2019,
        month = nov,
       volume = {158},
       number = {5},
          eid = {182},
        pages = {182},
          doi = {10.3847/1538-3881/ab3ebe},
archivePrefix = {arXiv},
       eprint = {1908.10988},
 primaryClass = {astro-ph.SR},
       adsurl = {https://ui.adsabs.harvard.edu/abs/2019AJ....158..182G}
}

@ARTICLE{burgasser2025,
       author = {{Burgasser}, Adam J. and {Schneider}, Adam C. and {Meisner}, Aaron M. and {Caselden}, Dan and {Hsu}, Chih-Chun and {Gerasimov}, Roman and {Aganze}, Christian and {Softich}, Emma and {Karpoor}, Preethi and {Theissen}, Christopher A. and {Brooks}, Hunter and {Bickle}, Thomas P. and {Gagn{\'e}}, Jonathan and {Artigau}, {\`E}tienne and {Marsset}, Micha{\"e}l and {Rothermich}, Austin and {Faherty}, Jacqueline K. and {Kirkpatrick}, J. Davy and {Kuchner}, Marc J. and {Andersen}, Nikolaj Stevnbak and {Beaulieu}, Paul and {Colin}, Guillaume and {Gantier}, Jean Marc and {Gramaize}, Leopold and {Hamlet}, Les and {Hinckley}, Ken and {Kabatnik}, Martin and {Kiwy}, Frank and {Martin}, David W. and {Massat}, Diego H. and {Pendrill}, William and {Sainio}, Arttu and {Sch{\"u}mann}, J{\"o}rg and {Th{\'e}venot}, Melina and {Walla}, Jim and {W{\k{e}}dracki}, Zbigniew and {Backyard Worlds: Planet 9 Collaboration}},
        title = "{New Cold Subdwarf Discoveries from Backyard Worlds and a Metallicity Classification System for T Subdwarfs}",
      journal = {\apj},
         year = 2025,
        month = apr,
       volume = {982},
       number = {2},
          eid = {79},
        pages = {79},
          doi = {10.3847/1538-4357/adb39f},
archivePrefix = {arXiv},
       eprint = {2411.01378},
 primaryClass = {astro-ph.SR},
       adsurl = {https://ui.adsabs.harvard.edu/abs/2025ApJ...982...79B}
}

@ARTICLE{stumpf2010,
       author = {{Stumpf}, M.~B. and {Brandner}, W. and {Bouy}, H. and {Henning}, Th. and {Hippler}, S.},
        title = "{2MASS J03105986 +1648155 AB - a new binary at the L/T transition}",
      journal = {\aap},
         year = 2010,
        month = jun,
       volume = {516},
          eid = {A37},
        pages = {A37},
          doi = {10.1051/0004-6361/200913711},
archivePrefix = {arXiv},
       eprint = {1003.3807},
 primaryClass = {astro-ph.SR},
       adsurl = {https://ui.adsabs.harvard.edu/abs/2010A&A...516A..37S}
}

@ARTICLE{dupuy2017,
       author = {{Dupuy}, Trent J. and {Liu}, Michael C.},
        title = "{Individual Dynamical Masses of Ultracool Dwarfs}",
      journal = {\apjs},
         year = 2017,
        month = aug,
       volume = {231},
       number = {2},
          eid = {15},
        pages = {15},
          doi = {10.3847/1538-4365/aa5e4c},
archivePrefix = {arXiv},
       eprint = {1703.05775},
 primaryClass = {astro-ph.SR},
       adsurl = {https://ui.adsabs.harvard.edu/abs/2017ApJS..231...15D}
}

@ARTICLE{konopacky2010,
       author = {{Konopacky}, Q.~M. and {Ghez}, A.~M. and {Barman}, T.~S. and {Rice}, E.~L. and {Bailey}, III, J.~I. and {White}, R.~J. and {McLean}, I.~S. and {Duch{\^e}ne}, G.},
        title = "{High-precision Dynamical Masses of Very Low Mass Binaries}",
      journal = {\apj},
         year = 2010,
        month = mar,
       volume = {711},
       number = {2},
        pages = {1087-1122},
          doi = {10.1088/0004-637X/711/2/1087},
archivePrefix = {arXiv},
       eprint = {1001.4800},
 primaryClass = {astro-ph.SR},
       adsurl = {https://ui.adsabs.harvard.edu/abs/2010ApJ...711.1087K}
}

@ARTICLE{liu2010,
       author = {{Liu}, Michael C. and {Dupuy}, Trent J. and {Leggett}, S.~K.},
        title = "{Discovery of a Highly Unequal-mass Binary T Dwarf with Keck Laser Guide Star Adaptive Optics: A Coevality Test of Substellar Theoretical Models and Effective Temperatures}",
      journal = {\apj},
         year = 2010,
        month = oct,
       volume = {722},
       number = {1},
        pages = {311-328},
          doi = {10.1088/0004-637X/722/1/311},
archivePrefix = {arXiv},
       eprint = {1008.2200},
 primaryClass = {astro-ph.SR},
       adsurl = {https://ui.adsabs.harvard.edu/abs/2010ApJ...722..311L}
}

@ARTICLE{burgasser2006c,
       author = {{Burgasser}, Adam J. and {Kirkpatrick}, J. Davy and {Cruz}, Kelle L. and {Reid}, I. Neill and {Leggett}, Sandy K. and {Liebert}, James and {Burrows}, Adam and {Brown}, Michael E.},
        title = "{Hubble Space Telescope NICMOS Observations of T Dwarfs: Brown Dwarf Multiplicity and New Probes of the L/T Transition}",
      journal = {\apjs},
         year = 2006,
        month = oct,
       volume = {166},
       number = {2},
        pages = {585-612},
          doi = {10.1086/506327},
archivePrefix = {arXiv},
       eprint = {astro-ph/0605577},
 primaryClass = {astro-ph},
       adsurl = {https://ui.adsabs.harvard.edu/abs/2006ApJS..166..585B}
}

@ARTICLE{liu2012,
       author = {{Liu}, Michael C. and {Dupuy}, Trent J. and {Bowler}, Brendan P. and {Leggett}, S.~K. and {Best}, William M.~J.},
        title = "{Two Extraordinary Substellar Binaries at the T/Y Transition and the Y-band Fluxes of the Coolest Brown Dwarfs}",
      journal = {\apj},
         year = 2012,
        month = oct,
       volume = {758},
       number = {1},
          eid = {57},
        pages = {57},
          doi = {10.1088/0004-637X/758/1/57},
archivePrefix = {arXiv},
       eprint = {1206.4044},
 primaryClass = {astro-ph.SR},
       adsurl = {https://ui.adsabs.harvard.edu/abs/2012ApJ...758...57L}
}

@ARTICLE{leggett2014,
       author = {{Leggett}, S.~K. and {Liu}, Michael C. and {Dupuy}, Trent J. and {Morley}, Caroline V. and {Marley}, M.~S. and {Saumon}, D.},
        title = "{Resolved Spectroscopy of the T8.5 and Y0-0.5 Binary WISEPC J121756.91+162640.2AB}",
      journal = {\apj},
         year = 2014,
        month = jan,
       volume = {780},
       number = {1},
          eid = {62},
        pages = {62},
          doi = {10.1088/0004-637X/780/1/62},
archivePrefix = {arXiv},
       eprint = {1311.2108},
 primaryClass = {astro-ph.SR},
       adsurl = {https://ui.adsabs.harvard.edu/abs/2014ApJ...780...62L}
}

@ARTICLE{bouy2003,
       author = {{Bouy}, Herv{\'e} and {Brandner}, Wolfgang and {Mart{\'\i}n}, Eduardo L. and {Delfosse}, Xavier and {Allard}, France and {Basri}, Gibor},
        title = "{Multiplicity of Nearby Free-Floating Ultracool Dwarfs: A Hubble Space Telescope WFPC2 Search for Companions}",
      journal = {\aj},
         year = 2003,
        month = sep,
       volume = {126},
       number = {3},
        pages = {1526-1554},
          doi = {10.1086/377343},
archivePrefix = {arXiv},
       eprint = {astro-ph/0305484},
 primaryClass = {astro-ph},
       adsurl = {https://ui.adsabs.harvard.edu/abs/2003AJ....126.1526B}
}

@ARTICLE{deacon2014,
       author = {{Deacon}, Niall R. and {Liu}, Michael C. and {Magnier}, Eugene A. and {Aller}, Kimberly M. and {Best}, William M.~J. and {Dupuy}, Trent and {Bowler}, Brendan P. and {Mann}, Andrew W. and {Redstone}, Joshua A. and {Burgett}, William S. and {Chambers}, Kenneth C. and {Draper}, Peter W. and {Flewelling}, H. and {Hodapp}, Klaus W. and {Kaiser}, Nick and {Kudritzki}, Rolf-Peter and {Morgan}, Jeff S. and {Metcalfe}, Nigel and {Price}, Paul A. and {Tonry}, John L. and {Wainscoat}, Richard J.},
        title = "{Wide Cool and Ultracool Companions to Nearby Stars from Pan-STARRS 1}",
      journal = {\apj},
         year = 2014,
        month = sep,
       volume = {792},
       number = {2},
          eid = {119},
        pages = {119},
          doi = {10.1088/0004-637X/792/2/119},
archivePrefix = {arXiv},
       eprint = {1407.2938},
 primaryClass = {astro-ph.SR},
       adsurl = {https://ui.adsabs.harvard.edu/abs/2014ApJ...792..119D}
}

@ARTICLE{burgasser2005b,
       author = {{Burgasser}, Adam J. and {Kirkpatrick}, J. Davy and {Lowrance}, Patrick J.},
        title = "{Multiplicity among Widely Separated Brown Dwarf Companions to Nearby Stars: Gliese 337CD}",
      journal = {\aj},
         year = 2005,
        month = jun,
       volume = {129},
       number = {6},
        pages = {2849-2855},
          doi = {10.1086/430218},
archivePrefix = {arXiv},
       eprint = {astro-ph/0503379},
 primaryClass = {astro-ph},
       adsurl = {https://ui.adsabs.harvard.edu/abs/2005AJ....129.2849B}
}

@ARTICLE{zhang2020,
       author = {{Zhang}, Zhoujian and {Liu}, Michael C. and {Hermes}, J.~J. and {Magnier}, Eugene A. and {Marley}, Mark S. and {Tremblay}, Pier-Emmanuel and {Tucker}, Michael A. and {Do}, Aaron and {Payne}, Anna V. and {Shappee}, Benjamin J.},
        title = "{COol Companions ON Ultrawide orbiTS (COCONUTS). I. A High-gravity T4 Benchmark around an Old White Dwarf and a Re-examination of the Surface-gravity Dependence of the L/T Transition}",
      journal = {\apj},
         year = 2020,
        month = mar,
       volume = {891},
       number = {2},
          eid = {171},
        pages = {171},
          doi = {10.3847/1538-4357/ab765c},
archivePrefix = {arXiv},
       eprint = {2002.05723},
 primaryClass = {astro-ph.SR},
       adsurl = {https://ui.adsabs.harvard.edu/abs/2020ApJ...891..171Z}
}

@ARTICLE{kirkpatrick2001,
       author = {{Kirkpatrick}, J. Davy and {Dahn}, Conard C. and {Monet}, David G. and {Reid}, I. Neill and {Gizis}, John E. and {Liebert}, James and {Burgasser}, Adam J.},
        title = "{Brown Dwarf Companions to G-Type Stars. I. Gliese 417B and Gliese 584C}",
      journal = {\aj},
         year = 2001,
        month = jun,
       volume = {121},
       number = {6},
        pages = {3235-3253},
          doi = {10.1086/321085},
archivePrefix = {arXiv},
       eprint = {astro-ph/0103218},
 primaryClass = {astro-ph},
       adsurl = {https://ui.adsabs.harvard.edu/abs/2001AJ....121.3235K}
}

@ARTICLE{faherty2010,
       author = {{Faherty}, Jacqueline K. and {Burgasser}, Adam J. and {West}, Andrew A. and {Bochanski}, John J. and {Cruz}, Kelle L. and {Shara}, Michael M. and {Walter}, Frederick M.},
        title = "{The Brown Dwarf Kinematics Project. II. Details on Nine Wide Common Proper Motion Very Low Mass Companions to Nearby Stars}",
      journal = {\aj},
         year = 2010,
        month = jan,
       volume = {139},
       number = {1},
        pages = {176-194},
          doi = {10.1088/0004-6256/139/1/176},
archivePrefix = {arXiv},
       eprint = {0911.1363},
 primaryClass = {astro-ph.SR},
       adsurl = {https://ui.adsabs.harvard.edu/abs/2010AJ....139..176F}
}

@ARTICLE{bardalez2014,
       author = {{Bardalez Gagliuffi}, Daniella C. and {Burgasser}, Adam J. and {Gelino}, Christopher R. and {Looper}, Dagny L. and {Nicholls}, Christine P. and {Schmidt}, Sarah J. and {Cruz}, Kelle and {West}, Andrew A. and {Gizis}, John E. and {Metchev}, Stanimir},
        title = "{SpeX Spectroscopy of Unresolved Very Low Mass Binaries. II. Identification of 14 Candidate Binaries with Late-M/Early-L and T Dwarf Components}",
      journal = {\apj},
         year = 2014,
        month = oct,
       volume = {794},
       number = {2},
          eid = {143},
        pages = {143},
          doi = {10.1088/0004-637X/794/2/143},
archivePrefix = {arXiv},
       eprint = {1408.3089},
 primaryClass = {astro-ph.SR},
       adsurl = {https://ui.adsabs.harvard.edu/abs/2014ApJ...794..143B}
}

@ARTICLE{manjavacas2019,
       author = {{Manjavacas}, Elena and {Apai}, D{\'a}niel and {Zhou}, Yifan and {Lew}, Ben W.~P. and {Schneider}, Glenn and {Metchev}, Stan and {Miles-P{\'a}ez}, Paulo A. and {Radigan}, Jacqueline and {Marley}, Mark S. and {Cowan}, Nicolas and {Karalidi}, Theodora and {Burgasser}, Adam J. and {Bedin}, Luigi R. and {Lowrance}, Patrick J. and {Kauffmann}, Parker},
        title = "{Cloud Atlas: Hubble Space Telescope Near-infrared Spectral Library of Brown Dwarfs, Planetary-mass Companions, and Hot Jupiters}",
      journal = {\aj},
         year = 2019,
        month = mar,
       volume = {157},
       number = {3},
          eid = {101},
        pages = {101},
          doi = {10.3847/1538-3881/aaf88f},
archivePrefix = {arXiv},
       eprint = {1812.03963},
 primaryClass = {astro-ph.SR},
       adsurl = {https://ui.adsabs.harvard.edu/abs/2019AJ....157..101M}
}

@ARTICLE{best2015,
       author = {{Best}, William M.~J. and {Liu}, Michael C. and {Magnier}, Eugene A. and {Deacon}, Niall R. and {Aller}, Kimberly M. and {Redstone}, Joshua and {Burgett}, W.~S. and {Chambers}, K.~C. and {Draper}, P. and {Flewelling}, H. and {Hodapp}, K.~W. and {Kaiser}, N. and {Metcalfe}, N. and {Tonry}, J.~L. and {Wainscoat}, R.~J. and {Waters}, C.},
        title = "{A Search for L/T Transition Dwarfs with Pan-STARRS1 and WISE. II. L/T Transition Atmospheres and Young Discoveries}",
      journal = {\apj},
         year = 2015,
        month = dec,
       volume = {814},
       number = {2},
          eid = {118},
        pages = {118},
          doi = {10.1088/0004-637X/814/2/118},
archivePrefix = {arXiv},
       eprint = {1612.02824},
 primaryClass = {astro-ph.SR},
       adsurl = {https://ui.adsabs.harvard.edu/abs/2015ApJ...814..118B}
}

@ARTICLE{bravo2023,
       author = {{Bravo}, Alexia and {Schneider}, Adam C. and {Bardalez Gagliuffi}, Daniella and {Burgasser}, Adam J. and {Meisner}, Aaron M. and {Kirkpatrick}, J. Davy and {Faherty}, Jacqueline K. and {Kuchner}, Marc J. and {Caselden}, Dan and {Sainio}, Arttu and {Hamlet}, Les and {Backyard Worlds: Planet 9 Collaboration}},
        title = "{An Investigation of New Brown Dwarf Spectral Binary Candidates From the Backyard Worlds: Planet 9 Citizen Science Initiative}",
      journal = {\aj},
         year = 2023,
        month = dec,
       volume = {166},
       number = {6},
          eid = {226},
        pages = {226},
          doi = {10.3847/1538-3881/acffc1},
archivePrefix = {arXiv},
       eprint = {2310.06957},
 primaryClass = {astro-ph.SR},
       adsurl = {https://ui.adsabs.harvard.edu/abs/2023AJ....166..226B}
}

@INPROCEEDINGS{burgasser2017,
       author = {{Burgasser}, A.~J. and {Splat Development Team}},
        title = "{The SpeX Prism Library Analysis Toolkit (SPLAT): A Data Curation Model}",
    booktitle = {Astronomical Society of India Conference Series},
         year = 2017,
       series = {Astronomical Society of India Conference Series},
       volume = {14},
        month = jan,
        pages = {7-12},
          doi = {10.48550/arXiv.1707.00062},
archivePrefix = {arXiv},
       eprint = {1707.00062},
 primaryClass = {astro-ph.SR},
       adsurl = {https://ui.adsabs.harvard.edu/abs/2017ASInC..14....7B}
}

@ARTICLE{gagne2018,
       author = {{Gagn{\'e}}, Jonathan and {Mamajek}, Eric E. and {Malo}, Lison and {Riedel}, Adric and {Rodriguez}, David and {Lafreni{\`e}re}, David and {Faherty}, Jacqueline K. and {Roy-Loubier}, Olivier and {Pueyo}, Laurent and {Robin}, Annie C. and {Doyon}, Ren{\'e}},
        title = "{BANYAN. XI. The BANYAN {\ensuremath{\Sigma}} Multivariate Bayesian Algorithm to Identify Members of Young Associations with 150 pc}",
      journal = {\apj},
         year = 2018,
        month = mar,
       volume = {856},
       number = {1},
          eid = {23},
        pages = {23},
          doi = {10.3847/1538-4357/aaae09},
archivePrefix = {arXiv},
       eprint = {1801.09051},
 primaryClass = {astro-ph.SR},
       adsurl = {https://ui.adsabs.harvard.edu/abs/2018ApJ...856...23G}
}

@ARTICLE{Schmidt2007,
       author = {{Schmidt}, Sarah J. and {Cruz}, Kelle L. and {Bongiorno}, Bethany J. and {Liebert}, James and {Reid}, I. Neill},
        title = "{Activity and Kinematics of Ultracool Dwarfs, Including an Amazing Flare Observation}",
      journal = {\aj},
         year = 2007,
        month = may,
       volume = {133},
       number = {5},
        pages = {2258-2273},
          doi = {10.1086/512158},
archivePrefix = {arXiv},
       eprint = {astro-ph/0701055},
 primaryClass = {astro-ph},
       adsurl = {https://ui.adsabs.harvard.edu/abs/2007AJ....133.2258S}
}

@ARTICLE{zapatero2014,
       author = {{Zapatero Osorio}, M.~R. and {B{\'e}jar}, V.~J.~S. and {Miles-P{\'a}ez}, P.~A. and {Pe{\~n}a Ram{\'\i}rez}, K. and {Rebolo}, R. and {Pall{\'e}}, E.},
        title = "{Trigonometric parallaxes of young field L dwarfs}",
      journal = {\aap},
         year = 2014,
        month = aug,
       volume = {568},
          eid = {A6},
        pages = {A6},
          doi = {10.1051/0004-6361/201321340},
archivePrefix = {arXiv},
       eprint = {1406.1345},
 primaryClass = {astro-ph.SR},
       adsurl = {https://ui.adsabs.harvard.edu/abs/2014A&A...568A...6Z}
}

@ARTICLE{zuckerman2006,
       author = {{Zuckerman}, B. and {Bessell}, M.~S. and {Song}, Inseok and {Kim}, S.},
        title = "{The Carina-Near Moving Group}",
      journal = {\apjl},
         year = 2006,
        month = oct,
       volume = {649},
       number = {2},
        pages = {L115-L118},
          doi = {10.1086/508060},
archivePrefix = {arXiv},
       eprint = {astro-ph/0609041},
 primaryClass = {astro-ph},
       adsurl = {https://ui.adsabs.harvard.edu/abs/2006ApJ...649L.115Z}
}

@ARTICLE{marley2002,
       author = {{Marley}, Mark S. and {Seager}, S. and {Saumon}, D. and {Lodders}, Katharina and {Ackerman}, Andrew S. and {Freedman}, Richard S. and {Fan}, Xiaohui},
        title = "{Clouds and Chemistry: Ultracool Dwarf Atmospheric Properties from Optical and Infrared Colors}",
      journal = {\apj},
         year = 2002,
        month = mar,
       volume = {568},
       number = {1},
        pages = {335-342},
          doi = {10.1086/338800},
archivePrefix = {arXiv},
       eprint = {astro-ph/0105438},
 primaryClass = {astro-ph},
       adsurl = {https://ui.adsabs.harvard.edu/abs/2002ApJ...568..335M}
}

@ARTICLE{zhang2021,
       author = {{Zhang}, Zhoujian and {Liu}, Michael C. and {Best}, William M.~J. and {Dupuy}, Trent J. and {Siverd}, Robert J.},
        title = "{The Hawaii Infrared Parallax Program. V. New T-dwarf Members and Candidate Members of Nearby Young Moving Groups}",
      journal = {\apj},
         year = 2021,
        month = apr,
       volume = {911},
       number = {1},
          eid = {7},
        pages = {7},
          doi = {10.3847/1538-4357/abe3fa},
archivePrefix = {arXiv},
       eprint = {2102.05045},
 primaryClass = {astro-ph.EP},
       adsurl = {https://ui.adsabs.harvard.edu/abs/2021ApJ...911....7Z}
}

@ARTICLE{zuckerman2004,
       author = {{Zuckerman}, B. and {Song}, Inseok and {Bessell}, M.~S.},
        title = "{The AB Doradus Moving Group}",
      journal = {\apjl},
         year = 2004,
        month = sep,
       volume = {613},
       number = {1},
        pages = {L65-L68},
          doi = {10.1086/425036},
       adsurl = {https://ui.adsabs.harvard.edu/abs/2004ApJ...613L..65Z}
}

@ARTICLE{casewell2008,
       author = {{Casewell}, S.~L. and {Jameson}, R.~F. and {Burleigh}, M.~R.},
        title = "{Proper motions of field L and T dwarfs - II}",
      journal = {\mnras},
         year = 2008,
        month = nov,
       volume = {390},
       number = {4},
        pages = {1517-1526},
          doi = {10.1111/j.1365-2966.2008.13855.x},
archivePrefix = {arXiv},
       eprint = {0808.1633},
 primaryClass = {astro-ph},
       adsurl = {https://ui.adsabs.harvard.edu/abs/2008MNRAS.390.1517C}
}

@ARTICLE{hurt2024,
       author = {{Hurt}, Spencer A. and {Liu}, Michael C. and {Zhang}, Zhoujian and {Phillips}, Mark and {Allers}, Katelyn N. and {Deacon}, Niall R. and {Aller}, Kimberly M. and {Best}, William M.~J.},
        title = "{Uniform Forward-modeling Analysis of Ultracool Dwarfs. III. Late-M and L Dwarfs in Young Moving Groups, the Pleiades, and the Hyades}",
      journal = {\apj},
         year = 2024,
        month = jan,
       volume = {961},
       number = {1},
          eid = {121},
        pages = {121},
          doi = {10.3847/1538-4357/ad0b12},
archivePrefix = {arXiv},
       eprint = {2311.04268},
 primaryClass = {astro-ph.SR},
       adsurl = {https://ui.adsabs.harvard.edu/abs/2024ApJ...961..121H}
}

@ARTICLE{faherty2013,
       author = {{Faherty}, Jacqueline K. and {Rice}, Emily L. and {Cruz}, Kelle L. and {Mamajek}, Eric E. and {N{\'u}{\~n}ez}, Alejandro},
        title = "{2MASS J035523.37+113343.7: A Young, Dusty, Nearby, Isolated Brown Dwarf Resembling a Giant Exoplanet}",
      journal = {\aj},
         year = 2013,
        month = jan,
       volume = {145},
       number = {1},
          eid = {2},
        pages = {2},
          doi = {10.1088/0004-6256/145/1/2},
archivePrefix = {arXiv},
       eprint = {1206.5519},
 primaryClass = {astro-ph.SR},
       adsurl = {https://ui.adsabs.harvard.edu/abs/2013AJ....145....2F}
}

@ARTICLE{liu2013,
       author = {{Liu}, M.~C. and {Dupuy}, T.~J. and {Allers}, K.~N.},
        title = "{Infrared parallaxes of young field brown dwarfs and connections to directly imaged gas-giant exoplanets}",
      journal = {Astronomische Nachrichten},
         year = 2013,
        month = feb,
       volume = {334},
       number = {1-2},
        pages = {85-88},
          doi = {10.1002/asna.201211783},
archivePrefix = {arXiv},
       eprint = {1305.3270},
 primaryClass = {astro-ph.SR},
       adsurl = {https://ui.adsabs.harvard.edu/abs/2013AN....334...85L}
}

@ARTICLE{stephens2009,
       author = {{Stephens}, D.~C. and {Leggett}, S.~K. and {Cushing}, Michael C. and {Marley}, Mark S. and {Saumon}, D. and {Geballe}, T.~R. and {Golimowski}, David A. and {Fan}, Xiaohui and {Noll}, K.~S.},
        title = "{The 0.8-14.5 {\ensuremath{\mu}}m Spectra of Mid-L to Mid-T Dwarfs: Diagnostics of Effective Temperature, Grain Sedimentation, Gas Transport, and Surface Gravity}",
      journal = {\apj},
         year = 2009,
        month = sep,
       volume = {702},
       number = {1},
        pages = {154-170},
          doi = {10.1088/0004-637X/702/1/154},
archivePrefix = {arXiv},
       eprint = {0906.2991},
 primaryClass = {astro-ph.SR},
       adsurl = {https://ui.adsabs.harvard.edu/abs/2009ApJ...702..154S}
}

@ARTICLE{radigan2014,
       author = {{Radigan}, Jacqueline and {Lafreni{\`e}re}, David and {Jayawardhana}, Ray and {Artigau}, Etienne},
        title = "{Strong Brightness Variations Signal Cloudy-to-clear Transition of Brown Dwarfs}",
      journal = {\apj},
         year = 2014,
        month = oct,
       volume = {793},
       number = {2},
          eid = {75},
        pages = {75},
          doi = {10.1088/0004-637X/793/2/75},
archivePrefix = {arXiv},
       eprint = {1404.3247},
 primaryClass = {astro-ph.SR},
       adsurl = {https://ui.adsabs.harvard.edu/abs/2014ApJ...793...75R}
}

@ARTICLE{ashraf2022,
       author = {{Ashraf}, Afra and {Bardalez Gagliuffi}, Daniella C. and {Manjavacas}, Elena and {Vos}, Johanna M. and {Mechmann}, Claire and {Faherty}, Jacqueline K.},
        title = "{Disentangling the Signatures of Blended-light Atmospheres in L/T Transition Brown Dwarfs}",
      journal = {\apj},
         year = 2022,
        month = aug,
       volume = {934},
       number = {2},
          eid = {178},
        pages = {178},
          doi = {10.3847/1538-4357/ac7aab},
archivePrefix = {arXiv},
       eprint = {2206.09025},
 primaryClass = {astro-ph.SR},
       adsurl = {https://ui.adsabs.harvard.edu/abs/2022ApJ...934..178A}
}

@ARTICLE{zuckerman2001,
       author = {{Zuckerman}, B. and {Song}, Inseok and {Bessell}, M.~S. and {Webb}, R.~A.},
        title = "{The {\ensuremath{\beta}} Pictoris Moving Group}",
      journal = {\apjl},
         year = 2001,
        month = nov,
       volume = {562},
       number = {1},
        pages = {L87-L90},
          doi = {10.1086/337968},
       adsurl = {https://ui.adsabs.harvard.edu/abs/2001ApJ...562L..87Z}
}

@ARTICLE{hsu2021,
       author = {{Hsu}, Chih-Chun and {Burgasser}, Adam J. and {Theissen}, Christopher A. and {Gelino}, Christopher R. and {Birky}, Jessica L. and {Diamant}, Sharon J.~M. and {Bardalez Gagliuffi}, Daniella C. and {Aganze}, Christian and {Blake}, Cullen H. and {Faherty}, Jacqueline K.},
        title = "{The Brown Dwarf Kinematics Project (BDKP). V. Radial and Rotational Velocities of T Dwarfs from Keck/NIRSPEC High-resolution Spectroscopy}",
      journal = {\apjs},
         year = 2021,
        month = dec,
       volume = {257},
       number = {2},
          eid = {45},
        pages = {45},
          doi = {10.3847/1538-4365/ac1c7d},
archivePrefix = {arXiv},
       eprint = {2107.01222},
 primaryClass = {astro-ph.SR},
       adsurl = {https://ui.adsabs.harvard.edu/abs/2021ApJS..257...45H}
}

@ARTICLE{liu2024,
       author = {{Liu}, Pengyu and {Biller}, Beth A. and {Vos}, Johanna M. and {Whiteford}, Niall and {Zhang}, Zhoujian and {Liu}, Michael C. and {Fontanive}, Cl{\'e}mence and {Manjavacas}, Elena and {Henning}, Thomas and {Kenworthy}, Matthew A. and {Bonavita}, Mariangela and {Bonnefoy}, Micka{\"e}l and {Bubb}, Emma and {Petrus}, Simon and {Schlieder}, Joshua},
        title = "{A near-infrared variability survey of young planetary-mass objects}",
      journal = {\mnras},
         year = 2024,
        month = jan,
       volume = {527},
       number = {3},
        pages = {6624-6674},
          doi = {10.1093/mnras/stad3502},
archivePrefix = {arXiv},
       eprint = {2311.07218},
 primaryClass = {astro-ph.EP},
       adsurl = {https://ui.adsabs.harvard.edu/abs/2024MNRAS.527.6624L}
}

@ARTICLE{wilson2014,
       author = {{Wilson}, P.~A. and {Rajan}, A. and {Patience}, J.},
        title = "{The brown dwarf atmosphere monitoring (BAM) project. I. The largest near-IR monitoring survey of L and T dwarfs}",
      journal = {\aap},
         year = 2014,
        month = jun,
       volume = {566},
          eid = {A111},
        pages = {A111},
          doi = {10.1051/0004-6361/201322995},
archivePrefix = {arXiv},
       eprint = {1404.4633},
 primaryClass = {astro-ph.SR},
       adsurl = {https://ui.adsabs.harvard.edu/abs/2014A&A...566A.111W}
}

@ARTICLE{radigan2014b,
       author = {{Radigan}, Jacqueline},
        title = "{An Independent Analysis of the Brown Dwarf Atmosphere Monitoring (BAM) Data: Large-amplitude Variability is Rare Outside the L/T Transition}",
      journal = {\apj},
         year = 2014,
        month = dec,
       volume = {797},
       number = {2},
          eid = {120},
        pages = {120},
          doi = {10.1088/0004-637X/797/2/120},
archivePrefix = {arXiv},
       eprint = {1408.5919},
 primaryClass = {astro-ph.SR},
       adsurl = {https://ui.adsabs.harvard.edu/abs/2014ApJ...797..120R}
}

@ARTICLE{sanghi2023,
       author = {{Sanghi}, Aniket and {Liu}, Michael C. and {Best}, William M.~J. and {Dupuy}, Trent J. and {Siverd}, Robert J. and {Zhang}, Zhoujian and {Hurt}, Spencer A. and {Magnier}, Eugene A. and {Aller}, Kimberly M. and {Deacon}, Niall R.},
        title = "{The Hawaii Infrared Parallax Program. VI. The Fundamental Properties of 1000+ Ultracool Dwarfs and Planetary-mass Objects Using Optical to Mid-infrared Spectral Energy Distributions and Comparison to BT-Settl and ATMO 2020 Model Atmospheres}",
      journal = {\apj},
         year = 2023,
        month = dec,
       volume = {959},
       number = {1},
          eid = {63},
        pages = {63},
          doi = {10.3847/1538-4357/acff66},
archivePrefix = {arXiv},
       eprint = {2309.03082},
 primaryClass = {astro-ph.SR},
       adsurl = {https://ui.adsabs.harvard.edu/abs/2023ApJ...959...63S}
}

@ARTICLE{gagne2015c,
       author = {{Gagn{\'e}}, Jonathan and {Burgasser}, Adam J. and {Faherty}, Jacqueline K. and {Lafreni{\'e}re}, David and {Doyon}, Ren{\'e} and {Filippazzo}, Joseph C. and {Bowsher}, Emily and {Nicholls}, Christine P.},
        title = "{SDSS J111010.01+011613.1: A New Planetary-mass T Dwarf Member of the AB Doradus Moving Group}",
      journal = {\apjl},
         year = 2015,
        month = jul,
       volume = {808},
       number = {1},
          eid = {L20},
        pages = {L20},
          doi = {10.1088/2041-8205/808/1/L20},
archivePrefix = {arXiv},
       eprint = {1506.04195},
 primaryClass = {astro-ph.SR},
       adsurl = {https://ui.adsabs.harvard.edu/abs/2015ApJ...808L..20G}
}

@ARTICLE{yang2016,
       author = {{Yang}, Hao and {Apai}, D{\'a}niel and {Marley}, Mark S. and {Karalidi}, Theodora and {Flateau}, Davin and {Showman}, Adam P. and {Metchev}, Stanimir and {Buenzli}, Esther and {Radigan}, Jacqueline and {Artigau}, {\'E}tienne and {Lowrance}, Patrick J. and {Burgasser}, Adam J.},
        title = "{Extrasolar Storms: Pressure-dependent Changes in Light-curve Phase in Brown Dwarfs from Simultaneous HST and Spitzer Observations}",
      journal = {\apj},
         year = 2016,
        month = jul,
       volume = {826},
       number = {1},
          eid = {8},
        pages = {8},
          doi = {10.3847/0004-637X/826/1/8},
archivePrefix = {arXiv},
       eprint = {1605.02708},
 primaryClass = {astro-ph.EP},
       adsurl = {https://ui.adsabs.harvard.edu/abs/2016ApJ...826....8Y}
}

@ARTICLE{apai2017,
       author = {{Apai}, D. and {Karalidi}, T. and {Marley}, M.~S. and {Yang}, H. and {Flateau}, D. and {Metchev}, S. and {Cowan}, N.~B. and {Buenzli}, E. and {Burgasser}, A.~J. and {Radigan}, J. and {Artigau}, E. and {Lowrance}, P.},
        title = "{Zones, spots, and planetary-scale waves beating in brown dwarf atmospheres}",
      journal = {Science},
         year = 2017,
        month = aug,
       volume = {357},
       number = {6352},
        pages = {683-687},
          doi = {10.1126/science.aam9848},
       adsurl = {https://ui.adsabs.harvard.edu/abs/2017Sci...357..683A}
}

@ARTICLE{gagne2018b,
       author = {{Gagn{\'e}}, Jonathan and {Allers}, Katelyn N. and {Theissen}, Christopher A. and {Faherty}, Jacqueline K. and {Bardalez Gagliuffi}, Daniella and {Artigau}, {\'E}tienne},
        title = "{2MASS J13243553+6358281 Is an Early T-type Planetary-mass Object in the AB Doradus Moving Group}",
      journal = {\apjl},
         year = 2018,
        month = feb,
       volume = {854},
       number = {2},
          eid = {L27},
        pages = {L27},
          doi = {10.3847/2041-8213/aaacfd},
archivePrefix = {arXiv},
       eprint = {1802.00493},
 primaryClass = {astro-ph.SR},
       adsurl = {https://ui.adsabs.harvard.edu/abs/2018ApJ...854L..27G}
}

@ARTICLE{radigan2012,
       author = {{Radigan}, Jacqueline and {Jayawardhana}, Ray and {Lafreni{\`e}re}, David and {Artigau}, {\'E}tienne and {Marley}, Mark and {Saumon}, Didier},
        title = "{Large-amplitude Variations of an L/T Transition Brown Dwarf: Multi-wavelength Observations of Patchy, High-contrast Cloud Features}",
      journal = {\apj},
         year = 2012,
        month = may,
       volume = {750},
       number = {2},
          eid = {105},
        pages = {105},
          doi = {10.1088/0004-637X/750/2/105},
archivePrefix = {arXiv},
       eprint = {1201.3403},
 primaryClass = {astro-ph.SR},
       adsurl = {https://ui.adsabs.harvard.edu/abs/2012ApJ...750..105R}
}

@ARTICLE{mclean2003,
       author = {{McLean}, Ian S. and {McGovern}, Mark R. and {Burgasser}, Adam J. and {Kirkpatrick}, J. Davy and {Prato}, L. and {Kim}, Sungsoo S.},
        title = "{The NIRSPEC Brown Dwarf Spectroscopic Survey. I. Low-Resolution Near-Infrared Spectra}",
      journal = {\apj},
         year = 2003,
        month = oct,
       volume = {596},
       number = {1},
        pages = {561-586},
          doi = {10.1086/377636},
archivePrefix = {arXiv},
       eprint = {astro-ph/0309257},
 primaryClass = {astro-ph},
       adsurl = {https://ui.adsabs.harvard.edu/abs/2003ApJ...596..561M}
}

@ARTICLE{apai2013,
       author = {{Apai}, D{\'a}niel and {Radigan}, Jacqueline and {Buenzli}, Esther and {Burrows}, Adam and {Reid}, Iain Neill and {Jayawardhana}, Ray},
        title = "{HST Spectral Mapping of L/T Transition Brown Dwarfs Reveals Cloud Thickness Variations}",
      journal = {\apj},
         year = 2013,
        month = may,
       volume = {768},
       number = {2},
          eid = {121},
        pages = {121},
          doi = {10.1088/0004-637X/768/2/121},
archivePrefix = {arXiv},
       eprint = {1303.4151},
 primaryClass = {astro-ph.EP},
       adsurl = {https://ui.adsabs.harvard.edu/abs/2013ApJ...768..121A}
}

@ARTICLE{burgasser2003e,
       author = {{Burgasser}, Adam J. and {Kirkpatrick}, J. Davy and {Liebert}, James and {Burrows}, Adam},
        title = "{The Spectra of T Dwarfs. II. Red Optical Data}",
      journal = {\apj},
         year = 2003,
        month = sep,
       volume = {594},
       number = {1},
        pages = {510-524},
          doi = {10.1086/376756},
archivePrefix = {arXiv},
       eprint = {astro-ph/0305139},
 primaryClass = {astro-ph},
       adsurl = {https://ui.adsabs.harvard.edu/abs/2003ApJ...594..510B}
}

@ARTICLE{cushing2006,
       author = {{Cushing}, Michael C. and {Roellig}, Thomas L. and {Marley}, Mark S. and {Saumon}, D. and {Leggett}, S.~K. and {Kirkpatrick}, J. Davy and {Wilson}, John C. and {Sloan}, G.~C. and {Mainzer}, Amy K. and {Van Cleve}, Jeff E. and {Houck}, James R.},
        title = "{A Spitzer Infrared Spectrograph Spectral Sequence of M, L, and T Dwarfs}",
      journal = {\apj},
         year = 2006,
        month = sep,
       volume = {648},
       number = {1},
        pages = {614-628},
          doi = {10.1086/505637},
archivePrefix = {arXiv},
       eprint = {astro-ph/0605639},
 primaryClass = {astro-ph},
       adsurl = {https://ui.adsabs.harvard.edu/abs/2006ApJ...648..614C}
}

@ARTICLE{leggett2007,
       author = {{Leggett}, S.~K. and {Marley}, M.~S. and {Freedman}, R. and {Saumon}, D. and {Liu}, Michael C. and {Geballe}, T.~R. and {Golimowski}, D.~A. and {Stephens}, D.~C.},
        title = "{Physical and Spectral Characteristics of the T8 and Later Type Dwarfs}",
      journal = {\apj},
         year = 2007,
        month = sep,
       volume = {667},
       number = {1},
        pages = {537-548},
          doi = {10.1086/519948},
archivePrefix = {arXiv},
       eprint = {0705.2602},
 primaryClass = {astro-ph},
       adsurl = {https://ui.adsabs.harvard.edu/abs/2007ApJ...667..537L}
}

@ARTICLE{line2017,
       author = {{Line}, Michael R. and {Marley}, Mark S. and {Liu}, Michael C. and {Burningham}, Ben and {Morley}, Caroline V. and {Hinkel}, Natalie R. and {Teske}, Johanna and {Fortney}, Jonathan J. and {Freedman}, Richard and {Lupu}, Roxana},
        title = "{Uniform Atmospheric Retrieval Analysis of Ultracool Dwarfs. II. Properties of 11 T dwarfs}",
      journal = {\apj},
         year = 2017,
        month = oct,
       volume = {848},
       number = {2},
          eid = {83},
        pages = {83},
          doi = {10.3847/1538-4357/aa7ff0},
archivePrefix = {arXiv},
       eprint = {1612.02809},
 primaryClass = {astro-ph.SR},
       adsurl = {https://ui.adsabs.harvard.edu/abs/2017ApJ...848...83L}
}

@ARTICLE{lodieu2017,
       author = {{Lodieu}, N. and {Espinoza Contreras}, M. and {Zapatero Osorio}, M.~R. and {Solano}, E. and {Aberasturi}, M. and {Mart{\'\i}n}, E.~L. and {Rodrigo}, C.},
        title = "{New ultracool subdwarfs identified in large-scale surveys using Virtual Observatory tools}",
      journal = {\aap},
         year = 2017,
        month = feb,
       volume = {598},
          eid = {A92},
        pages = {A92},
          doi = {10.1051/0004-6361/201629410},
archivePrefix = {arXiv},
       eprint = {1609.08323},
 primaryClass = {astro-ph.SR},
       adsurl = {https://ui.adsabs.harvard.edu/abs/2017A&A...598A..92L}
}

@ARTICLE{gonzales2018,
       author = {{Gonzales}, Eileen C. and {Faherty}, Jacqueline K. and {Gagn{\'e}}, Jonathan and {Artigau}, {\'E}tienne and {Bardalez Gagliuffi}, Daniella},
        title = "{Understanding Fundamental Properties and Atmospheric Features of Subdwarfs via a Case Study of SDSS J125637.13-022452.4}",
      journal = {\apj},
         year = 2018,
        month = sep,
       volume = {864},
       number = {1},
          eid = {100},
        pages = {100},
          doi = {10.3847/1538-4357/aad3c7},
archivePrefix = {arXiv},
       eprint = {1807.04794},
 primaryClass = {astro-ph.SR},
       adsurl = {https://ui.adsabs.harvard.edu/abs/2018ApJ...864..100G}
}

@ARTICLE{gizis2006,
       author = {{Gizis}, John E. and {Harvin}, James},
        title = "{Halo Stars near the Hydrogen-burning Limit: The M/L Subdwarf Transition}",
      journal = {\aj},
         year = 2006,
        month = dec,
       volume = {132},
       number = {6},
        pages = {2372-2375},
          doi = {10.1086/508514},
archivePrefix = {arXiv},
       eprint = {astro-ph/0608573},
 primaryClass = {astro-ph},
       adsurl = {https://ui.adsabs.harvard.edu/abs/2006AJ....132.2372G}
}

@ARTICLE{zhang2017b,
       author = {{Zhang}, Z.~H. and {Homeier}, D. and {Pinfield}, D.~J. and {Lodieu}, N. and {Jones}, H.~R.~A. and {Allard}, F. and {Pavlenko}, Ya. V.},
        title = "{Primeval very low-mass stars and brown dwarfs - II. The most metal-poor substellar object}",
      journal = {\mnras},
         year = 2017,
        month = jun,
       volume = {468},
       number = {1},
        pages = {261-271},
          doi = {10.1093/mnras/stx350},
archivePrefix = {arXiv},
       eprint = {1702.02001},
 primaryClass = {astro-ph.SR},
       adsurl = {https://ui.adsabs.harvard.edu/abs/2017MNRAS.468..261Z}
}

@ARTICLE{maldonado2010,
       author = {{Maldonado}, J. and {Mart{\'\i}nez-Arn{\'a}iz}, R.~M. and {Eiroa}, C. and {Montes}, D. and {Montesinos}, B.},
        title = "{A spectroscopy study of nearby late-type stars, possible members of stellar kinematic groups}",
      journal = {\aap},
         year = 2010,
        month = oct,
       volume = {521},
          eid = {A12},
        pages = {A12},
          doi = {10.1051/0004-6361/201014948},
archivePrefix = {arXiv},
       eprint = {1007.1132},
 primaryClass = {astro-ph.SR},
       adsurl = {https://ui.adsabs.harvard.edu/abs/2010A&A...521A..12M}
}

@ARTICLE{anosova1991,
       author = {{Anosova}, J.~P. and {Orlov}, V.~V.},
        title = "{The dynamical evolution of the nearby multiple stellar systems ADS 48, ADS 6175 (alpha Geminorum = Castor), alpha Centauri and ADS 9909 (ksi Scorpii).}",
      journal = {\aap},
         year = 1991,
        month = dec,
       volume = {252},
        pages = {123},
       adsurl = {https://ui.adsabs.harvard.edu/abs/1991A&A...252..123A}
}

@ARTICLE{brandt2014,
       author = {{Brandt}, Timothy D. and {Kuzuhara}, Masayuki and {McElwain}, Michael W. and {Schlieder}, Joshua E. and {Wisniewski}, John P. and {Turner}, Edwin L. and {Carson}, J. and {Matsuo}, T. and {Biller}, B. and {Bonnefoy}, M. and {Dressing}, C. and {Janson}, M. and {Knapp}, G.~R. and {Moro-Mart{\'\i}n}, A. and {Thalmann}, C. and {Kudo}, T. and {Kusakabe}, N. and {Hashimoto}, J. and {Abe}, L. and {Brandner}, W. and {Currie}, T. and {Egner}, S. and {Feldt}, M. and {Golota}, T. and {Goto}, M. and {Grady}, C.~A. and {Guyon}, O. and {Hayano}, Y. and {Hayashi}, M. and {Hayashi}, S. and {Henning}, T. and {Hodapp}, K.~W. and {Ishii}, M. and {Iye}, M. and {Kandori}, R. and {Kwon}, J. and {Mede}, K. and {Miyama}, S. and {Morino}, J. -I. and {Nishimura}, T. and {Pyo}, T. -S. and {Serabyn}, E. and {Suenaga}, T. and {Suto}, H. and {Suzuki}, R. and {Takami}, M. and {Takahashi}, Y. and {Takato}, N. and {Terada}, H. and {Tomono}, D. and {Watanabe}, M. and {Yamada}, T. and {Takami}, H. and {Usuda}, T. and {Tamura}, M.},
        title = "{The Moving Group Targets of the SEEDS High-contrast Imaging Survey of Exoplanets and Disks: Results and Observations from the First Three Years}",
      journal = {\apj},
         year = 2014,
        month = may,
       volume = {786},
       number = {1},
          eid = {1},
        pages = {1},
          doi = {10.1088/0004-637X/786/1/1},
archivePrefix = {arXiv},
       eprint = {1305.7264},
 primaryClass = {astro-ph.EP},
       adsurl = {https://ui.adsabs.harvard.edu/abs/2014ApJ...786....1B}
}

@ARTICLE{hog2000,
       author = {{H{\o}g}, E. and {Fabricius}, C. and {Makarov}, V.~V. and {Urban}, S. and {Corbin}, T. and {Wycoff}, G. and {Bastian}, U. and {Schwekendiek}, P. and {Wicenec}, A.},
        title = "{The Tycho-2 catalogue of the 2.5 million brightest stars}",
      journal = {\aap},
         year = 2000,
        month = mar,
       volume = {355},
        pages = {L27-L30},
       adsurl = {https://ui.adsabs.harvard.edu/abs/2000A&A...355L..27H}
}

@ARTICLE{bouy2022,
       author = {{Bouy}, H. and {Tamura}, M. and {Barrado}, D. and {Motohara}, K. and {Castro Rodr{\'\i}guez}, N. and {Miret-Roig}, N. and {Konishi}, M. and {Koyama}, S. and {Takahashi}, H. and {Hu{\'e}lamo}, N. and {Bertin}, E. and {Olivares}, J. and {Sarro}, L.~M. and {Berihuete}, A. and {Cuillandre}, J. -C. and {Galli}, P.~A.~B. and {Yoshii}, Y. and {Miyata}, T.},
        title = "{Infrared spectroscopy of free-floating planet candidates in Upper Scorpius and Ophiuchus}",
      journal = {\aap},
         year = 2022,
        month = aug,
       volume = {664},
          eid = {A111},
        pages = {A111},
          doi = {10.1051/0004-6361/202243850},
archivePrefix = {arXiv},
       eprint = {2206.00916},
 primaryClass = {astro-ph.SR},
       adsurl = {https://ui.adsabs.harvard.edu/abs/2022A&A...664A.111B}
}

@ARTICLE{vos2018,
       author = {{Vos}, Johanna M. and {Allers}, Katelyn N. and {Biller}, Beth A. and {Liu}, Michael C. and {Dupuy}, Trent J. and {Gallimore}, Jack F. and {Adenuga}, Iyadunni J. and {Best}, William M.~J.},
        title = "{Variability of the lowest mass objects in the AB Doradus moving group}",
      journal = {\mnras},
         year = 2018,
        month = feb,
       volume = {474},
       number = {1},
        pages = {1041-1053},
          doi = {10.1093/mnras/stx2752},
archivePrefix = {arXiv},
       eprint = {1710.07194},
 primaryClass = {astro-ph.EP},
       adsurl = {https://ui.adsabs.harvard.edu/abs/2018MNRAS.474.1041V}
}

@ARTICLE{schneider2025,
       author = {{Schneider}, Adam C. and {Vrba}, Frederick J. and {Bruursema}, Justice and {Munn}, Jeffrey A. and {Irwin}, Mike and {Read}, Mike and {Varricatt}, Watson and {Kerr}, Tom and {Hodapp}, Klaus and {Dye}, Simon and {Williams}, Stephen J. and {Cenko}, Andrew T. and {Tilleman}, Trudy M. and {Murison}, Marc A. and {Rothberg}, Barry and {Dahm}, Scott and {Dorland}, Bryan and {Lawrence}, Andy and {Chambers}, Kenneth C.},
        title = "{The Second and Third Data Releases from the UKIRT Hemisphere Survey}",
      journal = {\aj},
         year = 2025,
        month = aug,
       volume = {170},
       number = {2},
          eid = {86},
        pages = {86},
          doi = {10.3847/1538-3881/ade43c},
archivePrefix = {arXiv},
       eprint = {2506.14621},
 primaryClass = {astro-ph.GA},
       adsurl = {https://ui.adsabs.harvard.edu/abs/2025AJ....170...86S}
}

@ARTICLE{dupuy2013,
       author = {{Dupuy}, Trent J. and {Kraus}, Adam L.},
        title = "{Distances, Luminosities, and Temperatures of the Coldest Known Substellar Objects}",
      journal = {Science},
         year = 2013,
        month = sep,
       volume = {341},
       number = {6153},
        pages = {1492-1495},
          doi = {10.1126/science.1241917},
archivePrefix = {arXiv},
       eprint = {1309.1422},
 primaryClass = {astro-ph.SR},
       adsurl = {https://ui.adsabs.harvard.edu/abs/2013Sci...341.1492D}
}

@ARTICLE{martin2018,
       author = {{Martin}, Emily C. and {Kirkpatrick}, J. Davy and {Beichman}, Charles A. and {Smart}, Richard L. and {Faherty}, Jacqueline K. and {Gelino}, Christopher R. and {Cushing}, Michael C. and {Schneider}, Adam C. and {Wright}, Edward L. and {Lowrance}, Patrick and {Ingalls}, James and {Tinney}, C.~G. and {McLean}, Ian S. and {Logsdon}, Sarah E. and {Lebreton}, J{\'e}r{\'e}my},
        title = "{Y Dwarf Trigonometric Parallaxes from the Spitzer Space Telescope}",
      journal = {\apj},
         year = 2018,
        month = nov,
       volume = {867},
       number = {2},
          eid = {109},
        pages = {109},
          doi = {10.3847/1538-4357/aae1af},
archivePrefix = {arXiv},
       eprint = {1809.06479},
 primaryClass = {astro-ph.SR},
       adsurl = {https://ui.adsabs.harvard.edu/abs/2018ApJ...867..109M}
}

@ARTICLE{best2018,
       author = {{Best}, William M.~J. and {Magnier}, Eugene A. and {Liu}, Michael C. and {Aller}, Kimberly M. and {Zhang}, Zhoujian and {Burgett}, W.~S. and {Chambers}, K.~C. and {Draper}, P. and {Flewelling}, H. and {Kaiser}, N. and {Kudritzki}, R. -P. and {Metcalfe}, N. and {Tonry}, J.~L. and {Wainscoat}, R.~J. and {Waters}, C.},
        title = "{Photometry and Proper Motions of M, L, and T Dwarfs from the Pan-STARRS1 3{\ensuremath{\pi}} Survey}",
      journal = {\apjs},
         year = 2018,
        month = jan,
       volume = {234},
       number = {1},
          eid = {1},
        pages = {1},
          doi = {10.3847/1538-4365/aa9982},
archivePrefix = {arXiv},
       eprint = {1701.00490},
 primaryClass = {astro-ph.SR},
       adsurl = {https://ui.adsabs.harvard.edu/abs/2018ApJS..234....1B}
}

@ARTICLE{schneider2023,
       author = {{Schneider}, Adam C. and {Munn}, Jeffrey A. and {Vrba}, Frederick J. and {Bruursema}, Justice and {Dahm}, Scott E. and {Williams}, Stephen J. and {Liu}, Michael C. and {Dorland}, Bryan N.},
        title = "{Astrometry and Photometry for {\ensuremath{\approx}}1000 L, T, and Y Dwarfs from the UKIRT Hemisphere Survey}",
      journal = {\aj},
         year = 2023,
        month = sep,
       volume = {166},
       number = {3},
          eid = {103},
        pages = {103},
          doi = {10.3847/1538-3881/ace9bf},
archivePrefix = {arXiv},
       eprint = {2307.11882},
 primaryClass = {astro-ph.SR},
       adsurl = {https://ui.adsabs.harvard.edu/abs/2023AJ....166..103S}
}

@ARTICLE{vos2023,
       author = {{Vos}, Johanna M. and {Burningham}, Ben and {Faherty}, Jacqueline K. and {Alejandro}, Sherelyn and {Gonzales}, Eileen and {Calamari}, Emily and {Bardalez Gagliuffi}, Daniella and {Visscher}, Channon and {Tan}, Xianyu and {Morley}, Caroline V. and {Marley}, Mark and {Gemma}, Marina E. and {Whiteford}, Niall and {Gaarn}, Josefine and {Park}, Grace},
        title = "{Patchy Forsterite Clouds in the Atmospheres of Two Highly Variable Exoplanet Analogs}",
      journal = {\apj},
         year = 2023,
        month = feb,
       volume = {944},
       number = {2},
          eid = {138},
        pages = {138},
          doi = {10.3847/1538-4357/acab58},
archivePrefix = {arXiv},
       eprint = {2212.07399},
 primaryClass = {astro-ph.EP},
       adsurl = {https://ui.adsabs.harvard.edu/abs/2023ApJ...944..138V}
}
\bibliographystyle{aasjournal}

\end{document}